\documentclass[english, openany]{book}
\usepackage[utf8]{inputenc}
\usepackage[english]{babel}
\usepackage{import}
\usepackage{chapterbib}
\usepackage[nopostdot=true,xindy={language=english,codepage=utf8},nonumberlist,style=altlist,acronym,shortcuts,section=section]{glossaries}

\newglossaryentry{ABRAS}{
name={ABRAS},
description={Argentina-Brasil Astronomical Center},
first={Argentina-Brasil Astronomical Center (ABRAS)},
}
\newglossaryentry{ALICE}{
name={ALICE},
description={A Large Ion Collider Experiment},
first={A Large Ion Collider Experiment(ALICE)},
}
\newglossaryentry{ALPACA}{
name={ALPACA},
description={Andes Large-area PArticle detector for Cosmic-ray physics and Astronomy},
first={Andes Large-area PArticle detector for Cosmic-ray physics and Astronomy (ALPACA)},
}
\newglossaryentry{ANDES}{
name={ANDES},
description={Agua Negra Deep Experiment Site},
first={Agua Negra Deep Experiment Site (ANDES)},
}
\newglossaryentry{ANGRA}{
name={ANGRA},
description={Angra Dos Reis Nuclear complex},
first={Angra Dos Reis Nuclear complex (ANGRA)},
}
\newglossaryentry{ASTRI}{
name={ASTRI},
description={Astrofisica a Specchi con Tecnologia Replicante Italiana},
first={Astrofisica a Specchi con Tecnologia Replicante Italiana (ASTRI)},
}
\newglossaryentry{ATLAS}{
name={ATLAS},
description={A Toroidal LHC ApparatuS},
first={A Toroidal LHC ApparatuS (ATLAS)},
}
\newglossaryentry{AUGER}{
name={PAO},
description={Pierre Auger Observatory},
first={Pierre Auger Observatory (PAO)},
}
\newglossaryentry{BELLE2}{
name={BELLE 2},
description={B detector at KEK},
first={B detector at KEK},
}
\newglossaryentry{BINGO}{
name={BINGO},
description={BAO from Integrated Neutral Gas Observations},
first={BAO from Integrated Neutral Gas Observations (BINGO)},
}
\newglossaryentry{CEPC}{
name={CEPC},
description={Circular Electron-Positron Collider},
first={Circular Electron-Positron Collider (CEPC)},
}
\newglossaryentry{CLAS}{
name={CLAS},
description={CEBAF Large Acceptance Spectrometer},
first={CEBAF Large Acceptance Spectrometer (CLAS)},
}
\newglossaryentry{CMS}{
name={CMS},
description={Compact Muon Solenoid},
first={Compact Muon Solenoid},
}
\newglossaryentry{CNEATANDAR}{
name={TANDAR},
  description={Acelerador TANDAR, Comisión Nacional de Energía Atómica},  
}
\newglossaryentry{CONNIE}{
name={CONNIE},
description={Coherent Neutrino Nucleus Interaction Experiment},
first={Coherent Neutrino Nucleus Interaction Experiment (CONNIE)},
}
\newglossaryentry{CTA}{
name={CTA},
description={Cherenkov Telescope Array},
first={Cherenkov Telescope Array (CTA)},
}
\newglossaryentry{DARKSIDE}{
name={DARKSIDE},
description={Two phase TPC for Dark Matter Direct Detection},
first={Two phase TPC for Dark Matter Direct Detection (DARKSIDE)},
}
\newglossaryentry{DUNE}{
name={DUNE},
description={Deep Underground Neutrino Experiment},
first={Deep Underground Neutrino Experiment (DUNE)},
}
\newglossaryentry{FCC}{
name={FCC},
description={Future Circular Collider},
first={Future Circular Collider (FCC)},
}
\newglossaryentry{GRAND}{
name={GRAND},
description={Giant Radio Array for Neutrino Detection},
first={Giant Radio Array for Neutrino Detection (GRAND)},
}
\newglossaryentry{HAWC}{
name={HAWC},
description={High Altitude Water Cherenkov Gamma-ray Observatory},
first={High Altitude Water Cherenkov Gamma-ray Observatory (HAWC)},
}
\newglossaryentry{HL-LHC}{
name={HL-LHC},
description={High Luminosity Large Hadron Collider},
first={High Luminosity Large Hadron Collider (HL-LHC)},
}
\newglossaryentry{Hyper-K}{
name={Hyper-K},
description={Hyper Kamiokande},
first={Hyper Kamiokande},
}
\newglossaryentry{ILC}{
name={ILC},
description={International Linear Collider},
first={International Linear Collider (ILC)},
}
\newglossaryentry{JUNO}{
name={JUNO},
description={Jiangmen Underground Neutrino Observatory},
first={Jiangmen Underground Neutrino Observatory (JUNO)},
}
\newglossaryentry{KM3NET}{
name={KM3NeT},
description={Cubic Kilometre Neutrino Telescope},
first={Cubic Kilometre Neutrino Telescope (KM3NeT)},
}
\newglossaryentry{LAGO}{
name={LAGO},
description={Latin American Giant Observatory},
first={Latin American Giant Observatory (LAGO)},
}
\newglossaryentry{LAFNTANDEM}{
name={LAFN Tandem},
description={Tandem Accelerator at Laboratorio Aberto de Fisica Nuclear},
first={Tandem Accelerator at Laboratorio Aberto de Fisica Nuclear (LAFN Tandem)},
}
\newglossaryentry{LHCb}{
name={LHCb},
description={Large Hadron Collider beauty},
first={Large Hadron Collider beauty (LHCb)},
}
\newglossaryentry{LLAMA}{
name={LLAMA},
description={Large Latin American Millimetre Array},
first={Large Latin American Millimetre Array (LLAMA)},
}
\newglossaryentry{LSST}{
name={LSST},
description={Legacy Survey of Space and Time},
first={Legacy Survey of Space and Time (LSST)},
}
\newglossaryentry{NO-NU-A}{
name={No$\nu$A},
description={NuMI Off-axis $\nu$e Appearance},
first={NuMI Off-axis $\nu$e Appearance (No$\nu$A)},
}
\newglossaryentry{PHENIX}{
name={PHENIX},
description={Pioneering High-Energy Nuclear Interaction eXperiment},
first={Pioneering High-Energy Nuclear Interaction eXperiment (PHENIX)},
}
\newglossaryentry{PPS}{
name={PPS},
description={Precision Proton Spectrometer},
first={Precision Proton Spectrometer (PPS)},
}
\newglossaryentry{QUBIC}{
name={QUBIC},
description={Q$\&$U Bolometric Interferometer for Cosmology},
first={Q$\&$U Bolometric Interferometer for Cosmology (QUBIC)},
}
\newglossaryentry{RHIC}{
name={RHIC},
description={Relativistic Heavy Ion Collider},
first={Relativistic Heavy Ion Collider (RHIC)},
}
\newglossaryentry{SAGO}{
name={SAGO},
description={South American Gravitation Wave Observatory},
first={South American Gravitation Wave Observatory (SAGO)},
}
\newglossaryentry{SBND}{
name={SBND},
description={Short Baseline Neutrino Detector},
first={Short Baseline Neutrino Detector (SBND)},
}
\newglossaryentry{SENSEI}{
name={SENSEI},
description={Sub-Electron-Noise Skipper Experimental Instrument},
first={Sub-Electron-Noise Skipper Experimental Instrument (SENSEI)},
}
\newglossaryentry{STAR}{
name={STAR},
description={Solenoidal Tracker at RHIC},
first={Solenoidal Tracker at RHIC (STAR)},
}
\newglossaryentry{SWGO}{
name={SWGO},
description={Southern Wide-field-of-view Gamma-ray Observatory},
first={Southern Wide-field-of-view Gamma-ray Observatory (SWGO)},
}
\newglossaryentry{TAMBO}{
name={TAMBO},
description={Tau Air Shower Mountain-Based Observatory},
first={Tau Air Shower Mountain-Based Observatory (TAMBO)},
}
\newglossaryentry{TOROS}{
name={TOROS},
description={The Transient Optical Robotic Observatory of the South},
first={The Transient Optical Robotic Observatory of the South (TOROS)},
}
\newglossaryentry{NU-IOLETA}{
name={$\nu$IOLETA},
description={Neutrino Interaction Observation with a Low Energy Threshold Array},
first={Neutrino Interaction Observation with a Low Energy Threshold Array ($\nu$IOLETA)},
}

\makeglossaries

\usepackage[]{graphicx}
\usepackage[]{color}
\usepackage{alltt}
\usepackage[T1]{fontenc}
\usepackage[utf8]{inputenc}
\usepackage{mathptmx}
\usepackage{adjustbox}
\usepackage{amssymb}
\usepackage[utf8]{inputenc}
\usepackage{amsmath}
\usepackage{mathtools}
\usepackage{tabularx}
\usepackage{array}\newcolumntype{C}[1]{>{\centering\arraybackslash}p{#1}}
\usepackage{xcolor}
\usepackage{marvosym}
\usepackage[top=60pt,bottom=60pt,left=78pt,right=78pt]{geometry}

\usepackage{amssymb}
\usepackage[utf8]{inputenc}
\usepackage{amsmath}
\usepackage{mathtools}
\usepackage{tabularx}
\usepackage{xcolor}
\usepackage{marvosym}

\usepackage{graphicx}
\usepackage[amssymb]{SIunits}
\usepackage{cite}
\usepackage{graphicx}
\graphicspath{{figures/}}

\usepackage{subcaption}
\usepackage{hyperref}
\usepackage{listings}
\usepackage{marvosym}
\usepackage{mdframed}

\usepackage[official]{eurosym}

\usepackage{parskip}

\usepackage{subcaption}
\usepackage{lipsum}
\usepackage{booktabs}
\usepackage{multirow}
\usepackage[english]{babel}

\usepackage{fancyhdr}
\usepackage{titlesec}
\titleformat{\chapter}
   {\normalfont\LARGE\bfseries}{\thechapter.}{1em}{}
\titlespacing{\chapter}{0pt}{40pt}{2\baselineskip}

\usepackage{float}
\floatstyle{plaintop}
\restylefloat{table}

\usepackage[tableposition=top]{caption}

\usepackage[
type={CC},
modifier={by-nc-sa},
version={4.0},
]{doclicense}

\usepackage{hyperref}

\usepackage{authblk}
\usepackage{tcolorbox}

\usepackage{subfiles}

\usepackage{tikz}
\usetikzlibrary{calc,shapes.geometric,arrows,positioning,intersections}

\begin{document}
    
\begin{titlepage}
	\clearpage\thispagestyle{empty}
	\centering
	\vspace{1cm}

	{
		\textsc{Latin American Strategy Forum for Research Infrastructure - LASF4RI}
	}
		\vspace{2.5cm}

\normalfont

	\rule{\linewidth}{2mm} \\[0.5cm]
	{ \Huge \bfseries Latin American Strategy for Research Infrastructures\\ for High Energy, Cosmology, Astroparticle Physics\\
	\vspace{0.2cm}LASF4RI for HECAP} \\[0.5cm]
	\rule{\linewidth}{2mm} \\[3.4cm]

    { \Large \bfseries LATIN AMERICAN HECAP PHYSICS BRIEFING BOOK 2025}\\
    \vspace{1cm}
    { \Large \bfseries Preparatory Group}\\
    \vspace{0.5cm}
    \begin{table}[h]
        \centering {\small
        \begin{tabular}{l|l}
	    \textbf{Mario A. Acero - U. del Atlántico, Colombia}    &   \textbf{Fernando Monticelli - U. Nacional de la Plata, Argentina}
         \\
        \textbf{Alexis A. Aguilar-Arevalo - UNAM, México}  &       \textbf{Deywis Moreno - U. Antonio Nariño, Colombia}
        \\
        \textbf{Belén Andrada - CNEA-CONICET-UNSAM, Argentina}        &         \textbf{Martjin Mulders - CERN, Switzerland}
        \\
        \textbf{Andrés Baquero Larriva - U. del Azuay, Ecuador}        &         \textbf{Luis A. Núñez - U. Industrial de Santander, Colombia}
        \\ 
        \textbf{Mauro Cambiaso - U. Andrés Bello, Chile}        &         \textbf{Arturo S. Pineda - CC Venezuela \& CC Switzerland}
        \\ 
        \textbf{Edgar Carrera - U. San Francisco de Quito, Ecuador}        &         \textbf{Juan Ponciano - U. de San Carlos de Guatemala}
        \\
        \textbf{Melissa Cruz - U. Nacional Autónoma de Honduras}        &         \textbf{Farinaldo Queiroz - UFRN \& IIP, Brazil}
        \\
       \textbf{Lucía Duarte - U. de la República de Uruguay}     &       \textbf{Rogerio Rosenfeld - IFT-UNESP/ICTP-SAIFR, Brazil}
       \\
        \textbf{Juan Estrada - Brookhaven National Laboratory, USA}       &        \textbf{Sandro F. de Souza - DFNAE-UERJ, Brazil}
       \\
       \textbf{Alberto Gago - Pontifica U. Cat\'olica del Per\'u}       &        \textbf{Martin Alfonso Subieta Vasquez - U. Mayor de San Andr\'es, Bolivia}
       \\
      \textbf{Esteban Jimenez - U. de Costa Rica}       &        \textbf{Maria Elena Tejeda-Yeomans - U. de Colima, M\'exico}
       \\
       \textbf{Diana López Nacir - UBA-CONICET, Argentina}       &        \textbf{Luis Ureña - U. de Guanajuato, M\'exico}
       \\
       \textbf{José A. López - U. Central de Venezuela, Venezuela}       &        \textbf{Alfonso Zerwekh - U. T. Federico Santa Mar\'{i}a, Chile}
       \\
       \textbf{Marta Losada - New York University Abu Dhabi, UAE}       &        \textbf{}
       \\
       \textbf{}       &        \textbf{}
       \\
	\end{tabular} }
	\label{tab:my_label}
    \end{table}

\vspace{5cm}

\pagebreak

\end{titlepage}

{\hypersetup{linkcolor=black}

	\tableofcontents{}
}

\mainmatter

\normalfont

\chapter{Introduction}\label{chapt:intro}
It is widely recognized that large-scale long-term science collaborations deliver lasting benefits to
participating nations. Most of these collaborations require large infrastructures. 
Given the amount of human and financial resources involved, it is important to coordinate them in different
regions of the world. In particular, there have been efforts in the fields of High Energy, Cosmology, and 
Astroparticle Physics (HECAP) that resulted in documents summarizing the science drivers and how well the 
proposed projects will address them, taking into account the input of the scientific community involved.
These documents are used by policy makers and funding agencies to help inform their decision-making processes.

In the US, the most recent process was the US Community Study on the Future of Particle Physics (Snowmass 
2021), a grassroots study building on White Papers submitted by the community to plan for US particle 
physics in the decade 2025-2035. It was organized by the Division of Particle and Fields of the American 
Physical Society with ten working groups. After almost two years of intense activity, it led to a 700-page
report \cite{snowmass}. This so-called Snowmass report served to inform the Particle Physics Project Prioritization 
Panel (P5) in their recommendations to the funding agencies, which was concluded in 2023 \cite{p5}. The P5 is 
a scientific advisory panel tasked with recommending plans for US investment in particle physics research 
over the next ten years, on the basis of various funding scenarios. The P5 is a subcommittee of the High 
Energy Physics Advisory Panel (HEPAP), which serves the Department of Energy's Office of Science and the 
National Science Foundation.

In Europe, the third update of the European Strategy for Particle Physics (ESPP) was launched by the CERN
Council in 2024. There were 263 inputs submitted, and an open symposium was held in Venice in June 2025. The
Physics Briefing Book (PBB) of the ESPP was organized by the Physics Preparatory Group (PPG) and 
submitted to the European Strategy Group (ESG) by the end of September 2025 \cite{EPBB}. The Latin American HEP community represented by LAA-HECAP submitted a contribution to the ESPP \cite{surveylaa}.
The final drafting session of the Strategy update took place from 1 to 5 December, where the community recommendations were finalized. These will be presented to the CERN Council in March 2026 and discussed at a dedicated meeting of the CERN Council in May 2026.

The first process for the Latin American Strategy Forum for Research Infrastructure for High Energy, 
Cosmology and Astroparticle Physics (LASF4RI-HECAP) has come to a conclusion in October 2020. A Physics 
Briefing Book (PBB) was written by the Preparatory Group based on the 40 White Papers submitted by the 
community \cite{pbb-la}. The PBB served as the basis for the LASF4RI-HECAP Strategy Document that was 
endorsed in a letter by the High-Level Strategy Group. All documents can be found on the webpage {\tt 
lasf4ri.org}. The Strategy Document was submitted to the IV Iberoamerican Science and Technology Ministerial 
Meeting that took place in October 2020, where it was recognized in its Declaration.

This Physics Briefing Book is the result of the first update of LASF4RI-HECAP. The update process began with a call for White Papers from the HECAP community, issued in November 2023, with a submission deadline of July 26, 2024. The submitted contributions were presented at the “III LASF4RI for HECAP Symposium: Update of the Strategic Plan“, held at ICTP-SAIFR in São Paulo in August 26-29, 2024, with the participation of the Preparatory Group, High Level Strategy Group, Funding Agencies and representatives of similar efforts from around the globe.

In December 2024, a new Preparatory Group was formed to conduct the writing of the updated PBB, and seven working groups were established: 
Astronomy, Astrophysics and Astroparticle Physics; 
Cosmology;
Dark Matter;
Neutrinos;
Electroweak \& Strong Interactions, Higgs Physics, CP \& Flavour Physics and BSM;
Instrumentation and Computing;
Advanced Training and Capacity Building. The chapters in this document are the work of these seven working groups.

The main source of information for this PBB are the 46 White Papers listed in Appendix \ref{wp}. They are publicly available at this \href{https://drive.google.com/drive/folders/18ATto2vVbAKfB3M7hB6ty6wYWQ23f6Xl?usp=sharing}{link}.
In addition to the submitted inputs, the working groups were encouraged to gather more  
relevant information if deemed necessary to include in the PBB. The writing actually started in May 2025 with the creation of a 
shared Overleaf document. Although some overlap and differences in style among the chapters are unavoidable, we hope 
the PBB provides a smooth and informative reading.

Several developments in the Latin American HECAP community have happened since the end of 2020, when the first Strategy document was concluded. They are reported in this updated PBB. We want to highlight a few, in no particular order:
\begin{itemize} 
\item The creation of the Latin American Association for High Energy, Cosmology and Astroparticle 
Physics (LAA-HECAP \footnote{{\tt www.laa-hecap.org} } ) in 2021, following a recommendation of the High Level Strategy Group.
\item The C11 committee of IUPAP approved a proposal to organize the 43rd International Conference in High Energy Physics (ICHEP) in Natal, Brazil, in 2026. It is the first time that ICHEP will take place in Latin America.
\item Brazil became an Associate Member State of CERN in March 2024.
\item In May 2025, an agreement was signed admitting Chile as an Associate Member State of CERN and should be ratified in early 2026. 
\item International Cooperation Agreements (ICAs) were signed between CERN and Honduras (2021), Uruguay (2024 and Guatemala (2025).
\item Costa Rica became a full member of the \gls{LHCb} collaboration in 2022. Universidad Autónoma de Honduras also joined \gls{LHCb}.
\item The Brazilian Center for Research in Energy and Materials (CNPEM) has joined the ALICE 3 magnet project in 2025 with participation in its design and construction.
\item CNPEM (Brazil) signed in 2025 a cooperation agreement with CERN to be part of a feasibility study for the Future Circular Collider.
\item Brazilian funding agencies FAPESP and FINEP have approved in May 2024 a contribution of US\$ 36 million mainly towards the construction of a liquid argon purification system for the \gls{DUNE} experiment.
\item Brazilian funding agency FAPESP has approved in 2024 a contribution of US\$ 5 million towards the construction of the photon detection system for the \gls{DUNE} far detector. Production and assembly will be performed in Brazil. 
\item The ``Laboratorio Argentino de Mediciones de Bajo umbral de Detección y sus Aplicaciones'' (LAMBDA) was inaugurated in July 2022 at the University of Buenos Aires.
\item In July 2024, the Atacama Astronomical Park in Chile was selected as the site of the Southern Wide-field Gamma-ray Observatory (SWGO) project.
\item The Dark Energy Spectroscopic Instrument (DESI) collaboration, with participation from Mexico and Colombia, presented their first-year results in 2024 and the three-year results in 2025. 
\item Argentina and Mexico joined the Vera Rubin Observatory and the Legacy Survey of Space and Time (LSST).
\item The first LSST@LATAM meeting took place in La Serena, Chile, in June 2024 and the second one happened in Mexico City in December 2025. The third meeting will be at ICTP-SAIFR in São Paulo in 2026.
\item The first images of the Rubin Observatory were presented in June 2025 in the ``First Look'' event.
\item Representatives of 17 countries that are members of the Pierre Auger Observatory in Malargüe, Argentina, signed an agreement in November 2024 to support scientific research in the Observatory until 2034.
\item The Latin American Giant Observatory (LAGO), the world’s most geographically extended astroparticle
observatory, was granted a four-year CYTED grant, which supports the consolidation of the network and the long-term sustainability of the collaboration.
\item The Argentinian-based Q\&U Bolometric Interferometer for Cosmology (QUBIC) instrument has started its commissioning phase in 2025.
\item The Brazilian-based Coherent Neutrino-Nucleus Interaction Experiment (CONNIE), in collaboration with the Atucha-II experiment in Argentina, set the best-to-date individual and
combined limits in the parameter space of mass and charge fraction of millicharged particles from data obtained in 2021-2022 using skipper CCDs. 
\item Supported by a
2024 SECIHTI/CONAHCYT grant, a group in Mexico will manufacture the first full production batch of
19-channel multi-PMTs, with installation slated for the Hyper-Kamiokande 190-kt far detector during the
2027 integration campaign.
\item Chile led the design and manufacturing of 400 high-voltage splitters (HVS) and 200 under water boxes (UWB) for JUNO, successfully completed in late 2022.
\item The OpenIPMC project, spearheaded by SPRACE in Brazil, has provided open source hardware, firmware,
and software for more than 1000 electronics boards that were produced and assembled in Brazil and
shipped to CERN for the CMS tracker and high-granularity calorimeter upgrades. 
\item The 25th edition of the yearly international Conference on Particle Physics and Cosmology (COSMO'22) took place in Rio de Janeiro, Brazil, in August 2022. It was the first COSMO conference in Latin America.
\item The paper ``Statistical Analysis of Scientific Metrics in High Energy, Cosmology, and Astroparticle Physics in Latin America'' was uploaded to the arXives in March 2025\cite{statistics}.
\item There has been a steep increase in graduate-level schools, workshops and outreach activities in HECAP.
\item The construction of the \gls{ANDES} underground laboratory is in a standby status because the Agua Negra Tunnel project has been stalled since 2019. However, some \gls{ANDES}- related research activities are still being conducted. \gls{ANDES} was a top priority in the first Latin American Strategy.

The following chapters will delve into these and other major developments, providing an update of the HECAP landscape in Latin America.

\end{itemize}

\bibliographystyle{unsrt} 
\bibliography{intro/intro}

\chapter{Astronomy, Astrophysics and Astroparticle Physics}\label{chapt:astro}

\section{Introduction} 

Latin America plays an increasingly prominent role in the global ecosystem of High-Energy Cosmic and Astroparticle Physics (HECAP). Across a broad range of topics — from cosmic rays and gamma-ray astronomy to gravitational waves and multi-messenger astrophysics — the region contributes both scientific expertise and key infrastructure while developing programs that foster regional integration and train new generations of researchers.

In the 2021 version of this Briefing Book, the HECAP chapter outlined a landscape rich in scientific potential but facing challenges in coordination, access to infrastructure, and regional visibility~\cite{lasf4ri2021}. Key priorities were identified: strengthening Latin American participation in global collaborations, developing data infrastructure aligned with FAIR principles, and consolidating regional networks for education and outreach. In the years since, many of these goals have seen concrete progress, as detailed in the white papers collected for the current update.

The \gls{ANDES} Underground Laboratory project (Agua Negra Deep Experiment Site), a top priority in the 2021 Strategy Document, remains a key long-term infrastructure proposal in Latin America. While the associated Agua Negra Tunnel project is currently stalled and its timeline remains uncertain, \gls{ANDES} is not in complete standby. Collaboration between Argentina (ITeDA) and Chile (Universidad de La Serena) continues, with two innovative muon-veto telescopes built and being installed at underground sites in San Juan and Coquimbo. These prototypes provide both scientific value and training opportunities, preserving momentum while awaiting reactivation of the main tunnel initiative.

This chapter provides an updated snapshot of major experimental efforts involving Latin American groups. It focuses on flagship projects such as the Pierre Auger Observatory~\cite{auger2015,auger2023}, the Latin American Giant Observatory (LAGO)~\cite{lago2019status}, the Cherenkov Telescope Array Observatory (CTAO) and \gls{ASTRI} mini-array~\cite{cta2019science,astri2022overview}, and the Southern Wide-field Gamma-ray Observatory (SWGO)~\cite{swgocollaboration2025}, among others. Additional community-driven ERASMUs+ initiatives focused on regional capacity-building — such as LA-CoNGA (for Latin American Alliance for Capacity buildiNG in Advaced physics)~\cite{laconga2022}  and EL-BONGO (for E-Latin America Digital huB for OpeN Growing cOmmunities in physics)~\cite{elbongo2024}  — are also discussed.

We summarize recent milestones such as the renewal of the Auger collaboration agreement, the site selection for SWGO, and expansions of the \gls{LAGO} network. Where appropriate, we mention emerging efforts in instrumentation, data sharing, and open science practices.

Projects that are covered in more detail in other sections (such as neutrino and cosmology experiments) are only briefly mentioned here. 
Subsections also include qualitative summaries of country-level engagement and collaborative patterns. This reflects the broader LASF4RI goals of fostering integration, aligning strategies, and strengthening the region’s impact on global science.

\section{Cosmic Rays}  \label{sec:cr} 

Latin America has established itself as a key hub for cosmic ray research across a broad range of energies, built upon a foundation of global observatories and regional infrastructures. Its vast geography, scientific expertise, and institutional networks have enabled both leadership in flagship experiments and the development of innovative regional projects. The April 2021 Briefing Book underscored this potential by highlighting LATAM-led endeavours such as the Pierre Auger Observatory and LAGO, and proposed initiatives including \gls{ANDES} and BINGO, while calling for improved infrastructure, open-science practices, and regional coordination~\cite{lasf4ri2021}. Since then, the community has not only sustained its core contributions but also expanded its scientific and strategic footprint, positioning itself at the forefront of next-generation cosmic ray exploration.

\subsection{Key Projects with LATAM Leadership} 

\subsubsection{Pierre Auger Observatory} 

The Pierre Auger Observatory in Argentina is the world’s largest facility dedicated to ultra-high-energy cosmic rays (UHECR). Latin American countries, especially Argentina, Brazil, Mexico, and Colombia, are integral to the collaboration and have been central to its development since its inception.

The recent completion of the AugerPrime upgrade, including the installation of surface scintillator detectors, underground muon detectors, and advanced electronics, marks a critical enhancement of the observatory’s ability to determine primary mass composition and explore new physics at extreme energies. These upgrades extend the scientific reach of the Auger Observatory to address key open questions regarding cosmic-ray sources, acceleration mechanisms, and propagation models~\cite{auger2015, auger2023}.

The February 2025 release of the Pierre Auger Observatory’s open data, which includes an expanded dataset covering up to 30\% of cosmic-ray events above $2.5\times10^{18} $ eV (spanning 2004–2022) and complete atmospheric monitoring data, demonstrates the Collaboration’s commitment to transparency and global involvement. This initiative provides researchers worldwide with unprecedented access to high-quality data and analysis tools, enhancing accessibility for scientific and educational purposes~\cite{auger_open_data2025}. Equally significant, the renewal of Auger’s international collaboration agreement in 2024 and the full deployment and commissioning of the AugerPrime upgrade ensure that Latin American contributions will remain pivotal in advancing UHECR science well into the next decade~\cite{augerprimestatus2025}.

\subsubsection{\gls{LAGO} (Latin American Giant Observatory)} 

The Latin American Giant Observatory is the world’s most geographically extended astroparticle observatory, operating a distributed network of Water Cherenkov Detectors (WCDs) from Mexico to Antarctica. Managed by a decentralized collaboration of over 29 institutions from ten Latin American countries and Spain, \gls{LAGO} plays a dual role in advancing scientific research and fostering capacity building across the region. Originally designed in 2006 to study the high-energy component of gamma-ray bursts (GRBs), the project has since evolved into a multi-purpose infrastructure for research on space weather, cosmic rays, and atmospheric radiation. Its scientific objectives now span from astrophysical phenomena to interdisciplinary applications, including muography, soil moisture studies, and high-performance simulations.

The network currently includes WCDs at diverse altitudes and geomagnetic latitudes—key for mapping space weather and secondary cosmic-ray variations—such as those at Sierra Negra (Mexico), Chimborazo (Ecuador), and Antarctica (Marambio and San Martín bases). A new standardized WCD design is under deployment to enhance inter-calibration and network uniformity, incorporating recent advances in 3D printing, DAQ systems, and autonomous power solutions. Detector development remains a core strength of the collaboration, with innovations such as the C-Arapuca photodetector from UNICAMP (Brazil) offering low-cost, high-efficiency alternatives to traditional PMTs. \gls{LAGO} also plays a prominent role in regional training efforts, actively participating in educational programs such as the EU-funded LA-CoNGA and EL-BONGO platforms. These initiatives promote open science practices, physics education, and software skills through structured curricula and hands-on detector training, while reinforcing inter-institutional ties across the continent.

Institutional participation spans the entire region. Argentina contributes through multiple deployments, including the planet's southernmost WCDs. Brazil leads instrumentation research and contributes to the development of detector infrastructure. Colombia drives environmental and muography applications, while Peru and Ecuador focus on autonomous detectors at high altitudes. These efforts are now bolstered by a four-year CYTED grant (\gls{LAGO} INDICA), which supports the consolidation of the network and the long-term sustainability of the collaboration.

Beyond its scientific and technological goals, \gls{LAGO} embodies a regional model of collaborative, inclusive science. Its open data policies, decentralized management, and integration with academic curricula have made it a vehicle for expanding participation in astroparticle physics. As the project matures, \gls{LAGO} continues to exemplify how Latin American science can simultaneously address global research questions and local development needs~\cite{lago2024}.

\subsection{Emerging and Complementary Initiatives} 

Latin America continues to generate ambitious, forward-looking initiatives that complement flagship observatories and aim to explore new frontiers in cosmic-ray and multi-messenger science. Two particularly notable efforts are TAMBO and GRAND, both of which feature substantial Latin American involvement and have evolved significantly since the 2021 LASF4RI Briefing Book. These projects will be discussed in detail in Chapter 5.

\subsection{Regional Progress and Strategic Outlook} 

Latin America’s diverse geography, spanning high-altitude regions, the tropics, and subpolar areas, provides unique observational conditions for cosmic-ray research. Projects like LAGO and TAMBO take advantage of this diversity with distributed detector arrays that enable multi-site observations across energy and latitude ranges~\cite{lago2024, tambo2024}. These initiatives also foster academic engagement and infrastructure development across the region.

Collaboration and coordination are central features of the region’s scientific landscape. LAGO, TAMBO, and GRAND~\cite{grand2024}  promote open, modular participation models and emphasize training, inclusion, and cross-border cooperation. In parallel, the recent renewal of the Pierre Auger Observatory’s international collaboration agreement underscores the region’s long-term commitment and leadership in high-energy cosmic-ray physics.

As global astroparticle physics becomes more integrated, Latin America is increasingly recognized not only for its scientific contributions but also for its collaborative, inclusive frameworks.

\section{Gamma-Ray and Multi-Messenger Astronomy} \label{sec:gamma_mm} 

Gamma-ray and multi-messenger astrophysics offer powerful tools to explore some of the most energetic and enigmatic phenomena in the Universe, including black holes, supernova remnants, gamma-ray bursts, and active galactic nuclei. Latin America’s geographic location, technical expertise, and collaborative networks position the region as a key contributor to this frontier, both through participation in global observatories and through the development of regionally led initiatives.

The 2021 Briefing Book highlighted CTAO as the flagship next-generation gamma-ray facility, with Argentina and Brazil already actively participating in the project’s design and science preparation~\cite{lasf4ri2021}. Since then, the regional community has made significant advances: Brazil has consolidated its role in CTAO through instrument development and science organization, and Argentina has been involved in site qualification, atmospheric characterization, and data analysis frameworks. The Italian-led \gls{ASTRI} Mini-Array, a precursor to CTAO, has also become a channel for early science and regional capacity building~\cite{astri2024}. Meanwhile, the SWGO is emerging as a Latin American-led effort to establish a next-generation wide-field observatory in the Andes, bringing together partners from across the continent and beyond~\cite{swgo2024}.

In parallel, regional groups are increasingly involved in broader multi-messenger efforts, connecting gamma-ray observations with data from gravitational waves, neutrinos, and cosmic rays. These synergies are fostering cross-collaborative networks and helping to build a more integrated approach to high-energy astrophysics in Latin America.
\subsection{The Cherenkov Telescope Array Observatory (CTAO)} 

The CTAO is set to become the leading global observatory for very-high-energy gamma-ray astronomy in the coming decade and beyond \cite{CTAConsortium:2017dvg,Acharyya_2019,CTAConsortium:2023tdz} . Its scientific scope is remarkably broad, encompassing studies of the role of relativistic cosmic particles and searches for dark matter. CTAO will investigate the extreme universe, probing environments that range from the immediate vicinity of black holes to the vast scales of cosmic voids.

With an energy coverage spanning from 20 GeV to 300 TeV, CTAO represents a significant advancement over current instruments across all performance metrics. Its wide field of view and enhanced sensitivity will enable sky surveys at rates several hundred times faster than those of previous TeV telescopes. At the highest energies, the angular resolution of CTAO will be the finest resolution ever achieved by an instrument operating above the X-ray band. This capability will interestingly allow detailed imaging of a wide range of gamma-ray sources.

The \gls{CTA} collaboration has put forth a series of sensitivity studies to map the \gls{CTA} potential to probe the universe via high-energy gamma-rays, such as dark matter \cite{CTAConsortium:2023yak,CherenkovTelescopeArray:2023aqu,CTAO:2024wvb,Abe:2025qsi}  and fundamental cosmological properties \cite{CTA:2020qlo,CTA:2020hii}. Brazil, Chile and Mexico are part of the CTAO Consortium and Vitor de Souza, from Brazil, is the current chair of CTAO Consortium Board.

\subsection{The Southern Wide-field Gamma-ray Observatory (SWGO)} 

\begin{figure}[h!] 
        \centering 
        \includegraphics[width=0.8\textwidth]{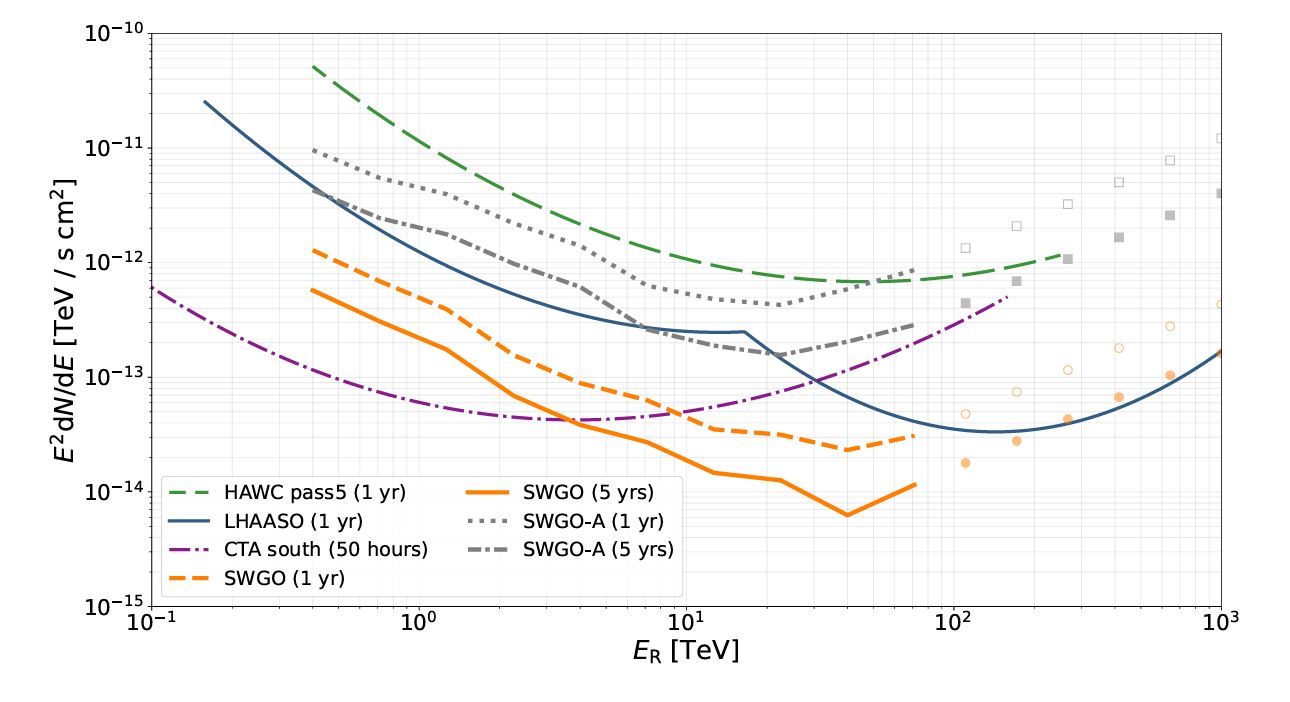}
        \caption{Differential point source sensitivity for SWGO and SWGO-A compared to CTA South, HAWC, and LHAASO. While the continuous lines of SWGO show the expected sensitivity, as calculated from simulations, the light markers denote extrapolations. At altitudes above 4,400 meters, where joint environmental and infrastructure studies have been conducted, America plays a central role in SWGO’s development. The collaboration has over 130 scientists from more than 40 institutions, including strong participation from Argentina, Brazil, Bolivia, Chile, Colombia, Mexico, and Peru. These countries contribute across all fronts: from site characterization and detector R\&D to simulation frameworks, prototype instrumentation, and science case development. The proposed sites are located in the Andes at altitudes above 4,400 meters, where joint environmental and infrastructure studies have been conducted to assess feasibility. Ulisses Barres, from Brazil, is one of the current vice-spokespersons}.
	\label{fig:swgo}
\end{figure}

SWGO is a proposed ground-based experiment designed to detect very-high-energy gamma rays using a water Cherenkov detector array at high altitude in South America. The project builds on the success of current-generation observatories like \gls{HAWC} and LHAASO, but offers a unique southern hemisphere view of the Galactic Center and large portions of the Milky Way not visible from northern facilities. \gls{SWGO} aims to operate in an energy range complementary to Imaging Atmospheric Cherenkov Telescopes (IACTs), enabling wide-field surveys, continuous monitoring of transient sources, and searches for signatures of dark matter, primordial black holes, or Lorentz invariance violation.
Figure \ref{fig:swgo} shows the sensitivity of SWGO compared to other gamma-ray experiments.

In recent years, the collaboration has reached several key milestones: completion of a technical design report, deployment of water Cherenkov and RPC prototypes in Bolivia and Peru, and refinement of a rich science case encompassing Galactic and extragalactic astrophysics, fundamental physics, and multi-messenger opportunities. The 2024 white paper highlights how \gls{SWGO} will provide continuity and expansion of regional expertise, reinforcing the legacy of past wide-field experiments such as LAGO and connecting with global efforts in gamma-ray and cosmic-ray astrophysics~\cite{swgo2024}.
In 2024, the Atacama Astronomical Park in Chile was selected as the site of the \gls{SWGO} project.

\subsection{Latin American Prospects in ASTRI} 
The \gls{ASTRI} Mini-Array, currently under deployment at the Teide Observatory in the Canary Islands, is a pathfinder for the Cherenkov Telescope Array Observatory (CTAO) and a standalone facility with unique capabilities above 1 TeV. It employs a novel dual-mirror Schwarzschild–Couder optical design and silicon photomultiplier (SiPM) cameras, enabling wide field-of-view observations and improved angular resolution. Once operational, the array of nine telescopes will enhance our understanding of very-high-energy (VHE) gamma-ray sources, including pulsar wind nebulae, active galactic nuclei, and the Galactic center.

While its deployment is led by INAF (Italy), the Brazilian astronomical community has engaged with \gls{ASTRI} to develop the technological and scientific expertise necessary for future participation in large-scale gamma-ray facilities. Brazilian researchers are developing SiPM-based instrumentation and simulation frameworks and promoting ASTRI-related activities as part of a national strategy to strengthen high-energy astrophysics in Latin America. This aligns with regional efforts to build critical mass in VHE gamma-ray astronomy and to foster training for students and early-career scientists.

\subsection{Outlook and Additional Efforts} 
Latin American groups are contributing across a growing range of gamma-ray and multi-messenger projects. These include the development of simulation and analysis tools, technical and scientific coordination, and significant involvement in R\&D for future facilities.

The \gls{SWGO} Collaboration plans to finalize the conceptual design in the coming years, laying the groundwork for a future proposal. In parallel, the \gls{ASTRI} Mini-Array is preparing to begin science operations from the Teide Observatory in Tenerife, with strong involvement from Brazilian and Argentinian institutions. These efforts offer unique opportunities to train new generations of researchers and strengthen regional capacities in high-energy astrophysics.

Latin American researchers have historically participated in international gamma-ray observatories, contributing to the field's development and building regional expertise in instrumentation and data analysis. These experiences help anchor the current wave of regional-led initiatives and promote greater international integration. 

\section{Collaborative Synergies and Joint Efforts} 

Several regional initiatives in Latin America are playing a central role in strengthening collaboration, training, and infrastructure development across multiple areas of HECAP. These efforts not only foster cross-border scientific networks but also ensure knowledge transfer, capacity building, and long-term sustainability of the region’s contributions to global science.

The Latin American Giant Observatory (LAGO), detailed in Section~\ref{sec:cr}, exemplifies such synergy with its decentralized structure and coordination among more than 30 institutions across 10 countries. \gls{LAGO} combines HEP and astroparticle research with strong educational and outreach goals~\cite{lago2024}.

LA-CoNGA physics (Latin American Alliance for Capacity Building in Astroparticle Physics) was launched in 2020 as an Erasmus+ capacity-building project involving France, Colombia, Ecuador, Peru, and Venezuela. It aims to develop joint virtual and local Master-level training programs in HEP and astroparticle physics. Since 2021, LA-CoNGA has consolidated its academic network, delivered regular courses and hands-on data analysis training, and contributed to the integration of Latin American early-career scientists into international research programs~\cite{laconga2022}.

EL BONGO (E-Latin America Digital huB for OpeN Growing cOmmunities in physics) ~\cite{elbongo2024}, is an Erasmus initiative that empowers the High-Energy Physics (HEP) community in Latin America by fostering digital transformation in education. It builds virtual, collaborative research-learning communities to democratize physics training. Key aims include developing practical digital skills and promoting open science across strategic domains, including HEP, astroparticle physics, geophysics, and AI. Starting in 2021 with a previous initiative, LA-CoNGA physics \cite{laconga2022}, both ERASMUS+ projects have trained over 200 students across the region, with strong institutional support and continuity. EL BONGO also interfaces with regional infrastructure efforts discussed in Chapter~\ref{chapt:instru}.
Efforts in capacity building will be further discussed in Chapter~\ref{chapt:training}.

Finally, we note that multiple HECAP projects in Latin America are embedded in large-scale global initiatives, as will also be clear in the following Chapters.

\bibliographystyle{unsrt} 
\bibliography{astro/astro}

\chapter{Cosmology}\label{chapt:cosmo}

\section{Introduction} \label{sec:intro} 
Cosmology lies at the intersection of theoretical physics, particle physics, and astrophysics. Among the most fundamental hypotheses of the concordance cosmological model are the validity of  Einstein's theory of general relativity and the Standard Model of particle physics, with possible and necessary extensions needed to account for the existence of a dark sector.  Cosmology maintains deep and exciting connections with fundamental physics.   Evidence for a dark sector composed of dark matter (DM) and dark energy (DE) relies on their gravitational effects. Despite decades of experimental progress and theoretical efforts dedicated to understanding the dark sector, its nature remains a mystery.

Since the previous Latin American (LA) Strategy, the LA  community  working on gravitation and cosmology   has continued contributing to the  main science drivers, which, as summarized in the   LA  HECAP Physics Briefing Book,  concern the  following questions:
\begin{itemize} 
\item What are the nature, properties and origin of the dark components (i.e., Dark Matter and Dark Energy)?
\item  What is the origin of matter-antimatter asymmetry?
\item  What are the nature and the properties of neutrinos?
\item  What are the nature and properties of Black Holes? Do primordial Black Holes exist? Are
they (part of) the Dark Matter?
\item  What is the origin and nature of the primordial perturbations? Are there primordial gravitational waves?
\item  Are the standard assumptions wrong? Is General Relativity the applicable theory of gravity? Is Lorentz symmetry violated?
\end{itemize}  
The key point of the previous LA Strategy for gravitation and cosmology were:  to promote the integration of the LA community working in astrophysics and cosmology; to join efforts for the development of cutting-edge experiments in the region, local facilities and resources;  to foster the LA participation in experiments installed in the region, their synergies and collaborative work; and to boost capacity building to increase the number of qualified human resources with the required experience.

The projects summarized in the previous LA  HECAP Physics Briefing Book are the following:   BAO from Integrated Neutral Gas Observations (BINGO, a Brazilian-led experiment); Macon Ridge Astronomical Site  (with    \gls{TOROS} project and The 
\gls{ABRAS} project ); Q\&U Bolometric Interferometer for Cosmology (QUBIC, a global international project in Alto Chorrillos site next to the Large Latin American Millimeter Array experiment); South American Gravitational-Wave Observatory (SAGO) (a next-generation gravitational wave detector proposed for the long-term future);  Vera Rubin Observatory’s Legacy Survey of Space and Time (LSST)  in Chile, which involves  the project of the implementation of an independent Data Access Center (iDAC) in Brazil;  Latin American PhD program (the project to create a collaborative LA PhD program in the area, as a network of institutions from different LA countries). These projects imply the development of new facilities and resources (including human resources) in the region and represent an opportunity for the LA community to participate in data analysis and simulations, as well as to contribute to electronics, software design, and technological development.

Observational cosmology is entering a new era, historically known as the stage IV cosmology. Within a few years the results of the next generation of ground and space based telescopes (such as Euclid\cite{Euclid:2024yrr} , the Dark Energy Spectroscopic Instrument (DESI) \cite{DESI:2022xcl} , LSST, Roman Space Telescope\cite{Spergel:2015sza} , the Simons Observatory (SO) \cite{Mangu:2024ouy}  and  the Lite (Light) spacecraft for the study of B-mode polarization and Inflation  from cosmic background Radiation Detection (LiteBIRD) \cite{LiteBIRD:2024wix}) will add crucial insights on  the current cosmological model.

Measurements of the Cosmic Microwave Background (CMB) radiation and its anisotropies provide  very valuable information about different parameters that describe the global properties of the Universe on large scales.  Primary CMB temperature and polarization  data from Planck 2018 \cite{Planck:2018vyg}, ACT  \cite{ACT:2025fju}   and SPT-3G \cite{SPT-3G:2025bzu}  have  achieved sub-percent precision on the six main parameters of the spatially flat Lambda Cold Dark Matter ($\Lambda$CDM) cosmological model. From these precise measurements one can infer tight constraints on the Hubble constant  $H_0$, on the relative abundance of matter, radiation, DM and  DE (which is described by the cosmological constant $\Lambda$), as well as on other parameters  determining the primordial perturbations that characterize the initial conditions for the cosmological perturbations  (i.e., the perturbed geometry of the universe, matter and radiation perturbations).  The minimal $\Lambda$CDM  model assumes adiabatic, Gaussian initial conditions and a single power-law form for the power spectrum of primordial  perturbations  \cite{Planck:2018jri} .

For the sensitivity level that the new generation of experiments  (such as QUBIC, the Simons Observatory\cite{Mangu:2024ouy} , LiteBIRD \cite{LiteBIRD:2024wix} ) aim to reach, not only  instrumental systematic effects   represent a challenge to CMB measurements, but also  the possibility to remove foregrounds,  particularly for the measurement of primordial gravitational waves via the polarization B-modes 
\cite{Planck:2018gnk} .  The latter can, in principle, be accounted for through their angular power spectra and frequency dependence, distinct from that of the CMB.

Beyond the  primary CMB anisotropies,  there are  known secondary anisotropies such as   gravitational lensing (which  refers to the fact that  CMB photon paths are deflected by gravitational potentials induced by matter overdensities,  for a review, see \cite{Lewis:2006fu} ), and 
the thermal and kinetic Sunyaev-Zel’dovich effect (caused by the fact that  CMB photons scatter off the free-electron gas in galaxies and clusters, allowing the  use  of CMB as a backlight to probe the gas in and around low-redshift galaxies,  see for  a review  \cite{Mroczkowski:2018nrv} ) that are measured with increasing precision (see, for instance 
  \cite{Planck:2018vyg,ACT:2023dou,ACT:2023kun,Liu:2025zqo, RiedGuachalla:2025byu} ). 
There are other important cosmological probes, such as cosmic shear (which refers to the coherent distortions of galaxy images caused by weak gravitational lensing due to intervening large-scale structure; see, e.g., \cite{DodelsonGL}).  These probes are sensitive to the cosmological perturbations on different  range of redshifts and  angular scales. Auto-correlations  and cross-correlations of these CMB  signals and galaxy or quasar  surveys are beginning to be detected at increasing signal-to-noise, hence their ability to provide useful constraints (see for instance  \cite{ACT:2023skz,Villagra:2025ztn} ).

Large photometric and spectroscopic surveys of galaxies have recently begun to place constraints on some cosmological parameters comparable in precision to those from CMB predictions \cite{DESI:2024hhd}. Optical spectroscopy provides crucial information on structure formation.  After 2dF and SDSS maps  have led to the first measurement of the baryon  acoustic peak  (which is a well understood feature in the correlation function of a tracer of the large-scale structure) in the large scale correlation function \cite{SDSS:2005xqv, 2dFGRS:2005yhx},   multi-object spectroscopy has allow for expansion rate measurements using the Baryon Acoustic Oscillation (BAO) method, using both galaxies and the Lyman-$\alpha$ forest as tracers, and  also  the  study of redshift-space distortions (RSD),  providing complementary cosmological probes. These maps can, in turn, be combined with weak lensing and CMB maps to capture physics at a wider range of scales, thus providing an accurate reconstruction of the power spectrum encoding the initial conditions of the universe produced by inflationary physics.  Using galaxies and the Lyman-$\alpha$ forest, the Dark Energy Spectroscopic Instrument (DESI) released  BAO measurements indicating signs of tension with current CMB data \cite{DESI:2025zgx,DESI:2025zpo} . Large-scale spectroscopic surveys require imaging surveys with sufficient sky coverage, wavelength range, and depth. 

Photometric surveys, such as the Dark Energy Survey, which has Brazilian participation and is completing the final analyses of the full data set in 2025, and the upcoming Vera Rubin Observatory’s Legacy Survey of Space and Time (LSST) will also contribute to our understanding of the cosmological model. In fact, the standard $\Lambda$CDM model is currently under stress, with accumulating data revealing tensions in $H_0$ measurements and evidence for dark energy evolution.

Regarding the power spectrum of primordial perturbations, in the minimal $\Lambda$CDM model, Planck CMB measurements have placed tight constraints on large scales (e.g., \cite{Planck:2018jri}). On smaller scales, constraints can be derived from high-multipole CMB data (e.g., from ACT\cite{ACT:2025tim} ), Lyman-$\alpha$ \cite{Rogers:2023upm} , $\mu$-distortions \cite{Nakama:2017ohe,Chluba:2019kpb} , and future 21cm experiments \cite{Naik:2025mba} .

An alternative approach to   mapping  the Universe through analysis of galaxy surveys consists of measuring  the collective 21 cm emission of the
underlying hydrogen distribution around large concentrations of galaxies, through a technique known as Intensity Mapping (IM).  
Using radio telescopes with a very large collecting
area  the technique allows  fast surveying of a  broad sky patch. Measurements of intensity fluctuations allows the   use of HI as a tracer of the 3-D large-scale
structure of the Universe.

Fast radio bursts (FRBs),  transient signals most of which were confirmed to come from cosmological distances, are dispersed by the ionized medium along their propagation path (mainly by the inhomogeneous intergalactic medium (IGM)), and in principle could provide valuable information of the Hubble rate and DE (see for instance \cite{Zhou:2014yta} ). While thousands of FRBs have been detected overall, most lack precise localisation (fewer than a hundred have been localised with a host and redshift).  FRBs are  already being used in combination with other datasets to  constrain the equation of state of DE \cite{Wang:2025ugc} .

Gravitational waves (GW), as predicted by General Relativity, 
are ripples in the space-time geometry propagating with the speed of light and produced
by the accelerated motion of massive astrophysical objects or by cosmological sources.
Indirect detection of GW was asserted since the 1970’s with the continuous observation of
PSR1913+16, the direct detection of  GWs  at high frequencies ($\sim 10$ Hz to a few kHz) by interferometers\cite{LIGOScientific:2016sjg,KAGRA:2020agh,VIRGO:2014yos}   and the more recent detectors based on pulsar timing that have shown convincing evidence of  a the stochastic GW background (SGWB)  in the frequency range of nanohertz  through various collaborations \cite{NANOGrav:2023gor,Xu:2023wog,EPTA:2023sfo,EPTA:2023akd,EPTA:2023fyk,Zic:2023gta,Reardon:2023gzh} ,  have sparked increased interest in  GW astronomy,  allowing for the possibility to constrain fundamental physics and understand the nature of gravity.    Signals from black hole coalescence events have been detected from nearly 100 mergers, including a spectacular event involving two merging neutron stars, which enabled the first coincident detection by electromagnetic observatories, ushering in the era of multi-messenger astronomy involving gravitational waves.
 The next decade will mark the arrival of new third-generation ground-based gravitational-wave observatories (Einstein Telescope and Cosmic Explorer), as well as the space-based detector Laser Interferometer Space Antenna, known as LISA\cite{LISA:2024hlh}, which opens the millihertz frequency window and is scheduled to launch in
 2035. 
So far, gravitational waves generated only by binary systems of black holes and neutron stars have been detected; however, waves originating from exotic astrophysical objects, such as supernovae, have not yet been detected. These astrophysical objects are of particular interest because their electromagnetic radiation, neutrino emission, and gravitational activity will provide new insights into the universe and advance multi-messenger astronomy. The  search  of primordial   gravitational waves, ripples in spacetime created in the very early universe, are a key prediction of inflationary cosmology and a major focus of scientific research. The primary method involves studying the polarization of the CMB,  as primordial gravitational waves are predicted to leave a specific B-mode polarization pattern in the CMB. In summary, the search for gravitational wave signals encompasses a frequency range spanning several orders of magnitude, diverse production mechanisms, and complementary modeling and data analysis techniques, all of which are necessary to maximize the probability of detecting signals of different shapes, intensities, and durations. The scientific knowledge expected from such intensive research is broad, from fundamental physics to astrophysics and cosmology.

\section{Experiments and infrastructure on Gravitation and Cosmology  with LA participation } 

\subsection{ Vera Rubin Observatory’s Legacy Survey of Space and Time (LSST)} 
After 10 years of construction, the first images produced by the \gls{LSST} 3.2 Gigapixels camera from the light collected by the 8.4 meter Simonyi Survey Telescope were presented to the world on June 23, 2025. The survey is now in the commissioning phase and will begin its 10-year observations. Using images from six filters to estimate the photometric redshift, a catalogue of billions of objects will be produced, along with millions of nightly alerts for transient phenomena. 

The four primary science goals of the Rubin Observatory are:\\
\begin{itemize}
 \item Probing dark energy and dark matter\\
 \item Mapping the Milky Way \\
 \item Taking an inventory of the Solar System \\
 \item Explore the transient optical sky
\end{itemize}

The science will be conducted by eight science collaborations, independent communities of scientists worldwide, and self-organized based on their research interests
and expertise. They are:\\
\begin{itemize}
  \item  Galaxies Science Collaboration (GSC) \\
  \item  Stars, Milky Way and Local Volume Science Collaboration (SMWLV) \\
  \item  Solar System Science Collaboration (SSSC) \\
  \item  Dark Energy Science Collaboration (DESC)\\
  \item  Active Galactic Nuclei Science Collaboration (AGN) \\
  \item  Transients and Variable Stars Science Collaboration (TVS) \\
  \item  Strong Lensing Science Collaboration (SLSC) \\
  \item  Informatics and Statistics Science Collaboration (ISSC)\\
\end{itemize}

{\bf Current LA Involvement: }  The Latin American community's participation has increased since the last Strategy
Planning, with Argentina and Mexico's entry into LSST. This community is involved in all science
collaborations, developing specialized software, and establishing the so-called independent
data access centers (IDACs), which are regional computing facilities for processing and
accessing the large amount of data to be produced. This community organized the first “\gls{LSST} Latin American Meeting (LSST@LATAM): Catalyzing Research Collaborations”, in June 2024, in La Serena, Chile. The second edition of this meeting will take place in Mexico City in December 2025, and a third one is planned to take place at ICTP-SAIFR in Sau Paulo in 2026. 

At this moment, there are the following Principal Investigator (PI) positions through in-kind contributions:
12 from Argentina, 29 from Brazil (24 from LIneA and 5 from FAPESP) and 20 from Mexico. Each PI can have up to four junior associates. As a host country, Chile can have an unlimited number of PIs.
  
{\bf  Computing requirements:}  
The in-kind contribution from Brazil includes an Independent Data Access Center (IDAC) with the requirements:
1000 processing cores (~100 Tflops), 5 PB storage, 500 TB database, Internet connection with speed greater than 40 Gbps. 
In addition, there will be contributions to photo-z infrastructure and pipeline scientists. The total contribution was valued at US \$ 4.4 million. The funds for the hardware are already secured, but there is a need to find resources for the operation of the IDAC for more than 10 years, in addition to a requirement to update the CPUs in 5 years.

The Mexican IDAC has secured funding for 1.6 petabytes, 264 cores, and 20 million CPU hours, as well as for the development of a user front-end to store the \gls{LSST} catalogues, and possibly some added-value image products for the use of Rubin/\gls{LSST} researchers. An extension to the contribution, including GPU support, is being considered.

\subsection{  Dark Energy Spectroscopic Instrument (DESI)  and its Extension} 
The Dark Energy Spectroscopic Instrument (DESI), located at the Mayall 4-meter telescope, is a Stage IV dark energy experiment, currently the world's foremost instrument in the world dedicated to spectroscopic observations of large-scale structure. DESI was designed to study the matter and energy contents of the universe by measuring the expansion history with unprecedented precision using the Baryon Acoustic Oscillation (BAO) and redshift-space distortions (RSD) techniques. The first-year results of DESI were presented in 2024, and its BAO data provide near-per cent level precision of cosmic distances in seven bins over the redshift range $z = 0.1-4.2$, which indicate tantalizing hints of models of evolving dark energy in contrast to a cosmological constant. The second data release presented this year, consisting of data from the first three years of operations, provided tighter constraints on parameters of the LCDM model and a statistical preference for an evolving dark energy model over a cosmological constant, with significances ranging from $\sim 3 \sigma$  to $\sim4 \sigma$, depending on the combination of datasets used for the analysis \cite{DESI:2025zgx} . The next analysis of the DESI main survey will include five years of data collection, with more DESI BAO data, as well as DESI measurements of redshift-space distortions and peculiar velocities, which can also test cosmic growth and gravity. 

The DESI extension consists of extending operations by 2.5 years (from June 2026 to December 2028), expanding the survey footprint (14,000 deg2 to 17,000 deg2) and improving the observational strategy and methodology. This would increase the total number of spectroscopically confirmed galaxies and quasars from 39 million to more than 63 million, improve the inverse variance of the BAO measurements by 22\% across all redshifts, and maximize synergies with other cosmological probes. 

{\bf Current LA Involvement: } 
The DESI Collaboration comprises hundreds of scientists from institutions worldwide. Participation in DESI is classified into Principal Investigators (PIs) and Sponsored Members (SMs), including postdoctoral researchers, students, and technical or other research staff with a short-term contracts. Latin American participation in DESI includes:
\begin{itemize} 
\item Universidad de los Andes (Uniandes), Colombia. 1 PI and 5
SM.
\item  Mexico Regional Participation Group (RPG) that includes the following
institutions and number of participants:
\begin{itemize} 
\item Centro de Investigaci\'on y de Estudios Avanzados del Instituto
Polit\'ecnico Nacional (CINVESTAV-IPN). 1 PI and 1 SM.
\item  Universidad de Guanajuato (Divisi\'on de Ciencias e Ingenier\'ias,
UG). 2 PI and 8 SM.
\item Universidad Nacional Aut\'onoma de M\'exico (UNAM-Instituto de
F\'isica, UNAM-Instituto de Astronom\'ia, UNAM-Instituto de Ciencias
F\'isicas). 4 PI, 15 SM and 1 external collaborator.
\item Instituto Nacional de Investigaciones Nucleares (ININ): 1 PI and 1 SM.
\end{itemize} 
\end{itemize} 
LATAM participants have played a very important role in
achieving DESI's milestones, not only contributing to the daily operations,
the data team, and/or the science analysis, but also assuming leadership roles.   
LATAM institutions are represented on the Institutional Board, and in particular, Mexico's RPG has one vote in DESI decisions. Finally, there are currently six LATAM DESI members who have obtained the builder status, as a recognition of a long engagement and a significant contribution to the collaboration infrastructure and service work.
 
{\bf  Computing requirements: } 
The extended survey will significantly increase the data volume and computational
requirements. Key needs include:  Enhanced data processing capabilities, increased storage capacity, improved network infrastructure for data transfer, and 
advanced analysis tools for cosmological measurements.  The DESI collaboration will leverage existing High-Performance Computing (HPC) resources and explore new technologies to meet these demands.

{\bf Technological advances, advanced training and connection with Industries:  } 
Former LA participants in DESI have made important contributions to DESI
and their participation in this collaboration has been crucial in advancing
their careers both inside and outside of academia. Those who left academia have been able to continue their careers in different work contexts,
as data scientists, for instance. This highlights the importance of basic science
research, and in particular in cosmology, as key to the workforce preparation needed in today's and future contexts. Moreover, there is a significant participation of mainly sponsored members worldwide, whose academic formation and first contact with DESI began in LATAM countries and/or are of LATAM origin. 
LA members have been part of the education and outreach DESI committees. Thanks to the LATAM contributions, DESI has public pages in Spanish, including some with outreach content like DESIHigh and Meet a DESI member

{\bf Construction and operational costs:} 
The extension of DESI will require approximately $30$ million in additional
funding over 2.5 years. In the LATAM context, participation in the DESI extension does not entail additional costs beyond the initial buy-in. However  the
community will need continuous funding for maintaining the active participation. It is necessary to increase the junior participation, postdoctoral, and Ph.D. students in particular, in order to provide more support to DESI for the operations and in particular in data and cosmological analysis. It would be desirable to increase the Full Participant members from LATAM through new buy-ins. 

{\bf Synergy: }  DESI will maximize synergies  across multiple cosmological probes   (including weak
lensing, CMB lensing, and galaxy-galaxy lensing) by increasing overlap with other surveys, such as the Rubin
Observatory's Legacy Survey of Space and Time (LSST) and  with CMB experiments, such as the Atacama Cosmology
Telescope (ACT) and the Simons Observatory (SO).

\subsection{Participation in Gravitational Wave projects and the South American Gravitational-Wave Observatory (SAGO) proposal } 

\gls{SAGO} is a so-called third-generation (3G) GW detector proposed by the Latin America community  to boost   their participation and  their impact in the field of gravitational wave astronomy,  particularly  on multi-messenger cosmology (see previous LA  HECAP Physics Briefing Book).

\begin{figure} [htbp]
\begin{center} 
\includegraphics[width=10cm]{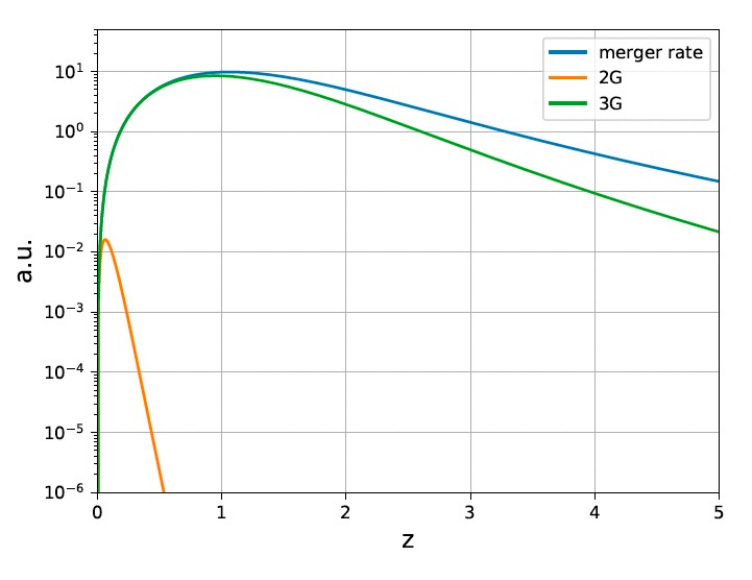}  
\caption{Extended redshift distribution of detections of GW from compact binary coalescences from second (2G) and third (3G) generation detectors, compared to expected merger rate inferred from star formation rate. Adapted from \cite{Leandro:2021qlc} . } 
\label{GWFoM} 
\end{center} 
\end{figure}

The current global network of detectors comprises the two Advanced LIGO (aLIGO) detectors in the United States, Advanced Virgo (aVirgo) in Italy and KAGRA in Japan.
In the near future, these detectors (which are close to the end of their fourth observing run (O4))   will be observing during one more period (O5), and it is likely that three detectors of the third generation (3G)  (with sensitivities about 10 times higher than aLIGO and aVirgo) will step in by 2030: Einstein Telescope, located in Europe, Cosmic Explorer, located in United States, and the Australian project NEMO.  The detection rate is expected to increase by several orders of magnitude as we move from second- to third-generation detectors and reach a higher redshift sources, as illustrated in Figure \ref{GWFoM}.  These three (3G) detectors will need a fourth partner for triangulation of arrival times in order to determine the source position in the sky more precisely and its polarization. To leverage this opportunity, the Latin American community has proposed the SAGO project. The first priority steps of the LA community have been to build a critical mass of researchers with technical expertise and to perform R\&D and site determination studies.  
For instance, in Mexico, the Guadalajara group GWDAMX   is proposing the construction of a control center (in the near future) for the search and parameter estimation of the GW generated by astrophysical objects.
 The main scientific goals highlighted in the GWDAMX proposal are to study the fundamental physics of formation and evolution of Core Collapse Supernovae and the associated mechanisms for generating and emitting GW, EM radiation, and neutrino emission, as well as to develop novel methods for the search, characterization, and parameter estimation of GW based on computational intelligence techniques.

{\bf  Current LA Involvement:  } 
There are in total about 50 LA people in LA or Caribbean countries that are already working or  have worked in gravitational wave detection collaborations with the laser interferometer technique,  other people interested and/or already working in Gravitational Wave Astrophysics, and many  researchers and professors working on GR, gravitational waves and astronomy/astrophysics. 
 
There are five groups in Latin America and Caribe (Anguilla (1), Brazil (2), Colombia (1), and Mexico (1)) involved in the LIGO Scientific Collaboration (LSC).  There is one person from Mexico at KAGRA working on the vibration-isolation subgroup.
In LATAM, the ET Collaboration involves six LA institutions, most of whom belong to an officially recognized research unit based in the region: Chile (1), Mexico (1),  Brazil (4). 
The LISA collaboration includes four LATAM institutions:  Brazil (3) Chile (1). 
The NANOGrav collaboration has  two members working in Brazil.

{\bf  Timeline and major milestones: } 
 The LA community   is planning  to   reach a critical mass of experimentalists already educated in the 3G technology by 2030.  The aim is to double the current  LA  involvement  having about  50 more people dedicated  mainly to instrumental science, detector characterization, and calibration/commissioning, which are the most vacant areas of know-how.  It is then timely to start involving Latin American students and post-docs in the 3G technology now.  At this stage, it is crucial to have fellowships and research funding to create a critical mass of researchers. This will also attract researchers already involved in 3G detectors to visit LA institutions regularly and participate in the training project. The plan involves conducting seismic studies to identify potential sites in South America. 

In order to grow, the LATAM GW community plans to continue   building bridges with
other well-established communities (theoretical physicists, astronomers, cosmologists) that are already present and active in this region.


After  these goals are achieved  a proposal for the construction of \gls{SAGO} can be made to all possible Latin American and Caribbean science foundations in collaboration with the other 3G projects.

{\bf  Advanced Training:}  The   plan is to involve the students, post-docs  and early-career-stage professors in the 3G R\&D, which is divided in the following topics: 
Facilities \& Infrastructure, 
Core Optics, 
Coatings, 
Cryogenics, 
Newtonian Noise, 
Light Sources, 
Quantum Enhancements, 
SAS \& SUS, 
Auxiliary Optics, 
Simulation and Controls, 
Calibration. 

To move forward with this, the LA community can leverage the fact that LIGO has adopted a very welcome measure to attract researchers from less developed countries: substantially discounted participation fees for collaboration memberships and collaboration meeting fees. As well, there are opportunities to participate in LIGO-related laboratories for limited time periods (e.g., the SURF Program or the Visitor Program).

{\bf  Construction and operational costs: } 
The budget for \gls{SAGO} can be roughly estimated at US$1–2$ billion. 

For the {\it near future objectives}, the expected costs related to data analysis, simulations and/or theoretical investigations, are the following (estimates including all LATAM countries, per year):

\begin{itemize}  \item Visit expenses to/from collaborating groups. Estimate (around 30 visits): USD 100,000.00.
 \item Open access publication fees. Estimate (around 100 papers): USD 200,000.00.
 \item Fellowships (M.Sc., Ph.D. Post-docs). Estimate (around 500): USD 5,000,000.00.
 \item  Conference participation (including travel, hotel, meals, conference fee). Estimate
(around 50 conferences): USD 100,000.00.
 \item Physical connection to established computational backbones (construction). Estimate:
USD 5,000,000.00.
  \item The estimated cost of the construction of the control center proposed by the  GWDAMX  is 100,000 USD. 
\end{itemize} 

{\bf Computing requirements: } For SAGO, these are going to be given by the present Cosmic Explorer and ET R\&D studies.

  For the  {\it near-future objectives} ,  estimates including all LATAM countries per year are:
\begin{itemize}  \item 
 Notebooks (500): USD 1,500,000.00.
\item Computer accessories (500): USD 50,000.00.
\item Fast, wide band Internet connection (some countries): USD 200,000.00.
\item Software license (300 users): USD 300,000.00.
\item Workstations (100): USD 300,000.00.
\end{itemize} 

progress.   

{\bf Synergy} :   As a crucial test of General Relativity and alternative theories of gravity, GW observations are clearly important for cosmology and astrophysics. Cosmology with GWs is a  field that has been evolving rapidly in recent years, mainly after the detection of the GW from the binary neutron star merger GW170817 and its electromagnetic counterpart.  
The detection of the GW signal from a compact binary allows us to determine the luminosity distance of the source, and, if additional information on the source redshift is provided, we can infer a value for the cosmological expansion rate. Several methods in addition to using direct electromagnetic counterparts have been purposed to study cosmology with GWs (see, for instance \cite{Mastrogiovanni:2024mqc} ).  For example, galaxy surveys, combined with GW detections from a compact binary, can be used to identify possible host galaxies of the binary system.   In the multi-messenger astronomy approach, GW observations can be combined with electromagnetic and neutrino observations to provide a more comprehensive picture of GW sources and their environments.

\subsection{BAO from Integrated Neutral Gas Observations (BINGO)} 
The \gls{BINGO}  project is a double-dish radio telescope
designed to observe the 21cm line emission from neutral hydrogen (HI) in a redshift interval 
$0.127\leq z\leq0.445$\  (corresponding to a frequency interval 980 MHz to 1260 MHz). As described in  the previous LA  HECAP Physics Briefing Book, BINGO's  main scientific goal is to claim the first BAO detection in radio in this redshift range, mapping the 3-D distribution of HI,   using   a technique known as
Intensity Mapping (IM) and with observations spanning several years.
High-quality measurements of BAO  in such a redshift range will   
be extremely useful to set constraints on DE models, since it  covers the transition from a matter-dominated Universe to a dark-energy (DE) dominated Universe
(transition occurred at $z\sim 0.3-0.4$). \gls{BINGO}  will also be capable to detect   and characterize Fast Radio Bursts (FRB) and other transients, and to characterize synchrotron polarization in the frequency range $980 \,{\rm MHz} <  f < 1260\,{\rm MHz} $.

{\bf Current LA Involvement: }  At this moment, only  Brazil is participating in BINGO.  \gls{BINGO} is open to involve  LA participation  in some specific areas
related to data analysis and pipeline preparation. 
 
 The \gls{BINGO} consortium is coordinated by the Universidade de S\~ao Paulo - USP, and co-led by Instituto Nacional de Pesquisas Espaciais - INPE and Universidade Federal de Campina Grande - UFCG (Brazil) and has Yang Zhou University and Shanghai Astronomical Observatory (China) as the main foreign partners.  
There are currently 7 scientists in the \gls{BINGO} steering committee:  Elcio Abdalla (USP); Amílcar Rabelo de Queiroz (UFCG); Carlos Alexandre Wuensche (INPE); Filipe Batoni Abdalla (USTC); Thyrso Villela (CGEE/INPE); Bin Wang (Yang Zhou University, China); Jiajun Zhang (Shanghai Astronomical Observatory, China).

{\bf Facilities and resources:}   \gls{BINGO} is being built in Paraíba, Northeastern Brazil, with a strong support of state authorities. There is also support from the Ministry of Science and Technology and FINEP, a related funding agency.

  {\bf  Technological advances:}      
a) Receiver construction, operating at room temperature, but with a careful front-end design,
delivering a very low system temperature and very good stabilization, due to the coupling to a
Peltier cooler. A Phase switch on a protoboard was designed, built and successfully tested at
UFCG. This subsystem is under testing and can be considered a technological advance.
b) The optical design was done in Brazil, with a complex focal plane distribution of the horns.
\gls{BINGO} is probably the largest example of a ''single dish, many horns''  telescope. The fabrication
of the two 40m dishes and their supporting structures, which were manufactured by CETC54, the same company that built the FAST telescope dishes and structures.
c) The prototypes of the horn, transitions, polarizer, magic tees and rectangular-to-coax transitions
were designed and fabricated in Brazil and successfully passed the electromagnetic tests.

{\bf  Advanced training:}  There are currently some Brazilian postdocs and a number of M.Sc. and Ph.D. students who are deeply involved with aspects of the mission preparation, including development of the
necessary instrumentation for the telescope. They are either leading or participating in the second series of papers on the instrument, to be released during 2025.

  {\bf Connection with Industries:}  Parts built by local industries/new industry efforts led by workforce trained
in physics: the construction of the instrument is led by Brazilian institutions and is being carried out by Brazilian companies. By March 2025, there were no spin-offs or start-ups from BINGO.
 
{\bf Timeline and Major Milestones:}   By March 2025, the dishes, horns, front and backend were already fabricated,
waiting for the civil structure to be completed; the terrain to receive the instrument was already
prepared, and the funding for the civil construction was already granted (they were waiting for the company selection process to be completed);  a series of new papers on  the  data analysis pipeline and science forecasts (following the eight
+2  published in 2022 and 2023) was being produced.  The commissioning phase is expected to start towards  the end of 2025. The major milestones are the
liberation of additional funds (FINEP, MCTI and approval of a new FAPESP project); the completion of the telescope assembly (foundations and metal support structure); and
a successful commissioning phase.

{\bf Synergy: }  
\gls{BINGO} will   have a strong synergy with a number of optical and
radio instruments, such as the LSST, MeerKat, ASKAP and Euclid, that will visit parts of the 5400 square
degrees observed by \gls{BINGO} in the southern sky. This will be superb for an immediate multiwavelength
approach for \gls{BINGO} data analysis.

\subsection{Q\&U Bolometric Interferometer for Cosmology (QUBIC)} 
\gls{QUBIC} is an experiment designed to measure the polarization of the CMB \cite{QUBICWP2020} . The main scientific goal is the measurement of a primordial B-modes signal. This is a smoking gun for primordial gravitational waves whose relative strength is given by $r$. The current bound is $r < 0.06$, which already disfavours the simplest inflationary models of the early Universe. In the near future, experiments are expected to achieve a sensitivity to $r$ of order $\sigma(r)= 0.01$, with improvements up to a few times $0.001$. \gls{QUBIC} \cite{QUBICWP2020}  is one of the international efforts currently pursuing this goal, and is using a novel kind of instrument, which combines the extreme sensitivity of two arrays of bolometric detectors, operating respectively at 150 GHz and 220 GHz, with the control of the systematics offered by the interferometric operation of the instrument.
The \gls{QUBIC} Collaboration reports that, using spectro-imaging capabilities, it is possible to reconstruct frequency sub-bands and obtain 5 maps used to estimate the parameter r, as well as a three-parameter B-mode dust emission model (from the latest Planck model). Not only should the reachable sensitivity be taken into account, but also the control of systematic effects and the possibility of removing foregrounds. In this sense, \gls{QUBIC} has a unique status due to its particular architecture as a bolometric interferometer.

{\bf  Current LA involvement:}  \gls{QUBIC} International Collaboration is integrated by France, Italy, the UK, Ireland, the USA and Argentina. In Argentina, 49 members (researchers, engineers, technicians and students) are part of different working groups. The involved institutions are ITEDA-CNEA (CAC-ITeDA, CAB-IB, Regional NOA (Salta) and Cuyo (Mendoza)), as well as CONICET, UNLP, and IAR. As part of the local work, 4 PhD theses are in progress. For the first module, Argentina is the only Latin American country involved.
 
{\bf  Local facilities and resources:}  Development of Alto Chorrillo site (a new scientific/astronomical pole in Salta, Argentina, where \gls{LLAMA} will also be located). This site is 180 km from the Chajnantor site in Chile, where other millimetre-wave experiments and observatories are located (ALMA, APEX, Advanced ACTPol, POLARBEAR, and CLASS), and offers similar atmospheric conditions. This location has been characterized for several years within the framework of the site selection process for the \gls{CTA} project and for the \gls{LLAMA} project, which has atmospheric requirements similar to those of QUBIC. The specific contributions from Argentina include roads, energy, telecommunications, qualified human resources, and community engagement.

{\bf  Scientific advances:}  The contributions of the Argentine \gls{QUBIC} collaboration can be grouped in simulations for the data acquisition, map-making process and component separation to clean the CMB and obtain forecasts on the parameter r. In particular, the study of the angular resolution of the reconstructed CMB maps obtained with the simulation pipeline for different sub-frequencies within the 150GHz band, taking into account the spectro-imaging capabilities of QUBIC. Another important contribution is the improvement of the end-to-end pipeline and the simulation of reconstructed maps, both of which are useful for studying instrumental noise. The development of a component separation framework to be able to separate the CMB signal from the foregrounds, and studying the ability of \gls{QUBIC} for distinguishing between different dust models in the data analysis procedure to obtain r and the dust model parameters. Advancements in machine learning tools are underway to develop new methods for map-making and component separation.

{\bf  Technological advances: } No new technological advances are being developed at the moment. Advanced training: The collaboration has carried out a series of advanced tutorials for young
scientists and engineers to help them start working with the simulation pipelines (qubicsoft). Connection with industries: There is no direct involvement of the local industry in the project.
Timeline and major milestones: The prototype instrument with 1/4 of the focal plane was installed at the observation site, and after a vacuum system failure was repaired, commissioning began in early 2025. First sky measurements are underway alongside in-situ calibration. It is expected that the mount's full angular system will be completed by mid-2025, and full-sky scanning will commence in the second semester of the current year.

{ \bf Computing requirements: } For the data analysis, several Tb of storage will be needed, which will be provided at NERSC computing center, where the final analysis will be performed.

\subsection{ A Latin-American network on Astrophysics, Cosmology and Gravitation: Latin American PhD program } 

The proposal is to create a graduate program in Astrophysics and Cosmology, in the form of a network, based on the PPGCosmo experience (a Brazilian PhD program based at UFES), with participation from universities and scientific institutions across Latin America. Students in the program could, for example, have two co-advisors from different countries, thereby increasing scientific cooperation between the participating institutions. The primary objective is to undertake a collaborative PhD program that fosters in-depth scientific collaboration among all participating institutions and research groups, with extensive utilisation of the continent's existing astronomical facilities. 

PPGCosmo started its activities in 2016. The research topics range from theoretical to observational aspects of Astrophysics, Cosmology and Gravitation, including participation in large collaborations such as Euclid, J-PAS, DES and LIGO-Virgo. Thematic areas include: Dark Matter, Astroparticle Physics, Cosmology, Gravitational Waves, Dark Energy, High Energy Theory, Astronomy and Astrophysics. In the original configuration of the PPGCosmo, students were from the beginning of their courses associated to an advisor at Brazil and a co-advisor outside Brazil. 
PPGCosmo initially involved researchers from 6 Brazilian institutions and four international research groups. Students are selected after a careful selection process that includes analysis of CVs, projects, and reference letters, as well as an interview. Each selection process had candidates from different continents. Once accepted, students should start their research project at UFES but their mobility between associate institutes is highly encouraged. The thesis should be written in English, which is also the language of their public defences. Until now, more than 20 PhD thesis have been defended.

Given the PPGCosmo example, we discuss how a similar graduate school can be established to bring together Latin American research groups in the thematic areas cited above. Such an initiative can effectively advance scientific collaboration across the Continent.

{\bf Advanced training: } The student's training could cover theoretical aspects of Astrophysics and Cosmology. The use of the astronomical/experimental structures installed in the South of the Continent would be a priority in the PhD network program, but this does not preclude theoretical research themes: In Physics and Astronomy, Theory and Observations/Experiments, which must be developed together. 
The PhD student will develop their project thesis under the supervision of at least two advisors from at least two institutions in different countries.  The formation of students, even if it is primarily conducted at a given institution, must include an internship at another institution within the network, thereby integrating the Latin American community working in Astrophysics and Cosmology. 

{\bf Timeline and major milestones:} The goal is to create a collaborative LA PhD program in Astrophysics, Cosmology, and Gravitation, as a network of institutions across the continent.
Firstly, it is essential to establish an advisory scientific committee comprising leading scientists from various countries specialising in the field. Institutions and research groups must explicitly express their interest in joining the proposed program and specify the terms of the necessary agreement. Hence, the necessary scientific and administrative personnel, as well as the infrastructure, are in principle already available. The project's significant new contributions would include a substantial number of PhD fellowships from various countries, as well as dedicated financial support for scientific missions across the continent. It is important that those fellowships have a fixed, common value for all students.  The value of the fellowship, for example, may enable a student from a given country to spend a long period (typically one year) in another country. To simplify bureaucratic processes, PhD certificates would be issued by a single host institution and immediately recognised and validated by all participating institutions.

\section{Conclusions} \label{sec:conclusions} 
Currently, most of the previous strategic objectives remain to be achieved. There is at least one example, the development of the Macon  site, that has been stalled. However, the LA community working on gravitation and cosmology, although relatively small (compared, for instance, to that in Europe and the US) and sparse, has successfully moved towards integrating into joint projects. Indeed, since the last strategic plan, the participation of research groups in international cosmological experiments located in the Latin American region, such as \gls{LSST}, has increased and their strengths and expertise have been consolidated. The number of researchers dedicated to different aspects of GW physics has increased, more LA groups have joined experiments with second-generation detectors, and there are already members working in the research units of the consortium for third-generation detectors.  The social context of science and societal priorities has brought to the forefront the importance of addressing not only capacity building but also retaining and attracting human resources with the pertinent knowledge in LA countries.

\bibliographystyle{plain} 
\bibliography{cosmo/cosmo}

\chapter{Dark Matter}\label{chapt:dm}


\section{Introduction}  \label{sec:dmintro} 

\begin{figure} [h!]
    \centering
    \includegraphics[width=0.85\textwidth]{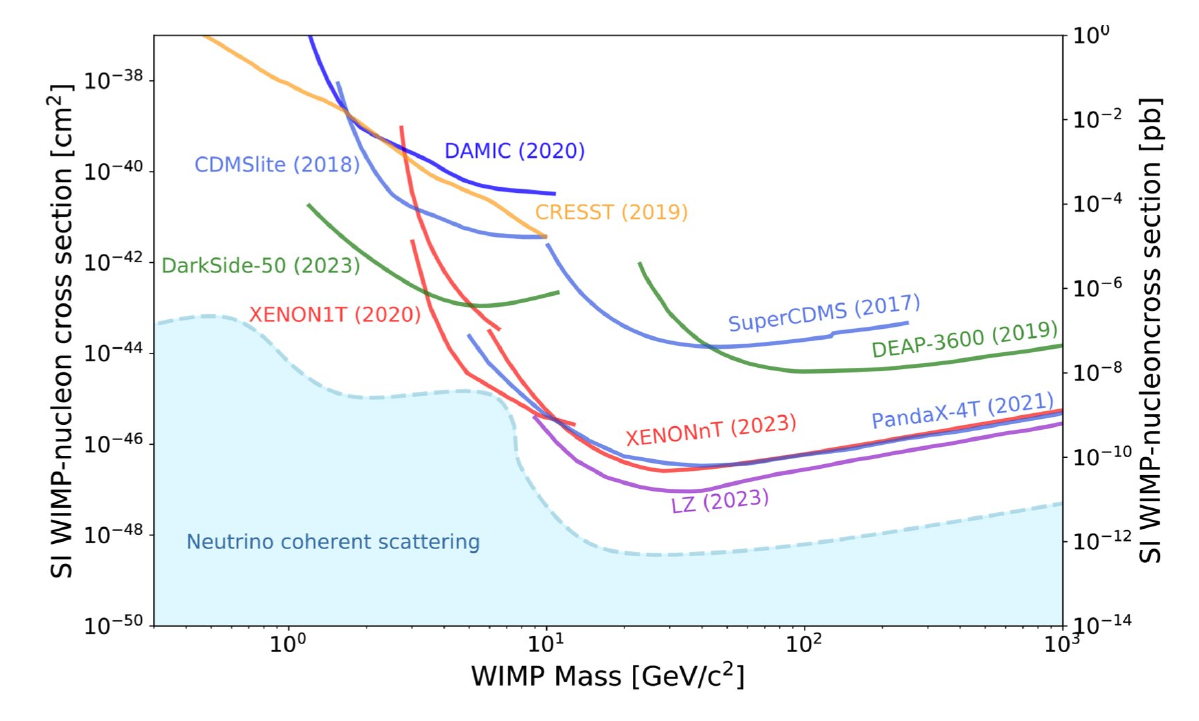} 
    \caption{ Upper limits on the Spin Independent DM -nucleon cross section as a function of DM mass \cite{pdg:2023} .} 
    \label{fig:dmlimits:2023} 
\end{figure}

Strong astronomical and cosmological evidence across many scales, ranging from velocities of stars in ultra-faint galaxies to the Cosmic Microwave Background, indicates that more than 80\% of the matter in our Universe consists of non-baryonic matter known as “Dark Matter” (DM). This evidence has been steadily collected over the last eight decades and suggests that DM consists of one or more new particles not contained within the Standard Model (SM) of Particle Physics. From these observations, we can set some basic requirements for any DM candidate:
(i) it should yield the correct relic density; (ii) should be non-relativistic at matter-radiation equality to form structures in the early Universe in agreement with the observation; (iii) it should be effectively neutral otherwise it would form unobserved stable charged particles; (iv) and last but not least, the dominant part of DM should be cosmological stable with a lifetime much larger than the age of the Universe to be consistent with cosmic rays and gamma-rays observations.
From these requirements, we can infer some of the general properties of DM. The formation of consistent cosmological structures sets a limit on the strength of DM interactions, both with the SM and with itself. DM production in the early Universe may occur through a wide range of channels, but many of these require additional non-gravitational interactions in some form (See \cite{Arcadi:2024ukq}  for an extensive review). Several dark matter scenarios are consistent with these observations. In terms of elementary particles, masses from $10^{-22} $~eV all the way up to the Planck scale are allowed in principle.
Figure \ref{fig:dmlimits:2023}  presents the current dark matter limits obtained for several experiments (ca. 2023).

In what follows, we describe some of the efforts of the Latin American community in the global search for DM. We address direct, indirect, and collider searches for DM, and mention some complementary probes and theoretical efforts, and their alignment with experiments. Cosmological facilities also provide useful information on dark matter, and we refer the reader to the LA participation discussed in the previous chapter.
The LA community has grown and consolidated, actively participating in large-scale and long-term international collaborations. A clear example is the Brazilian DM community, as detailed in the submitted white paper \cite{wp:brazilliandmreport2024}. \footnote{The structure of this section has been adapted directly from that of Ref. \cite{wp:brazilliandmreport2024} , with more detailed summaries of a few key efforts.}

\section{Direct Detection}  \label{sec:dmdirect} 

Data from galaxy rotation curves show the presence of a dark matter halo surrounding galaxies. Due to Earth's relative motion with respect to this stationary dark-matter halo, an incoming flux of dark particles is expected to impinge on our detectors. Underground laboratories were built to observe low-energy nuclear or electron recoils arising from dark matter scatterings. By measuring the energy of the recoiling particles, which is how this energy is deposited, as well as the shape of the scattering rate as a function of deposited energy, one may discriminate dark matter signals from potential background sources through different readout techniques.

Latin American participation in direct-detection experiments has increased over the last five years. We will detail this participation below.

\subsection{The DarkSide experiment}  \label{sec:darkside} 

\begin{figure} 
\centering
    \includegraphics[width=0.6\textwidth]{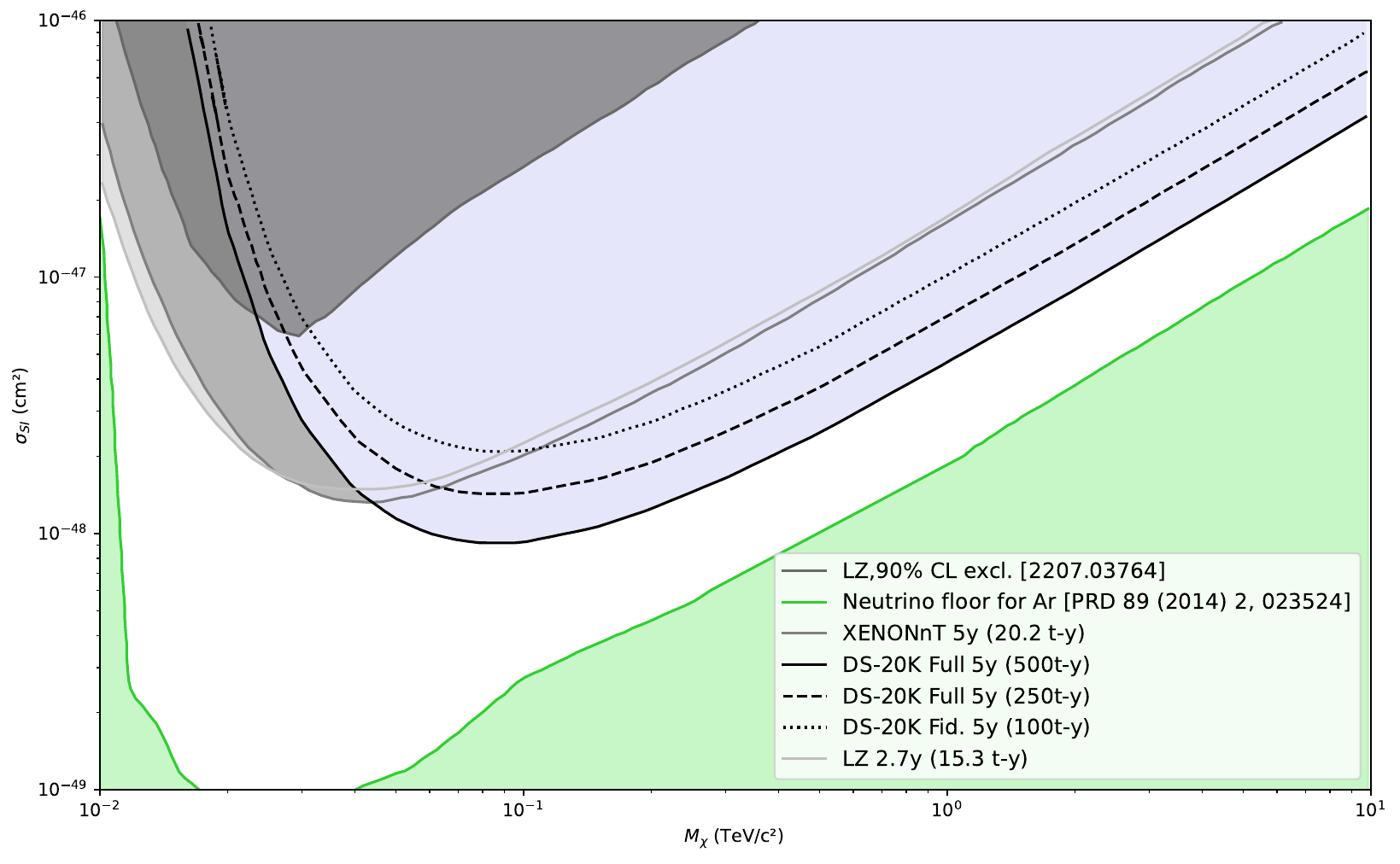} 
    \caption{Projected sensitivity of the DarkSide experiment on the dark matter--nucleon spin-independent scattering cross-section \cite{DarkSide-20k:2017zyg}.} 
    \label{fig:darkside-limits}    
\end{figure}

The DarkSide experiment \cite{DarkSide-20k:2017zyg}  at the Laboratori Nazionali del Gran Sasso (LNGS), employs Liquid Argon (LAr) to search for dark matter particles through scintillation and ionization readout techniques. When dark matter interacts with Argon, a prompt scintillation light ($S_1$ signal) is produced, along with an ionization process. During ionization, electrons drift towards the anode of the time projection chamber, reaching a gas phase of Argon where they emit electroluminescence light ($S_2$ signal). By leveraging the high trigger efficiency of the $S_2$ signal even for low-energy recoils, researchers can use the $S_2$ signal alone to constrain dark matter-electron scattering and establish restrictive bounds in the low dark matter mass regime \cite{DarkSide:2022knj}  (see Fig.~\ref{fig:darkside-limits} ). 
The DarkSide-20k detector, currently under construction, aims to search for WIMP Dark Matter particles that scatter with liquid Argon nuclei in a 20 t fiducial volume. It expects to reach a sensitivity to WIMP-nucleon cross-section of $7.4\times 10^{-48} \textrm{cm} ^2$ for a TeV WIMP mass at the 90\% C.L with a 200 t.yr exposure \cite{Agnes:2023izm}  (see Fig.~\ref{fig:darkside-limits} ).  

The DarkSide group at USP has grown and is now consolidated. As the traditional WIMP search for masses larger than $10$~GeV has excluded a large fraction of its region of interest, the USP group has contributed significantly to light dark matter probes, as well as to constraints in leptophilic dark matter and bounds to the Migdal effect. 

\subsection{The Oscura experiment}  \label{sec:oscura} 

This summary is based on refs. \cite{oscura:2022},  \cite{oscura:2023},  and private communication with collaboration members. \\

Oscura is a next-generation direct dark matter (DM) search experiment designed to probe low-mass DM particles using state-of-the-art Skipper Charge Coupled Devices (Skipper-CCDs)\cite{holland:2003} . It will achieve unprecedented sensitivity to DM-electron interactions in the sub-GeV mass range \cite{essig:2012}, filling a gap in the largely unexplored parameter space of direct searches below 1 GeV. With a projected total exposure of 30 kg-year, it will be the most sensitive experiment in the sub-GeV DM mass range.

\begin{figure}[h!]
    \centering
    \includegraphics[width=0.9\textwidth]{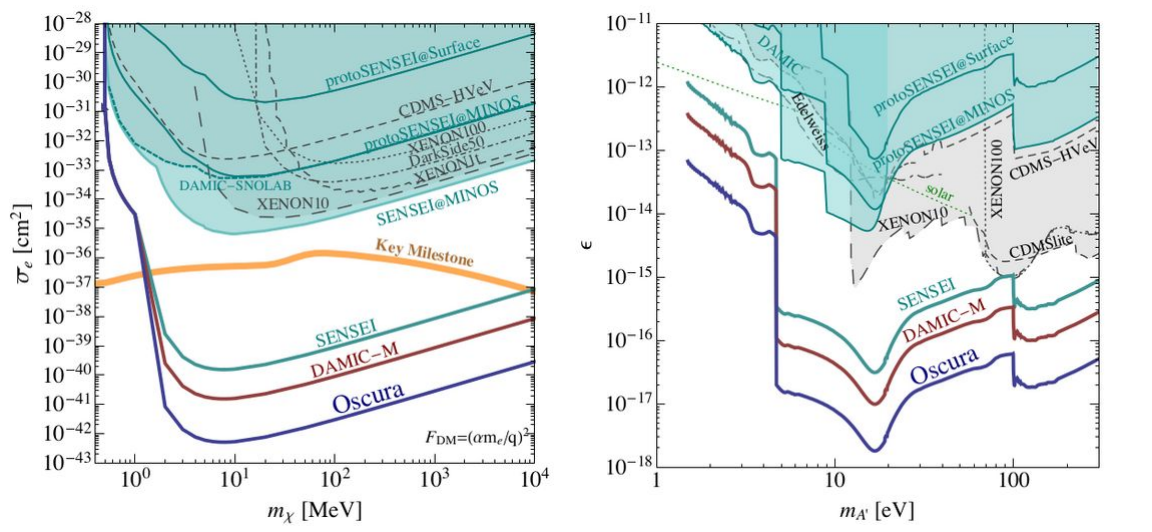} 
    \caption{Projected sensitivity of the Oscura experiment with a 30 kg-yr exposure to dark matter-electron scattering via a light mediator (left) and bosonic dark matter absorption (right) compared to constraints from other experiments \cite{oscura:2022} . } 
    \label{fig:oscura-sens} 
\end{figure}

Oscura builds on the success of pioneering experiments like DAMIC \cite{damic:2019}  and \gls{SENSEI} \cite{adari:2025} , which demonstrated the feasibility of using Skipper-CCDs for direct detection of sub-GeV DM. With its much larger target mass of 10 kg, Oscura will increase this sensitivity significantly studying a variety of processes including:
\textit{ i) DM-electron scattering:}  Oscura will search for signals of dark matter scattering off electrons, depositing enough energy to ionize one or more electrons in the detector, reaching sensitivity to DM particle masses as low as 500 keV,
\textit{ ii) Absorption of bosonic dark matter:}  The experiment will probe bosonic dark matter candidates, such as dark photons \cite{bloch:2017} , with masses as low as 1 eV.
\textit{ iii) Migdal effect:}  Oscura will also be sensitive to nuclear recoils, leveraging the Migdal effect \cite{ibe:2018}  to detect ionization electrons produced by dark matter-nucleus interactions.
Figure \ref{fig:oscura-sens}  shows the projected sensitivity of Oscura  to DM-electron scattering and bosonic dark matter absorption \cite{oscura:2022} , comparing it with other experiments using Skipper CCDs such as \gls{SENSEI} \cite{crisler:2020}  and DAMIC-M \cite{amquist:2023} .

\begin{figure} [h!]
    \centering
    \includegraphics[width=0.85\textwidth]{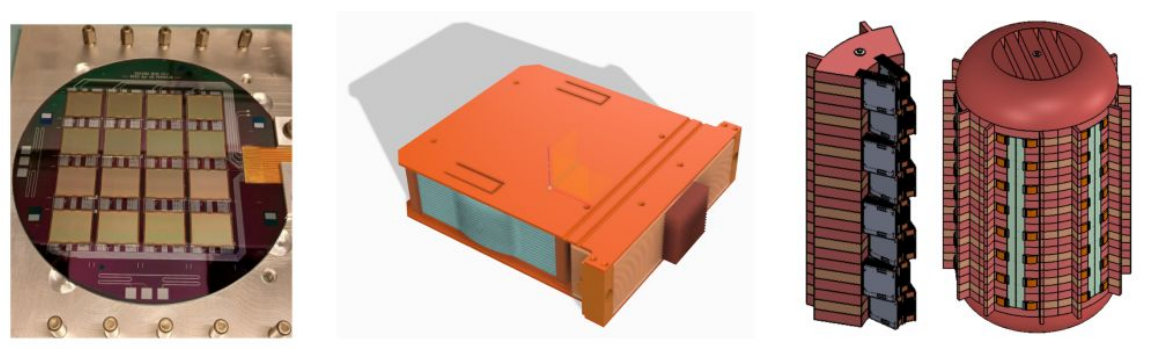} 
    \caption{Design for a Oscura Multi-Chip Module (left) Super Module (center) and full detector array with about 20,500 Skipper-CCDs (right) \cite{oscura:2023} .} 
    \label{fig:oscura-det-design} 
\end{figure}

The proposed active mass of 10 kg in Skipper CCDs with standard pixel size of 15~$\mu$m $\times$ 15~$\mu$m and 725~$\mu$m substrate thickness corresponds to 26 gigapixels. To reach its scientific goals Oscura requires to achieve a background rate of less than 1 event between 2e- and 10e- over the full 30 kg-yr exposure (equivalent to 0.01 events/keV/kg/day), as well as a dark count rate of $10^{-6} $ e-/pixel/day. The sensors must cooled down to a temperature between 120 and 140 K, which will be achieved by submerging the full detector array in a Liquid Nitrogen (LN2) bath at a vapor pressure of 450 psi. 
The Oscura design is based on 1.35 Mpix sensors (1278 × 1058 pixels) packaged on a compact Multi-Chip Module (MCM), see figure 2 (left). Each MCM consists of 16 sensors epoxied to a 150 mm-diameter silicon wafer fabricated by {\it Microchip Technology, Inc.}, with traces connecting the sensors to a low-radiation background flex cable. MCMs will be integrated into Super Modules (SMs), with each SM containing 16 MCMs and using a custom, ultrapure, electro-deposited copper support and shielding structure. Oscura requires roughly 80 SMs containing a total of 1280 MCMs (see Figure \ref{fig:oscura-det-design} ).

{\bf Timeline and costs.}  The following assumes funding for the project constrution acquired in 2025. Delays should shift the schedule accordingly.
{\it Phase 1: Detector Development and Testing (2024-2026)} .
The first stage of Oscura involves completing the design and testing of prototype Skipper-CCDs. During this phase, already underway, the main areas of focus are:
{\it 1)}  Testing the charge resolution and noise characteristics of the detectors.
{\it 2)}  Developing low-background electronics and packaging solutions.
{\it 3)}  Integrating the detectors into a liquid nitrogen (LN2) cryogenic system for operation at 120-140K.
By 2026, a working prototype will be operational at SNOLAB, providing the first science data with a subset of the detector array.
{\it Phase 2: Full Detector Construction and Installation (2026-2028)} .
The full 10 kg detector array will be constructed over a two-year period, consisting of 28 gigapixels spread across Multi-Chip Modules (MCMs). Each module will be supported by copper shielding to reduce background radiation. The LN2 cryogenic system will be scaled up to accommodate the entire detector, and readout electronics will be optimized to minimize noise and maximize data throughput.
{\it Phase 3: Science Data Collection and Analysis (2028-2030).} 
Following the detector's full integration, Oscura will enter its primary data-taking phase. The goal is to collect up to 30 kg-year of exposure by 2030, with sensitivity to dark matter interactions down to 500 keV. Data will be continuously analyzed to refine background estimates and ensure that the experimental thresholds are met.
Construction of a 10 kg Skipper-CCD detector is estimated around 7.7 million US dollars, where 78\% is for the production of the CCD wafers and the rest is distributed between packaging, electronics, cryogenics, vessel, shielding, and other expenses. The detector will operate at SNOLAB.

{\bf Latin American involvement.}  

Latin American institutions have played a vital role in the development of Skipper-CCD technology. To enhance this collaboration further, several initiatives have been put forward:
{\it Expanding Regional Expertise.} 
Institutions such as Centro Atómico Bariloche (Argentina) and Universidad Nacional Autónoma de México (UNAM), have been key players in Oscura. Expanding collaboration with other regional research centers can provide access to additional resources and expertise. This will strengthen the region's ability to lead cutting-edge experiments.
{\it Training and Educational Programs} .
A critical component of strengthening Latin American participation is developing educational programs focused on Skipper-CCD technology and dark matter research. These programs could include workshops and hands-on training at universities like UFRJ in Brazil, fostering a new generation of scientists.
{\it Funding and Support} .
Securing long-term funding for Latin American participants is essential. National science organizations such as CONICET (Argentina) and CNPq (Brazil) should be approached to support regional involvement in Oscura. Strengthening ties with international partners, including Fermilab, can also provide financial and technical support for participants from Latin America.
{\it Outreach and Public Engagement} .
Public outreach programs in Latin America will raise awareness about dark matter research and highlight the critical role of regional scientists in global collaborations. This will inspire future generations of scientists and broaden the public's understanding of the importance of experiments like Oscura.

In conclusion the Oscura experiment represents a major leap forward in the search for low-mass dark matter. By leveraging the groundbreaking capabilities of Skipper-CCD technology, Oscura will open up new avenues for discovery in particle physics. 
Latin American institutions and scientists will continue to play an essential role in this project, contributing both technically and scientifically. By enhancing collaboration and fostering educational initiatives, we can ensure a strong Latin American presence in Oscura and future dark matter experiments.

\subsection{The CYGNO experiment}  \label{sec:cygno} 

Conventional direct detection experiments will be challenged by the neutrino floor, and new avenues need to be explored, such as directionality \cite{Amaro:2024prf}. The CYGNO project plans to develop a high-precision optical Time Projection Chamber (TPC) for directional dark matter search and solar neutrino spectroscopy to be hosted at the LNGS. CYGNO's distinctive features include using scientific CMOS cameras and photomultiplier tubes coupled with a Gas Electron Multiplier for amplification within a helium-fluorine-based gas mixture at atmospheric pressure. The primary objective of this project is to achieve three-dimensional tracking with head-tail capability and to enhance background rejection down to the keV energy range. It is expected to yield competitive limits on the spin-independent dark matter-nucleon scattering cross-section for sub-GeV WIMPs and competing bounds on the spin-dependent scattering for GeV masses. Brazilian involvement in CYGNO represents a new addition compared to the previous LASF4RI report.

\subsection{LAMBDA: A World-Class Particle Physics Laboratory
in South America} \label{sec:lambda} 

This summary is based on ref. \cite{wp:lambda} . \\

The {\it Laboratorio Argentino de Mediciones de Bajo umbral de Detección y sus Aplicaciones}  (LAMBDA) was created to investigate and develop the capabilities of Skipper-CCD technology \cite{dm:tiffenberg:2017}  for dark matter searches, neutrino physics experiments, and searches for physics BSM. It is the result of the strong collaboration between the Physics Department of the University of Buenos Aires (UBA) and Fermilab in the USA.
Skipper-CCDs are pixelated silicon detectors with a spatial resolution of 15 $\mu$m and the capability of measuring the deposited charge per pixel with sub-electron readout noise, which permits the counting of individual electrons hence reaching the lowest possible detection energy threshold of an ionization detector.
The following experiments using Skipper-CCDs are LAMBDA users:
SENSEI: running at SNOLAB \cite{adari:2025} , uses Skipper CCDs to search for dark matter with masses from MeV to tens of GeV. \gls{CONNIE}: A Skipper CCD experiment 30 m away from the Angra-II reactor in Brazil, looking for CE$\nu$NS interactions of the neutrinos emitted by the nuclear reactor, to be discussed in the next chapter. ATUCHA-II \cite{depaoli:2024}: A Skipper CCD experiment 12 m away from the core of the 2 GW Atucha II reactor in Argentina, looking for CE$\nu$NS and exotic interactions. OSCURA (\autoref{sec:oscura}): A future experiment with 10 kg of Skipper-CCDs, aimed at searching for light dark matter.
LAMBDA users performed the most precise measurements to date of the energy required to create electron-hole pairs and the Fano factor in silicon at energies below 1 keV \cite{rodriguez:2021} .
Under the umbrella of the LAA-HECAP a fruitful exchange between the \gls{CONNIE} and Atucha-II collaborations took place at the LAMBDA laboratory in 2023 that resulted in a joint analysis searching for mCPs with data from the two experiments, producing world-leading limits on mCPs produced at nuclear reactors, complementary to those imposed by \gls{SENSEI} \cite{barak:2024} .
It should be noted that the installation of skipper-CCD detector at Atucha-II was mainly driven by LAMBDA members.
Skipper-CCDs also have the potential to exploit the quantum nature of light, using quantum entanglement of photons as a tool to search for dark photons. In a pioneering
work, LAMBDA members have demonstrated the experimental challenges required to conduct a competitive experiment using correlated photons and single-photon-counting technologies \cite{estrada:2021}.
Other studies related to quantum imaging and interferometry with low photon numbers have been performed at LAMBDA \cite{pears:2023, pears:2024}.

{\bf Latin American involvement.} 
The following institutions are currently participating or interested in LAMBDA: from Argentina Universidad de Buenos Aires, Comisión Nacional de Energía Atómica, Universidad Nacional de San Martin, Laboratorio de Óptica Cuántica, DEILAP, UNIDEF, Instituto de Investigaciones en Ingeniería Eléctrica. From Brazil, Universidade Federal de Rio de Janeiro is involved. 

{\bf Infrastructure investment.} 
The laboratory was innaugurated in July 2022 in a 100 m$^2$ area in the Physics Department at UBA, with an equipment investment of approximately 74,000 USD.

\subsection{Cryogenic Rare
Event Search with Superconducting Thermometers (CRESST)}  \label{sec:qsltd} 

This summary is based on ref. \cite{wp:qsltd} . \\

Dark matter direct searches in the sub-GeV range have gained significantly attention in the last decade. Several experiments employing different techniques are in operation or under development. Among the most promising ones are those with cryogenic calorimeters operated at milli-Kelvin temperatures, which have demonstrated unprecedented efficiency and sensitivity. Using quantum sensors known as transition edge sensors (TES), the CRESST experiment is leading direct dark matter searches in the sub-GeV range.
Milli-Kelvin calorimeters operated at 10-30 mK are excellent particle detectors for rare event searches \cite{fiorini:1984} . 
A particle crossing a crystal deposits part or the totality of its energy, producing an increase in temperature of the order of $\Delta T \propto \Delta E/C$, where $\Delta E$ is the amount of thermal energy deposited and $C$ is the thermal capacitance of the crystal.
In general, the thermal capacity of a crystal at low-temperatures given by $C \propto T^3$ . At room temperature, this increase in temperature is negligible, but for crystals operated at 10 mK, particles of < 3 MeV produce an increase of $\sim\mu$K; which is measurable.
Operating these detectors requires a stable temperature of {\cal O} (mK) for long periods, high power to cool down large samples, ultra-low vibration systems, and super-sesitive quantum sensors to measure the small variations of temperatures corresponding to energy depositions of < 100 eV. 
Transition Edge Sensors (TES) and Neutron Transmutation Doped Sensors (NTDS) are examples of ultra-sensitivs sensors that can measure $\mu$K variations at a base temperature of mK. 

The Quantum Sensors and Low-Temperature Detectors (QSLTD) group at the University of São Paulo is specialized in using Germanium NTDs (Ge-NTD) and Tungsten TES (W-TES). Transition Edge Sensors explore superconducting phase transition to measure a temperature variation. At the critical temperature $T_c$ , the resistance of the TES shows a sharp response to the temperature, which allows a precise measurement of the energy of the particle. The $T_c$ of these sensors depend strongly on the parameters of their deposition. The CRESST group at Max Planck Institute for Physics, led by Dr. F. Petricca, has a tradition of almost 30 years in the development and operation of such sensors. W-TES works with voltages of mV and currents of nA, which requires quantum amplifiers like SQUIDs to read the signal with low noise. With these sensors, the CRESST collaboration was able to produce detectors with a threshold of 6.7 eV \cite{cresst:2024}  and single-photon counting. As a disadvantage, their dynamic range is small, not being suitable for energies > MeV.
The Quantum Sensors and Low-Temperature Detector (QSLTD) Group at the University of São Paulo aims to
develop novel cryogenic calorimeters that can advance the state-of-art of particle physics. In particular, the QSLTD group will be working on the CRESST experiment, developing new strategies to probe or exclude light dark matter in the sub-GeV/c 2 range through elastic dark-matter-nucleus scattering.

WIMPs represent one of the most promising theoretical models to explain dark matter together with axions, but no direct evidence has yet been found \cite{goodman:1985, wasserman:1986, lee:1977, peccei:1977} . Only indirect cosmological evidence has been observed \cite{kochanek:2003, allen:2002} . In the range > 1 GeV/c$^2$ , experiments like DarkSide and Xenon1T have established stringent limits on spin-independent dark matter interactions (cross-section) of $<10^{-43} $ cm$^2$ \cite{agnes:2018, agnes:2018:2, xenon:2023} . Since nothing has yet been found, the search for light WIMPs (< 1 GeV/c 2 ) has become a hot topic, where CRESST was the first and still the best experiment to search for them, together with others \cite{angloher:2022, rothe:2018, cdms:2010} . Due to low threshold, particle discrimination, high efficiency, and high resolution, CRESST-III was able to set exclusion limits $>10^{-5} $ pb for spin-dependent and independent dark matter interactions of 0.1 GeV/c 2. CRESST-III uses small calorimeters of a a few tens of grams in coincidence with light detectors operated with transition edge sensors (TES). The goal is to measure the elastic scattering nucleus-dark matter. Different target materials have been tested, like CaWO$_4$ , sapphire (Al$_2$O$_ 3$) and diamond. A lot of effort has been put by CRESST experiment to reduce the detector threshold by optimizing TES, which allows to search for lighter dark matter. This led to the observation of low-energy excess, which is one of the challenges, \cite{angloher:2023} . 
%
%
%

The goal of the QSLTD group is to test and produce new detectors that can improve the capability of de-
tection of dark matter in CRESST. The group expects to have the first part of the laboratory fully commissioned in the first two years (two cryostats with DAQ installed and tested, and a chemistry lab). The anticipated costs are around $\sim 4$ Mi-Euro.
In the first two years, the QSLTD Group at Universidade de São Paulo should have 3 full-time post-doctoral researcher working on that, plus 6 Ph.D/Master researchers and one faculty. From the Universidade de Campinas the expectation is to have one post-doctoral researchers plus 2 Ph.D./Master researcher and one permamnent professor	
The QSLTD group at Universidade de São Paulo was born from an international collaboration from researchers
of different institutions around the world. In particular, they will maintain a strong and close relation with the Max-Planck-Institut für Physik (MPP) at Munich, Germany, led by Dr. Federica Petricca. MPP is one of the researcher centers leading the direct dark matter searches in the light dark matter range (with CRESST) and in the axion range (with MADMAX). They expect to have not only a marginal contribution in the searches of dark matter and neutrinoless double beta decay, but to be comparable to the best existing centers.

\subsection{PICO, DEAP and BULLKID-DM}  \label{sec:qsltd} 

A group at Instituto de Física, UNAM in Mexico participates in dark matter detection experiments
hosted at underground laboratories such as SNOLAB and Gran Sasso. It contributes to the PICO \cite{pico:2025} , DEAP \cite{deap:2023}  and BULLKID-DM experiments, focusing on detailed Monte Carlo simulations to model detector response and background sources, which are essential for data interpretation and signal optimization. The group leads the physics interpretation and phenomenology of experimental results in the PICO experiment using non-relativistic effective field theory (NREFT), studying photon-mediated interactions, inelastic dark matter scenarios, fermionic dark matter absorption, among others.

\section{Indirect Detection} \label{sec:dmindirect}

Indirect dark matter detection relies on observing the flux of stable particles such as electrons, protons, neutrinos, and gamma rays produced by dark matter annihilation or decay in dense astrophysical environments like dwarf spheroidal galaxies and the Galactic Center. For instance, the flux of gamma rays from dark matter annihilation is proportional to the dark matter annihilation cross-section, the energy spectrum (photons produced per annihilation), and the dark matter density along the line of sight, while being inversely proportional to the dark matter mass. Knowing the dark matter density, one can estimate the gamma-ray flux from dark matter annihilation into a given final state. 
In the past, Latin American involvement in indirect dark matter searches has been represented by Brazilian participation in the AMS-02 and H.E.S.S. telescopes. However, the Brazilian community has shifted its efforts to CTAO and the Southern Wide-field Gamma-ray Observatory (SWGO), which are the main gamma-ray instruments in the next decades.
It is worth mentioning that Brazil, Chile and Mexico have participating institutions in the CTAO consortium. In fact, the current chair of the consortium board is from a Brazilian institution. 
The HAWK observatory has conducted several searches for dark matter candidates in collaboration with several Mexican institutions. Gamma-ray observatories were already described in Chapter 2. Here, we concentrate on their dark matter aspects.

\subsection{Dark Matter with the CTAO}  \label{sec:ctaodm} 

The CTAO will be the main observatory for high-energy gamma rays in the near future. The broad scientific agenda of CTAO includes dark matter as a key component \cite{CherenkovTelescopeArray:2024osy}. With instrumental improvement, a wider field of view, and a large energy window covering gamma-rays from $20$~GeV to $300$~TeV, CTAO will surpass its predecessors in many ways, and it will be sensitive to dark matter interactions more than one order of magnitude better than current instruments. We highlight that the dark matter annihilation cross-section ($\sigma v = 3\times 10^{-26} \, {\rm cm} ^3{\rm s} ^{-1} $) within reach of CTAO is natural in several dark matter models \cite{Balazs:2017hxh} . The observatory will operate arrays at sites in both the Northern and Southern hemispheres to provide full-sky coverage and maximize the discovery potential for the rarest phenomena. In summary, CTAO has the potential to detect dark matter through the observation of high-energy gamma rays, and for this reason, it is one of the flagship experiments. It is worth noting that Brazilian researchers have been actively working on dark matter research within the scope of the CTAO.

\subsection{Dark Matter with the SWGO}  \label{sec:sugodm} 

This summary is based on ref. \cite{wp:swgo} . \\

Searching for dark matter in the Galactic Halo is one of the key science goals of the proposed South Wide field Gamma-ray Observatory (SWGO). It plans to accomplish this with a densely populated core with ~4,000 detection units, which will drive the low-energy performance of the observatory, surrounded by a sparser outer array of the order of 1,000 detection units and reaching a coverage area of 1 km$^2$ to reach PeV detection capabilities.
\gls{SWGO} is fundamentally different from Imaging Atmospheric Cherenkov Telescopes (IACTs) such as H.E.S.S. and CTAO. IACTs observe the Cherenkov radiation produced in the atmosphere by particle cascades initiated by incident high-energy gamma rays, while \gls{SWGO} is planned to be an Extensive Air Shower (EAS) array, able to collect data directly from particles in the shower that reaches the Earth’s surface continuously, even during the day.

Thermal WIMPs with masses between $\sim 2$ TeV and $\sim 100$ TeV remain still viable dark matter candidates \cite{smirnov:2019} . Such heavy dark matter can only be probed by astrophysical experiments similar to SWGO, by looking for high energy gammas from its annihilation or decay. \gls{SWGO} will be most sensitive to nearby extended sources of gamma rays from dark matter interactions in the inner Galactic halo, and will overlap in the region between 1 and 10 TeV with the sensitivity of the Cherenkov Telescope Array Observatory (CTAO) \cite{ctao} , as well as with the energy range of future observatories. 
Because of its wide field of view and southern latitude, \gls{SWGO} will rutinely monitor a large fraction of the entire sky, making it capable of constraining widly varying models of the dark matter distribution. As a consequence, \gls{SWGO} will have no need to optimize its observing strategy for different dark matter profiles, as opposed to pointing instruments.
\gls{SWGO} will be able to test numerous new ultra-faint dwarf spheroidal galaxies that are promising targets for the production of VHE gamma rays by dark matter annihilation, that have been recently identified in deep surveys, mostly in the Southern hemisphere (see e.g. \cite{bechtol:2015} ). The observatory will also test models in which dark matter trapped in the Sun annihilates into a metastable mediator that decays to gamma rays outside of the Sun (see eg. \cite{batell:2010,leane:2017} ).

\begin{figure} 
    \centering
    \includegraphics[width=0.4\textwidth]{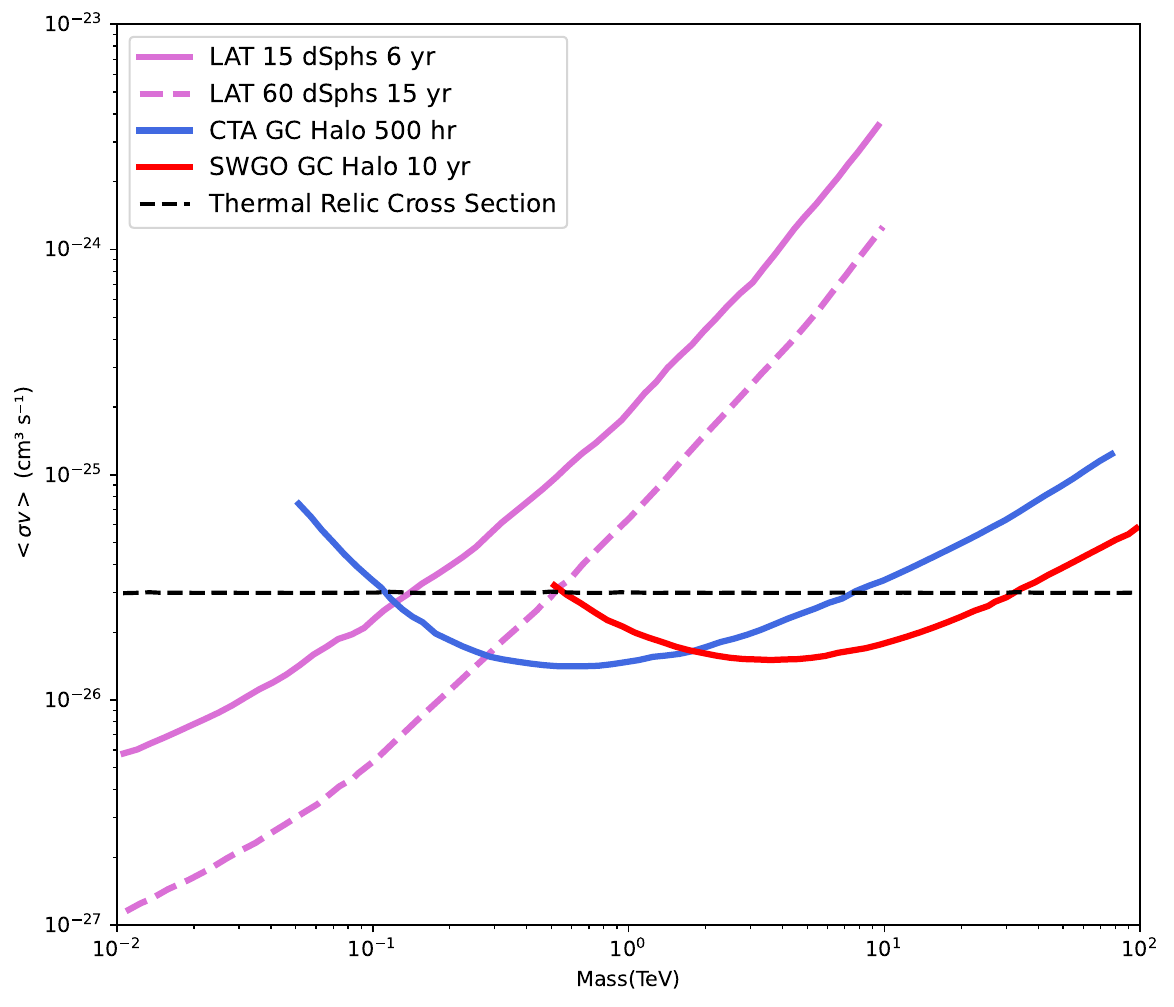} 
    \includegraphics[width=0.4\textwidth]{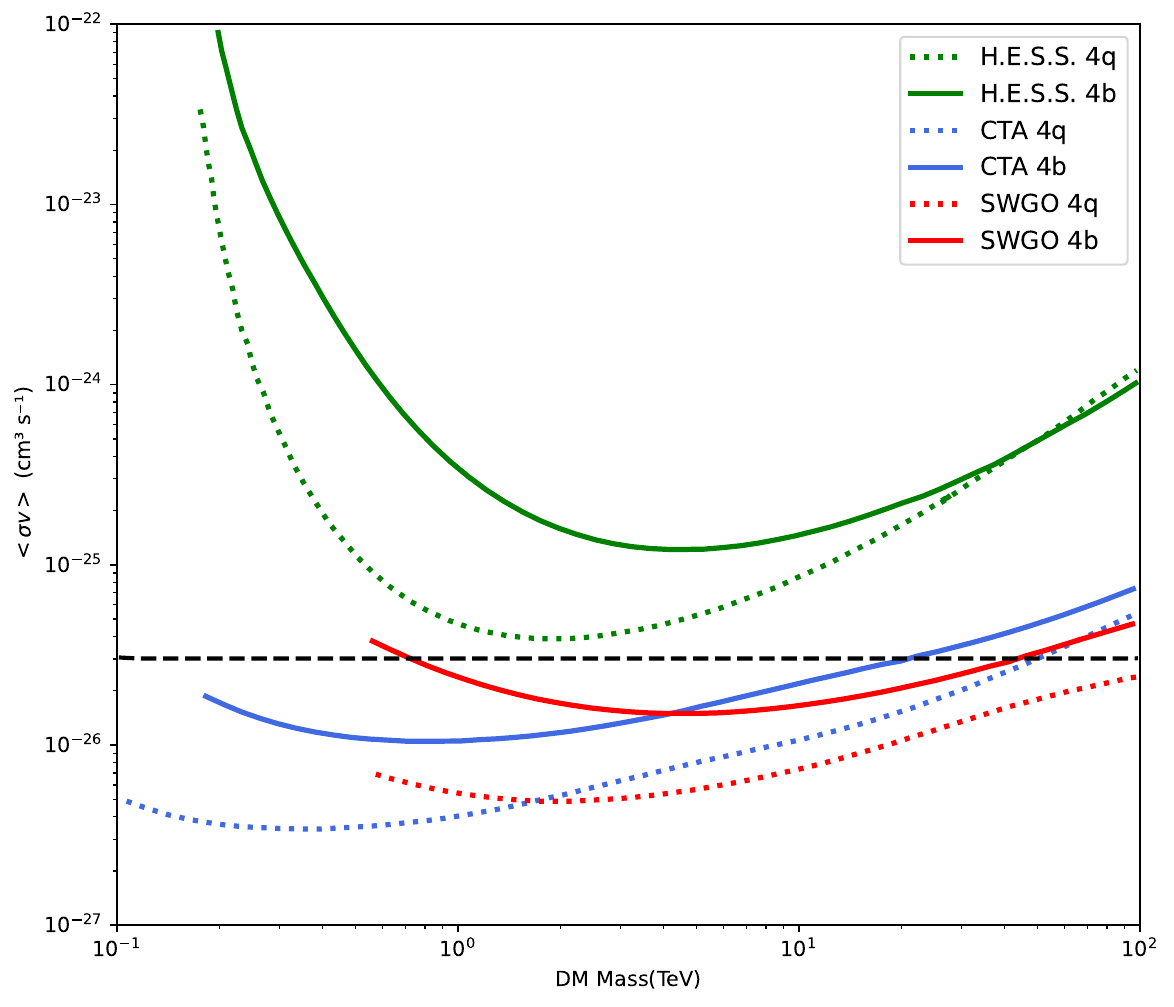}
    \caption{Projected limits on the dark matter annihilation into $b\bar{b} $ (left-panel) \cite{Viana:2019ucn}  and into dark vector boson that eventually decay into quarks (right-panel) \cite{NFortes:2022dkj}. Works from Brazilian groups.} 
    \label{fig:swgo-limits} 
\end{figure}

The left panel of Fig.\ref{fig:swgo-limits} shows the CTAO and \gls{SWGO} sensitivity to dark matter annihilation into quarks \cite{Viana:2019ucn}, and in the right panel, their sensitivity to secluded dark matter \cite{NFortes:2022dkj}, where dark matter annihilates into dark vector bosons which later decay into quarks. It is visible from the plots that there is an interesting complementarity between the CTAO and SGWO telescopes in the TeV mass range. 

Latin American participation in \gls{SWGO} includes 7 institutions from Argentina, 8 from Brazil, 11 from Chile, 14 from Mexico, and 7 institutions from Peru.

\section{Dark matter at colliders and accelerators}  \label{sec:dmcollider} 

The Large Hadron Collider (LHC) probes the fundamental interactions through proton-proton collisions at high energies. Photons, charged and strongly interacting particles, produce signals at the detectors and are used as triggers to search for dark matter particles. As dark matter particles interact weakly with matter and leave no trace at the detectors, their presence is inferred from energy-momentum reconstructions. Generally speaking, at colliders, missing-momentum searches are dark matter searches, but there is also an effort to reinterpret other searches in terms of an extended dark sector, in which DM would be just a component. In this extended panorama, particle colliders play a crucial role in the search for dark sectors.

At colliders, dark-sector searches typically involve the pair production of dark-matter particles, which yields a distinctive signature of missing transverse momentum. The community has developed simplified benchmark models to guide these searches~\cite{Abercrombie:2015wmb}, providing experimental targets and guidelines for comparing collider findings with those from direct and indirect detection experiments. These benchmark models typically include a dark matter candidate and a mediator particle, which may also be a particle beyond the Standard Model. The ability of colliders to explore the properties of mediator particles and their interactions enhances our potential to uncover new physics, underscoring the importance of collider experiments in the broader search for dark sectors.

The design, deployment, and monitoring of the trigger algorithms is a fundamental to the LHC experiments due to the high instantaneous luminosity delivered by the accelerator. Dark sectors pose challenges to standard trigger and data-acquisition setups, which in turn require alternative strategies from the experiments~\cite{CMS:2024zhe}.

\subsection{Dark Matter searches at the LHC}  \label{sec:lhcdm} 

The following describes efforts to search for DM in the LCH experiments, with participation from Latin American groups, particularly Brazilian institutions.

\paragraph{ATLAS.} 

Designed to be a general-purpose experiment, it has been capable of probing several dark sectors (see \cite{wp:brazilliandmreport2024}  for references). Machine learning (ML) methods have improved the event background rejection. In particular, the Neural-Ringer algorithm (NR) \cite{DaFonsecaPinto:2777971}, developed by Brazilian researchers, is currently used in the community to probe vector dark matter particles.

\paragraph{CMS.} 

 It has a broad physics program which also features dedicated searches for dark sectors (see \cite{wp:brazilliandmreport2024}  for references). In addition to searches for transverse momentum imbalances, the \gls{CMS} collaboration also reinterprets many of its other analyses as searches for the dark sector. Those include measurements of the Higgs boson invisible decay width, which can constrain a posited H\,$\to$\,DM decays; searches for dijet resonances, reinterpreted as dark-sector mediators; and searches for long-lived particles, which are present in many models that lead to DM candidates. A comprehensive report on searches for the dark sector with the \gls{CMS} experiment is available in Ref.~\cite{CMS:2024zqs} .

\paragraph{LHCb.} 

 It is experiment dedicated to b-physics with the capability to search for dark sectors in the low mass regime. In general, \gls{LHCb} targets low mass and short-lived particles because of its relatively low energy threshold and integrated luminosity compared to the general purpose detectors \gls{CMS} and ATLAS. In 2018, \gls{LHCb} implemented electron identification in the first high-level trigger which permitted the search for dark particles decaying into $e^+e^-$ \cite{Craik:2022riw}. In the next LHC run, \gls{LHCb} will continue the search for dark matter covering dark mediators masses ranging from  $10^{-2} $~GeV to $50$~GeV.

\subsection{Dark Matter searches in future colliders}  \label{sec:futurecollidersdm}

Naturally, groups involved in the LHC detectors are also participating in detector upgrades needed for the High Luminosity-LHC (HL-LHC) that is planned to achieve an integrated luminosity of $\mathcal{L} =3$~ab$^{-1} $ operating with a center-of-mass energy of $14$~TeV. The \gls{HL-LHC} is a future accelerator that has been approved and it will be a reality in the upcoming years. In the process of building an $100$~TeV future accelerator as the Future Circular Collider (FCC), a Large Hadron electron Collider (LHeC) could be constructed offering a complementary probe to dark sectors \cite{Huang:2022ceu} . In the context of the FCC, there are two Brazilian groups (CBPF anf UFPel) with signed a MoU (Memorandum of Understanding) who are actively participating in the \gls{FCC} feasibility study \cite{RebelloTeles:2023uig} . 
The Latin American community, with a relevant contribution from Brazil,  plays an important role in developing collider physics and is involved in science projects crucial to the search for dark matter.

\subsection{Dark Matter searches at other facilities}\label{sec:acceleratorsdm}

Beyond traditional particle colliders, other high-intensity facilities can contribute to the search for dark sectors. A proposal to search for dark matter at the Brazilian synchrotron light source has been put forth \cite{Duarte:2022feb}. The European Spallation Source (ESS), a next-generation neutron source based on a high-power proton accelerator, also offers new opportunities to explore physics beyond the Standard Model.\\

\paragraph{SIRIUS.} 
A proposal to search for dark matter at the Brazilian synchrotron light source has been put forth \cite{Duarte:2022feb}. An idea to repurpose UVX, the predecessor of SIRIUS, showed that a 1-3 GeV positron beam impinging on a target, followed by a missing mass spectrum event reconstruction, could cover an unexplored region of parameter space of the dark photon model. A similar proposal has been done for axion-like particles \cite{Angel:2023exb}. New experimental strategies to search for dark sectors are being explored without interfering with SIRIUS's original purpose. 

\paragraph{HIBEAM-NNBAR at ESS.} 

The ESS-driven experiment HIBEAM/NNBAR is designed to investigate baryon-number violation by searching for neutron conversions into sterile states or antineutrons. These processes could be mediated by hypothetical bosons associated with the dark sector and baryogenesis, providing an alternative pathway to uncover new interactions and possible connections to dark matter. Latin American participation is represented by the Brazilian groups in the HIBEAM-NNBAR experiment \cite{Santoro:2024lvc}, focused on the development of the tracking detector and associated simulations, with recent work devoted to the calorimeter fast-simulation and optimization of the Time Projection Chamber configuration, and simulations of the readout electronics.

\paragraph{Cohereht CAPTAIN-Mills (CCM).} 
It is a 10 ton liquid argon (LAr) scintillation detector located 26~m away from the production target of the Lujan spallation facility in the Los Alamos National Laboratory in the USA. CCM has performed sensitive searches for sub-GeV and leptophobic dark matter. The upgraded CCM200 detector expects to collect about 1.7$\times10^{22} $~POT by the end of 2025 to explore uncharted territory in the parameter space of various models, including axion-like particles and some that would explain the anomalies reported in short baseline neutrino experiments (see \cite{ccm:2024}  and references therein). CCM will also perform crucial measurements of neutrino-argon cross sections in support of the broader USA neutrino physics program. Mexico is participating in $\nu_e-Ar$ CC cross-section measurements and multi-purpose calibration efforts.

\section{Complementary probes for DM}  \label{sec:dmcomplementary} 

There are complementary probes that do not fall neatly into direct, indirect, or collider searches for dark matter. Nevertheless, they can probe various properties of dark matter. They include facilities in neutrino physics and cosmology, and therefore, there is an overlap with the corresponding Chapters in this Briefing Book. They will be briefly discussed below.

\subsection{Probes from Neutrino Physics}  \label{sec:neutrinosdm}

\paragraph{CONNIE. } 

\begin{figure}[h!] 
    \centering
    \includegraphics[width=0.95\linewidth]{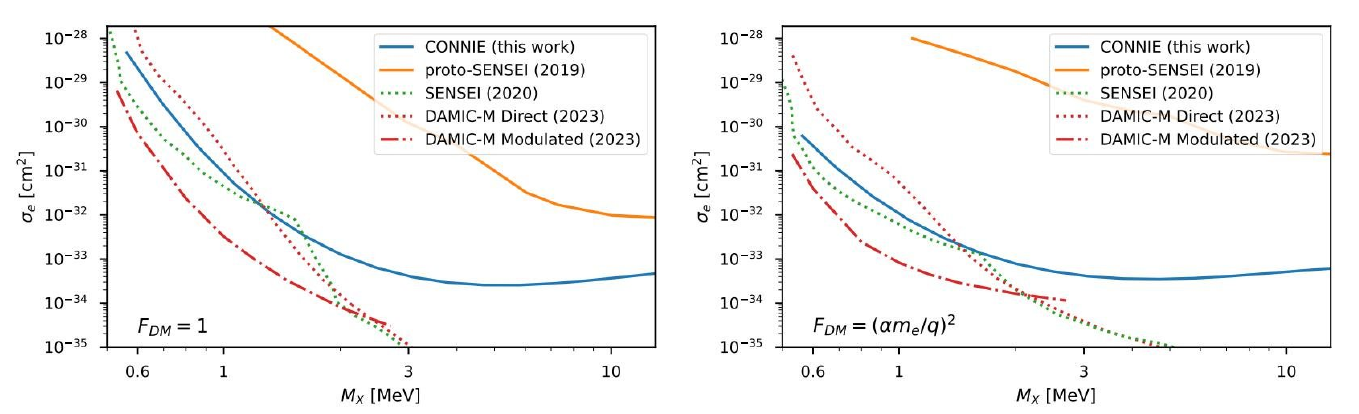} 
    \caption{\gls{CONNIE} limits on DM-e interactions mediated by a heavy dark photon (left) and an ultralight dark photon (right), compared to other experiments. See \cite{connie:2024b}  for details.}  
    \label{fig:connie-diurnaldm} 
\end{figure}

Described in detail in \autoref{sec:lambda}, CONNIE was the first experiment to use silicon charge-coupled devices (CCD) at a nuclear reactor to look for coherent elastic neutrino-nucleus scattering (CE$\nu$NS) and to impose competitive constraints on BSM physics related to dark sectors.  In addition to setting world-leading limits on milli-charged particles from its first run with Skipper CCDs, it set competitive constraints, for a surface experiment, on DM-electron interactions by searching for a diurnal modulation of its single-electron rate \cite{connie:2024b} , shown in Fig. \ref{fig:connie-diurnaldm} .
After the completion of a recently commissioned upgrade to Oscura-design Multi-Chip-Modules (\autoref{sec:oscura}), and with the use of newly available multiplexing electronics, \gls{CONNIE} can host several such modules, therefore, we expect improved constraints on dark sectors in the future.
It is important to re-emphasize here that \gls{CONNIE} is a largely Latin American effort in terms of overall participants, with key contributions from Fermilab. 

\paragraph{SBND. } 

The Short-Baseline Neutrino Detector presents an exciting opportunity to probe dark sectors. Light mediators that mix with photons can be produced in neutrino beams through meson decays and subsequently detected. Although similar to CONNIE, these neutrino detectors do not directly probe dark matter particles. They can provide complementary and crucial information on viable dark-matter interactions within a dark sector and on other new-physics signals. The signals from the dark sector are similar to those produced by neutrino coherent scattering, making neutrinos the primary background. However, spectral information and triggers for highly off-axis beams can help manage the neutrino background, thereby enhancing the ability to detect dark-sector interactions. This highlights the importance of neutrinos as a probe of dark sectors, offering a unique and valuable perspective in the quest to understand dark matter and its interactions \cite{Acevedo:2024wmx}. 

\paragraph{DUNE} 

The Deep Underground Neutrino Experiment (see section \ref{sec:dune}) will allow to probe dark matter scenarios such as boosted dark matter, which relies on the presence of a small and relativistic component of dark matter that can be detected via its interactions with the Standard Model particles. Signals from boosted dark matter can in principle be separated from potential background sources (neutrinos) and thus turn neutrino experiments into dark sector detectors, as already happens with dark matter detectors \cite{Acevedo:2024wmx} .

\subsection{Probes from cosmology and astrophysics}  \label{sec:cosmoastropdm} 

There are several facilities for cosmological and astrophysical surveys such as DES, DESI, BINGO and LSST, previously discussed in this Briefing Book, that can test properties of dark matter through modifications in the process of structure formation in the Universe. We refer the reader to Chapter 3 for a description of these facilities and the participation of Latin American scientists in them.

\section{Theory efforts} \label{sec:dmtheory}

There is an active Latin American phenomenological community  that has produced important results on theoretical aspects of dark matter on several fronts.

The Brazilian groups working on dark sectors from a particle physics perspective have grown in the past years. Their work includes signatures of heavy unstable dark matter particles and dark matter annihilation in gamma-ray observatories and the IceCube detector \cite{Esmaili:2021yaw,Bhattacharya:2019ucd,Angel:2023rdd}, the effect of a large non-thermal production on the limits derived from direct detection experiments \cite{Dutra:2021lto}, using neutron stars to probe light dark matter \cite{Maity:2021fxw}, signatures at colliders from heavy mediators coupling the SM with dark matter particles via the conversion-driven freeze-out mechanism \cite{Heisig:2024xbh}, detailed computations of the relic density with near-resonance production, relevant for current and future colliders \cite{daSilveira:2024tpy}, studies of the potential to use the UVX sinchrotron light source to produce and detect dark photons \cite{Duarte:2022feb} or axion-like particles \cite{Angel:2023exb}, calculations of the sensitivity of a future hadron-electron collider (LHeC) to WIMPS \cite{Huang:2022ceu} and dark photons \cite{Oliveira:2022ypu} in leptophilic scenarios. On the side of cosmology, studies of the effects of a dark sector on the abundance of light elements \cite{Alves:2023jlo}, alleviating the $H_0$ tension with non-thermal production of dark-matter \cite{daCosta:2023mow} , and imprints of thermal dark matter on cosmological observables \cite{Rodrigues:2023xqu}. Also work on model building with important implications to the axion-photon coupling \cite{Dias:2021lmf}. Part of the Brazilian community has been involved in the design of machine learning algorithms to reconstruct dark states in high-energy collisions~\cite{Alves:2024sai} to characterize events with essential information lost through neutrinos and/or dark matter.

Theoretical groups in Mexico are actively investigating extensions of the Standard Model that provide well-motivated candidates for dark matter. At UNAM, research has explored radiative neutrino mass models, where a stable particle arises as a dark matter candidate, as well as scenarios involving extra U(1)' gauge symmetries with new extra gauge interactions and/or scalar portals \cite{delaVega:2024, Bonilla:2023}. These studies combine model building with phenomenological analyses, confronting collider, flavor, and direct detection constraints. Collaborative work with groups in Mexico (Colima, Guanajuato and Sonora) and Latin America, such as in Chile , has further developed dark sector phenomenology and its interplay with neutrino physics.
Other researchers have studied dark matter models with discrete modular flavour symmetries, and those from Puebla, Hidalgo, and Tlaxcala have performed theoretical calculations of cross sections for fermionic dark matter within extensions of the SM, with magnetic and electric dipole moments \cite{arellano:2022}. 

%
%
The Brazilian community engaged in cosmology and astrophysics research is very active, but most studies are not focused on dark matter physics. Dark matter is an interesting connection, with the main focus on dark energy, modified gravity, and other cosmological observables.
Their work ranges from investigating the connection between the dark matter and dark energy equation of state\cite{vonMarttens:2020apn}, to parameterization of the dark matter density in in cosmological models \cite{Pinto-Neto:2021gcl,Pereira:2022cmu}, to studies of primordial black holes as dark matter \cite{Capanema:2021hnm,Magana:2022cwq}, and studies of the dark matter density evolution law \cite{Santana:2023gkx}, and of the dark matter density profile in galaxies \cite{vonMarttens:2021fmj}, 
For a comprehensive list of contributions to Dark Matter theoretical studies from Brazillian researchers see the ``Theoretical Efforts" section from Ref. \cite{wp:brazilliandmreport2024} from which this section has been adapted.

In Chile, groups at U. Católica Norte and San Sebastián have studied dark matter in non standard cosmologies \cite{gonzalez:2024}. At PUC, researchers have been mostly focused on long-lived particles that depending on the model can have implications for dark matter \cite{Arbelaez:2024lcr,Cottin:2025avd,Bertholet:2025lar}. Others have studied primordial black holes as the sole component of dark matter, and theoretical studies of dark matter with spin on a faraday-like effect on gravitational waves propagating through dark matter \cite{barriga:2025}. At Universidade Técnica Federico Santa María (UTFSM) and at SAPHIR, simplified dark matter models have been investigated. Either in the context vector dark matter \cite{Escalona:2024zkb,Benitez-Irarrazabal:2024ich} and singlet fermion \cite{DiazSaez:2021pmg}. 

Groups at Cinvestav, Chiapas, Guanajuato, and UNAM in México have explored the cosmological implications of a bosonic dark matter equation of state \cite{padilla:2021}. At UNAM, researchers have also investigated neutrino coherent scattering and possible implications to dark matter \cite{delaVega:2021wpx}. This is a light of research that has not been explored much in the Latin American community. Other works are in the context of the well-studied Z' portal \cite{delaVega:2022uko} and neutrino masses \cite{delaVega:2024tuu}
In Colombia, researchers have been working on theoretical studies of multi-component dark matter \cite{Yaguna:2021vhb,Yaguna:2021rds}. Some astrophysical aspects have been explored in connection with gravitational waves and primordial black holes \cite{Bernal:2020ili,Barman:2023ymn}.


\section{Conclusions}  \label{sec:dmconclusions} 

The Latin American community of scientists working on dark matter-related topics, either on the experimental or theoretical side, has grown significantly since the previous iteration of this document in 2021. Particularly notable is the case of Brazil, whose scientific output and number of experiments with Brazilian participation have more than doubled in the period. Overall, there is a healthy level of participation in experimental collaborations and in the development of theoretical work. 

As the searches for the SUSY-motivated standard WIMP and related thermally produced dark matter candidates continue to yield negative results, the focus of research in Latin America has followed the global trend to shift towards the exploration of dark matter candidates from dark sector and other BSM contexts, with different mechanisms that set their relic density, and possible connections to other sectors, such as neutrino physics. This is a rich field whose reach is yet to be understood. 

Latin American scientists are, in many cases, at the forefront of the development of particular experimental techniques, as with CCD detector technology and other quantum sensor technologies. Dedicated laboratories based in Latin American countries have begun to push the frontier in detector development. Many theoretical ideas originating in Latin American institutions guide experimental efforts and advance hand in hand with the field.

In the coming years, the advancement in our understanding of dark matter will have a very important contribution from Latin America.

\bibliographystyle{unsrt} 
\bibliography{dm/dm}

\chapter{Neutrinos}\label{chapt:neutrinos}
\section{Introduction} \label{sec:intro} 
\noindent In our understanding of matter and the subatomic world, Latin America's participation has been steadily growing in the last decades, not only in international neutrino experiments (see Secs.~\ref{sec:v-angra}, \ref{sec:connie}, \ref{sec:hyperk}, \ref{sec:tambo}, \ref{sec:grand}, \ref{sec:novaLA}, \ref{sec:dune}, \ref{sec:junoLA}), but also contributing to the development of phenomenology and theory studies, also promoting and strengthening collaboration and research networks in the region (see Secs.~\ref{sec:ifunam}, \ref{sec:conhep}, \ref{sec:GI-FT}).

The neutrino puzzle is still not fully solved. However, potential groundbreaking applications (still unpractical) such as Earth tomography and nuclear activity monitoring have recently attracted the attention of the academic and experimental community and beyond, representing a potential opportunity for technology hubs, spin-offs, investment funds, and venture capital from the private sector \cite{ref11}\cite{ref12}\cite{ref13}\cite{ref14}\cite{ref15}. 

Current efforts on neutrino physics include a precision measurements of neutrino oscillation parameters, with particular attention to the $\theta_{23}$-mixing angle and the leptonic CP-violation phase ($\delta_{CP}$), the neutrino mass ordering, their Dirac or Majorana  nature, the absolute neutrino mass scale, possible astrophysical sources, and new neutrino states (such as a sterile neutrino). 

\subsection{Neutrino oscillations, mass hierarchy and leptonic phase}\label{sec:1a}

\noindent 
Before introducing relevant Latin American contributions, let us describe some of the basic aspects of neutrino physics.

Neutrino oscillations are described by the Pontecorvo–Maki–Nakagawa–Sakata (PMNS) mixing matrix, which relates the flavor eigenstates ($\nu_{e}$, $\nu_{\mu}$, $\nu_{\tau}$) to mass eigenstates ($\nu_{1}$, $\nu_{2}$, $\nu_{3}$)  \cite{ref27}, as shown in Equation (\ref{eq01}) for the Dirac neutrino case:

\begin{eqnarray}\label{eq01}
\begin{bmatrix*}[c]
    \nu_{e}  \\
    \nu_{\mu} \\
    \nu_{\tau}  
\end{bmatrix*}
= 
\begin{bmatrix*}[c]
    U_{e1} & U_{e2} & U_{e3} \\
    U_{\mu1} & U_{\mu2} & U_{\mu3} \\
    U_{\tau1} & U_{\tau2} & U_{\tau3}    
\end{bmatrix*}
\begin{bmatrix*}[c]
    \nu_{1}  \\
    \nu_{2} \\
    \nu_{3}  
\end{bmatrix*}.
\end{eqnarray}
\\
The simplest approximation to the oscillation probability between two neutrino states (2$\nu$), having a neutrino state $\nu_{\alpha}$ detected as $\nu_{\beta}$ is:

\begin{equation}\label{eq02}
P_{2\nu}(\nu_{\alpha} \rightarrow \nu_{\beta}) = \sin^{2}\left(2\theta\right)\sin^{2}\left(\frac{\Delta m_{ji}^{2}L}{4E_{\nu}}\right),
\end{equation}
with $\Delta m_{ji}^{2} = m_{i}^{2}-m_{j}^{2}$ ($m_{j}$ is the mass of the \textit{j}-th neutrino mass eigenstate). Equation (\ref{eq02}) is able to explain almost all the data available on neutrino oscillation parameters (except Short Baseline (SBL) anomalies) even in a 3-flavor neutrino (3$\nu$) scenario \cite{ref28}. The PMNS matrix can be parametrized  as the product of three independent rotations, through three mixing angles ($\theta_{23}$, $\theta_{13}$, $\theta_{12}$), where the second (unitary) rotation depends on a phase ($\delta_{CP}$), and of a diagonal matrix of phases (\textit{P}):

\begin{eqnarray}\label{eq03}
\newcommand\tab[1][1cm]{\hspace*{#1}}
\begin{aligned}
& U_{PMNS} = 
\begin{bmatrix}
    1 & 0 & 0 \\
    0 & \cos\theta_{23} & \sin\theta_{23} \\
    0 & -\sin\theta_{23} & \cos\theta_{23}    
\end{bmatrix}   
\begin{bmatrix}
    \cos\theta_{13} & 0 & e^{i\delta_{CP}}\sin\theta_{13} \\
    0 & 1 & 0 \\
    -e^{i\delta_{CP}}\sin\theta_{13} & 0 & \cos\theta_{13}
\end{bmatrix} 
\begin{bmatrix}
    \cos\theta_{12} & \sin\theta_{12} & 0 \\
    -\sin\theta_{12} & \cos\theta_{12} & 0 \\
    0 & 0 & 1
\end{bmatrix}P = \\ &
\tab[1.4cm]
\begin{bmatrix}
 c_{12}c_{13} & s_{12}c_{13} & e^{i\delta}s_{13} \\
 -s_{12}c_{23}-e^{i\delta}s_{13}c_{12}s_{23} & c_{12}c_{23}-e^{i\delta}s_{13}s_{12}s_{23} & s_{23}c_{13} \\
  s_{12}s_{23}-e^{i\delta}s_{13}c_{12}c_{23} & -c_{12}s_{23}-e^{i\delta}s_{13}s_{12}c_{23} & c_{23}c_{13}
\end{bmatrix}P 
\end{aligned}
\end{eqnarray}
\\
where the first, second, and third matrices represent the accessible regions to atmospheric, reactor, and solar neutrino fluxes, respectively, and \textit{P} is a unitary matrix for the Dirac neutrino case or a diagonal matrix (depending on two phases) for the Majorana neutrino case \cite{ref29}. The three mixing angles have all been measured and current efforts are focused on precision measurements, while $\delta_{CP}$ is still unknown. In addition to these parameters, there are two independent mass-squared difference scales present in the puzzle ($\Delta m^{2}_{21}$, $\Delta m^{2}_{32}$). 

\noindent 
Usually two possibilities are defined for the ordering of the neutrino masses: a ``Normal Ordering" (NO) with $m_3>m_2>m_1$ and an "Inverted Ordering" with $m_2>m_1>m_3$. In the latter case,  $\Delta m^{2}_{31}$ and $\Delta m^{2}_{32}$ are both negative and can be precisely measured in reactor experiments \cite{ref30}, and represent the mass difference between the heaviest and lightest neutrino or the opposite. The 3$\nu$ oscillation parameters from fits based on the most recent available data are summarized in Table \ref{tab0a} \cite{refosc1, refosc2}.

\noindent 
\begin{table*}[htp]
\caption{The 3$\nu$ scenario fit to global data (October 2024) obtained with the inclusion of the tabulated $\chi^{2}$ data on atmospheric neutrinos provided by the Super-K Collaboration. The best fit point (bfp) is shown as the central value.}\label{tab0a}
\centering
\begin{adjustbox}{width=\textwidth}
\begin{tabular}{ccccc}
\hline\noalign{\smallskip}
Parameter & bfp $\pm$ 1$\sigma$ (NO) & 3$\sigma$ range (NO) & bfp $\pm$ 1$\sigma$ (IO) & 3$\sigma$ range (IO) \\
\noalign{\smallskip}\hline\noalign{\smallskip}
$\sin^2\theta_{12}$ & $0.307_{-0.011}^{+0.012}$ & $0.275 \to 0.345$ & $0.308_{-0.011}^{+0.012}$ & $0.275 \to 0.345$
\\
$\theta_{12}/^{\circ}$ & $33.68_{-0.70}^{+0.73}$ & $31.63 \to 35.95$ & $33.68_{-0.70}^{+0.73}$ & $31.63 \to 35.95$ 
\\
$\sin^{2}\theta_{23}$ & $0.561_{-0.015}^{+0.012}$ & $0.430 \to 0.596$ & $0.562_{-0.015}^{+0.012}$ & $0.437 \to 0.597$
\\
$\theta_{23}/^{\circ}$ & $48.5_{-0.9}^{+0.7}$ & $41.0 \to 50.5$ & $48.6_{-0.9}^{+0.7}$ & $41.4 \to 50.6$ 
\\
$\sin^{2}\theta_{13}$ & $0.02195_{-0.00058}^{+0.00054}$ & $0.02023 \to 0.02376$ & $0.02224_{-0.00057}^{+0.00056}$ & $0.02053 \to 0.02397$ 
\\
$\theta_{13}/^{\circ}$ & $8.52_{-0.11}^{+0.11}$ & $8.18 \to 8.87$ & $8.58_{-0.11}^{+0.11}$ & $8.24 \to 8.91$ 
\\
$\delta_{CP}/^{\circ}$ & $177_{-20}^{+19}$ & $96 \to 422$ & $285_{-28}^{+25}$ & $201 \to 348$ 
\\
$\frac{\Delta m_{21}^{2}}{10^{-5}eV^{2}}$ & $7.49_{-0.19}^{+0.19}$ & $6.92 \to 8.05$ & $7.49_{-0.19}^{+0.19}$ & $6.92 \to 8.05$ 
\\
$\frac{\Delta m_{3l}^{2}}{10^{-3}eV^{2}}$ & $+2.534_{-0.023}^{+0.025}$ & $+2.463 \to +2.606$ & $-2.510_{-0.025}^{+0.024}$ & $-2.584 \to -2.438$ 
\\
\noalign{\smallskip}\hline
\end{tabular}
\end{adjustbox}
\end{table*}

Determination of $\delta_{CP}$ is being searched through enormous efforts of future upgrades to artificial neutrino sources in the main accelerator facilities worldwide at Fermilab, JPARC and Protvino. Outstanding synergies are taking place with leading expectations from the analysis of data collected by experiments such as NO$\nu$A and T2K \cite{refosc1}. Nonetheless, competitive sensitivity is also estimated by the next generation of experiments such as DUNE \cite{ref55} and KM3NeT-ORCA (the P2O side project \cite{refdelkm}), with planned upgrades to the beam at different power and exposures, in turn being suited for precision on oscillation parameters with NO. P2O, with a sufficiently long beam exposure ($\sim$ 4 year at 450 kW), can  reach a $2\sigma$ sensitivity to $\delta_{CP}$, comparable with the projected sensitivity of NO$\nu$A and T2K. An advanced phase of P2O would give a $6\sigma$ sensitivity to $\delta_{CP}$ after 10 years of operation at 450 kW, competitive with the projected sensitivity of the next generation experiments DUNE and T2HK. The best accuracy on $\delta_{CP}$ would be achieved for $\delta_{CP}$ = 0$^{\circ}$ and 180$^{\circ}$. 

\noindent The above equations refer to the standard 3-flavor neutrino scenario. However, for a (3+1)-flavor scenario, in which a hypothetical sterile neutrino is added, Equation~(\ref{eq01}) can be  modified accordingly by adding new flavor ($\nu_{s}$) and mass ($\nu_{4}$) eigenstates connected through the $[U_{s1}, U_{s2}, U_{s3}]$ row vector in the PMNS matrix, in order to give room to include a fourth flavor, hence, obtaining a $4\times4$ PMNS matrix and three independent squared-mass differences $\Delta m^{2}$ where $U_{e4}^{2}+U_{\mu4}^{2}+U_{\tau4}^{2}+U_{s4}^{2} = 1$, is a PMNS unitarity condition. 

\subsection{Neutrino masses and nature}\label{sec:1b}

\noindent Neutrinoless Double Beta Decay ($\beta\beta0\nu$) searches have the potential to solve the questions related to neutrino nature and absolute mass scale. This process requires the neutrino to be its own antiparticle, thus neutrinos are Majorana particles with $\nu_{eL}=\bar{\nu}_{eR}$, \textit{L} (\textit{R} subindeces denoting the Left- or Right-handed nature of neutrinos).  Experimental searches look for the so-called effective Majorana mass \cite{ref31}:

\begin{equation}\label{eq06}
\langle m_{\beta\beta} \rangle = \left\lvert \sum_{i=1}^{3}U^{2}_{ei}m_{i} \right\rvert .
\end{equation}

\noindent If $\beta\beta0\nu$ is mediated by light neutrino exchange, Equation (\ref{eq06}) can be linked to the corresponding isotope half-life ($T_{1/2}^{0\nu}$) as a function of either $\langle m_{\beta\beta} \rangle$ and the nuclear matrix element ($M^{0\nu}$) (describing all the nuclear structure effects):

\begin{equation}\label{eq07}
T_{1/2}^{0\nu} = G^{0\nu}(Q,Z) \left\lvert M^{0\nu} \right\rvert ^{2} \left(\frac{\langle m_{\beta\beta} \rangle}{m_{e}} \right) ^{2},
\end{equation}

\noindent where $G^{0\nu}(Q,Z)$ corresponds to the lepton phase-space (kinematic) factor (a function of the Q-value of the decay, charge, and mass of the final state nucleus).  

\noindent The ranges are obtained by projecting the results of the global analysis of oscillation data, with the exclusion of the tabulated $\chi^{2}$ data on atmospheric neutrinos provided by Super-K. The region for each ordering is defined with respect to its local minimum. To date, the best neutrino mass limit is set by the KamLand
Zen experiment, with $m_{\beta\beta} = 50 - 160$ meV, the best half-life sensitivity is set by the GERDA experiment, with $T_{1/2} > 11 \times 10^{25}$ years (90\% Confidence Level). 

\section{The Neutrinos-Angra Experiment} \label{sec:v-angra} 

This summary is based on ref. \cite{wp:v-angra} . \\

The Neutrinos-Angra ($\nu$-Angra) experiment is located at the Angra dos Reis nuclear power plant near Rio de Janeiro, Brazil, and is focused on non-proliferation safeguards studies through neutrino detection. 
The experiment aims to develop a reliable and cost-effective technology for monitoring nuclear reactors and potentially the spectral evolution of neutrinos, which could reveal changes in fuel composition, particularly plutonium fractions, crucial for diversion detection \cite{bowden:2008,carr:2018}.
Its objectives are: {\it i)}  non-invasive monitoring of reactor activity, {\it ii)} Estimate the thermal power produced in the reactor core, {\it iii)}  Development of new antineutrino detection techniques, {\it iv)} Contribute to the International Atomic Energy Agency (IAEA) safeguards and non-proliferation efforts \cite{Anjos:safeguards:2009}, and {\it v)} Integration of Latin American scientists and engineers into global scientific collaborations.

The experiment uses a gadolinium (Gd)-doped water Cherenkov detector to monitor the electron antineutrino ($\bar{\nu} _e$) flux generated by the Angra-II reactor (3.95 GW$_{\rm th} $ power) via the inverse beta decay (IBD) reaction: $\bar{\nu} _e + p \rightarrow e^+ + n$.
Monitoring the antineutrino flux allows to measure the fission rates and the reactor’s power.
In IBD, a $\bar{\nu} _e$ interacts with a proton ($p$), producing a positron ($e^+$) and a neutron ($n$). The positron produces a Cherenkov flash and annihilates with an electron, creating two photons of approximately 511 keV each: $e^+ + e^- \rightarrow \gamma + \gamma$, causing a light burst called the {\it prompt signal} . The neutron is captured by a Gd nucleus within about 12.3 $\mu$s, releasing 8 MeV in photons as the Gd atom deexcites, producing a {\it delayed signal}.

The $\nu$-Angra detector is located at a distance of 25 m from the reactor core in a standard shipping container shared with the \gls{CONNIE} experiment \cite{connie:2019}. It has a 1-ton target volume filled with GdCl$_3$-doped water, surrounded by 32 eight-inch photomultiplier tubes (PMTs) to detect Cherenkov radiation produced by IBD events. The detector also includes a Top VETO system and a Lateral VETO system with 4 PMTs each, to reduce background noise from cosmogenic muons. For a comprehensive description of the detector and its electronics, see \cite{angra-det:2019}.
The detector is installed at surface level, subject to the influx of cosmogenic particles, which impact its signal-to-noise ratio and pose an additional challenge to effectively monitoring the reactor activity. 
The High Voltage system and computer infrastructure are commercial off-the-shelf components, while  the Front End electronics \cite{angra:fe:2016}  and digitizers \cite{angra:daq:2014} were entirely designed and constructed by the collaboration.

Data from the detector is continuously written to local storage and subsequently transferred to primary and mirror data servers located at CBPF (Rio de Janeiro) and Unicamp (Campinas) for analysis.
Due to restrictions on using radioactive sources at the nuclear plant, detector calibrations rely on Monte Carlo Simulations of the expected prompt $e^+$ energy spectrum from the reactor $\bar{\nu} _e$ flux \cite{huber:2004}, and of the smeared 8 MeV capture gamma line \cite{santos:2023}.
IBD events are selected based on the reconstructed positron energy $E_{e+}$, the delayed capture gamma energy, and the time between the prompt and the delayed signals ({\it prompt-delay} pairs). 
The antineutrino energy is  related to the incident positron energy by $E_{e+} = E_{\bar{\nu} }  - \Delta$, where $\Delta = m_n - m_p = 1.293$~MeV is the neutron-proton mass difference. 
The detector time resolution enables the separation of muon-decay electrons, neutron captures, and noise.
Reactor monitoring is based on the comparison of the {\it prompt}  spectra of candidate {\it prompt-delay}  pairs from data acquired with the reactor ON and OFF.

\begin{figure} 
\includegraphics[width=0.42\linewidth]{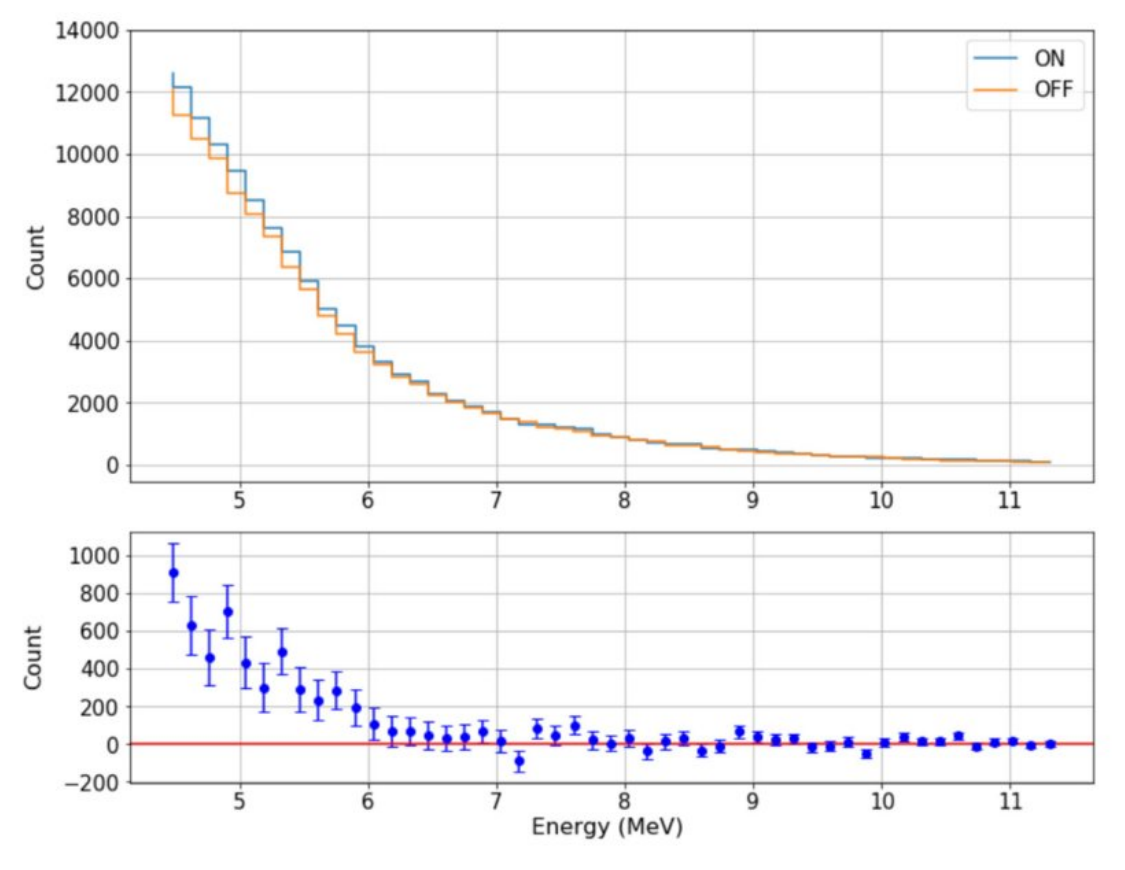}
\includegraphics[width=0.42\linewidth]{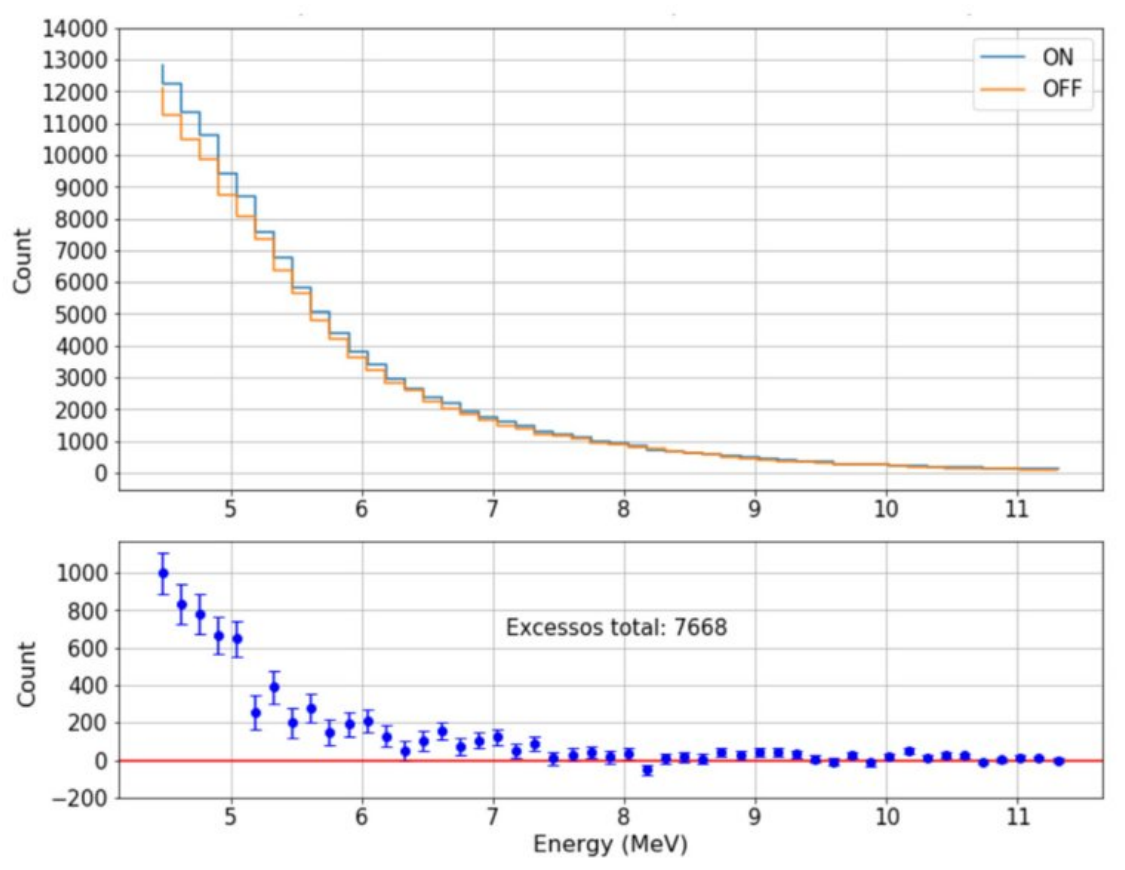}  
\centering
\caption{\small 
Neutrinos-Angra ON-OFF analysis results showing event rates comparison during two reactor ON and
OFF periods. A clear excess of events is observed during the ON1 (left) and ON2 (right) periods \cite{angra:2025}.
} 
\label{fig:angra-results} 
\end{figure}

Based on datasets acquired in August-September of 2020 (ON1), September-October of 2020 (ON2), and July-August of 2020 (OFF), the collaboration reports a clear and statistically significant increase in the number of events during the reactor ON period compared to the OFF period in the 4.5-6.5 MeV energy range expected for IBD events (see Figure \ref{fig:angra-results}). The background-only hypothesis was rejected in the ON1 (ON2) period with a $\chi^2/$dof$\sim13.4$ ($\sim17.3$). These excesses are attributed to the detection of reactor antineutrinos through the IBD process, validating the detector’s performance and its potential for non-intrusive reactor monitoring. The results have recently been published in \cite{angra:2025}.

{\bf Current status and proposed upgrades.} During the pandemic, maintenance visits became challenging, and loss of water transparency affected the detector response. Eventually, data collection was interrupted due to insufficient man-power to carry out operations and on-site maintenance. The detector was reactivated in July 2024 and has been collecting data steadily since then. 

Planned upgrades include: 
{\it i)}  Increase the Veto efficiency and sensitivity and use advanced algorithms for background rejection, 
{\it ii)}  Restore the stability and robust operation required to collect data before, during and after the annual reactor shutdown, demonstrating the non-intrusive monitoring of the reactor status, and performing safeguards-related measurements during the handling of nuclear fuel during maintenance periods.
{\it iii)}  Develop improved calibration simulations, use of portable calibration sources that comply with nuclear plant safety regulations, and integrate an LED-based energy scale calibration system. 
{\it iv)}  Improved data acquisition system and Front-End electronics to an FPGA-based system with higher efficiency and speed, and developed AI tools for real-time event identification. 
{\it v)}  Replace water with water-based liquid scintillator (WBLS) to increase light yield and timing response, while keeping Cherenkov light production. This will improve the energy resolution and potentially enable the measurement of the reactor fuel composition's evolution via changes in the antineutrino energy spectrum, resulting from uranium burn-up and plutonium build-up.

{\bf Cryogenic Calorimeter.} 
The collaboration plans to develop a cryogenic calorimeter that will eventually replace the water-based detector. MilliKelvin calorimeters have been successfully used in large-scale experiments for neutrinoless double-beta decay and searches for sub-GeV Dark matter (see, for example, \cite{cuore:2022} and \cite{cresst:2024}). By converting particle energy deposited in a crystal into phonons and measuring a temperature increase, these detectors can reach thresholds of the order of 10 eV, ideal for detecting low-energy depositions. Such an instrument will be able to detect neutrinos via Coherent Elastic Neutrino-Nucleus Scattering (CE$\nu$NS) and expand the experiment's reach to explore new physics phenomena.

{\bf Timeline} 
The collaboration has proposed a 5-year plan to implement the mentioned upgrades on the water-based detector and develop the cryogenic calorimeter in parallel. It plans to continue operating and collecting data with the water-based detector for reactor monitoring, while implementing system upgrades to improve sensitivity and data quality. 
In parallel, a feasibility study will be conducted to assess the potential and requirements for the cryogenic calorimeter, and to develop the detector design, as well as the DAQ and veto systems. In the later stage components will be procured and the system will be integrated for preliminary testing. The assembly and commissioning of the cryogenic calorimeter is envisioned post-year 5. 

{\bf Costs and computational requirements.}  The estimated costs for upgrading and running the water-based detector over the next 5 years are \$270,000 USD. The development of the cryogenic calorimeter is estimated to cost \$670,000 USD. The computational infrastructure of the institutions involved in the $\nu$-Angra experiment, primarily CBPF, UNICAMP, and USP, is expected to be excellent and fully capable of meeting the experiment's needs in the near future. Upgrades required for maintaining them are estimated at \$50,000 USD.

The $\nu$-Angra experiment was a pioneer in the detection of reactor neutrinos using a Cherenkov detector that runs at the surface level. The proposed technologies to be developed for the cryogenic calorimeter hold great potential for applications in other areas. It produced a significant number of Master's and doctoral theses, training human resources in hardware and software, preparing the next generation of scientists and engineers equipped with cutting-edge skills and knowledge. 

The $\nu$-Angra experiment, alongside CONNIE, demonstrates the successful partnership with Eletronuclear at Angra II. This collaboration highlights Brazil and Latin America’s capacity for advanced particle physics research without large-scale infrastructure investments. Using existing facilities, these experiments yield high-quality scientific results, showcasing the region’s potential for world-class research.
 
\section{The Coherent Neutrino-Nucleus Interaction Experiment (CONNIE)} \label{sec:connie} 

This summary is based on ref. \cite{wp:connie}. \\

\gls{CONNIE} uses low-noise fully depleted charge-coupled devices (CCDs) \cite{holland:2003}  with the goal of measuring low-energy recoils from coherent elastic scattering (CE$\nu$NS) of reactor antineutrinos with silicon (Si) nuclei and probing physics Beyond the Standard Model (BSM).
CE$\nu$NS was predicted in the 1970's \cite{freedman:1974,kopeliovich:1974}  and was detected for the first time in 2017 by the COHERENT collaboration at the Spallation Neutron Source (SNS) at the Oak Ridge National Laboratory in the U.S.A. 
CE$\nu$NS from solar, atmospheric and diffuse supernova neutrinos has long been identified as a limiting background for future dark matter (DM) searches \cite{anderson:2011}. The XENONnT and PandaX \cite{xenonnt:2024,pandax:2024}  collaborations recently announced the first measurements of low-energy nuclear recoils from solar neutrinos. The CONUS+ experiment has recently announced \cite{conus:2025}  a 3.7 $\sigma$ observation with Ge detectors at reactor in Brokdorf, Germany.
CE$\nu$NS also provides a new window into the low-energy neutrino sector, which has received increasing interest as a potential probe for new physics \cite{carey:2024}. Several nonstandard interactions of neutrinos predicted by extensions of the SM as well as other nonstandard neutrino properties can be probed at low energies from CE$\nu$NS \cite{denton:2018,miranda:2019,parada:2019}.

The \gls{CONNIE} experiment is located at about 30 m from the core of the 3.8 GW Angra-II nuclear reactor in Rio de Janeiro, Brazil, inside a refurbished standard shipping container, alongside the $\nu$-Angra experiment (see \autoref{sec:v-angra} ), receiving a neutrino flux of $7.8\times10^{12} $~$\nu/{\rm cm} ^{2} /{\rm s} $.
An engineering run was carried out in 2014-2015 \cite{connie:2016} , followed by a science run from 2016-2020 using 14 standard CCDs with 16 million pixels with 15$\mu$m$\times$15$\mu$m size and a substrate thickness of 675 $\mu$m. The sensors are operated at temperatures below 100 K to reduce dark current, and are installed in a copper box inside a copper vacuum vessel ($10^{-7} $~torr), shielded by a 15~cm layer of lead sandwiched between two 30-cm layers of polyethylene.

The energy of the recoiling Si nucleus is reconstructed from the ionisation charge it produces by means of the {\it quenching factor} , which is evaluated from a recent model developed at UNAM (Mexico) \cite{Sarkis:2023}  that extends the Lindhard theory \cite{lindhard}  down to the energies \gls{CONNIE} is sensitive to. 
To search for CE$\nu$NS, neutrino event selection criteria are applied to the reactor ON and OFF data and their spectra are subtracted. 

 For the 2016-2018 run, a total exposure of 3.7 kg-days from 8 CCDs (47.6 g active mass), no significant excess was observed from the ON-OFF subtraction, allowing us to place a 95\% C.L. limit in the CE$\nu$NS rate at a factor of 40 above the SM prediction for deposited energies of 100 eV (1 keV recoil).
 These results were used to obtain the, at the time, world-leading constraints on the parameter space of two simplified extensions of the Standard Model with light mediators \cite{connie:2020}.
An improved readout scheme using hardware binning was applied to the 2019-2020 run, allowing for a reduction of the threshold from 100 eV to 50 eV. However, while a stronger limit on the CE$\nu$NS rate was expected,  due a statistical fluctuation in the data a weaker limit was obtained \cite{connie:2022}.

\begin{figure} 
\scalebox{0.5} {\includegraphics{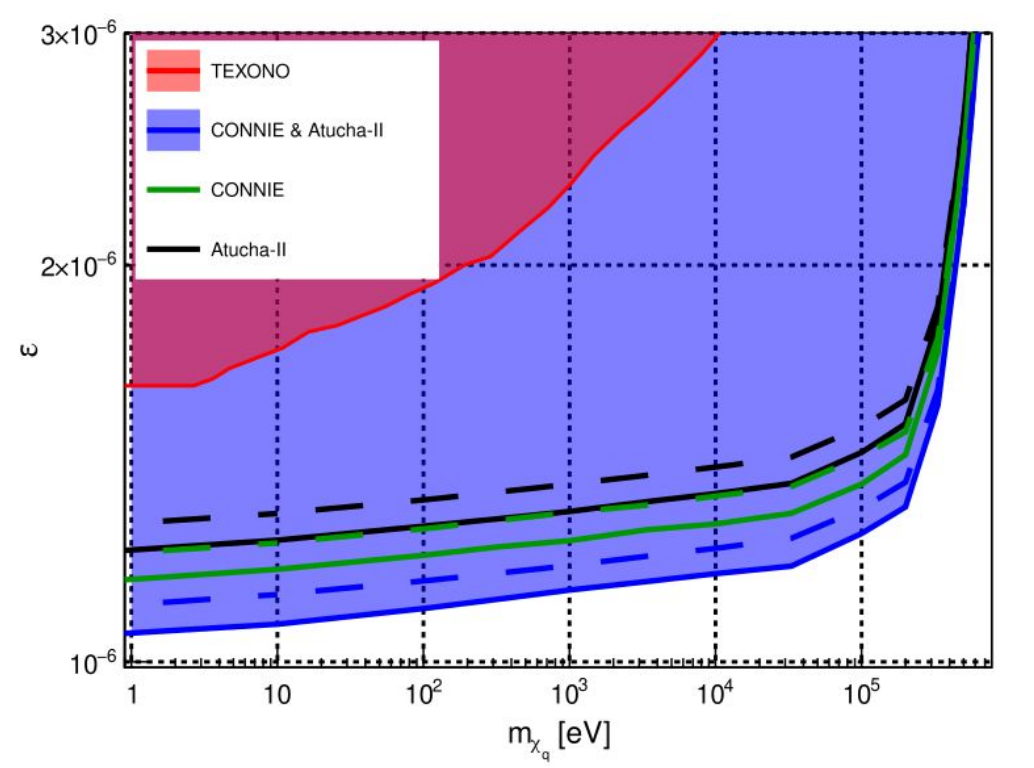} } 
\centering
\caption{\small \gls{CONNIE} and Atucha-II limits on the mass and charge fraction of millicharged particles obtained from the 2021-2022 run with two Skipper-CCDs of 0.5 g mass \cite{connie:2025}.
} 
\label{fig:connie-milliQ} 
\end{figure}

 In July 2021 two Skipper-CCDs with 0.5 g total mass were installed together with new dedicated Low Threshold Acquisition (LTA) electronics \cite{cancelo:2021}  speciallly developed by Fermilab and UNS (Universidad Nacional del Sur, Bahia Blanca, Argentina) for low-noise Skipper-CCD readout. In Skipper-CCDs, the readout stage is modified to allow multiple nondestructive samplings of each pixel, thus decreasing the noise to sub-electron level and allowing to count individual electrons in a pixel. 
 During the 2021-2023 run with Skipper-CCDs a readout noise of 0.15 e- was achieved, and a new analysis chain was developed reducing the background and reaching a threshold of 15 eV, albeit accumulating a small exposure of only 18.4 g-days. New limits on the CE$\nu$NS rate \cite{connie:2024} were obtained, comparable to those of the previous analysis, but with an exposure $10^3$ times smaller.
The limits on light vector-mediator models also improved relative to the previous analysis. A study of the diurnal modulation of dark matter (DM) interacting with electrons was carried out, yielding the best limits on the cross section for DM scattering with electrons, obtained from a surface experiment\cite{connie:2024}.
In addition, a search for relativistic millicharged particles produced by high-energy photons in the reactor core and interacting electromagnetically in the detector was performed. The statistical analysis was developed in collaboration with the Atucha-II experiment, which also uses Skipper-CCDs at a reactor, resulting in the best-to-date individual and combined limits in the parameter space of millicharged particle mass and charge fraction (see Figure \ref{fig:connie-milliQ}  \cite{connie:2025}).
 
The successful 2021-2023 Skipper run showed that the signal and background rates (29.3 and 4000 events/keV/kg/day, respectively) achieved in the lowest energy bin, from 15 to 215 eV, would allow a 1 kg detector to observe CE$\nu$NS at 90\% C.L. with approximately 200 days of reactor-on exposure and 50 days of reactor-off exposure. With the same operating conditions and background level but located at half the distance to the reactor core, hence quadrupling the antineutrino flux, the measurement would be achieved with only 13 days of reactor-on exposure. The collaboration is pursuing improvements in both fronts (larger mass, shorter distance).

{\bf Current status and plans.} 
Between May and October of 2024 the collaboration installed and commissioned 16 new Skipper-CCDs mounted in a compact single-board array called a Multi-Chip Module (MCM), designed for the Oscura DM experiment \cite{cervantes:2023}  reaching a mass of 8 g. Data taking started in November of 2024. An Oscura array of 16 MCMs and about 130 g of active mass, is called a Super Module.
With just one SuperModule of Skipper-CCDs at 15 m from the core under the operating conditions of 2021-2023, the CE$\nu$NS measurement would be accomplished in 100 days. Negotiations with Eletronuclear, the company that manages the reactor, are underway to relocate to a position within the reactor dome \cite{eletronuclear}. 

A broader Latin American effort is underway to construct the next-generation neutrino reactor experiment with Skipper-CCDs at several kg of mass. This requires integrating thousands of sensors and their electronics, along with robust packaging and stable cooling. Members of \gls{CONNIE} are leading the Oscura effort in the U.S.A. to develop a 10 kg Skipper-CCD experiment to search for DM. The technology developed for the DM experiment will be available for next-generation neutrino detectors using Skipper-CCDs.

{\bf Latin American involvement.} 
The collaboration is currently formed by 29 scientists and graduate students mostly from Latin America. Argentinian institutions include Instituto de Ciencias Físicas at ECyT-Universidad Nacional de San Martín, Instituto Balseiro at Centro Atómico Bariloche, as well as Universidad Nacional del Sur (UNS), Universidad Nacional de Córdoba and Universidad de Buenos Aires (host of the LAMBDA laboratory \ref{sec:lambda} ). Brazillian institutions include Universidade Federal de Rio de Janeiro, Centro Federal de Educação Tecnológica (CFET) Celso Suckow da Fonseca, Centro Brasileiro de Pesquisas Fisicas, Instituto Tecnologico de Aeronautica, Universidade Federal do ABC. Also participate Universidad Nacional de Asunción, Paraguay, and Universidad Nacional Autónoma de México (UNAM) from México. Also participate the Institute of Physics from the University of Zürich, Switzerland, and Fermilab from the United States. Skipper-CCD detector development is led by coleagues at Fermilab, UBA and UNS, with shared technical kowledge, both in hardware and analysis, distributed among the Latin American Institutions.

{\bf Timeline. } 
Operation with the current MCM will continue for a few years (2024-2027). In parallel, additional MCMs will be installed to achieve a mass of approximately 100 g (one Super Module) using a new vessel and cryogenics. New location opportunities with respect to proximity and shielding will be explored (2025-2028). It is expected that sensitivity to detect CE$\nu$S will be reached by 2027. The development of a 1-10 kg-scale detector will depend on the results of the previous phases and could take place between 2027 and the early 2030s.

{\bf Costs and computational requirements.} 
Construction of a 10 kg Skipper-CCD detector (Oscura design) is estimated around 7.7 million US dollars, where 78\% is for the production of the CCD wafers and the rest is distributed between packaging, electronics, cryogenics, vessel, shielding, and other expenses. The operational costs are expected to be small in comparison with construction, and include maintenance and technical visits as well as scientic visits between institutes and labs.
Support to develop an instrumentation lab at UFRJ in Rio de Janeiro for Skipper-CCD studies has been secured through three FAPERJ programs. A CCD lab is also under development at UNAM in Mexico.

\gls{CONNIE} data is stored, processed and analysed at the CHE cluster at CBPF, the only machine allowed to connect directly to the container at Angra-II. CHE has 100 TB of storage and 280 cores available for processing for CONNIE, and operates in a highly stable environment with fast internet connections. The planned upgrades will require processing and storage upgrades, in particular, a dedicated storage system for \gls{CONNIE} to accommodate the expected high volume of data. Both the physical infrastructure and the cluster hardware are ready to host the needed upgrades, so that only the purchase of new processing and storage machines will be required.
 
\section{Hyper-Kamiokande experiment} \label{sec:hyperk} 

Hyper-K, successor of Super-Kamiokande \cite{fukuda:2003}, consists of a next-generation underground water Cherenkov detector and the upgraded Japan Proton Accelerator Research Complex (J-PARC) neutrino beam. The detector will be 71 m high and 68 m diameter, and ﬁlled with 260 ktons of ultra pure water, corresponding to an order of magnitude larger ﬁducial mass (188 ktons) than Super-K. In January 2020, the \gls{Hyper-K} project ofﬁcially began, and the operation is expected to start in 2027. 
Equipped with newly developed high-sensitivity photosensors and a high intensity neutrino beam, it will have large potential to discover CP violation ($>5\sigma$ significance in 60\% of the parameter space) by observing neutrinos and anti-neutrinos from J-PARC, to investigate Grand Uniﬁed Theories by exploring proton decay via $p\rightarrow e^+ \pi^0$ with lifetimes larger than $10^{35} $~yr, and to determine the neutrino mass ordering by combining atmospheric and beam data. \gls{Hyper-K} will also have far better capabilities to observe $^8$B solar neutrinos and neutrinos from other astronomical sources than previous water-based detectors. It will also have capabilities to detect Supernova burst and Supernova relic neutrinos.

{\bf Latin American involvement.} 
Brazil is currently involved in \gls{Hyper-K} in physics studies of NSI, as well as possible contributions to GRID computing and data storage through Pontificia Universidad Catolica do Rio de Janeiro.
Since 2020, a consortium of four Mexican universities —Tecnológico de Monterrey (TEC), Universidad de Guadalajara (UdeG), Universidad Autónoma de Sinaloa, and Universidad Autónoma de Chiapas— has joined the \gls{Hyper-K} collaboration. The team co-leads the mechanical design, local prototyping, and low-power front-end electronics of the 19-channel multi-PMT optical modules, and it is also developing AI algorithms for neutrino classification in water-Cherenkov detectors, work highlighted in recent collaboration reports.
Supported by a 2024 SECIHTI/CONAHCYT grant, the group will manufacture the first full production batch of these mPMTs at TEC and commission them in Mexico, with installation slated for Hyper-K’s 190-kt far detector during the 2027 integration campaign. Large-scale Monte-Carlo production and AI-based reconstruction run on UdeG’s “Leo Atrox” supercomputer at the CADS centre, which is operated by the UdeG team. Mexican collaborators additionally contribute to Hyper-K’s underwater mechanical structures, electronic vessels, and transportation studies, and they hold coordination roles on the collaboration’s Steering/Resource, Publications, and Outreach Boards, while the “Outreach \gls{Hyper-K} México”
initiative organizes national training schools that broaden Latin-American engagement in neutrino physics.
 
\section{The Tau Air-shower Mountain-Based Observatory: TAMBO} \label{sec:tambo} 

Over the past ten years, neutrino astronomy, particularly through the IceCube experiment, has achieved several remarkable milestones: the discovery of a diffuse astrophysical neutrino flux~\cite{IceCube:2013low} ; the detection of neutrino emission from a distant blazar~\cite{IceCube:2018cha}  and from a nearby active galaxy~\cite{IceCube:2022der}; and the identification of neutrinos originating from the Galactic Plane~\cite{IceCube:2023ame}. More recently, the highest-energy cosmic neutrino to date has been observed by KM3NeT~\cite{KM3NeT:2025npi}. 

Moreover, neutrino astronomy has opened a new energy frontier, enabling stringent constraints on fundamental symmetries and probing physics beyond the Standard Model through high-energy neutrino observations~\cite{IceCube:2016rnb,IceCube:2016dgk,IceCube:2017qyp,IceCube:2021tdn,IceCube:2022ubv}. Despite these significant advances, many fundamental questions remain unresolved, and important expectations are still unfulfilled, such as identifying the origin of cosmic rays, detecting a broader population of neutrino point sources, and achieving higher-precision measurements of the astrophysical neutrino flux, among others.

{\bf \gls{TAMBO} experimental concept.} 
A key step toward improving our understanding of the astrophysical neutrino flux is a more accurate determination of its flavor composition, which in turn requires the unambiguous detection of tau-neutrino events. The Tau Air-shower Mountain-Based Observatory (TAMBO) aims to meet this goal by enabling the cleanest detection of cosmic tau neutrinos to date~\cite{TAMBO:2023plw}. 

\begin{figure} [htbp]
    \centering
    \includegraphics[width=0.8\textwidth]{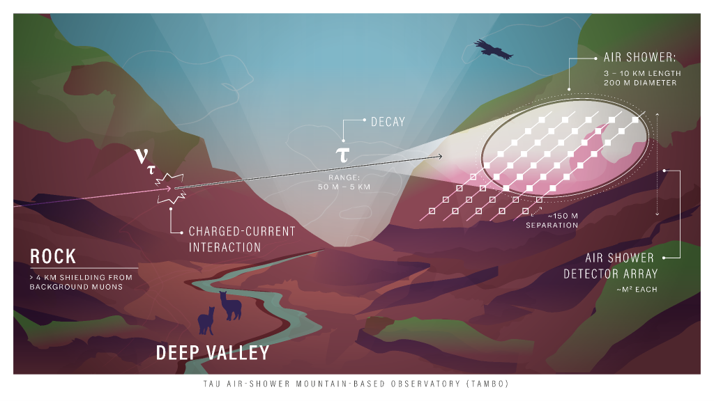} 
    \caption{\gls{TAMBO} detector concept.} 
    \label{fig:TAMdetConcp} 
\end{figure}

\gls{TAMBO} will be deployed in the Colca Valley, Arequipa, Peru, where air showers are induced by tau leptons emerging from the mountains. This process begins when cosmic tau neutrinos, upon arrival, traverse the Earth’s crust and interact via charged-current interactions with nuclei in the rock. Tau leptons created in interactions that occur near the surface of one of the valley’s faces are likely to escape the rock. These emerging tau leptons then decay in flight, initiating an extensive air shower that propagates across the valley and reaches the opposite face, where the detection units will be located. \gls{TAMBO} is projected to consist of approximately 5,000 detection units spaced 150 meters apart, which will be responsible for detecting the particle shower. In Fig.~\ref{fig:TAMdetConcp} , the Earth-skimming detection technique behind \gls{TAMBO} is illustrated, as described above.

{\bf \gls{TAMBO} main features.}  Given the Earth-skimming geometry, all potential background particles are absorbed by the rock. This implies that any horizontal air shower detected by \gls{TAMBO} can be unambiguously associated with a tau neutrino of cosmic origin. \gls{TAMBO} is expected to observe six $\nu_\tau$ events over ten years, within an energy window ranging from approximately $0.3$ to $1000$~PeV. It is also projected to attain sub-degree angular resolution, which would help pinpoint the location of neutrino point sources for follow-up observations with other neutrino telescopes. Another key feature is TAMBO’s ability to identify neutrino sources without electromagnetic counterparts.

{\bf \gls{TAMBO} scientific goals.}  These can be summarized as follows: the identification of new cosmic neutrino sources; the measurement of the diffuse flux of cosmogenic neutrinos in the energy range above 100 PeV, which remains inaccessible to current observations; the tracking of muon evolution from the lateral development of air showers to probe the observed muon excess; and last but not least, the testing of several beyond-the-standard model hypotheses at an unprecedented energy frontier.

\textbf{Latin America's participation.}  Peruvian institutions such as the Pontificia Universidad Católica del Perú (PUCP), UTEC, and Universidad del Pacífico are heavily involved. In particular, PUCP and UTEC are responsible for designing the data acquisition (DAQ) and synchronization systems, which are the cornerstones for determining TAMBO’s capability for shower identification and angular resolution—i.e., the experiment’s key measurements. Meanwhile, Universidad del Pacífico is primarily focused on Montecarlo simulations. 
\section{The Giant Radio Array for Neutrino Detection (GRAND)} \label{sec:grand} 

This summary is based on ref. \cite{wp:GRAND_LA}. \\

The Giant Radio Array for Neutrino Detection (GRAND) is a proposed large-scale observatory designed
to discover and study the sources of Ultra-High-Energy Cosmic Rays (UHECRs). \gls{GRAND} will detect the radio signals produced by the interaction of UHE cosmic rays, gamma rays, and neutrinos with the atmosphere. Its setup also enables studies of
fundamental particle physics, cosmology, and radioastronomy. UHECRs are atomic nuclei with energies exceeding $10^{18} $ eV whose origins
are a mystery. The mechanisms whereby they attain their extreme energies are not well understood. 
GRAND’s primary goal is to solve the long-standing mystery concerning the origins of the UHECRs. It will also
serve as a neutrino telescope to study point sources of UHE neutrinos.

{\bf List of interested scientists in the community}  

The \gls{GRAND} Collaboration has 120 members, comprising scientists and engineers from 14 countries and 45 institutions. A Memorandum of Understanding (MoU) is signed by 13 institutions, which coordinates scientific and technical efforts toward building, operating, and extracting scientific information from the experiment.
Argentinian and Brazilian scientists have been active in \gls{GRAND} since the early stages. In particular, Universidade Federal do Rio de Janeiro (UFRJ) and researchers at the Instituto de Física La Plata, ITeDA, and Centro Atómico de Bariloche, are participating in the collaboration. It is expected that Argentinian institutions sign the MoU sometime in the future.

Latin America has a strong tradition in cosmic-ray physics. For example, the prominent role played by many groups in the commissioning and operation of the Pierre Auger Observatory in Argentina. The theoretical framework for GRAND’s main science case (cosmogenic neutrinos) was developed mainly in Brazil by USP (São Paulo) and UFF, CEFET/RJ (Rio de Janeiro). Work on physics simulations, algorithm adaptations, outputs, and futures to meet \gls{GRAND} needs was carried out by the group at the Instituto de Física La Plata in Argentina.

Projects in the \gls{GRAND} Collaboration are organized at a higher level within forums. The CEFET/RJ group is responsible for the Analysis Forum, where activities related to data analysis and methodology development are coordinated through monthly meetings, providing a space for discussion and strategic planning of analysis tasks. The group is also responsible for organizing the analysis sessions at collaboration meetings
and other important gatherings. The UFRJ group is involved in the deployment and commissioning efforts of the GRAND@Auger prototype. A joint effort with Argentinian collaborators allows the deployment and maintenance of detection units.

The main analysis for the GRAND@Auger prototype is led by the UFRJ group, which is responsible for writing the Collaboration paper on the progress of the prototypes, providing data analysis, and coordinating the efforts.
The group is also developing algorithms to estimate the energy of the primary particle from the signals in the antennas. Another activity involves generating shower libraries using the CORSIKA code in parallel with other libraries generated using AIRES at IAP. The Argentinian group is currently leading the development of the event simulation chain and the production of event libraries for end-to-end validation.

{\bf Current status and expected challenges} 

The first stage of the experiment consists of three prototype arrays distributed in three continents, each of them with its corresponding goals: GRANDProto300 in China, GRAND@Auger in Argentina and GRAND@Nançay in France. All the prototype arrays were deployed in 2023 and have been in the commissioning phase since then.

GRANDProto300 is located in the Gobi Desert, Dunhuang province, China. Its construction is being done in phases. The 13-antenna first stage was deployed and is currently taking data. The prototype has already provided measurements that validate the \gls{GRAND} detection principle. The next 80-antenna phase was deployed by the end of 2024, allowing the start of physics-motivated runs. 

An agreement between the \gls{GRAND} and Pierre Auger Collaborations allowed for the deployment of a \gls{GRAND} prototype array, GRAND@Auger, in the Auger Engineering Radio Array (AERA) site, in Malargüe, Argentina. Ten AERA stations were converted into \gls{GRAND} Detection Units. Besides validating the \gls{GRAND} detection principle and allowing for cross-check results with GRANDProto300, GRAND@Auger will allow for the quantification of reconstruction performance through an event-by-event comparison with showers detected by Auger.
A small 4-antenna prototype array, GRAND@Nançay, was also deployed in the Nançay Radio Observatory, France. Its closeness to the European laboratories allows for test benching hardware, noise rejection and triggering under field conditions but the small size of the prototype limits the reliability to detect air showers.

GRANDProto300 (GP300) will comprise an array of 300 antennas, with their deployment currently underway in the Gobi Desert, covering over 200 $km^2$ in Dunhuang Province, China. Its main goal is to demonstrate the feasibility of autonomous detection of nearly horizontal EASs with high efficiency, and to reconstruct the properties of the primary particles with an accuracy similar to other techniques employed for cosmic-ray detection.  

{\bf Computational requirements.} 

One important component of the project is the simulation of cosmic rays. Their propagation from sources
to Earth is simulated with the CRPropa framework. It takes into account the primary physical processes that can potentially influence the propagation of cosmic rays through galactic and extragalactic environments.

\gls{GRAND} will collect huge volumes of data. In the next five years, nearly half a petabyte of data will be generated from the detection units alone. On top of that, roughly the same amount of data will be generated from shower simulations. It is anticipated a data stream of roughly 2.5 petabyte per year. To this end, GRANDlib was developed, a software tool specifically tailored to process such huge datasets 

The Argentinian collaborators in \gls{GRAND} from the Instituto de Fisica La Plata-CONICET are the developers of the ZHAireS simulation suite and use the computing resources made available by the computing center of IN2P3 in Lyon, France as well as the Universidad Nacional de La Plata (Departamento de Física) local cluster.

\begin{figure} [h]
\scalebox{0.7} {\includegraphics{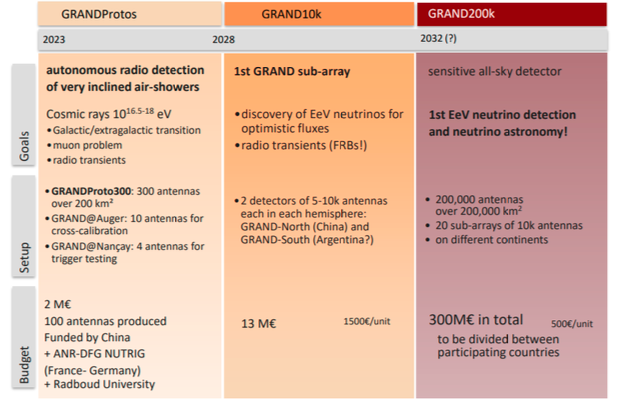} } 
\centering
\caption{\small Timeline of the \gls{GRAND} construction. \cite{connie:2025}.} 
\label{fig:grand-timeline} 
\end{figure}

{\bf Timeline and Costs} 

By 2028, it is expected to finalize the design of the detector units, paving the way for the construction of two
GRAND10k arrays, each containing 10000 units. China and Argentina are the leading candidates for hosting
the bases of GRAND-North and GRAND-South, respectively, ensuring comprehensive sky coverage. In the 2030s, GRAND10k replication is planned, with the goal of eventually creating twenty subarrays that will comprise the entire \gls{GRAND} project. Scaling up production to an industrial level will allow the front-end electronics to
transition to a fully integrated ASIC design. This will allow for reduced costs, improved reliability, and greater
reproducibility of individual units. In addition, the design of each subarray can be tailored to its specific
environment, topography, and scientific objectives. A detailed timeline is shown in \ref{fig:grand-timeline} 

  The hardware cost for the three prototype arrays was \euro 125k, mainly driven by the digitizers \euro100k and
the low-noise amplifiers \euro25k. The costs for GRANDProto300 will be around \euro1.2M, neither the deployment nor the computing costs are included in this value. Science costs are also not included in the estimate. Individual collaborators will apply for funds from their funding agencies to support data analysis, low-level monitoring, and calibration activities. The total cost of GRANDProto300 and GRAND10k will be funded by international collaboration. The exact cost-sharing has not yet been determined. 
\section{The NOvA Experiment} \label{sec:novaLA} 

This summary is based on ref. \cite{wp:novaLA}. \\

With its two functionally identical detectors exposed to neutrinos and antineutrinos from the NuMI beamline, the NOvA long-baseline neutrino experiment has measured neutrino oscillation parameters by observing $\nu_{\mu}  (\bar{\nu} _{\mu} )$ disappearance and $\nu_{e}  (\bar{\nu} _{e} )$ appearance.

\begin{figure} [h]
\centering
\includegraphics[width=0.42\textwidth]{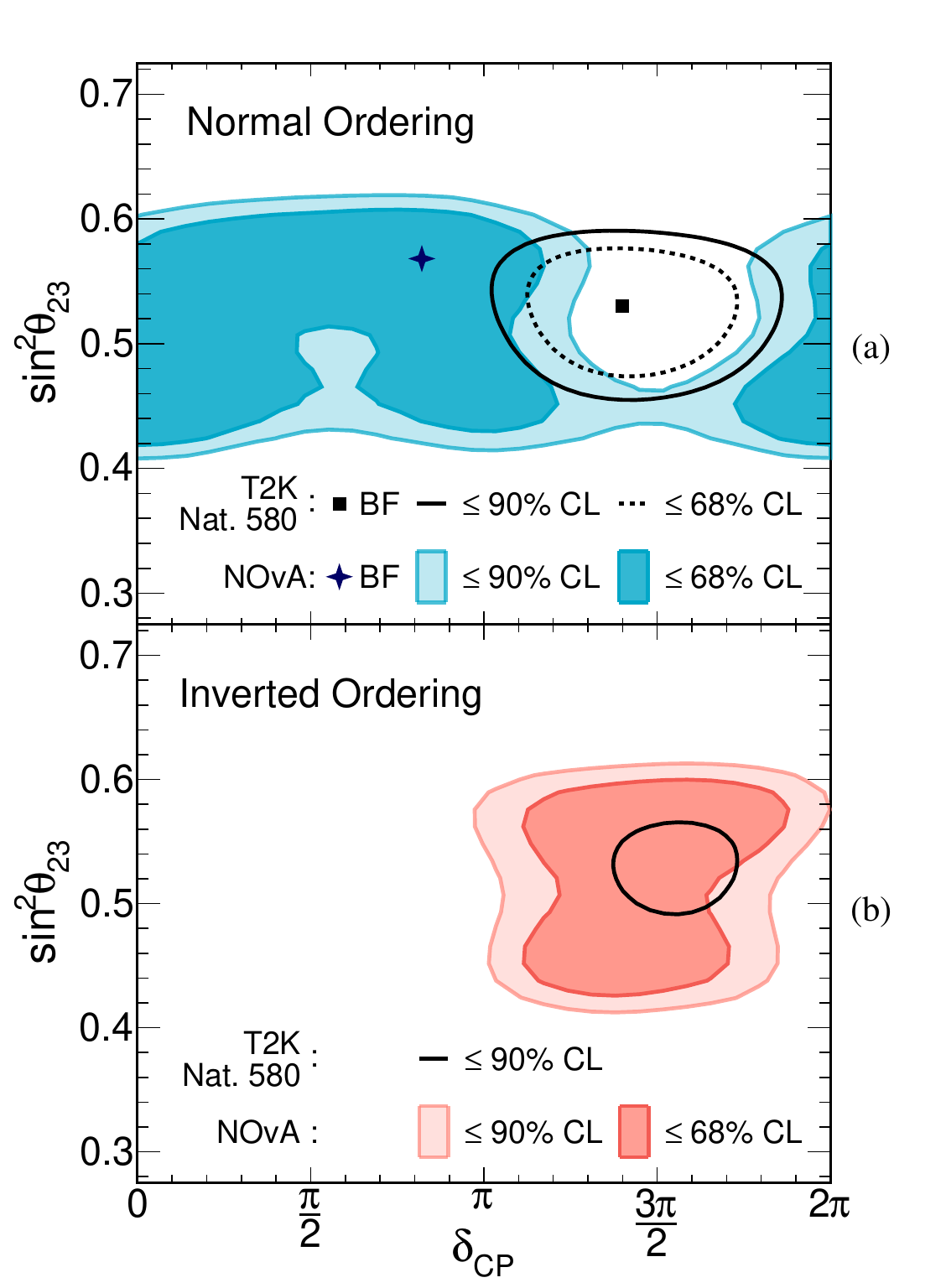} 
\includegraphics[width=0.4\textwidth]{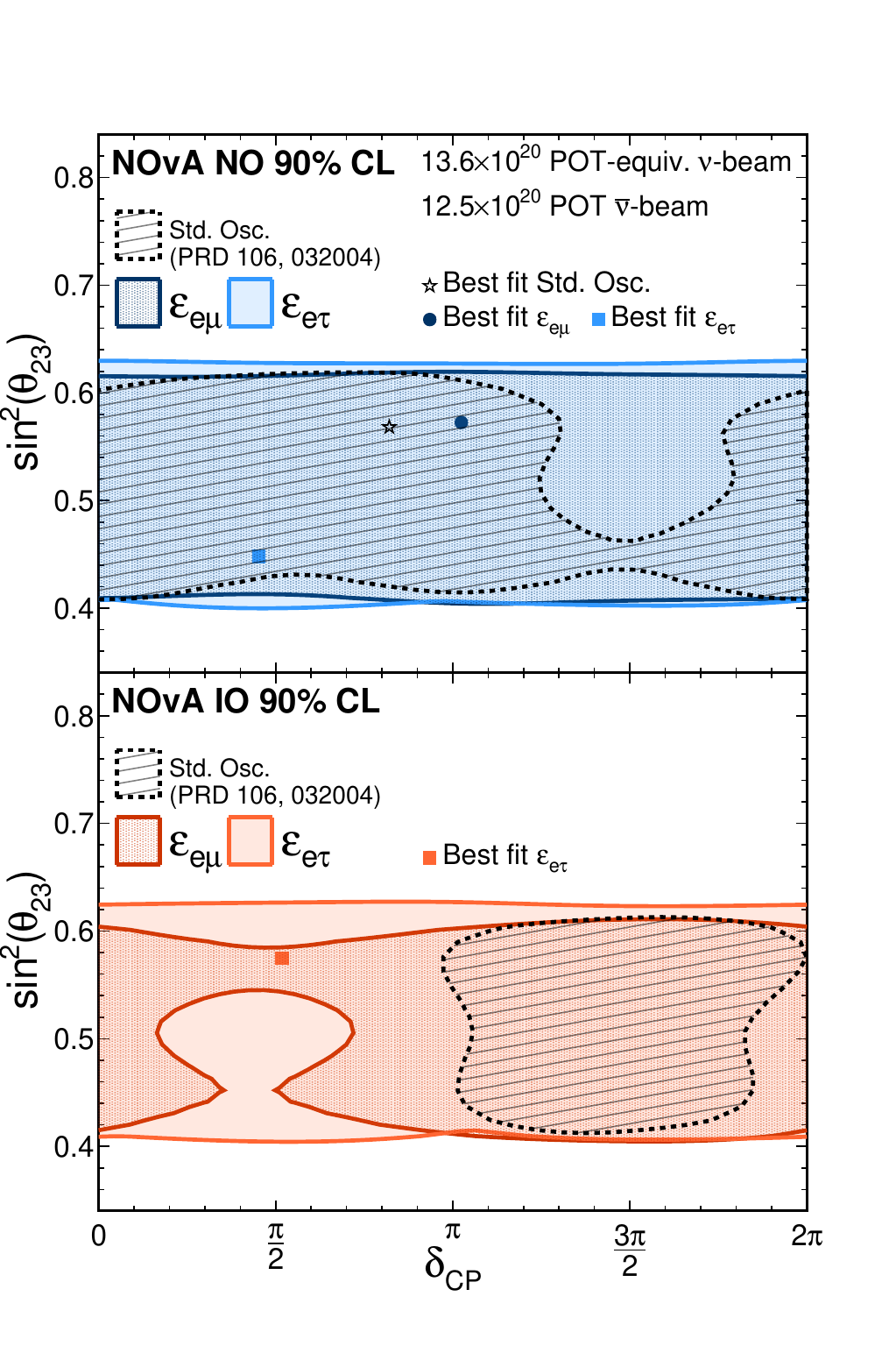} 
\caption{\small 
\emph{Left.}  Allowed regions and best-fit for $\sin^2\theta_{23} $ vs. $\delta_{CP} $ obtained by NOvA \cite{novaOsc:2022}. The results reported by T2K in Ref.~ \cite{t2k:2020}  are also shown for comparison. \emph{Right.}  Allowed regions and best-fit for $\sin^2\theta_{23} $ vs. $\delta_{CP} $ parameter space, for NO (top) and IO (bottom), comparing the NOvA standard oscillation result with cases when NSI sectors ($\varepsilon_{e\mu} $ or $\varepsilon_{e\tau} $) are included \cite{novaNSI:2024}.} 
\label{fig:nova-osc} 
\end{figure} 

After starting data collection in 2014, NOvA has regularly presented improvements in the measurements of the relevant oscillation parameters ($\Delta m_{32} ^2, \sin^2\theta_{23} ,\delta_{\rm{CP} } $), implementing different statistical analysis techniques \cite{novaOsc:2022,novaOsc:2024} , opening the possibility of also obtaining constraints on the mixing angle $\theta_{13} $ \cite{novaOsc:2024} , consistent with the results of the reactor experiments. In addition to neutrino oscillations, thanks to its design capabilities, NOvA has also reported important results on neutrino cross-section measurements, sterile neutrinos, non-standard interactions (NSI), cosmic rays, and multimessenger astronomy \cite{novaXS:2023A,novaXS:2023B,novaXS:2023D,novaXS:2020,novaXS:2019,novaST:2017,novaST:2021,novaNSI:2024,novaATM:2019,novaATM:2021,novaMM:2020A,novaMM:2020B}.

Regarding the neutrino oscillation parameters, the results obtained from analyzing the 2020 dataset are shown in the left panel of Fig.~\ref{fig:nova-osc}, where the T2K results are also presented for comparison. Although the best-fit points for NOvA and T2K \cite{t2k:2020}  are in the Normal Ordering (NO), T2K points to a region that NOvA disfavors, while for the Inverted Ordering (IO), the T2K allowed region is contained within the NOvA one. These results have been interpreted as an indication of possible beyond-standard model physics such as NSI \cite{Denton:2021,Chatterjee:2021}.

\begin{figure} [h]
\centering
\includegraphics[width=0.4\textwidth]{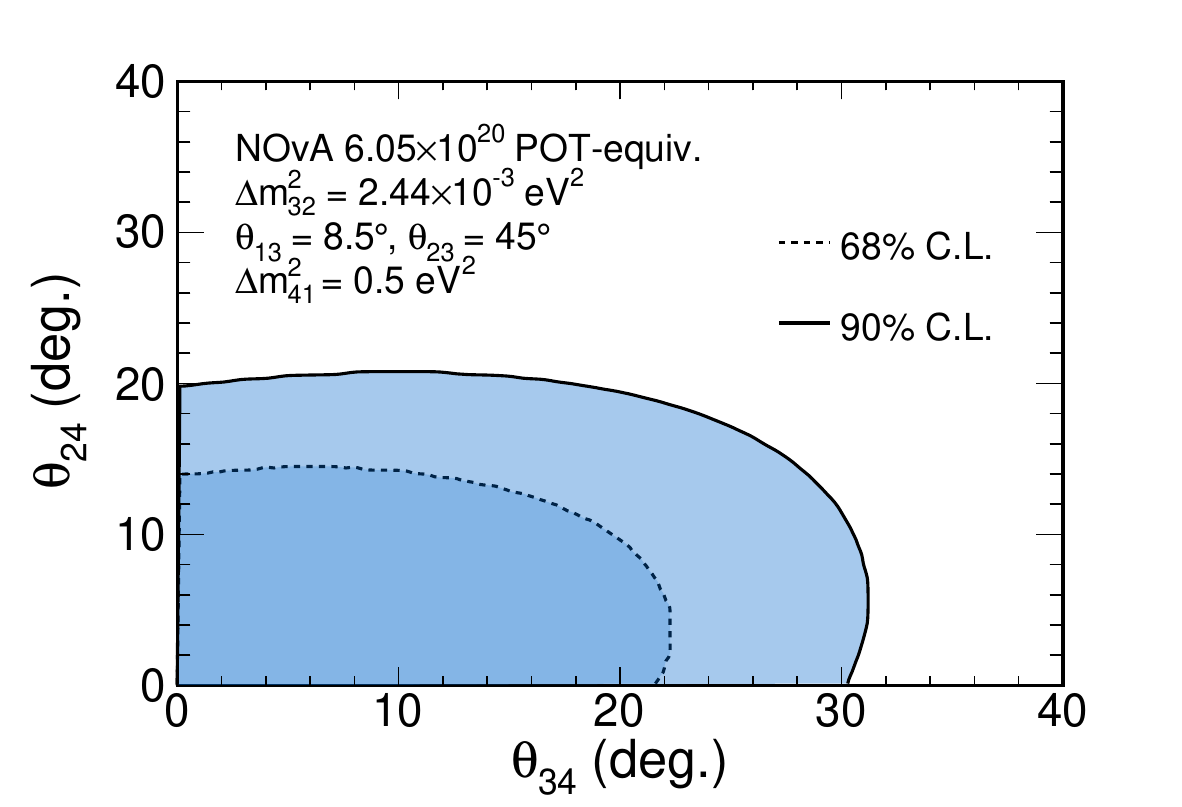} 
\includegraphics[width=0.46\textwidth]{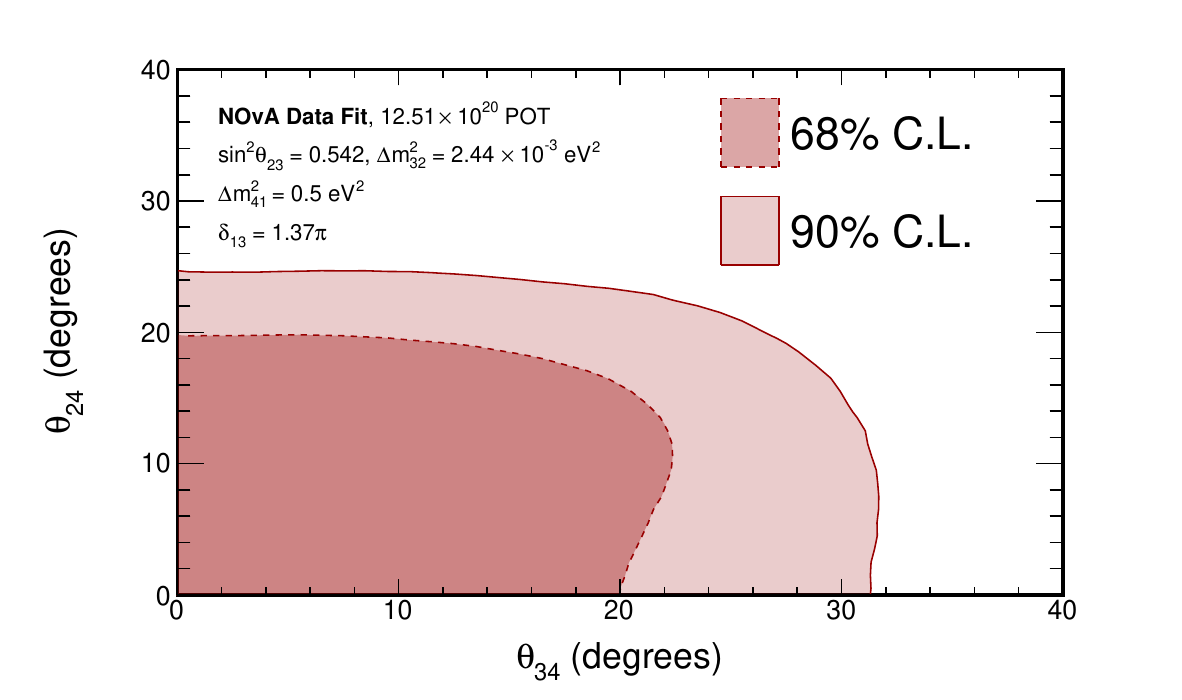} 
\caption{\small 
Allowed regions in parameter space defined by the mixing elements describing neutrino oscillations in a 3+1 model for neutrinos (left) \cite{novaST:2017}  and antineutrinos (right) \cite{novaST:2021}.} 
\label{fig:nova-nus} 
\end{figure}

Precisely, NOvA has recently studied the possibility of Neutral Current-like NSI (NC-NSI) of neutrinos with matter \cite{novaNSI:2024}. Analysis of NOvA FD data within a three-neutrino framework including NSI couplings allowed setting constraints on the NSI parameters considered ($\varepsilon_{e\mu} $ and $\varepsilon_{e\tau} $) finding that, since the NSI parameters add new CP-violation phases, the sensitivity to $\delta_{\rm{CP} } $ is weakened for both neutrino mass orderings, while the constraints on $\sin^2\theta_{23} $ are scarcely modified (right plot in Fig. \ref{fig:nova-osc}).

The search for sterile (anti)neutrinos in NOvA is carried out looking for a reduction in the rate of neutral-current interactions due to the mixing of sterile neutrinos with active neutrinos \cite{novaST:2017,novaST:2021}. So far, no evidence for $\nu_{\mu}  \to \nu_{s} $ oscillation has been observed. However, based on a $3+1$ model NOvA has set constraints on the mixing angles $\theta_{24} $ and $\theta_{34} $, and the difference of the quadratic masses $\Delta m^2_{41} $. Figure \ref{fig:nova-nus}  shows the corresponding contours for both neutrinos (left) and antineutrinos (right). In particular, the NOvA limits from Ref.~\cite{novaST:2021}  are the first set from studying an antineutrino-dominated sample, while new search of active to sterile neutrino transition was recently released, based on a most up-to-date dataset \cite{novaST:2024}.
 
 Within Latin America, three institutions contribute to the NOvA Collaboration, both in hardware (instrumentation) and software (data analysis and simulation): \emph{Universidade Federal de Goias}  (Brazil), \emph{Universidad del Atlantico}  (Colombia) and \emph{Universidad del Magdalena}  (Colombia). Scientists and students from these institutions have been involved in the NOvA experiment, and it is expected that this contribution will extend even after the NOvA data taking process comes to an end by 2027.

\textbf{Timeline.}  It is expected that NuMI will continue operations through 2027, determining the end of the NOvA data collection process. An important goal is to double the antinetrino-mode data in the final dataset, contributing to improve the NOvA results on neutrino oscillations and in a broader and exciting program (sterile neutrinos, NSI, cross-section measurements, cosmic ray physics, etc.). In conclusion, for a few years after the shutdown of the NuMI beam, NOvA will continue to work on physics analyses, serving as an excellent collaboration for students and researchers.
 
\section{The Deep Underground Neutrino Experiment (DUNE)} \label{sec:dune} 

This summary is based on ref. \cite{wp:DUNE_LA, wp:DUNE_Brazil}. \\

\begin{figure} [h]
\scalebox{0.2} {\includegraphics{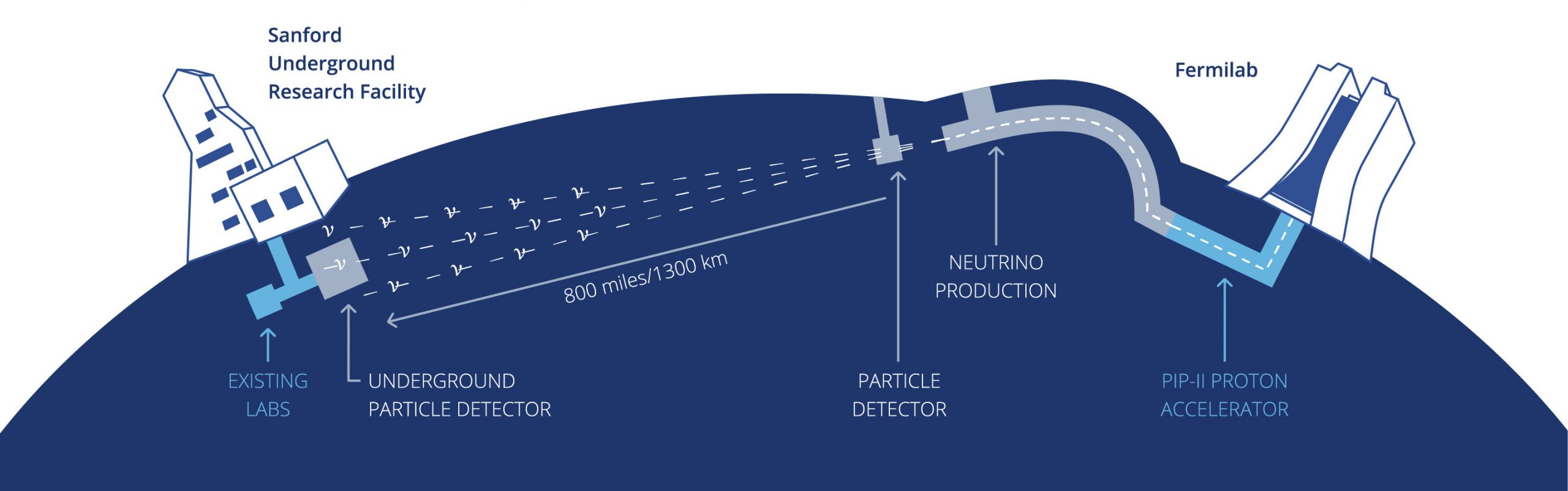} } 
\centering
\caption{\small \gls{DUNE} scheme. Taken from www.fnal.gov} 
\label{fig:dune-scheme} 
\end{figure}

The Deep Underground Neutrino Experiment (DUNE) is one of the most ambitious and promising experiments currently under construction. In the neutrino sector, \gls{DUNE} will look for signs of CP violation, put constraints on the neutrino mass hierarchy, and improve some of the parameters related to neutrino physics. For proton decay, \gls{DUNE} will have unprecedented sensitivity to the decay of a proton into a charged kaon and an antineutrino. However, other nucleon decay modes can also be studied. In addition, \gls{DUNE} has the potential to detect neutrinos produced in supernova events, thereby playing an important role in astrophysics and multi-messenger astronomy. The scheme of the experiment is shown in \ref{fig:dune-scheme} 

\gls{DUNE} has an ancillary program that goes beyond the goals above. Due to the great potential of the experiment, it is possible to study and constrain physics beyond the Standard Model (BSM)~\cite{conhep_dune1}. For instance, \gls{DUNE} could be sensitive to deviations of the neutrino standard interactions, which arise from BSM physics. Moreover, because of the high-intensity proton beam used in the experiment, BSM particles can be produced and detected in the near detector, and these particles could be part of a dark sector or themselves dark-matter candidates. 

\gls{DUNE} is a mega-science project that involves contributions from many countries worldwide. Currently, more than 1000 scientists from more than 30 countries are involved in the design and construction of the experiment. Latin American involvement in the experiment began with the \gls{DUNE} collaboration's adoption of a novel system for collecting photons produced by scintillating light, called ARAPUCA. Since then, the \gls{DUNE} Photon Detection System (PDS), based on the ARAPUCA concept, has been managed by Latin American institutions. The PDS system, in conjunction with the time projection chamber (TPC), will enable event time reconstruction, which is crucial for several of the experiment's scientific goals. 

{\bf List of interested scientists in the community} 

Latin American institutions play a key role in the \gls{DUNE} experiment. The involvement began with the crucial development of the ARAPUCA concept, designed by scientists at UNICAMP, Brazil. This novel technology focuses on detecting scintillation photons produced by the interaction of charged particles in liquid argon. An improved ARAPUCAs, named X-ARAPUCAS, has been tested in the first \gls{DUNE} prototypes at CERN and, in all cases, has demonstrated good performance and high efficiency in the harsh DUNE environment. An evolution of the concept is under testing and used at the second far detector module (vertical drift style), adding the reading and powering of the SiPMs of the ARAPUCAS using optical fibers. 

The Digitization of the X-ARAPUCAS signals is performed at room temperature by electronics boards designed by Colombian engineers in collaboration with Fermilab. The boards known as DAPHNE (Detector Electronic for Acquiring Photons from Neutrinos) will be in charge of the digitization, initial processing, and communication of the PDS with the DAQ system.

The Colombian institutions actively participating in the \gls{DUNE} Collaboration are:
\begin{itemize} 
\item Universidad del Atlántico
\item Universidad EIA 
\item Universidad Antonio Narino (UAN) 
\item Universidad de Antioquia
\item Universidad de Medellín 
\item Universidad del Magdalena. 
\end{itemize} 
Since the last update of the previous white paper (2020), three new institutions have joined the \gls{DUNE} experiment, reflecting an increased regional interest in this project.

Brazilian institutions involved in \gls{DUNE} are: 
\begin{itemize} 
\item Centro Brasileiro de Pesquisas Físicas (CBPF)
\item     Centro de Tecnologia da Informação Renato Archer
\item     Instituto Tecnológico de Aeronáutica
\item     Universidade Estadual de Campinas (UNICAMP)
\item     Universidade Estadual de Feira de Santana (UEFS; State University of Feira de Santana)
\item     Universidade Federal de Alfenas (UNIFAL-MG)
\item     Universidade Federal de Goiás (UFG)
\item     Universidade Federal de São Carlos
\item     Universidade Federal de São Paulo
\item     Universidade Federal do ABC (UFABC)
\item     Universidade Tecnológica Federal do Paraná
\item     Universidade Federal do Rio de Janeiro (UFRJ; University of Brazil)
\item     Universidade Federal Fluminense (UFF)
\item     Sao Paulo, Inst. Tech. Aeronautics (ITA)
\item Centro de pesquisas em materiais avancados e energia (UFSCAR)
\item Laboratorio Nacional de Astrofisica, Itajuba
\end{itemize} 

For a more detailed list of participating persons, see \cite{wp:DUNE_Brazil}. In addition to the institutions listed above, there are scientific groups working at \gls{DUNE} in countries such as Peru and Chile, performing activities that cover both instrumentation and theoretical aspects.

{\bf Current status and expected challenges} 

An ambitious mega-experiment such as \gls{DUNE} poses significant challenges across various areas to achieve its primary physics goals. For example, the experiment must be sensitive to a broad energy range, from MeV-scale supernova neutrinos to higher-energy neutrinos originating from the galactic core. The expected number of particles to be detected varies widely across processes, from only one in proton decay to millions in beam events, which demands an efficient trigger system with internal and external inputs. Reconstruction of liquid argon data is complex and requires novel machine-learning-based algorithms and substantial computational resources. 
For the Photon detector system, precise timing and energy calibration require fast electronics components on a board to meet budget constraints and the physics requirements. The dimmer light from the farthest region of the TPC must be collected; therefore, the PDS includes a light-collection system, light sensors, readout electronics, and monitoring systems. 
The PDS system used in \gls{DUNE} is based on the X-ARAPUCA modules. It is a development of the traditional ARAPUCA. This concept was conceived to reduce losses on the internal surfaces of the ARAPUCA by diminishing the average number of reflections before detection. The idea is to introduce a light-
guide plate in the reflective cavity of a standard ARAPUCA parallel to the filter and with the same dimensions. A dichroic filter with a WLS deposited on its external side to convert the 127 nm photons to longer wavelengths (below the filter cutoff) is the device’s window. Converted photons traverse the filter and, before reaching one of the inner surfaces of the box, find an acrylic plate that acts as WLS and as
a light guide. The acrylic plate has a WLS compound that converts the photon to
a wavelength above the filter cut-off. The new configuration of the X-ARAPUCA
allows for two different detection mechanisms:

\begin{itemize} 
\item the photon when entering the X-ARAPUCA is converted by the WLS deposited on the filter and again converted by the WLS inside the acrylic plate. Still, the plate itself does not guide it. In this case, the photon can be reflected
by the internal walls of the X-ARAPUCA until being detected by the SIPMs
installed at one end of the box, precisely in the same way as a traditional ARAPUCA

\item the photon is converted by the filter and again converted by the WLS inside the plate, the photon can be guided inside the plate by total internal reflection
if it is emitted with an angle greater than the critical angle of the interface
acrylic/LAr. The photon will propagate up to the boundaries of the sheet
where the array of SiPMs eventually detects it. This mechanism rep-
resents the main improvement when compared to a traditional ARAPUCA
\end{itemize} 

As mentioned DAPHNE boards will be responsible for the digitization, initial processing and communication of the PDS with the DAQ system, the boards has been developed as a joint effort of Colombian and Fermilab engineers. Currently, DAPHNE boards are undergoing testing at CERN, where two \gls{DUNE}- scale prototypes were built: one with a horizontal drift configuration and the other with a vertical drift configuration. The performance of the DAPHNE boards meets the expected specifications for signal-to-noise ratio and power consumption. The self-trigger algorithm for operating the board under Protodune conditions is under development and is a joint effort between Colombian institutions (EIA University) and European collaborators. In addition, the proposal for a testbench QA/QC of the future boards has been proposed by Colombian institutions, and its formulation is ongoing, the latest one an initiative of researchers at the Antioquia University. 
More recently, new Colombian institutions have become involved in PDS activities. In particular, the research experimental group at the University of Medellin, in collaboration with the South Dakota School of Mines and Technology (SDSMT), is working on the PDS calibration and monitoring system. In the PDS, a pulsed UV-light monitoring and calibration system is incorporated to determine the gains, crosstalk, linearities, and timing resolution of the photosensors, as well as to monitor the system's response stability over time. The system hardware consists of both warm and cold components; it is expected to contribute to the design and production of new parts for the system, in the production of optical fibers, diffusers, and other parts that will be placed inside and outside the cryostat. Also, the QA/QC of the components produced, assembled, and tested for one of the future \gls{DUNE} modules will play an important role, led by the Colombian institution. Besides the FD1 contributions, it is expected that the University of Medellin takes part in QA/QC on the Power over Fiber technology, a novel application designed to provide power to PDS of the \gls{DUNE} VD in cryogenic environments.

Finally, the University of Atlantico is participating actively in the design of the calibration system for the Muon Spectrometer (TMS). The TMS is an 850 ton magnetized steel range stack that will play different roles in the \gls{DUNE} experiment. The primary role of TMS is to determine the kinetic energy and charge of muons exiting the back of ND-LAr. This lepton information is a crucial component of event reconstruction in ND-LAr for a wide range of neutrino interactions relevant to long-baseline oscillation analysis. This ND-LAr-TMS partnership is needed for the experiment to reach phase-1 physics goals.

To achieve the full potential of \gls{DUNE} with increasing statistics, it is anticipated that TMS will be replaced by a More Capable Near Detector (MCND). The MCND will both measure muons exiting ND-LAr, as with TMS, and perform detailed neutrino-interaction measurements that will help constrain and reduce systematic errors. A possible scenario, as put forth in the \gls{DUNE} near detector (ND) CDR, is one where TMS is replaced by a ND-GAr, which is a high-pressure gaseous argon time projection chamber (TPC) in a magnetic field

\gls{DUNE} is the first science project of this scale in the USA that will be built under an international collaboration scheme. This opens a series of opportunities for countries worldwide to participate in the design and construction of specific components, gaining experience and strengthening relationships with local industry. In particular, Colombian institutions are planning to produce an unspecified number of DAPHNE boards, as well as testbench boards for DAPHNE testing. It faces numerous challenges, including a funding shortfall, difficulties with export and import, limited access to information and infrastructure, and high personnel turnover; nevertheless, Colombian institutions remain committed to participating in the \gls{DUNE} experiment.

Participation in the \gls{DUNE} experiment presents an unprecedented opportunity to make a significant contribution of great responsibility at the regional level and to understand the challenges associated with community work. We must identify mechanisms to link basic science research to the productive and industrial sectors through concrete strategies that assign responsibilities to both sectors. Until now, the X-ARAPUCA and DAPHNE boards represent the most relevant commitment in hardware of
the Brazilian and Latin American groups in the \gls{DUNE} experiment, and it is a
major contribution since the PDS is one of the most important subsystems of the
experiment.

Having a long-term budget is a general objective for the scientific community to assume long-term responsibilities in extensive experiments without the migration of qualified personnel. Enhancing communication channels among groups involved in the same experiment is mandatory for the allocation and management of available resources.

{\bf Timeline and Costs} 

Currently, the second \gls{DUNE} scale prototype (ProtoDUNE-VD) located at CERN is ready for commissioning. The previous year a successful  operation of ProtoDUNE-HD was reported, allowing for a fine-tuning of the technologies to be used and a test of the hardware performance under real conditions.
The expected general timeline is:
\begin{itemize} 
\item 2024: Testing of DAPHNE boards at ProtoDUNE-HD 
\item 2025: Testing of DAPHNE boards at ProtoDUNE-VD 
\item 2026: Start of hardware production
\item 2028:  Beginning of installation of first FD module
\item 2029:  Beginning of installation of second FD module 
\end{itemize} 

The final design of the \gls{DUNE} experiment is still under discussion. In particular, the Photon detection system is planned to have a total production cost of 17 million U.S dollars. The digitalization electronics of the photon detector system are calculated to be on the order of 400000 US dollars. A common fund is defined as a fee corresponding to \$3,000 per year per member. 

In May 2024 Brazilian funding agencies FAPESP and FINEP have approved a contribution of US\$ 36
million mainly towards the construction of a liquid argon purification system for DUNE. In addition,
FAPESP has approved in 2024 a contribution of US\$ 5 million towards the
construction of the photon detection system for the DUNE far detector. Production and assembly will
be performed in Brazil.

\section{The Jiangmen Underground Neutrino Observatory (JUNO)} \label{sec:junoLA} 

This summary is based on ref. \cite{wp:junoLA}. \\

\textbf{Overview.}  With a 20 kton liquid scintillator detector, currently under construction in southern China, 52.5 km from eight nuclear reactor cores in the Taishan and Yangjiang nuclear power plants, \gls{JUNO} expects to reach an unprecedented energy resolution and exceptional transparency of the liquid scintillator, aiming to achieve a median sensitivity significance of $3\sigma$ for the determination of the neutrino mass ordering with an exposure of approximately 6.5 years $\times$ 26.6 GW thermal power. Its detection system includes 17,612 20-inch photomultiplier tubes and a secondary array of 25,600 3-inch photomultiplier tubes, implementing a novel ``double calorimetry” technique to enhance energy resolution and control systematic uncertainties. 

The physics objectives of \gls{JUNO} include the precision measurement of neutrino oscillation parameters, detection of galactic supernova-burst neutrinos, and observation of the diffuse supernova neutrino background. As reported in \cite{wp:junoLA} , Latin American institutions have played an important role in the \gls{JUNO} project, with contributions that include hardware and software.

The physics contribution of reactor neutrino experiments reached a maximum with Daya Bay, Double Chooz, and RENO, successfully measuring the last mixing angle, $\theta_{13} $. Notably, Latin American institutions have been actively participating in this research area since the 2000s, contributing to the Daya Bay \cite{dayabay:2012}  and Double Chooz experiments \cite{dooblechooz:2012}. 
\\

\textbf{Objectives.}  The \gls{JUNO} observatory has several important physical objectives, including \cite{juno:2022} :
\begin{itemize} 
\item Neutrino Mass Ordering (NMO) \cite{juno:2024}.
\item Precision measurement of oscillation parameters \cite{juno:2022osc}.
\item Detection of Supernova Neutrino Burst: With its unique combination of size, energy resolution, and timing capability, \gls{JUNO} can play a crucial role in facilitating follow-up multi-messenger observations of the next galactic or nearby extragalactic core-collapse supernova \cite{juno:2024}.
\item Diffuse Supernova Neutrino Background (DSNB). \gls{JUNO} can detect the DSNB with a 3$\sigma$ significance within three years of data collection and surpass 5$\sigma$ significance after ten years \cite{juno:2022dsn}.
\item Solar Neutrinos. \gls{JUNO} may improve the best current measurements on $^{7} $Be, pep, and CNO solar neutrino fluxes \cite{juno:2023sun}. 
\item Atmospheric neutrinos. \gls{JUNO} is able to reconstruct the energy spectrum of atmospheric neutrinos in the energy range [100 MeV - 10 GeV], offering valuable data for the study of neutrino oscillations, and providing complementary information on the NMO \cite{juno:2021atm,juno:2024atm}.
\item Geo-neutrinos: The detection of about 300-500 geo-neutrinos per year is expected, resulting in an expected precision of 8\% in 10 years \cite{juno:2024geo}.
\item Nucleon decay: \gls{JUNO} has the potential to provide important information on proton \cite{juno:2023pdk}  and neutron decay \cite{juno:2024ndk}  on decay channels such as $P \to K + \nu$ or neutron invisible decay modes.
\item Other physics: \gls{JUNO} has the potential to address a wide range of open questions \cite{juno:2022} , including the Majorana nature of neutrinos, the existence of Dark Matter (DM), light sterile neutrinos, non-unitarity of the neutrino mixing matrix, and non-standard neutrino interaction.
\end{itemize} 

\textbf{Methodology.}  The \gls{JUNO} detector is located in an underground laboratory in Guangdong Province, China. It consists of a 20 kton liquid scintillator target designed to achieve an energy resolution better than 3\% at 1 MeV \cite{juno:2024ene} , crucial for NMO determination. Thanks to its proximity to the Yangjiang (six reactor cores with a thermal power of 2.9 GW) and Taishan (two reactor cores each generating 4.6 GW) Nuclear Power Plants (NPP) ($\sim$52.5 km away), the \gls{JUNO} detector will observe a high flux of reactor antineutrino events. The experimental site is located under Dashi Hill, providing a vertical overburden of approximately 650 m, reducing the cosmic muon rate.

\begin{figure} [h]
\centering
\includegraphics[width=0.8\textwidth]{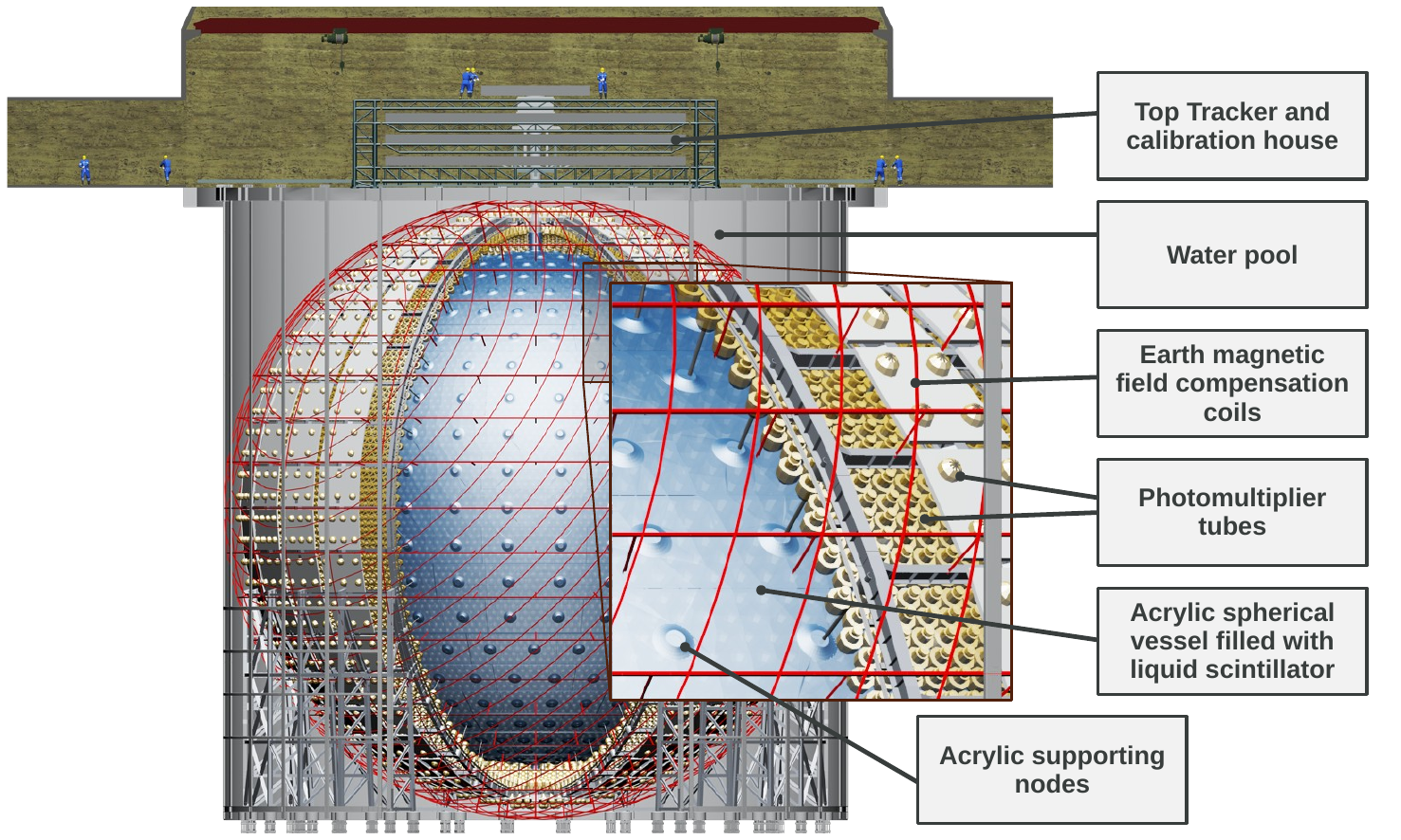} 
\caption{\small 
Schematic view of the \gls{JUNO} detector \cite{juno:2022osc}.} 
\label{fig:juno-det} 
\end{figure} 

Figure \ref{fig:juno-det}  shows the \gls{JUNO} detector system which is made up of four main components:
\begin{itemize} 
\item The Central Detector (CD) is a large acrylic sphere filled with liquid scintillator, surrounded by large 20-inch photomultiplier tubes (LPMTs) and small 3-inch photomultiplier tubes (SPMTs), providing a total photo-cathode coverage of 78\%. The acrylic vessel is supported by a stainless steel (SS) structure using connecting bars and photomultiplier tubes (PMTs) are installed on the inner surface of the SS structure.
\item The water Cherenkov detector, surrounding the CD, is equipped with PMTs to detect Cherenkov light from cosmic muons, acting as a veto to reduce background noise. The water pool also provides shielding against external radioactivity and helps minimize the impact of the Earth’s magnetic field on the PMTs through compensation coils.
\item A Top Tracker (TT) located on top of the water pool is a plastic scintillator array designed to accurately measure muon tracks to suppress accidental backgrounds caused by high radioactivity from the surrounding rock.
\item A chimney connects the CD to the outside, allowing for calibration operations. The calibration systems are managed from the Calibration House, which includes special shielding and a muon detector to ensure accurate measurements and maintain the integrity of the detector performance.
\end{itemize} 

\textbf{Latin America Involvement.}  
\begin{itemize} 
\item Chile. Rafael Herrera, Pablo Walker, Ignacio Jeria, Giancarlo Troni, Angel Abusleme, Sergey Kuleshov - PUC (Pontificia Universidad Católica de Chile), SAPHIR and UCI (University of California, Irvine) led the design and manufacture of the HVS and the UWBs in partnership with France and China. Those hardware responsibilities have now been completed. The UCI group also contributes to the leadership of the experiment, actively participates in preparations for physics with reactor antineutrinos, co-leads studies on using the sPMT system for LPMT calibration, and is involved in the installation and commissioning of the sPMT system.

PUC has left the collaboration following the completion of its hardware responsibilities, but SAPHIR remains active and is increasingly involved in analysis processes, focusing on reactor antineutrinos. Chile's funding originates from ANID, the Chilean National Agency of Research and Development.

\item Brazil. Hiroshi Nunokawa and Pietro Chimenti - Discussions, writing, edition, and revision of several papers for the Collaboration. Also studying the physics potential of \gls{JUNO} from a phenomenological perspective and testing non-standard neutrino properties as \gls{JUNO} begins collecting real data. Contributions also to the initial simulation of the sPMT system and studies of the sensitivity for precision measurements and neutrino mass ordering.

\gls{JUNO} has received partial funding from CNPq through
research grants coordinated by Prof. H. Nunokawa and Prof. P. Chimenti. 
\end{itemize}  
\section{Particle Physics Experiments with Low-Temperature Detectors at the University of S\~ao Paulo} \label{sec:qsltdBrazil} 

This summary is based on ref. \cite{wp:qsltd}. \\

Neutrinoless double beta ($0\nu\beta\beta$) decay has gained significant attention in the last decade, with several experiments employing different techniques currently in operation or in development. In particular, those with cryogenic calorimeters operated at milli-Kelvin temperatures have demonstrated unprecedented efficiency and sensitivity. Among the most promising experiments for elucidating the nature of neutrinos through $0\nu\beta\beta$ decay are the CUORE/CUPID ton-scale calorimeter experiment.

The Quantum Sensors and Low-Temperature Detectors (QSLTD) Group at the University of S\~ao Paulo (USP) aims to participate in these projects, working with other interested groups to develop low-temperature detectors for dark matter, neutrinos, and other rare event searches.

One of the objectives of the QSLTD group is to develop novel cryogenic calorimeters that can advance the state-of-the-art of particle physics, looking for contributing to the search of answering important open questions in the Standard Model of Particle Physics (SM) such as the nature of dark matter and the origin of the neutrino mass.

To do so, the QSLTD group is expected to contribute to the $0\nu\beta\beta$ decay searches on $^{130} $Te and $^{100} $Mo through the Cryogenic Underground Observatory for Rare Events (CUORE) experiment and its planned upgrade called CUPID (CUORE Upgrade with Particle Identification).

\begin{figure} [h]
\centering
\includegraphics[width=0.7\textwidth]{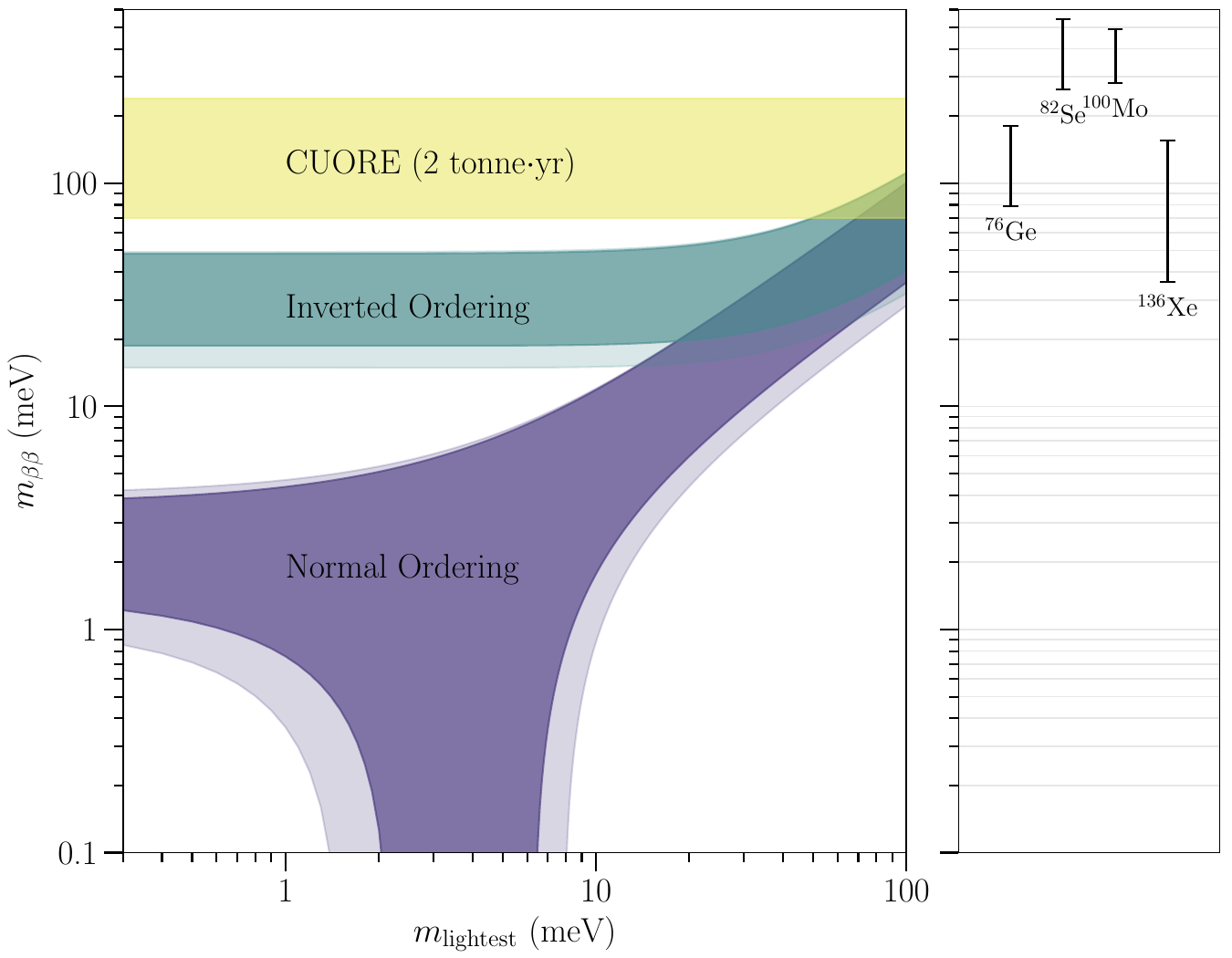} 
\caption{\small Allowed parameter space as a function of the lightest neutrino mass in the case of inverted (normal) ordering is shown in green (purple). The lighter shaded areas correspond to the $3\sigma$ uncertainties on the oscillation parameters. The yellow band corresponds to the limit obtained from the analysis in Ref. \cite{qsltd:cuore}.} 
\label{fig:cuore-mbb} 
\end{figure} 

\textbf{Double Beta Decay Searches.}  One of the most puzzling questions in neutrino physics is their nature: are neutrinos Dirac or Majorana particles? In order to answer this question, different experimental techniques have been employed to search for $0\nu\beta\beta$ decay. It is well established that its observation would provide evidence of a Majorana neutrino, together with a measurement of its mass \cite{qsltd:0nbb}.

Most of the recent experimental advances in $0\nu\beta\beta$ have been obtained by employing high-purity germanium detectors (LEGEND experiment), but also with large calorimeters at mK temperatures (CUORE experiment). CUORE, using 988 TeO$_2$ crystals of $\sim$750 g each, set a limit of $2.2\times10^{25} $ years for $^{130} $Te, which corresponds to a limit for the neutrino mass of $m_{\beta\beta}  < 70-240$ \cite{qsltd:cuore2015,qsltd:cuore2022,qsltd:cuore}. CUPID, its next upgrade, aims to increase the sensitivity by adding light detectors to the crystals, and use scintillating crystals of Li$_2$MoO$_4$. These new detectors have an energy resolution of of $\mathcal{O} $(7 keV) at the Q-value of 3034 keV. This upgrade will also improve the discrimination efficiency and background suppression. CUORE have placed stringent . It is expected that the CUPID experiment will be able to probe the full inverted hierarchy space, \cite{qsltd:cupid2022}. Figure \ref{fig:cuore-mbb}  shows the neutrino mass already probed by CUORE.

The goal of the QSLTD group is to test and produce new detectors that can improve the capability of detection $0\nu\beta\beta$ decay in CUPID and other rare event searches, like sterile neutrinos, solar axions and majorons. With these main goals, the group will have two main lines of research: novel low-temperature detectors and sensors; low-/high-level analysis and machine learning methods.

\textbf{International Collaboration.}  The QSLTD group at \emph{Universidade de S\~ao Paulo}  was born from an international collaboration of researchers from different institutions around the world. The plan is to keep an active exchange of permanent and non-permanent researchers from Germany, Italy, United States, and Brazil that are members of the CRESST, CUORE and CUPID experiments.
In particular, the group will keep a strong and close relationship with the \emph{Max-Planck-Institut f\"ur Physik}  in Munich, Germany.

\textbf{Perspectives.}  In the next 20 years it will be possible to measure neutrinoless double beta decay. The QSLTD group expects to contribute in the searches of this phenomena. The group will work in close collaboration with researchers around the world by receiving them in S\~ao Paulo or sending researchers from S\~ao Paulo to other centers. 
\section{Neutrino Physics at the Instituto de Física at UNAM, Mexico} \label{sec:ifunam} 

At \emph{Instituto de Física, UNAM}  (IFUNAM), in Mexico, a number of researchers are involved 
in neutrino physics through their participation in coherent elastic neutrino-nucleus scattering experiments for reactor neutrinos. Their work on neutrino phenomenology has focused on CE$\nu$NS \cite{ifunam1}, and on key studies on the physics reach and sensitivity to new physics of the SBC experiment, a low-threshold scintillating argon bubble chamber designed to operate at nuclear reactors \cite{ifunam2}. 

Furthermore, they have contributed to recent analyses of reactor antineutrino data from the CONUS+ experiment. Complementing these activities, some researchers are working on developing a gadolinium-loaded liquid scintillator detector with two 100-liter modules near the ININ reactor to characterize background, assessing reactor neutrino detection via inverse beta decay, and exploring the potential for deploying CE$\nu$NS experiments.

There are also contributions directed to addressing the question of the origin of neutrino masses and their connection to new physics at the electroweak scale, particularly with the dark sector. In the model building for neutrino mass generations, radiative neutrino mass mechanisms, inverse, and linear seesaw realizations are used in conjunction with flavor- symmetry- based frameworks, leading to predictions for CP violation and lepton number and flavor violations. An essential aspect of this work is the link to low-energy precision probes, including coherent elastic neutrino–nucleus scattering (CE$\nu$NS) \cite{ifunam3,ifunam4,ifunam5}. 

Active collaboration characterizes the work in Mexico, including researchers from IFUNAM, CINVESTAV, Universities in Zacatecas and Pachuca in Mexico, and in Antofagasta, Chile.

\section{Colombian Network on High Energy Physics: Input on Neutrino Physics} \label{sec:conhep} 

This summary is based on ref. \cite{wp:conhep}. \\

The Colombian community presents some of the activities related to high energy physics that drive its research. In addition to synergistic efforts done in neutrino physics, paramount progress on strengthening the community through the workshops and meetings developed during the last decade has been made. This section summarizes the current status and interests of the different groups working on neutrino physics in Colombia and their future perspectives (see \cite{wp:conhep}  for additional details).
\\
Among other important aspects, the reseach of an important number of members of CONHEP aims to simulate events observed at neutrino experiments including neutrino oscillations, neutrinoless double beta decay, direct measurement of the absolute neutrino mass, CE$\nu$NS, etc., as a calibration tool of the theoretical inputs and the analysis framework needed, for each observable. Also, efforts are also directed to probe well-motivated BSM scenarios, resulting, for instance, from models tackling most of the open problems, using the observables detailed above.

Working on neutrino oscillations, researchers carry out analytical and numerical calculations of fluxes of neutrinos at the source, neutrino cross sections at the detectors, and mainly neutrino oscillation probabilities in vacuum and matter. Experimental information like detector energy resolution or the smearing matrices is necessary to account for the mapping between true vs. reconstructed neutrino energy. This is implemented in calculating the expected (un)oscillated events at near and far detectors (the same procedure is followed for the calculation of the background events). Then, systematic uncertainties are included in the analysis and a statistical procedure should be followed, usually maximizing some likelihood function, in order to produce regions of confidence for the free parameters considered in the analysis. When observational data are available and included in the analysis, the described procedure corresponds to a statistical data analysis approach that is, in general, frequentist. For forecast studies, when there is no experimental information, it is also possible to simulate the data following the procedure for the expected events. In this case, some set of `true parameters' have to be assumed in advance, and the results have to be checked for all possible values that these true parameters might have. Statistical fluctuations can be added to the fake data in order to produce realistic sensitivity regions. The whole procedure can also be adopted for BSM studies, affecting any of the stages production, propagation, and/or detection. 

\textbf{NuCo (\emph{Neutrinos en Colombia} ) Workshop.}  Colombian researchers who participate and develop projects on various topics of neutrino physics gather to communicate their results and to foster the creation of new and strengthening the existing national collaborations. A large number of researchers contribute to this community, working on areas like neutrino oscillations, neutrino masses and interactions, BSM physics (e.g., sterile neutrinos, non-standard interactions), from theoretical, experimental, and phenomenological approaches. In 2023 \href{https://indico.cern.ch/e/nuco2023} {NuCo}  took place at \emph{Universidad Antonio Nariño}  (Santa Marta, Magdalena). Since 2024, we decided to join efforts and increase the impact of this activity and organized the new \href{https://indico.cern.ch/event/NeMO-C_2024} {NeMO-C}  (\emph{Neutrinos y Materia Oscura en Colombia} ) at \emph{Universidad de Medellín} , a workshop to talk about neutrinos and DM, considering the large relationship between these areas, as well as the active participation of many researchers in both of them.
\\

\textbf{Current status and expected challenges}  The following are the some contributions by Colombian scientists to different efforts studying neutrino physics.
\begin{enumerate} 
    \item The three active neutrino oscillation framework
    \begin{itemize} 
        \item Although current experiments like T2K, NOvA (at which the Colombian community contributes~\cite{conhep_nova1,conhep_nova2,conhep_nova3} ) provide some preference for certain CP-violating values, there is not a concluding measurement. In addition, there is no conclusive information on the neutrino mass ordering~\cite{conhep_nmo}. For this reason, future neutrino facilities like JUNO, \gls{DUNE} and \gls{Hyper-K} are being planned to measure the current unknowns and to improve the precision of the neutrino oscillation parameters already measured. 
    \end{itemize} 
    \item Beyond the three-active neutrino oscillation framework
    \begin{itemize} 
    \item Neutrinos may have interactions that are not considered in the SM. For phenomenological studies, a general framework that generalises the Fermi four-fermion interactions have been considered increasing the couplings to neutrinos in a week interaction relative to the Fermi SM interaction. This is the so-called Nonstandard Neutrino Interactions (NSI) which can be a result of an effective field theory framework (SMEFT). The NSI can modify any of the three stages involved in a neutrino oscillation experiment (production, propagation, and/or detection). There are plenty of phenomenological works constraining the magnitude of these new couplings. Also, NSI couplings provide new sources of CP violation. It is therefore interesting to establish the robustness of the six neutrino oscillation parameters under the NSI and whether the undetermined Dirac CP violation can have a nonstandard source~\cite{conhep_nsi}. Let us stress here that Colombian researchers took part of the recent data analysis by the NOvA Collaboration regarding, precisely, the search for evidences of NSI CP-violation \cite{novaNSI:2024}.
    \item Sterile neutrinos is another recurrent subject within the neutrino community. In general, the mass of the extra neutrinos can span different energy scales (from eV to TeV) therefore having unequal phenomenological consequences. In particular, motivated by several anomalies, the hypothesis of an sterile neutrino at the eV scale is under investigation. Searches for a light sterile neutrino, in short and long baseline neutrino experiments, have been performed without any positive result, therefore reducing the parameter space for the simplest 3+1 scheme. It is expected that a definitive test of this hypothesis would come from the results of future facilities like the short-baseline neutrino program at FERMILAB. Synergies with long-baseline facilities, like DUNE, or from searches at different neutrino sources such as the case of atmospheric neutrino experiments like IceCUBE deep-core, will be also relevant. Efforts on this area have been done by the researchers from the Colombian community, for instance, looking for the possible constrains that could be obtained by the \gls{DUNE} experiment to the additional mixing angles that appear in the case of a 3+1-neutrinos framework, through the observation of neutral current events at the far detector (see Ref.~\cite{conhep_sterile}  and references there included). 
    \item Right handed neutrinos, which are singlets, are allowed to travel in Extra Dimensions. In these models, this right handed neutrinos couple to the SM three-active neutrinos only through Yukawa couplings, which exist to generate Dirac neutrino masses for the active neutrinos. It is assumed that one of the extradimension is large than the others such that the compatification is asymmetric and leads to potential signatures at neutrino oscillation experiments. In Large Extra Dimensions (LED), after compactification, at low energies in four dimensions, neutrinos appear as a tower on infinite modes that modify the three-active neutrino oscillation pattern in a unique way. Therefore, as a consequence, it is expected an spectral distortion of the neutrino energy events that can be used to test this LED hypothesis at both, near and far detectors simultaneously. The Colombian neutrino community have been participating on LED forecast studies at DUNE~\cite{conhep_dune1,conhep_dune2}. Posterior studies in the subject of this kind of searches at neutrino oscillation experiments appear in Refs.~\cite{conhep_sbl,conhep_led}.
 \end{itemize} 
\end{enumerate} 
 
\section{Theoretical Physics Group (GI-FT) at Escuela Politécnica Nacional} \label{sec:GI-FT} 

\textbf{Overview.}  Ecuador’s higher education system originated in the 16th century with the establishment of religious institutions. Over time, particularly by the 20th century, it evolved into a more inclusive academic model. However, the proliferation of private universities since the 1990s has raised concerns regarding educational quality. Throughout this period, the Escuela Politécnica Nacional (EPN) has maintained a leading role in scientific education and research 
\cite{Beltran2021}.

\textbf{Theoretical Physics Group (GI-FT) 
description.}  The GI-FT 
at EPN is a research initiative created by the Physics Department, bringing together Ecuadorian researchers working in the areas of dark matter, cosmology, high-energy theory, astronomy and astrophysics, and instrumentation and computing.

\textbf{Main objectives of GI-FT.}  Its goals include: promoting the study of theoretical high-energy physics and gravitation in Ecuador; encouraging the use of open-access data from high-energy observatories; strengthening collaboration with international institutions such as the IAU and HECAP-ICT; and fostering regional scientific events and joint research initiatives. The group currently comprises seven faculty members from EPN and six collaborators from other Ecuadorian universities, with research spanning quantum field theory, general relativity, astrophysics, and complex systems.

\textbf{GI-FT’s research.}  The research of the GI-FT  focuses on the mathematical structure of topological field theories, particularly the non-Abelian Chern-Simons-Wong theory. These models describe link invariants using perturbative techniques \cite{Anda2023,Leal2007}.

The group is also actively involved in gamma-ray astronomy. Their studies analyze peculiar gamma-ray bursts (GRBs), such as GRB070724A and GRB090426, observed by the \textit{Swift}  telescope. Temporal and spectral analyses of these events reveal deviations from standard GRB classifications, providing new insights into these highly energetic phenomena~\cite{Zhang2018,Lien2016}. In parallel, the group seeks to integrate machine learning into GRB analysis and train students in modern data science techniques.

\textbf{Current and upcoming activities.}  The GI-FT has hosted the National Physics Meeting annually since 1989. In 2025, the group will co-organize the \textbf{46th International School for Young Astronomers (ISYA)}. Looking ahead, they also plan to establish a computational laboratory for complex systems to enhance interdisciplinary collaboration.

\section{Conclusions} \label{sec:conclusions} 
The participation and contribution of Latin American scientists on different topics of neutrino physics keep growing steadily and strongly. Robust and stable collaborations have been established and/or strengthened within the region and around the world, demonstrating the relevance and impact of this community on advances in our understanding of the physics of neutrinos. 

\noindent In addition to participating in large international collaborations such as DUNE, GRAND, Hyper-K, JUNO, and NOvA, experimental facilities are being developed and proposed to be based in LA ($\nu$-Angra, CONNIE, TAMBO), covering a variety of topics of neutrino physics, including the evolution of novel experimental techniques (quantum sensors and low-temperature detectors).

\noindent Furthermore, an important number of LA researchers are involved in theoretical and phenomenological aspects of neutrino physics, supporting experimental projects (contributing to data analysis and simulations, for instance) and studying models of particle physics to address relevant questions (such as neutrino nature and the origin of the neutrino mass). 

\noindent Overall, scientists in and from Latin America have built a strong, integrated community working on neutrino physics, promoting not only the progress in different research projects, but also a constant communication through frequent regional and national meetings, and supporting educational and outreach activities.

\bibliographystyle{unsrt} 
\bibliography{neutrinos/neutrinos}

\chapter{Electroweak \& Strong Interactions, Higgs Physics, CP \& Flavour Physics and BSM}\label{chapt:colliders}

\section{Introduction} \label{sec:intro} 

\begin{figure} 
     \centering
    \includegraphics[width=0.8\textwidth]{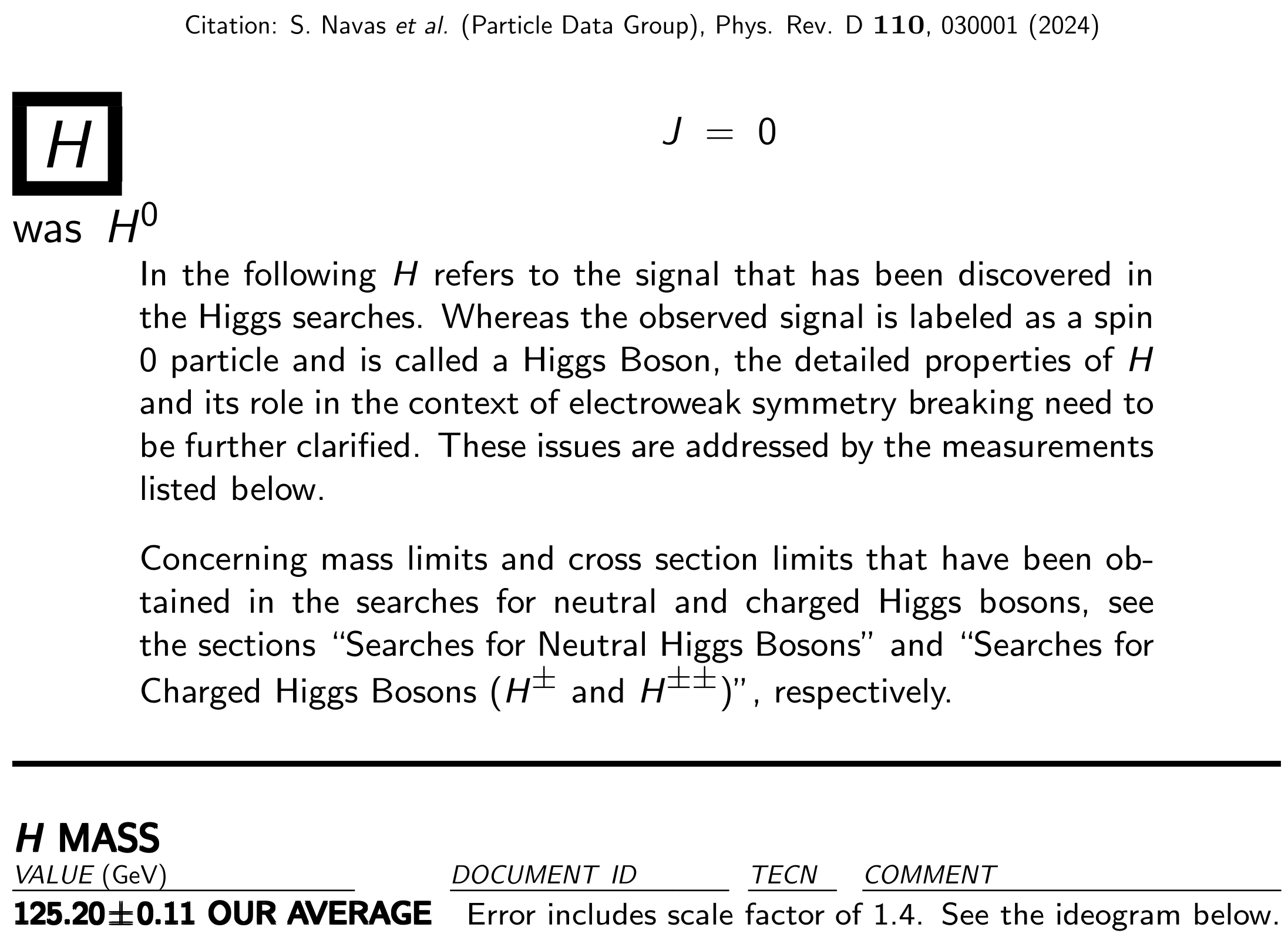} 
     \caption{First page of the Higgs boson entry in the 2024 PDG~\cite{ParticleDataGroup:2024cfk} .} 
     \label{fig:HiggsPDG} 
 \end{figure}

Figure \ref{fig:HiggsPDG}  shows the first lines of the Higgs boson entry in the 2024 PDG~\cite{ParticleDataGroup:2024cfk} . Notice the mass average of $m_H = 125.20 \pm 0.11$ - a measurement of $0.1 \%$! 
The couplings of the Higgs boson are also in very good agreement with the SM, as reported in the 10-year reviews from CMS ~\cite{CMS:2022dwd} and ATLAS~\cite{ATLAS:2022vkf}.

\begin{figure} 
     \centering
    \includegraphics[width=0.8\textwidth, scale=0.5]{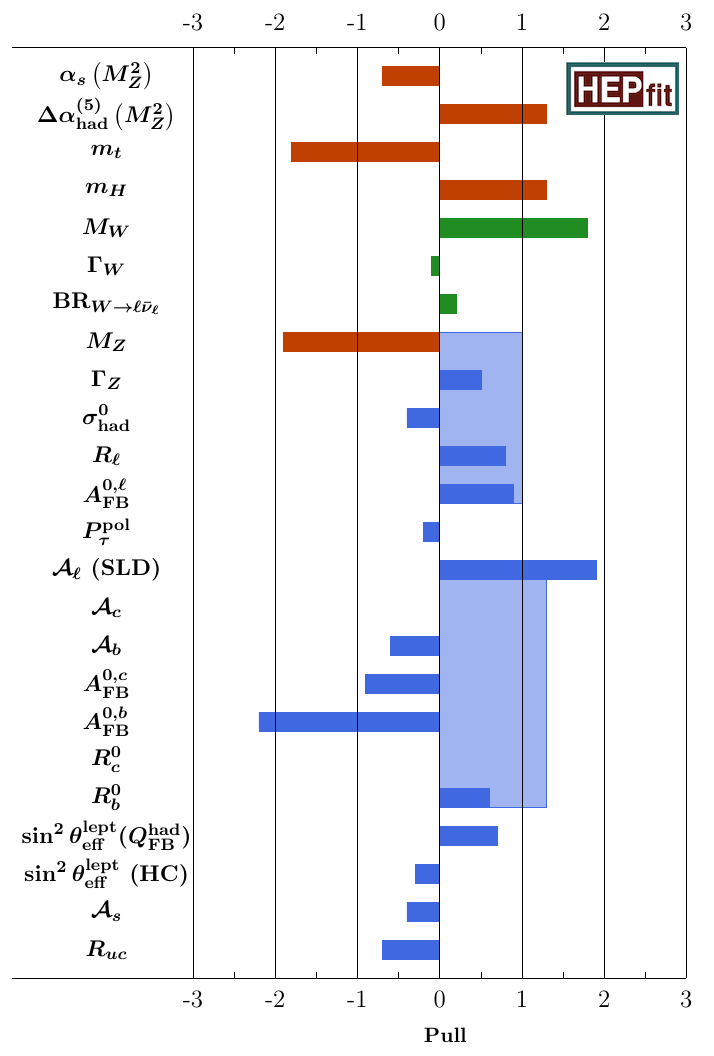} 
     \caption{Comparing fit results with direct measurements: 1D pulls between the observed experimental values and the SM predictions for the different electroweak precision observables (SM input parameters) considered in the fit, i.e. deviations between experimental measurements and theoretical calculations in units of the experimental uncertainty. The different colors in the figure are simply used to distinguish the SM inputs [orange], charged-current observables [green], and neutral-current observables [blue]~\cite{PhysRevD.106.033003} .}      
     \label{fig:EWPrecision} 
 \end{figure}

In Figure \ref{fig:EWPrecision} , the agreement between precision measurements of electroweak observables and theoretical calculations is shown. Each prediction is obtained by removing the corresponding observable from the fit. The transparent bars represent the corresponding nD pulls for groups of correlated observables. We can see a clear consistency between the measurements of all electroweak precision observables and their SM predictions~\cite{PhysRevD.106.033003}.

Despite its remarkable success in accurately describing a large range of phenomena observed in HEP experiments, the SM is widely recognized as a highly effective but not quite complete theory of fundamental interactions. Its precision and predictive power have made it a cornerstone of modern physics. At the same time, the SM opens exciting avenues for further inquiry and discovery that point toward new physics and a deeper understanding of the Universe, such as
\begin{itemize} 
    \item the origin of its 19 free parameters, closely related to open questions in flavour physics and CP violation;
    \item the mechanism behind the generation of neutrino masses;
    \item the nature and constituents of dark matter;
    \item the origin of dark energy;
    \item the hierarchy problem and the related question of the naturalness of the SM parameters;
    \item the inclusion of gravity in the SM framework.
\end{itemize} 

Surprisingly, the SM accounts for only about 5\% of the Universe's known content. Hence, 95\% of its contents is in the form of dark matter and dark energy. In addition, the discovery that neutrinos have mass also lies beyond the current formulation of the SM. These open challenges are not shortcomings but opportunities that drive global scientific efforts to extend the SM and explore what lies beyond, using new experimental setups. In the following, we describe the participation of the Latin American community in several High Energy Physics (HEP) activities related to tests of the Standard Model (SM) and beyond (BSM).

\section{Participation of LA groups in HEP Activities} \label{sec:lahep} 

\subsection{Large Hadron Collider (LHC) at the European Organization for Nuclear Research (CERN)} 

There is a large and growing engagement of the HEP Latin American community in CERN experiments. Table \ref{tab:GreyBook} shows
the participation of scientists and engineers from institutes in the Latin American region in CERN experiments, comparing September 2020 to September 2025. In the following we will briefly describe some activities that were reported in the submitted white papers.

   \begin{table} [h!]
       \centering
       \begin{tabular} {l|lcccc} 
 Country	&	Involvement	& Participants 2020	&	( $\supset$ Authors 2020 ) &  Participants 2025 & ( $\supset$ Authors 2025 )\\
 \hline
 Argentina	&	ATLAS	&	23	&	5 & 22 & 10 \\
 Brazil	&	ALICE	&	50	&	7 & 37 & 20 \\
 	&	ATLAS	&	67	&	8 & 103 & 24 \\
 	&	CMS	&	70	&	32 & 93 & 34 \\
 	&	LHCb	&	39	&	19 & 60 & 29 \\
 	&	ALPHA	&	5	&	3 & 3 & 3 \\
 	&	ISOLDE	&	12	&	4 & 7 & 6 \\
 	&	AMS	&	2	&	0 & - & - \\
 	&	ProtoDune	&	6	&	0 & 9 & 3 \\
 	&	RD51	&	5	&	1 & 4 & 1 \\
    &   DRD1    & - & - & 2 & 2 \\
 Chile	&	ATLAS	&	70	&	9 & 60 & 18 \\
    &   \gls{CMS} & - & - & 11 & 4 \\
    &	CLIC	&	2	& -	& 1 & - \\
 	&	ISOLDE	&	5	&	1 & 1 & - \\
 	&   LHCB & - & - & 3 & 1 \\
 	&	NA64	&	7	&	2 & 18 & 5 \\
 	&	SHiP	&	6	& -  & 21 & 3 \\
    &   SND     &  - & - & 26 & 12 \\
 Colombia	&	 ATLAS	&	7	&	2 & 19 & 3 \\
 	&	CMS	&	21	&	6 & 26 & 6 \\
 	&	ISOLDE	&	2	&	2 & - & - \\
 	&	LHCB	&	3	&	3 & 5 & 3 \\
 	&	RD51	&	4	&	3 & - & - \\
    &   Proto\gls{DUNE} & - &- & 4 & 1 \\
 Costa Rica	&	ISOLDE	&	1	&	1 & - & - \\
     & LHCB & - & - & 23 & 7 \\
Cuba	&	ALICE	&	5	&	1 & 3 & 3 \\
 Ecuador	&	CMS	&	9	&	0 & 21 & 1 \\
 Mexico	&	ALICE	&	82	&	14 & 51 & 30 \\
    &   AMS  & - & - & 2 & 1 \\
 	&	CMS	&	41	&	18 & 60 & 19 \\
 	&	NA62	&	5	&	2 & 9 & 6 \\
    & OTHER & - & - & 5 & 1 \\
 Peru	&	ALICE	&	12	&	3 & 3 & 3 \\
 \hline
 \hline
 LA Region	&	ALICE	&	108	&	29	& 93 & 56 \\
 	&	ATLAS	&	167	&	24	& 204 & 55 \\
 	&	CMS	&	141	&	56	& 211 & 64 \\
 	&	LHCB	&	42	&	22	& 91 & 40 \\
 	&	OTHER	&	103	&	15	& 112 & 44 \\
 	&	\textbf{TOTAL} 	&	\textbf{561} 	&	\textbf{146} 	& \textbf{(+26.7\%) 711} & \textbf{(+77.4\%) 259}  \\
 	\hline
       \end{tabular} 
      \caption{Participation of scientists and engineers from institutes in the Latin American region in CERN experiments, comparing September 2020 to September 2025. Participants are included if they are members of a research team officially associated with the experimental collaboration. Authors are participants who can sign physics publications of the corresponding experiment. [Source: CERN Greybook database 
    \tt{https://greybook.cern.ch/greybook/}]. 
    } 
       \label{tab:GreyBook} 
   \end{table}

\subsubsection{A Toroidal LHC ApparatuS (ATLAS)} 
Latin American groups participating in ATLAS, particularly in Brazil and Argentina, are developing ML-based tools for calorimeter signal processing, electron and jet triggering, and cross-talk mitigation. Both Argentinean and Brazilian groups are deploying algorithms on FPGA-based systems to support low-latency inference, and applying blind source separation and signal deconvolution to improve event reconstruction in real-time on an upgraded version of the \gls{ATLAS} trigger for the HL-LHC, also for offline environments.

\noindent In {\bf Argentina} , HEP groups have been members of \gls{ATLAS} since 2006.
Argentina is committed to the \gls{ATLAS} upgrade for the \gls{HL-LHC} with contributions to the trigger electronics, Data Acquisition, High-Level trigger software and hardware, and photon/electron/tau/jet reconstruction.
The responsibilities in \gls{ATLAS} have included so far
participating in the online testing of data taking, the development of the programs for the
online selection of events, the calibration of detectors, the measurement and improvement of
their performance, as well as the pursuit of multiple lines of physics analysis, several of them
as leaders of the respective analysis teams.

\noindent The main interests have concentrated on event triggering, reconstruction and physics analyses based on photons and/or jets. These have included SM precision measurements and searches for supersymmetry and other BSM models. Several PhD thesis in the group have addressed these analysis.


\noindent The Argentinian institutions are involved in the {\bf design, construction, and testing of the Global Trigger} for ATLAS. This is the first step of the online Trigger system, a hardware-based Level-0, with the following commitments: \\
\begin{itemize} 
  \item  Design, production, testing, installation and commissioning of the Global Trigger for ATLAS: a common hardware module with different functionality implemented in firmware, composed of ATCA blades and FPGAs with many multi-gigabit transceivers; 
  \item  Development of algorithmic firmware and software; 
  \item  Design, production, testing, installation and commissioning of Fibre Management; 
  \item  Procurement of general infrastructure.
\end{itemize}

\noindent A {\bf HEP laboratory}  has been set up at Instituto de Fisica La Plata (IFLP, UNLP-CONICET). It currently consists mainly of an electronics laboratory for the development and deployment of hardware for high-speed, real-time signal processing. It also has the tools and equipment to build prototype PCBs in-house.
Some of the
equipment that this lab will have when fully operational includes: \\
\begin{itemize} 
    \item FPGA evaluation kits 
    \item High-performance computers for firmware synthesis 
    \item Soldering station and tools for handling fine pitch PCBs SMD components
    \item  High speed oscilloscopes and high speed function generators 
    \item  ATCA shelf 
    \item High-speed optical fibres management. 
    \item CNC machine for drilling and milling for PCB making and equipment prototyping
\end{itemize} 

\noindent The
Argentine \gls{ATLAS} groups are also involved in applying Machine Learning techniques in current and near future
analyses (for instance, in jet tagging, Tau identification, fast simulation, calorimeter calibration, and more), and
foresee a much heavier involvement and usage in Artificial Intelligence
tools. This requires preparation, both in terms of peripheral computing and software power, as well as in education and supporting human resources.
However, one difficulty mentioned in the white paper is the {\bf current lack of computer resources}.

{\bf Brazil} has been participating in ATLAS since its conception, starting in
1988, through a pioneering initiative by UFRJ, which allowed the formation
of the so-called ATLAS/Brazil Cluster, composed by physicists, engineers and
computer technicians from UERJ, UFBA, UFJF, UFRJ, and USP.
The Brazilian group is involved in the muon trigger assisted by the hadronic calorimeter.
An advanced electronic interface system for this purpose (called TMDB) was 
designed, produced and
commissioned by the ATLAS/Brazil Cluster and has been operating on ATLAS since Run2
in the context of the Tilemuon project.
The Cluster is responsible for advanced signal processing techniques used
for TileCal signal detection. It also identifies bunch crossings and establishes
coincidence with the muon TGC through high-speed optical links. Currently,
a matched filter detector is used for both signal detection and bunch crossing
identification.
Concerning the electromagnetic calorimeter (LAr), activities in the Cluster are mainly related
to the implementation of the Super Cells, which refer to a new trigger piece of
hardware. Additionally, signal processing and machine learning techniques are
being applied for cross-talk mitigation.
For the triggering system, the Cluster is participating in the development of new computational
intelligence-based techniques for the calorimeter High-Level Trigger
(HLT) system, which are also to be extended to the offline filtering.
In addition, efforts are being made to improve the trigger calibration for
electrons. Possible extensions of the machine learning techniques developed
for HLT to level-one triggering (embedded electronics) are also being pursued
(FPGA-based designs).

The ATLAS experiment is preparing the construction of a new high-segmented
detector subsystem able to provide timing information to the reconstructed
tracks with a resolution of 30 ps and covering the region between 2.4 < |$\eta$| < 4.0.
This new detector, called the High Granularity Timing Detector (HGTD), will use 
Low-Gain Avalanche Diodes (LGAD) sensors. Given the novelty of the sensors, one key aspect of high interest is the radiation
hardening of the LGAD sensors in order to maximize the lifetime of the
sensors in the highest radiation areas. The presence of
several irradiation facilities in S\~ao Paulo provides support for this research.
ATLAS USP group has projected and constructed a dedicated facility for
the sensors R\&D and has a compromise with the qualification, installation and
commissioning of the detector layers of HGTD, as well as providing part of the
LGAD sensor arrays are needed for the innermost rings of HGTD. 
It is foreseen for the next years a contribution
to the qualification and construction of the electronic boards (Peripheral Electronic
Boards) that will interface the front-end electronics and the trigger and
data acquisition system in this region. The installation (stave loading, module
assembly and peripheral electronics integration) will be done at a CERN facility
assembled exclusively for this task with the participation of members of ATLAS
USP and UERJ groups, which have also participated in the test beams at SPS
(CERN) and in the construction of a small section of HGTD (demonstrator) for
testing purposes at CERN.

The Brazilian group is heavily involved in the double Higgs analyses in the channel $HH \rightarrow bb \tau \tau$.


\subsubsection{The Compact Muon Solenoid (CMS)} 

In total 14 groups from Latin America ({\bf Chile, Colombia, Ecuador, Brazil, Mexico}) are contributing to CMS with about 60 scientists. Latin American countries have so far awarded around 75+ PhDs and MSC-level degrees for research connected to CMS, and work is in progress for many more. The collaboration with CMS was instrumental to establish the Master- and the PhD programme in some institutes. There are also many engineering and computing students/specialists that are working in CMS, bringing the total number of active participants to well over 200.

Latin American researchers have made significant, and in many cases highly visible, contributions to CMS across the spectrum of activities: detector hardware, building, commissioning, data taking at CERN, operation, maintenance and upgrades, physics analysis, software development in support of physics analysis, data processing including validation and certification, computing with the operation of Tier-1, Tier-2 and Tier-3 sites, and central software developments. High-level management roles in these areas have also been filled by individual researchers from Latin-American institutes, thus making key contributions to the success of the collaboration. Several authors from Latin America who have been member of CMS from the start of data taking have signed over 1450 CMS journal publications to date.

In the next five years the CMS detector will undergo an ambitious set of upgrades in order to be ready for data taking at the High-Luminosity phase of the LHC. A finely segmented High-Granularity calorimeter in the forward region, ultra-fast timing layer, and a powerful upgrade of the trigger system allowing tracking and pattern recognition with ML-based algorithms at 40 MHz will vastly extend the capabilities of the detector beyond what is currently possible. 

It is noted that the use of AI (ML techniques) is becoming increasingly sophisticated and ubiquitous at many levels in the experiment. CMS groups in the region are applying ML techniques in fast simulation, jet tagging, calorimeter calibration, and physics object reconstruction. Efforts are underway to integrate these tools into both online and offline pipelines, often constrained by limited access to GPU clusters. Some work focuses on simulation-based inference and model interpretability for high-precision physics analyses.


{\bf Ecuador} submitted a white paper highlighting 10 years of participation in the \gls{CMS} experiment,  through groups at the Escuela Politécnica Nacional (EPN) and the Universidad San Francisco de Quito (USFQ), contributing to data analysis, detector operations, and software development. Ecuadorian researchers are active in collider physics, focusing on Higgs boson studies, dark matter searches, and new phenomena, and they have historically contributed to the High Level Trigger, while more recent efforts have centered on the operations and upgrades of the BRIL subsystem. During the difficult months of the COVID pandemic and the Long Shutdown 2, Ecuadorian engineers provided essential in situ support to ensure the successful restart of the experiment, and the groups are now engaged in the preparations for the High Luminosity LHC through contributions to BRIL and to the testing of acquisition software for newer methods for luminosity measurements. In parallel, they have made substantial contributions to \gls{CMS} open data initiatives and training programs, and their HECAP groups aim to strengthen software development and open data tools to further support fundamental physics research in collaboration with CMS. In total, around ten participants from Ecuador, including students, professors, and engineers, are involved in these efforts.

The 200+ scientists, engineers and students from {\bf Brazil} who are a member of the CMS collaboration play a crucial role in the experiment, with important contributions to hardware, software and computing operations, the ongoing Phase-II detector upgrades, physics analysis and various management leadership roles. The four Brazilian groups (GECMS at the Federal University in Rio Grande do Sul, SPRACE at São Paulo State University, COHEP at Centro Brasileiro de Pesquisas Físicas in Rio de Janeiro and the  CMS UERJ group at the State University of Rio de Janeiro) cover a broad spectrum of physics interests, from measuring Higgs boson properties to photon-photon induced electroweak physics with forward proton tagging with the Precion Proton Spectrometer (PPS), studying properties of the quark-gluon plasma in heavy-ion collisions, exploring flavour physics and rare decays, and searching for dark matter or magnetic monopoles and other exotic physics processes beyond the Standard Model of particle physics. Details can be found in the submitted CMS-Brazil White Paper.

The OpenIPMC project, spearheaded by SPRACE, has provided open source hardware, firmware, and software for 1000+ electronics boards that were produced and assembled in Brazil and shipped to CERN for the CMS tracker and high-granularity calorimeter upgrades. This development will be used in ATCA electronics boards for ATLAS and CMS upgrades and other experiments and has attracted interest for applications outside particle physics. Brazilian groups also play a key role in the optimization of calibration procedures and operations as part of the CMS ECAL upgrade, and a strong national programme is underway for advanced R{\&}D in gaseous detectors, relevant in particular for RPC technology used in various experiments including CMS. A cutting-edge R{\&}D programme in diamond sensor technology focuses on providing ultra-fast timing solutions for HL-LHC environments and aims at further strengthening Brazil's role in providing timing instrumentation for the CMS phase-II upgrade. 

UERJ and SPRACE host Tier-2 computing centers with 100+ computer nodes and several PB of disk storage space, connected via fast internet connections to the Worldwide LHC Computing Grid. In 2024 alone, the SPRACE cluster executed more than 150'000 computer jobs for CMS. SPRACE also plays a key role in preparing for future computing developments to meet unprecedented CMS computing demands, and is involved in the NGT project aiming at developing technologies that go beyond the baseline HL-LHC plans in CMS heterogenous computing, fast triggering and real-time data reconstruction and prompt calibration for physics analysis.  These advances in Instrumentation and Computing will be further addressed in chapter 7.

\subsubsection{The Large Hadron Collider beauty (LHCb)} 
{\bf Costa Rica} became a full member of the \gls{LHCb} collaboration in 2022 through the Consejo Nacional de Rectores (CONARE), which represents the country's five public universities. The Costa Rican group is contributing to analyses of rare decays and to R\&D projects on uncertainty estimation in machine learning models, electronics, and parallel programming. Guatemala and Honduras are also actively preparing for integration into LHCb. Guatemala has initiated discussions to formalize an International Cooperation Agreement with CERN, while Honduras signed such an agreement in 2021 and is seeking a Memorandum of Understanding with LHCb. Both countries are focusing on building the human and technical infrastructure to support long-term participation in the experiment.

{\bf Brazilian} researchers have been part of the \gls{LHCb} collaboration since 1998 \cite{LHCb:2000xej} , contributing to the experiment's design, construction, and operation. Since 2010, the detector has performed precise measurements, including studies of CP-violating phases, lepton flavor violation, QCD/EW probes, and searches for new physics. The \gls{LHCb} detector has undergone Upgrade I-a for higher luminosity, with a minor Upgrade I-b planned for 2025 and a major Upgrade II proposed for 2030. These upgrades are motivated to provide complementary approaches of the energy and intensity frontier in the search for physics beyond the Standard Model \cite{LHCb:2018roe, Aaij:2244311} .
 Currently, there are 19 permanent members in several institutes with several contributions, including work on detector readout electronics, multi-wire proportional chamber design, solid-state sensors characterization,
and real-time analysis development. The Brazilian team is involved in the operation of three detector projects of the current experiment: VELO, SciFi and RTA. The timeline for Brazilian participation in the \gls{LHCb} experiment is as follows:
\begin{itemize} 
  \item {\bf Until 2041:}  Planned participation in experiment operations.
  \item  {\bf Post-2041 (with \gls{LHCb} Upgrade II approval):}  The timeline would be extended by at least five years if \gls{LHCb} Upgrade II is approved to collect data during the High Luminosity LHC (HL-LHC) operational period.
\end{itemize} 
The construction and operational costs for Brazil's participation in the \gls{LHCb} experiment are summarized as follows:
\begin{itemize}   
\item  {\bf Operation Cost (M$\&$O):}  Approximately 5 kCHF/year per PhD researcher. For the current permanent scientists (19 individuals), this amounts to approximately 95,000 CHF per year. With the addition of one post-doctoral researcher, the cost increases to 100 kCHF/year.
\item  {\bf \gls{LHCb} Upgrades:} 
   \subitem {\bf Upgrade I:}  Completed last year, with additional financial support expected from institutes.
     \subitem {\bf Upgrade II:}  Expected to be approved soon. A significant contribution to \gls{LHCb} upgrades is proposed to be one-third of the M$\&$O over 15 years, totaling 500 kCHF.
\item {\bf Mission Trips to CERN:}  This is a crucial part of the upgrade and operational costs. It is assumed that 10 trips of 10 days per year are needed, costing 20 kCHF per year (about 20$\%$ of the M$\&$O cost). The document highlights the lack of formal financial support for these trips since 2015, which has negatively impacted Brazilian participation.
\item   {\bf Total Estimated Cost:}  To maintain a fruitful collaboration with the \gls{LHCb} experiment, Brazilian institutes would need **2300 kCHF over the next 15 years**.
\end{itemize} 
Brazilian institutes have made significant contributions to the \gls{LHCb} experiment, but they face challenges. A major issue is the lack of long-term financial commitment from funding agencies, unlike in Europe and the USA, hindering sustainable participation and upgrades. This financial instability also impedes detector research and development (R$\&$D) due to insufficient infrastructure and engineering capabilities, making it difficult to properly train students. Furthermore, the absence of postdoctoral positions and the current funding strategy prevent Brazilian institutes from attracting international researchers, an issue that could be resolved with long-term grants that enable competitive salaries and CERN-based opportunities. 

\subsubsection{A Large Ion Collider Experiment (ALICE)} 

{\bf Brazilian} participation in the ALICE-LHC experiment, focusing on the study of the Quark-Gluon Plasma (QGP), a state of matter where quarks and gluons are deconfined. The study is crucial for understanding Quantum Chromodynamics (QCD) and is conducted through heavy-ion collisions at the LHC. The main areas are Strangeness, QGP Tomography with Hard-Probe Production, and Ultra-peripheral Collisions. The Brazilian groups aim to contribute to physics analysis for Run 3, as well as to instrumentation development, specifically working on the FoCal project, which involves the readout system and physics simulations, and training future scientists. For Run-5 LHC dataken, the upgrade program from \gls{ALICE}  experiment it calls, ALICE3 will focus on heavy-ion collisions with ultra-low mass silicon trackers and high-resolution timing detectors. Brazilian groups are contributing to the design of the time-of-flight system. Development of Advanced Particle Detectors: Contributions to the upgrade of the \gls{ALICE}  experiment, including the Time Projection Chamber (TPC) with GEM detectors and the SAMPA chip, which enhances tracking and particle identification capabilities. 
Machine Learning Applications: Use of machine learning algorithms to improve reconstruction techniques for strange baryons, photon identification, and background reduction in heavy-ion collision data. 
FoCal Calorimeter: Development of the Forward Calorimeter (FoCal) with silicon sensors for high spatial resolution, enabling measurements at very low Bjorken-x values and distinguishing between single and double photon showers.
High-Performance Electronics: Contributions to the readout electronics of the FoCal system, including the configuration of the HGCROC chip and firmware development for the Common Readout Unity (CRU). 
Time-of-Flight System for ALICE3: Participation in the development of ultra-high timing resolution detectors ($\approx$ 20 ps) for particle identification, testing advanced technologies like MadPix and LGADs. 
Computing Infrastructure: Maintenance and expansion of the SAMPA cluster for data processing and storage, supporting the computational demands of the \gls{ALICE}  experiment. Timeline and Major Milestones:
\begin{itemize} 
    \item  LHC Run 3 (2023–2025): Ongoing physics analyses using the new O2 framework. 
Optimization of algorithms for Sigma baryon and Lambda polarization measurements. 
Studies on heavy flavor jets and charm fragmentation functions. 
\item Long Shutdown 3 (2026–2028): Preparation for LHC Run 4. Contributions to the FoCal calorimeter upgrade, including simulations and hardware development.
\item LHC Run 4 (2028–2030): Deployment and operation of the upgraded FoCal detector.
Continued studies on gluon saturation effects and heavy quarkonia production. 
\item LHC Run 5 (Post-2030): Development of ALICE3, focusing on heavy ion collisions and soft electromagnetic radiation. Contributions to the design and construction of the time-of-flight system with ultra-high timing resolution. 
\item Computing Infrastructure: Annual 15$\%$ increase in computing power and storage to meet growing data demands. 
\end{itemize} 
The summary about Maintenance and Operation (M$\&$O):
\begin{itemize} 
\item M$\&$O-A: General costs shared among all PhD authors in the collaboration.  Expected contribution from Brazilian groups in 2024: 109,017 CHF for 9 scientists. 
\item M$\&$O-B: Specific costs for detectors contributed by Brazilian groups (TPC, MFT, and FoCal).  Expected annual contribution: $\approx$40,000 CHF. 
\end{itemize} 

\vspace{0.5cm}


\noindent The {\bf Mexican} ALICE group at Instituto de Ciencias Nucleares (ICN) of Universidad Nacional Aut\'onoma de M\'exico (UNAM), is focused on high-energy heavy-ion and small-system collisions to study QCD matter at extreme temperature and energy density, with emphasis on the properties of the strongly interacting quark--gluon plasma, the QCD phase structure at (approximately) zero baryon chemical potential, collectivity and strangeness enhancement in small systems (p+p, p-Pb) and parton energy loss and jet quenching. The contributions of the ICN-UNAM group within ALICE are~\cite{Dainese:2925455,MunozMendez:2025ttk,Mendez:2025dqz,Alfaro:2024sxc,Ortiz:2024ndh,ALICE:2024vaf,ALICE:2023ulm,ALICE:2023plt,ALICE:2023csm,ALICE:2022wpn,ALICE:2022wwr,ALICE:2022qxg,Ortiz:2022mfv,Bencedi:2021tst,ALICE:2020fuk,ALICE:2019dfi,ALICE:2019hno,Ortiz:2017jaz,Ortiz:2017cul,ALICE:2016dei,Bala:2016hlf,Ortiz:2015ttf,ALICE:2015xmh,ALICE:2015dtd,ALICE:2014juv,OrtizVelasquez:2013ofg,ALICE:2013hur,ALICE:2012cor,Ortiz:2011pu}:
\begin{itemize}
  \item Precision measurements of identified-hadron ($\pi^\pm$, K$^\pm$, p, $\bar{\mathrm{p}}$) spectra in Pb--Pb and pp collisions at $\sqrt{s_{\rm NN}} = 2.76$ and 5.02~TeV, including centrality-dependent yields and particle ratios.
  \item Observation and systematic study of the baryon-to-meson enhancement at intermediate $p_{\rm T}$ exemplified by:
    \[
      \frac{p + \bar{p}}{\pi^+ + \pi^-}, \quad
      \frac{p + \bar{p}}{2\phi}
    \]
    indicating a strong interplay between radial flow and hadronization mechanisms (e.g.\ recombination).
  \item Measurements of the nuclear modification factor $R_{\rm AA}$ for identified hadrons, demonstrating strong suppression at high $p_{\rm T}$ and a mass-ordered peak at intermediate $p_{\rm T}$ consistent with strong collective flow.
  \item Complementary comparisons with isolated photon measurements (from CMS) showing $R_{\rm AA} \simeq 1$ for colorless probes, supporting the interpretation of parton energy loss in a colored medium.
  \item Systematic investigations of QGP-like signatures in p+p and p+Pb:
    \begin{itemize}
      \item Event-shape observables (transverse sphericity, spherocity).
      \item Underlying-event studies and relative transverse activity classifiers.
    \end{itemize}
  \item Demonstration that apparent ``jet-quenching''-like features in small systems are largely driven by event-selection biases rather than genuine medium effects, via measurements of jet-like yields and the $I_{\rm AA}$ observable in different event-activity classes.
  \item Development and experimental application of a new event classifier, ``flattenicity'', to quantify local multiplicity fluctuations and disentangle soft multi-parton interaction effects from geometric biases in pp and p--Pb collisions.
  \item First measurements of a $Q_{pp}$ observable (pp analogue of $R_{\rm AA}$) for identified hadrons in classes of flattenicity and multiplicity, providing stringent constraints on models (PYTHIA 8 with/without color reconnection, EPOS LHC, etc.).
\end{itemize}

The next-generation ALICE 3 detector (LHC Run~5, 2036--2041) includes a Muon Identification Detector (MID) in the forward region. The Mexican ICN-UNAM group plays a leading role in the conceptual design and performance studies of a MID based on plastic scintillator bars with wavelength-shifting (WLS) fibers and SiPM readout, together with the construction and characterization of small-area prototypes (e.g.\ $21.4\times21.4$~cm$^2$ chamber) with two orthogonal scintillator layers, tested with pion/muon beams at CERN T10. The team needs to demonstrate a time resolution of $1.6$--$2.0$~ns, uniform detection efficiency close to $100\%$ along $\sim 1$~m scintillator bars, and pion misidentification efficiencies at the few-per-cent level after machine-learning-based selection. In order to achieve this the ICN-UNAM team develops and deploys GEANT4-based simulations, including detailed modeling of primary and secondary particles in the iron absorber and machine learning workflows (e.g.\ Boosted Decision Trees) for muon/pion discrimination using time-over-threshold, time-of-arrival and hit-position information in the MID. This group has contributed to simulation and analysis frameworks for small-system event classifiers (flattenicity, sphericity, spherocity), as well as to underlying-event and jet-like correlation analyses.

Timeline and major milestones of the ALICE ICN-UNAM group can be summarized as follows:
\begin{itemize}
  \item \textbf{2009 - 2018}: ALICE Run~1 and Run~2 at the LHC; first heavy-ion and pp results on identified hadrons, $R_{\rm AA}$, and collectivity.
  \item \textbf{2015 - now}: ICN joins and consolidates its role in ALICE masterclasses for high-school students.
  \item \textbf{2018 - 2024}: Series of ICN-led analyses on small systems (event shapes, underlying event, flattenicity) and high-$p_{\rm T}$ particle production.
  \item \textbf{2020 - now}: Particle Therapy masterclass initiated at ICN; expansion to more countries and schools.
  \item \textbf{2024}: Construction and beam test of the first MID small-area prototype for ALICE~3 at CERN T10.
  \item \textbf{2022 - 2033}: LHC Run~3 and Run~4; continued ALICE data taking and physics program.
  \item \textbf{2036 - 2041}: LHC Run~5 with ALICE~3; full deployment of the MID subsystem and extended heavy-ion and small-system program.
\end{itemize}

The ICN-UNAM ALICE team has a continuous plan for training of undergraduate, MSc and PhD students in experimental high-energy nuclear physics, detector technologies, and data analysis; the formation of a regional expertise hub in QCD matter and collider physics within Latin America and the participation of early-career researchers in CERN test beams and analysis coordination tasks. This allows for many technological advances, such as the development and validation of fast, efficient plastic scintillator-based muon detectors with SiPM readout, the integration of ML techniques into detector operation and offline analysis chains, and the transfer of know-how in photodetectors, front-end electronics, and fast-timing systems to local laboratories. On the other hand, the team identifies challenges such as scaling up from small-area prototypes to full-size MID modules while maintaining time resolution, efficiency and uniformity; ensuring radiation tolerance and long-term stability of scintillators, SiPMs and electronics and upgrading local infrastructure (labs, clean areas, test equipment) to support serial production and QA.
Some of the challenges on the formal approvals and governance matters are the alignment of national funding cycles and priorities with the long-term ALICE~3 construction and operation timeline; the administrative and legal frameworks for in-kind contributions from Latin American institutions to CERN-based experiments and the coordination with governmental agencies for international agreements and customs/logistics associated with hardware shipment.

The ICN-UNAM ALICE group organises outreach and educational activities such as the ALICE Particle Physics Masterclass (since 2015) and the Particle Therapy Masterclass (since 2020) which allows for initial training for young people and continuous engagement with schools and the general public, promoting STEM careers and highlighting the role of Latin American scientists in large international collaborations.

\subsection{Belle II at the High Energy Accelerator Research Organization (SuperKEKB)}

\noindent{\bf Current Latin American involvement } 

The Mexican participation in the Belle~II experiment focuses on advancing the understanding of flavor physics at the intensity frontier. The program aims to search for potential deviations from the Standard Model through precision measurements of rare $B$, $D$, and $\tau$ decays, where new physics phenomena such as charged lepton flavor violation (cLFV), lepton number violation (LNV), or non-standard CP violation could emerge.

Mexico joined the Belle~II Collaboration in July~2013 and is the only Latin American country currently participating. The group integrates theoretical and experimental expertise to maximize the physics reach of data collected at SuperKEKB. Current efforts include searches for new physics in $\tau$ decays, precision electroweak measurements, detector operation, simulation, and software development. The group is also involved in phenomenological studies, such as the use of advanced reconstruction methods for invisible $\tau$ decays~\cite{DeLaCruz2020}.

These activities align with the long-term physics goals defined in the \textit{Belle~II Physics Book}~\cite{BelleIIPhysicsBook}.

The Mexican group is an active team in the Belle~II Collaboration and the only one from Latin America. The program is executed through a coordinated effort between Cinvestav, UNAM, and UAS, involving five senior researchers, two postdoctoral fellows, and about ten graduate students.

The team contributes to several Belle~II physics and technical groups:
\begin{itemize}
  \item $\tau$ and charm physics analysis groups,
  \item Detector operation and on-site shifts at KEK,
  \item Grid computing and software validation,
  \item Theoretical interpretation of experimental results.
\end{itemize}

\noindent{\bf Timeline and major milestones} 

Key milestones in the participation of Mexico within Belle~II include:

\begin{itemize}
  \item \textbf{2013 (July):} Mexico officially joined the Belle~II Collaboration, establishing the Mexican group at Cinvestav --- the only Latin American country participating.
  \item \textbf{2019:} Graduation of the first PhD student trained within the Belle~II program, consolidating the first generation of Mexican researchers.
  \item \textbf{2020 (February):} First Belle~II luminosity measurement, reported in \textit{Chin. Phys. C} \textbf{44} (2020) 021001, DOI: 10.1088/1674-1137/44/2/021001.
  \item \textbf{2020 (April):} First Belle~II physics result --- search for $Z'$ bosons in $e^+e^- \to \mu^+\mu^- +$ missing energy, published in \textit{Phys. Rev. Lett.} \textbf{124} (2020) 141801.
  \item \textbf{2022:} Development of beam diagnostics with the LABM using machine-learning techniques~\cite{LABM2022}.
  \item \textbf{2023 (August):} Precision measurement of the $\tau$ lepton mass in collaboration with DESY, published in \textit{Phys. Rev. D} \textbf{108} (2023) 032006~\cite{TauMass2023}.
  \item \textbf{2023 (December):} First publication led by the Mexican group within Belle~II --- search for lepton-flavor-violating $\tau$ decays to a lepton and an invisible boson~\cite{LFV2023}.
  \item \textbf{2025–2027:} Execution of next-generation $\tau$ physics analyses and upgrade of the LABM with radiation-hard sensors.
\end{itemize}

\noindent{\bf Funding  } 

National funding supports:
\begin{itemize}
  \item Travel and collaboration activities at KEK,
  \item LABM hardware maintenance and upgrades,
  \item Operation and upgrade of the JAGUAR computing cluster,
  \item Student and postdoctoral training.
\end{itemize}

Operating costs are moderate and primarily associated with personnel mobility, equipment maintenance, and computing infrastructure.  
The upcoming phase requires investment to expand JAGUAR’s storage and processing capacity to meet the computing demands of Belle~II Run~3.

\noindent{\bf Areas of expertise} 

\begin{itemize}
  \item Searches for lepton-flavor-violating $\tau$ decays,
  \item Precision measurement of the $\tau$ lepton mass,
  \item Advanced kinematic reconstruction and machine learning,
  \item Development of beam diagnostics instrumentation (LABM),
  \item Distributed computing and data processing.
\end{itemize}

\noindent{\bf Technological impact  } 

Mexico plays a leading and active role in the development and operation of the Large Angle Beamstrahlung Monitor (LABM) at SuperKEKB---a unique instrument designed to measure beamstrahlung light from the Interaction Point (IP). The system provides real-time sensitivity to the shape, position, and optical parameters of colliding beams.

The Mexican group co-leads the LABM upgrade program in collaboration with KEK and Wayne State University, contributing to:
\begin{itemize}
  \item Hardware upgrades (radiation-hard cameras, acquisition electronics, vacuum mirror system),
  \item Machine-learning-based beam-parameter reconstruction,
  \item Detector operation during beam commissioning and physics runs,
  \item Integration with Belle~II online and offline monitoring systems.
\end{itemize}

Recent work demonstrated that LABM data, analyzed using neural-network regression, reproduces key beam parameters (vertical size, luminosity proxies) with a few-percent precision~\cite{LABM2022}. This validates the LABM as an independent beam monitor for luminosity optimization. The effort represents the flagship technological contribution of Latin America to Belle~II.

Additionally, Mexico operates and maintains the JAGUAR computing cluster at Cinvestav, which supports remote data reconstruction and simulation, strengthening national computing autonomy within the collaboration.

\subsection{The Electron-Ion Collider (EIC) at Brookhaven National Laboratory (BNL) and the Thomas Jefferson National Accelerator Facility (JLab)} 

\noindent{\bf Current Latin American involvement:  } 

Currently these efforts involve 16 of scientists from institutions in M\'exico, Brazil, Colombia and Argentina and with support from Latin American scientists all around the world. At this stage, efforts are focused on early stages of EIC physics as reported by PIs from Latin American institutions such as Universidad Michoacana de San Nicolás de Hidalgo, Universidad de Sonora, Universidad Nacional Autónoma de México, Universidad de las Américas
Puebla, Universidad Autónoma de Sinaloa, Universidade Estadual de Campinas, Universidade Federal de São Paulo, Instituto Tecnológico de Aeronáutica, Universidade Federal de Pelotas, Universidade Estadual Paulista, Universidade Federal do Rio Grande do Sul, Universidade Cidade de São Paulo, Instituto Tecnológico de Aeronáutica, Universidad de Sucre, Universidad Nacional de San Mart\'in and Universidad de Buenos Aires. These scientists have played a crucial role in the exploration of hadron structure, hadron spectroscopy, and hadron phenomenology, alongside theoretical efforts on hadrons at finite temperature and in dense matter, functional approaches to nonperturbative QCD, light-front quantum field theory, and a wide variety of QCD-inspired effective theories. To summarize
\begin{itemize} 
\item M\'exico: Continuum studies through Schwinger-Dyson equations in QCD to predict observables for JLab and the EIC; effective field-theory models of QCD to compute the mass spectrum, decay constants, electromagnetic form factors, transition form factors and charge radii of pseudoscalar and vector meson form factors of light, heavy as well as heavy-light mesons; studies of light-front wave functions of unflavored vector mesons, as well as electromagnetic form factors, parton distribution functions and generalized parton distributions of heavy-light pseudoscalar mesons; calculations of distribution functions of diquarks and nucleons with the Schwinger-Dyson equation; phenomenological analyses of electromagnetic and two-photon transition form factors of pseudoscalar mesons to predict the pion and kaon form factors in the projected EIC and JLab $Q^2$ range; studies of entanglement entropy in inclusive and diffractive DIS; global analyses on hadron spectroscopy and structure with predictions for the EIC~\cite{ayala_quark_2012,albino_electron-photon_2022,bashir_collective_2012,roberts_dyson-schwinger_1994,
aguilar_schwinger_2023,aguilar_quark_2018,aguilar_infrared_2024,aguilar_gluon_2016,rojas_quark-gluon_2013,
albino_transverse_2019,rojas_exciting_2014,aguilar_pion_2019,ayala_quark_2012-1,albino_electron-photon_2022-1,
bermudez_quark-gluon_2017,gutierrez-guerrero_pion_2010,roberts_abelian_2010,roberts__2011,bedolla_charmonia_2015,
bedolla__2016,gutierrez-guerrero_masses_2019,gutierrez-guerrero_mesons_2021,hernandez-pinto_electromagnetic_2023,
paredes-torres_first_2024,miramontes_pion_2022,raya_partonic_2017,ding__2019,miramontes_timelike_2023,hentschinski_evidence_2022,
hentschinski_probing_2023,hentschinski_maximally_2022,bautista_bfkl_2016,arroyo_garcia_qcd_2019,hentschinski_exclusive_2021,
alcazar_peredo_ratio_2024,courtoy_extraction_2022,courtoy_parton_2023,kotz_analysis_2024,ablat_new_2024,
ramirez-uribe_rewording_2024,aguilera-verdugo_open_2020,ochoa-oregon_using_2024,raya_structure_2016,ding_drawing_2020,
raya_revealing_2022,xu_pion_2024,xu_empirical_2023,lu_pion_2024,raya_contribution_2020,albino_pseudoscalar_2022} .

\item Brazil: studies of non-perturbative phenomena with Schwinger-Dyson equations to compute the non-perturbative QCD propagators and vertices; studies of hadron mass generation due to dynamical chiral symmetry breaking in the QCD gap equation; light-front projection of Bethe-Salpeter amplitudes in order to compute the parton distribution amplitude, the parton distribution function and the transverse momentum distribution of mesons including heavy pseudoscalar mesons and quarkonia; pion targets are an experimentally challenging task at JLab and EIC so indirect approaches are explored with the Sullivan process; pion-structure studies, off-shell effects in electromagnetic form factors of light pseudoscalar mesons, in view of the pion and kaon elastic form factors which will be measured at large momenta and small virtualities at the EIC; explorations of heavy quarkonia interacting with atomic nuclei, which allows to access a matrix element related to the QCD trace anomaly, a quantum effect that is key to our understanding of the origin of the proton’s mass and its distribution within the hadron; research efforts into a dynamical continuum formulation of generalized transverse momentum distribution in Minkowski space, solving the Bethe-Salpeter equation; looking for constraints on the description
of the exclusive $J/\psi$ and $\Upsilon$ photo-production at low energies, since theoretical studies indicate that the near-threshold production of heavy quarkonium is sensitive to the trace anomaly contribution to the nucleon mass; an improved description of the photo-production of exotic states and the emerging predictions for their production, considering ep and eA collisions at the EIC; exclusive particle production and diffractive deep-inelastic scattering at the EIC; entanglement entropy in ep and eA collisions at small-$x$ ~\cite{souza_pseudoscalar_2020,segovia_completing_2015,chen_structure_2018,mojica_mass_2017,choi_pion_2019,
leao_off-shell_2024,burkert_precision_2023,tarrus_castella_effective_2018,krein_nuclear-bound_2018,
de_paula_parton_2022,de_paula_unpolarized_2023,de_paula_observing_2021,duarte_dynamical_2022,
castro_exploring_2023,oliveira_quark-gluon_2019,oliveira_exploring_2018,oliveira_soft-gluon_2020,
goncalves_searching_2000,goncalves_nuclear_2004,kugeratski_probing_2006,kugeratski_saturation_2006,
cazaroto_constraining_2008,cazaroto_could_2009,goncalves_exclusive_2009,goncalves_heavy_2010,
cazaroto_exclusive_2011,carvalho_nuclear_2013,goncalves_deeply_2015,goncalves_diffractive_2016,
goncalves_probing_2018,bendova_diffractive_2021,goncalves_exclusive_2020,goncalves_coherent_2022,
bendova_deeply_2022,xie_exclusive_2022,xie_investigating_2024,fagundes_asymptotic_2023,
peccini_exclusive_2022,peccini_exclusive_2021,peccini_investigating_2020,goncalves_coulomb_2018,
ramos_investigating_2020,el-bennich_dressed_2022,goncalves_meson_2013,goncalves_deciphering_2014,
moreira_production_2016,goncalves_probing_2019,babiarz_exclusive_2023,francener_photoproduction_2024} .
\item Colombia: currently exploring hadron three-dimensional structure distributions; light-front wave functions of pseudoscalar mesons; transverse momentum distributions and parton distribution functions of $D$ and $B$ mesons~\cite{albino_impact_2021,lessa_gauge_2023,serna_distribution_2020,serna_d_2022,da_silveira_strong_2023,serna_parton_2024} .
\item Argentina: spin physics and hadronization at the EIC; computation of EIC observables beyond next-to-leading order approximation in QCD; impact studies based on the foreseen experimental precision at the EIC~\cite{borsa_nnlo_2023,borsa_full_2022,borsa_jet_2020,borsa_revisiting_2020,aschenauer_semi-inclusive_2019,buonocore_precise_2023,mazzitelli_next--next--leading_2021,borsa_inclusive-jet_2021,borsa_parton-shower_2024,borsa_next--next--leading_2024} .

\end{itemize} 

\begin{figure} 
    \centering
    \begin{tikzpicture} [auto,node distance=1cm,thick,
main node/.style={black,draw,minimum width=1.5cm,minimum height=0.75cm} ]
\node[main node] (L)                 {FF, radii} ;
\node[main node] (F) [left = of L  ] {SDE, D$\chi$SB} ;
\node[main node] (B)    [right = of L  ] {PDF, TMD, BSA} ;
\node[main node] (H)  [right = of B  ] {Spectroscopy} ;
\node[main node] (G) [right = of H] {Phenomenology} ;
\node[main node] (C) [teal,above = of L  ] {M\'exico} ;
\node[main node] (A) [olive,above = of H  ] {Brazil} ;
\node[main node] (AP)    [orange,below = of B  ] {Colombia} ;
\node[main node] (BR)  [violet,below = of H] {Argentina} ;
\path[every node/.style={font=\sffamily\small} ]
(L)      edge [teal] node [right]                   {}  (C)
(F)      edge [teal] node [right]                   {}  (C)
(B)      edge [teal] node [right]                   {}  (C)
(H)      edge [teal] node [right]                   {}  (C)
(G)      edge [teal] node [right]                   {}  (C)
(L)      edge [olive] node [right]                   {}  (A)
(F)      edge [olive] node [right]                   {}  (A)
(B)      edge [olive] node [right]                   {}  (A)
(H)      edge [olive] node [right]                   {}  (A)
(G)      edge [olive] node [right]                   {}  (A)
(B)      edge [orange] node [right]                   {}  (AP)
(B)      edge [violet] node [right]                   {}  (BR)
(G)      edge [violet] node [right]                   {}  (BR)
(F)      edge node [right]                   {}  (L)
(L)      edge node [right]                   {}  (B)
(B)      edge node [right]                   {}  (H)
(H)    edge node [right]                   {}  (G);
\end{tikzpicture} 
\caption{Hadron physics and phenomenology efforts for the EIC in M\'exico, Brazil, Colombia and Argentina.} 
    \label{fig:enter-label} 
\end{figure}
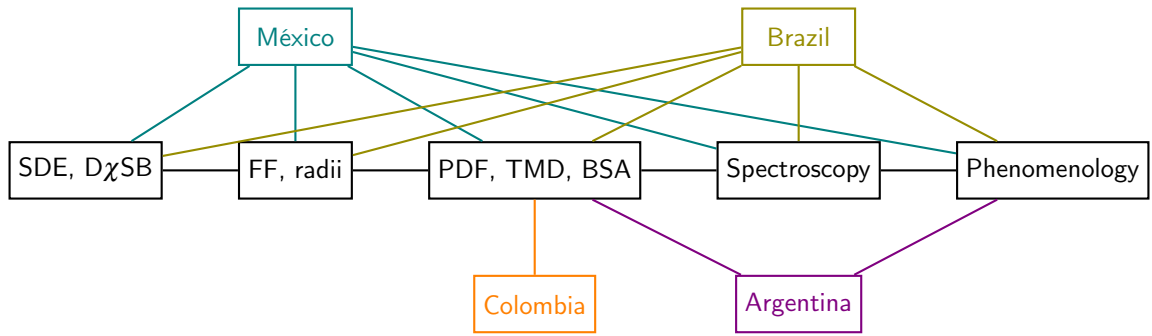 

\subsection{The Joint Institute for Nuclear Research (JINR)} 

\noindent{\bf Current Latin American involvement:  } 

The Joint Institute for Nuclear Research (JINR) is an international scientific research organization with Latin American involvement through Cuba as a Member State and, more recently with Mexico and Brazil as international cooperation partners~\cite{JINR,NICA} .
\begin{itemize}

\item M\'exico

Since 2016, a group of Mexican scientists (professors, researchers, engineers, postdoctoral researchers, undergraduate and graduate students) from seven national institutions have been part of the collaboration building the Multi Purpose Detector (MPD) at the Nuclotron Ion Collider fAcility (NICA). Back in 2019, a Memorandum of Understanding was signed between JINR and Mexican higher education institutions and centers to consolidate efforts for joint participation, cooperation, and collaboration in the NICA project. NICA has opened opportunities for bilateral cooperation between Mexico and the JINR, not only in science but also in industry and technology transfer. The laboratory directors have invited other Mexican communities of scientists and technologists to establish initiatives supported by the Mexican National Council of Science, Humanities, Technology and Innovation (SECIHTI), as well as with the Mexican diplomatic representation in Russia, to enhance Mexico cooperation at JINR. The areas of opportunity focus on the use of the facilities not only for basic research in frontier nuclear physics but also in the study of radiation effects for the development of advanced materials, in the design of electronic components, in modern medicine and biomedical research, in the safe disposal of nuclear waste, in cancer studies, among others. Currently, there are 18 participants (PIs, postdocs and technicians) from six Mexican institutions involved at MPD-NICA: Instituto de Ciencias Nucleares, Universidad Nacional Aut\'onoma de M\'exico (ICN-UNAM); Facultad de Ciencias F\'isico Matem\'aticas, Universidad Aut\'onoma de Sinaloa (FCFM-UAS); Instituto de F\'isica y Matem\'aticas, Universidad Michoacana de San Nicol\'as Hidalgo (IFM-UMSNH); Departamento de Física, Universidad Aut\'onoma de M\'exico - Iztapalapa (DF-UAM-I); Facultad de Ciencias, Universidad de Colima (FC-UCOL) and Facultad de Ingeniería Eléctrica, Universidad Michoacana de San Nicol\'as Hidalgo (FIE-UMSNH)~\cite{Kado:2020evi,Herrera:2024ymg} .
 
\item Brazil

Recently, JINR and the Ministry of Science, Technology and Innovation of the Federative Republic of Brazil (MCTI) signed a Memorandum of Understanding aimed at strengthening cooperation between the parties in fundamental and applied research. The agreement provides for the creation of a JINR–Brazil Joint Coordinating Committee designed to establish a structured method of selecting and implementing collaborative projects, particularly those that involve using the large research infrastructure of JINR and Brazil to achieve the national goals of scientific and technological development of Brazil and the JINR Member States. The Minister of Science, Technology, and Innovation of Brazil Luciana Barbosa de Oliveira Santos expressed a wish to establish closer contact between JINR and Russia’s Ministry of Science and Higher Education with organizations subordinate to the Ministry of Science, Technology and Innovation of Brazil: the Brazilian Centre for Research in Physics (CBPF), the Brazilian Centre for Research in Energy and Materials (CNPEM), and the National Institute of Pure and Applied Mathematics (IMPA). The areas for cooperation between JINR and Brazilian research laboratories and universities that have been targeted are: 
theoretical physics (quantum field theory; the interface between QFT, cosmology, and general relativity; lattice quantum chromodynamics, theory of hadronic matter under extreme conditions, topological superconductivity, and theory of atomic nuclei); nuclear physics (low energy nuclear reactions, nuclear structure and clusterisation in nuclei, and production cross sections of superheavy elements); applied nuclear research (materials and life sciences, ecology, and radiobiology); condensed matter physics (atomic and magnetic structure of novel materials and materials in extreme conditions, soft matter studies, and structural biophysics); computing and IT (distributed data processing and machine learning for data analysis) and educational programmes (training and exchange of highly qualified personnel)~\cite{BrazilJINR} .

\item Cuba

Cuba has been a member state since 1976. Currently, cuban scientists are involved in several JINR research programs, including the development of new semiconductor detectors for fundamental and applied research, the physics programme of the SPD and MPD detectors at NICA, and the modernization of the LINAC-800 linear electron accelerator. More than 20 young Cuban scientists have already participated in the educational programmes of JINR. Starting in 2014, Cuban students have continuously participated in the JINR Summer Student Programmes. Since 2015, Cuba has also participated in the International Student Practice (ISP). In the period 2015 – 2019, nine graduate students of InSTEC have completed internships at the DLNP JINR, which have allowed them to complete their Bachelor's theses. Since 2017, annually, Cuba has also participated in the International Training Program for Decision Makers in Science and International Scientific Cooperation of the JINR (JEMS), and in the Helmholtz International Summer School on the theme “Nuclear Theory and Astrophysical Applications”~\cite{CubaJINR} .
\end{itemize}

\subsubsection{Multi Purpose Detector (MPD) at the Nuclotron Ion Collider fAcility (NICA)} 

The Multi Purpose Detector (MPD)is an experiment at one of Nuclotron-based Ion Collider fAcility (NICA) interaction points, designed to deep-dive into the baryon-rich region of the QCD phase diagram by colliding heavy nuclei at $\sqrt{s_{NN} } =4-11$ GeV. The general structure of the MPD is a central barrel organized in a shell-like structure surrounding the interaction point. In general, it will reconstruct particle traces in the pseudorapidity range $\eta \leq 1.5$. The MPD also has two end caps to detect particles with larger pseudorapidity and in the central barrel the Time Projection Chamber (TPC) and the Time of Flight (TOF) systems will be the main subsystems. The Mexican collaboration \textit{MexNICA}  is working on a trigger subsystem called \textit{minibebe}, which must be efficient for low-multiplicity p + p, p + A, and A + A events, as well as have a fast response. The minibebe has been optimized in size, thickness and number of SiPMs of the detector cells, to achieve a fast response signal of order 20-30 ps. If this fast response is combined with fast read-out electronics with a response time of about 20 ps, it is then conceivable that the designed detector can serve as a good TOF trigger, provided it is efficient for low multiplicity events. Current efforts in the Mexican group are dedicated to both improving and testing the proposed electronics and DAQ, and in collaborations with the MPD inner tracking system working group to develop the mechanical structure for minibebe~\cite{MPD,MPD:2022qhn,MPD:2025jzd} .

\subsection{The High Intensity Baryon Extraction and Measurement (HIBEAM) and the Neutron-to-Neutron-Antineutron (NNBAR) experiment at the European Spallation Source (ESS)}

The HIBEAM/NNBAR experiment is one of the experiments of the ESS, currently under construction in Lund, Sweden. It is a two-stage experiment at the ESS to search for baryon number violation. The experiment will conduct high-sensitivity searches for baryon number–violating processes, either by one or two units. The ultimate sensitivity increases by three orders of magnitude the bounds from earlier work. 
The experiment addresses open questions such as baryogenesis and dark matter, and it is sensitive to a scale of new physics substantially beyond that available at colliders. 

Currently, the experiment comprises around 100 scientists from over 50 institutes in more than 10 countries, including Sweden, Denmark, Finland, the USA, and Japan. Projects under development are creating new research partnerships that are becoming firmly established, expanding the collaboration beyond the US and European borders and creating a HIBEAM/NNBAR collaboration in South America, with a special focus on Brazil.
The Brazilian members of the HIBEAM/NNBAR experiment come from four institutions: Universidade Federal Fluminense (UFF), Universidade do Estado do Rio de Janeiro (UERJ), Universidade Estadual de Campinas (Unicamp), and Universidade Estadual de Feira de Santana (UEFS). Together, they comprise six scientists as well as several undergraduate and graduate students.

The Brazilian group has been actively engaged in detector simulations, track-detector development, and electronics. The simulation effort plays a central role in ensuring the accuracy and reliability of the experiment’s results. The team is also deeply involved in the use of machine learning to achieve signal-background separation, in the development of TPC prototypes, contributing to advance detector technology and enhancing measurement precision. In addition, they are working to improve the electronic systems required for data acquisition and signal processing, further strengthening the experimental capabilities of the project. Studies will also be conducted to assess the feasibility of using gas-argon scintillation light from the TPCs as an additional input to track reconstruction.

\subsection{Future Colliders} 
Costa Rica has stated its intention to contribute to \gls{LHCb} Upgrade II, particularly through detector R\&D projects, including scintillating materials and ASIC development. These projects align with LHCb’s future needs in handling increased radiation and track densities. Guatemala and Honduras also aim to contribute to future upgrades through their proposed integration into the collaboration.

ML will be central to the physics and computing programs of future colliders. Latin American researchers are involved in the development of scalable architectures, including graph neural networks and transformer models, for tracking and calorimetry. Resource limitations — especially access to centralized GPU clusters — are a current barrier to further progress. Strategic investment in shared infrastructure is needed to prepare for \gls{HL-LHC} and beyond.

\subsubsection*{Brazilian plans} 
Young scientists from Brazil have reinforced their interest in contributing to the next generation of colliders in an organized central effort. Since the last report, and with the goal of establishing a organized community in Brazil, the National Network
for High Energy Physics (Rede Nacional de Física de Altas Energias - RENAFAE) has submitted a proposal to the initiative of National Institutes promoted by CNPq. The objective is the consolidation of hardware development, mobility, and the promotion of training human resources, as well as the broadening of High-Energy Physics in Brazil. As the results of synergies of the HEP community contributing to CERN collaborations, it was created the 'INCT CERN Brasil', the  CERN National Institute of Science and Technology of Brazil. Its main goal is to establish concrete advancements in R\&D detectors within a 5-year timeline and the consolidation of Brazilian participation in CERN experiments. In this sense, the CERN collaboration in Brazil has been restructured to facilitate faster communication and bring together different research groups, thereby increasing their impact.

In March of 2024, Brazil became an associate member of CERN, which lead to opening opportunities with Brazilian industry. Through this milestone for the Brazilian CERN High Energy Communities, it is expected the trigger of new collaborations, particularly in R\&D for new detector for the \gls{HL-LHC} using common technologies.  

One of the main challenges foreseen is identifying a stable funding line, given that INCT CERN-Brazil has a limited duration of 5 years. This has become particularly critical as discussions on the next-generation collider is starting
to converge. One way forward is to follow the example of the Brazilian Center for Research in Energy and Materials (Centro Nacional de Pesquisa em Energia
e Materiais – CNPEM). A similar strategy could be opted for the creation of a  National Laboratory for HEP, in a similar legal framework, supported by a series of interlinked projects that would lay the foundational structure required for the initial laboratory setup.

\subsubsection*{Scientific Context} 

The unprecedented amount of data collected by the LHC since 2022, with operations set to end in 2025, has led researchers to focus intensely on investigating parameters of the Standard Model (SM) at high precision and searching for new physics. The next upcoming data period, called Phase 2 or High Luminosity LHC (HL-LHC), is projected to start in 2029 and collect nearly 10 times more data than Phase 1. This will enable us to conduct more refined searches for new physics and measurements on the precision frontier. The LHC experiments have been planning their upgrades for many years and will start the installation and commissioning during Long Shutdown 3, aimed to start in 2026. Beyond the LHC, there are projects for future accelerators to be built around the world, namely:

\begin{itemize} 
    \item Circular Electron Positron Collider (CEPC): to be 100 Km $e^+ e^-$ collider to be built in China to start data taking in 2040, aimed to be a Higgs Factory.

    \item International Linear Collider (ILC): planned to be based in Japan, foresee as 20 Km linear  $e^+ e^-$ collider of 1 TeV energy.

    \item Compact Linear Collider (CLIC): similar to \gls{ILC} but of 50 Km based at CERN and achieving energies of 1.5 TeV and 3.0 TeV in two operation phases.

    \item Future Circular Collider (FCC): 90 Km $e^+ e^-$, $eh$, $hh$ circular collider based at CERN with up to center-of-mass energy of 100 TeV. Fisrt phase projected by 2048 and $hh$ projected by 2070.

    \item Electron-Ion Collider (EIC): project already approved by DoE and to start construction at Brookhaven National Laboratory. It will has the capability to collide electrons and ions at energies ranging from 20 GeV to 140 GeV.
  
\end{itemize} 

From all of them, only the EIC experiment is planned to start data taking by 2030; all others are scheduled to start by 2030 or later.

Among the physics motivation and objectives by the Brazilian HEP community, it includes:

\begin{enumerate} 
    \item {\bf Central Exclusive Production:}  motivated in the search for New Physics, particularly in the coupling of new particles to the electroweak sector. It provides an opportunity for searches for new particles at the LHC. The experimental signature of {\it Central Exclusive Production} , is the presence of events with intact protons in the final state interaction and low hadronic activity due to the exchange of colorless particles. This results in a gap in the pseudorapidity in the detector between the protons and the produced central system. Near-beam detectors are required to observe the intact protons (so-called {\it proton tagging} , which are scattered at very small angles ($\sim$ mrad)

    \item {\bf Heavy ion Physics:}  The heavy ion physics program in Run 3 and Run 4 at LHC will receive great benefits from detector upgrades, higher luminosities, to new data analysis techniques. All four main experiments at LHC are strongly involved in these advancements. Heavy ion physics places special importance on the understanding of the thermodynamic and transport properties of the Quark-Gluon Plasma (QGP), a deconfined state of quarks and gluons created in heavy ion collisions. This area, reaching a precision era, is expected to shed light on the following topics:
    \begin{itemize} 
        \item Search for the onset of parton saturation.
        \item Search for the onset of collective effects in small colliding systems, characterization of the nuclear parton distribution functions, and modeling of ultrahigh energy (cosmic ray) phenomena.

        \item Search for new physics in the low-mass range, understanding of heavy-flavour production, among others.

        \item  Better understanding of the hard probe sector. 

        \item Study of the 3D nucleus/proton structure, which can be accessed by Multiple Parton Interactions (MPI) observables.
    
    \end{itemize} 

  \item {\bf Higgs boson pair production and Higgs boson self-coupling:} 
Since the discovery of the Higgs boson in 2012, its properties has been widely investigated. Until now, all measurements of its quantum numbers and couplings are in agreement with the SM.  However, the verification of properties of interactions involving several Higgs boson is planned. Even more, the observation of the production of Higgs boson pairs ({\it HH} ) would  represent the next milestone in the physics program of the LHC, due to their role in cosmological theories involving the vacuum stability among others.
The one key measurements to be made, and within reach of the HL-LHC, is the trilinear term of the Higgs self-interaction. Deviations of this trilinear self-interaction coupling from the SM would shed light and play fundamental role in BSM physics theories, being of them the Electroweak Baryogenesis. The most sensitive test of Higgs boson self-interaction comes from processes of {\it HH}  production via gluon-gluon fusion (ggF) and vector-boson fusion (VBF). The small SM production cross-section of these processes implies that the main experimental signatures will come from the decay process where at least one of the two Higgs bosons decays into a final state with large branching ration, e.g $H \to b\bar{b} $. The most sensitive signatures would be $H \to b\bar{b} b\bar{b} $, $H\to b\bar{b} \tau^{+} \tau{-} $ and $H\to b\bar{b} \gamma \gamma$, with branching ratios of 33.9\%, 7.3\% and 0.26\%, respectively.
Although the study of the $HH$ production has been considered out of the LHC reach, but given the spectacular improvement of $HH$ analyses over the past decays, many progress has been made. Upper limits of $HH$ production, stringent constraints on new physics are rapidly approaching to SM prediction sensitivity with recent results with Run 2 data by \gls{CMS} and \gls{ATLAS} experiments. Many other improvements, including analysis techniques, are foresee with the Run 3 data-taking period. 

The goal for future machines beyond the \gls{HL-LHC} should be to probe the Higgs potential quantitatively. This will possible particularly through the measurement of $e^+e^-$ production at lepton machines at energies above 500 GeV and at hadron machines (FCC-hh). On the other hand, \gls{HL-LHC} constraints on the scales factor, related to couplings of $HH$ production via ggF or VBF, will have an unmatched precision for an very long time.

\end{enumerate}  
    
\subsubsection*{Constributions for hardware development} 

\begin{itemize} 
    \item {\bf Ultra-fast semiconductor detectors CMS-PPS}  \\
    Semiconductor particle detectors are among the most advanced detector technologies in HEP and have been employed extensively due to their precision and favorable characteristics. They are usually the detectors of choice in tracking and primary and secondary vertex reconstruction. Frontiers are being pushing in time resoution in order to develop ultra-fast  sensors and their corresponding electronics on one hand, and pushong for ever lower detection threshold to achieve quantum detectors on the other. Also, given their use in increasingly more intense radiation enviroments, motivates the search for radiation hard devices and improvement in cooling technologies.
    \begin{itemize} 
        \item {\it \bf Silicon detectors} : Because of the challenges presented by the LHC experiments, it has motivated an intense R\&D program on semiconductor sensors for radiation detection towards smaller, faster, and more radiation hard devices for tracking, timing and energy measurements. The Ultra-fast Silicon Detectors (UFSD) are capable of timing resolution of the order of few picoseconds. This came with the introduction of Low Gain Avalanche Diodes (LGAD), a structure pionereed by the RD50 collaboration. The device fullfill the performance requirements for \gls{HL-LHC} experiments and beyond.

        \item {\it \bf Synthetic diamond detectors} : diamond sensors as ultra-fast detectors of high-energy proton has been used in the \gls{CMS} and TOTEM collaborations since 2016 in PPS. It provides the possibility to access to exclusive events at the LHC, turning it inyo photon collider. synthetic diamond sensors are wide band-gap (WBG) semiconductors featuring reduced leakage current and hence low noise but with smaller generated signals.
        Teams from \gls{CMS} experiment at UFRGS (Universidade Federal do Rio Grande do Sul) is working in developing the metallization process in Brazil with diamonds acquired from Brazilian company. The local production can secure the preparation and characterization of new sensors to \gls{PPS} in Run-3 and HL-LHC.
        \item {\it \bf 3D Pixel Detectors}  3D pixel sensors achieve excellent spatial resolution due to small pixel sizes. They are radiation tolerant sustaining up to $\mathcal{O} $(16)n$_{eq} $/cm$^2$ and feature small inactive edge.        
    \end{itemize} 

\vskip 0.3 cm
    \item  {\bf HGTD for \gls{ATLAS} Phase-II Upgrade} 
In the plans of the \gls{ATLAS} experiment it lays the construction of a new high segmented detector subsystem able to provide timing information to the reconstructed tracks with resolution of 30 ps and covering the region between $2.4 < |\eta| < 4.0$. This detector is called High Granularity Timing Detector (HGTD). This device is a solution that has been sought not only by \gls{ATLAS} but also by \gls{CMS} and \gls{LHCb} collaborations as one of the most promising technologies to deal with the increased pileup in HL-LHC. Among the Brazilian participation there is the \gls{ATLAS} USP group, who has projected and constructed a dedicated facility for the sensors R\&D. They have the compromise  about the qualification, installation and commissioning of the detectors layers of HGTD, as well as providing part of the LGAD sensors arrays needed for the innermost rings for the HGTD. The R\&D and qualification of the sensors has been done at IFUSP, partnered with the engineering groups of EPUSP and FEI. It is also foreseen, for the upcoming years, contributions to the qualification and construction of the electronic boards (Peripheral Electronic Boards) that will interface the front-end electronics and the trigger and data acquisition system in the region. 

\vskip 0.3 cm
    \item  {\bf Gas Based Detectors: From the RPC to the Future} 
The RPC or Resistive Plate Chamber is a gas detector technology which allows for small localized region of dead time, resulting in excellent time resolution. In preparation for the \gls{HL-LHC} era, one of the main challenges for the RPC technology is the developing of new gas mixture with lower experimental impact while maintaining cost-effectiveness and performance for current and future HEP experiments. Through the INCT CERN Brasil initiative, Brazilian laboratories associated with \gls{CMS} RCP system, are working towards advancement in gaseous detector devices as well as in the upgrading aiming for the development of eco-friendly gas mixtures, gas recirculation and deeper understanding on how avalanches form. Partnership has been form with industry, specifically with Recigases, where the development of a regeneration system for a type of RCP is taking place.

\vskip 0.3 cm
    \item  {\bf High-Speed Electronics for Triggering and Energy Estimation} 

    The INCT-CERN Brazil is aiming to support the development of trigger systems in FPGA, in demand of the needed scientific and technological development for the HL-LHC. The combined of knowledge from the Electronic Engineering, Physics and Computational Intelligence is very welcome in this kind of projects, in the several tasks with luminosity and pileup conditions foreseen for Phase-II. Efforts are mainly concentrated in the UFBA and UFRJ institutes. The NeuralRinger algorithm, Brazilian method used as standard during the end of Run-2 and the whole of Run-3 for online electron identification in the \gls{ATLAS} High Level Trigger (and commissioned for photon triggers) is being implemented in simulation in the environment planned for Run-4, and is being designed for its operation in hardware.

\vskip 0.3 cm
    \item  {\bf The Mighty tracker detector for \gls{LHCb} Experiment upgrade-II} 

    During the so-called upgrade II, scheduled for the long shut down (LS4) 2033-2034, the \gls{LHCb} will replace a significant portion fo detectors and data acquisition systems. The purpose is to address the obsolescence or substantial damage caused by exposure to ionizing radiation and to handle the challenges of collecting data with even higher beam luminosity. The Mighty tracker is a planned system to upgrade the downstream tracking system of the \gls{LHCb} experiment.
    
\end{itemize}

\subsection{Theory}

Guatemalan and Honduran research groups are engaged in theoretical work focused on physics beyond the Standard Model (BSM), dark matter phenomenology, and cosmology. Specific efforts include quantum gravity and supergravity, relativistic kinetic theory in curved spacetime, helicity methods in scattering amplitudes, and studies of cosmic inflation and perturbations. These lines of research are contributing to capacity-building at the national level, and forming the foundation for deeper experimental engagement.


The higher education system in Ecuador has had structural problems, such as low-quality education system, violence, and social inequality. In this context and with the understanding that improving education helps with these problems, albeit indirectly and in the long-term, there have been actions proposed to mitigate these problems. The Group of Theoretical Physics (GI-FT) of the Escuela Politecnica Nacional (EPN) coordinates actions with a research agenda in high energy theory and gamma-ray astronomy; the promotion of the use of open databases of high-energy observatories; consolidate an institutional role in national and international training (Ecuadorian annual meeting); and the implementation of a computational laboratory for complex systems to catalyze multidisciplinary work. This group, which has supervised 40 undergraduate and graduate graduation projects, consists of collaborators affiliated to an Ecuadorian institution as follows:
\begin{itemize} 
\item Researchers: 
     \begin{itemize} 
     \item 7 faculty of the Physics Department from EPN and GI-FT members.
     \item 6 collaborators from other Ecuadorian universities, members of the GI-FT.
     \end{itemize} 
\item Institutions:
\begin{itemize} 
    \item Escuela Polit\'ecnica Nacional (EPN).
    \item Universidad de las Am\'ericas.
    \item Yachay Tech University.
    \item Universidad del Azuay.
    \item Universidad de las Fuerzas Armadas ESPE.
    \item Escuela Superior Polit\'ecnica del Chimborazo.
\end{itemize} 
\end{itemize} 

This Ecuatorian GI-FT group consolidates areas of expertise such as: 
high-energy theory, QFT and topological field theories, knot and theoretical particle physics, astrophysics and gamma ray astronomy. Some of the specific scientific goals of this group include encouraging the study of theoretical high energy physics and gravitation among local universities; promoting the use of open databases of high-energy observatories; engaging with local scientific communities through regional events; proposing joint projects with regional research centers and strategic allies as IAU. In the context of non-Abelian topological field theories, this group promotes the development and use of perturbative methods in non-Abelian Chern-Simons-Wong (CSW) classical theory to identify each term in the series expansion of the on-shell action to link invariants \cite{anda2023} , such as the Gauss linking number (GLN), and the third Milnor coefficient (TMC). Also, they propose to reproduce the non-trivial link invariants different from the GLN, an ``Intermediate Abelian Theory'' is proposed. This intermediate Abelian theory is one that is related to both the non-abelian CSW and a particular Abelian CS theory \cite{leal2008topological}  whose on-shell action,  coupled to a suitable current, reproduces the GLN invariant only. In gamma-ray astronomy, the group proposes to analyze peculiar GRBs (GRB070724A, GRB070429B, GRB090426, GRB120804A) using temporal estimators (emission time, spectral lag) \cite{lien2016third}  from the Swift catalog, with interest in machine-learning techniques and teaching data analysis, that could also increase the job opportunities for our students. The  GI-FT proposes the implementation of the computational laboratory for complex systems to promote multidisciplinary collaborations. An immediate theoretical impact would be the link-invariant program (GLN, TMC, four-component links) indicates a pathway to systematic constructions of higher-order invariants and a geometric interpretation via surface integrals, suggesting broader applications across knot/link problems in theoretical particle physics. On the technology front, the implementation of a computational laboratory for complex systems “with considerable computing power” is positioned to support GI-FT’s main areas and promote multidisciplinary collaboration; teaching data analysis and using open observatory databases expand workforce-ready skills.



The Mexican group MexNICA focuses on theoretical and phenomenological aspects relevant to heavy-ion physics and collider experiments, where the QGP and other phenomena can be studied. 

The MexNICA group has 18 researchers and currently has 2 postdocs and 5 postgraduate students from the following mexican institutions:
\begin{itemize}
    \item Instituto de Ciencias Nucleares, Universidad Nacional Aut\'onoma de M\'exico (ICN-UNAM)
    \item Facultad de Ciencias F\'isico Matem\'aticas, Universidad Aut\'onoma de Sinaloa (FCFM-UAS). 
    \item Instituto de F\'isica y Matem\'aticas, Universidad Michoacana de San Nicol\'as Hidalgo (IFM-UMSNH).
    \item Departamento de F\'isica, Universidad Aut\'onoma de M\'exico - Iztapalapa (DF-UAM-I).
    \item Facultad de Ciencias, Universidad de Colima (FC-UCOL).
    \item Facultad de Ingenier\'ia El\'ectrica, Universidad Michoacana de San Nicol\'as Hidalgo (FIE-UMSNH).
\end{itemize}

Their current efforts focus on meson and baryon production from the hadronic medium, pion production and femptoscopy, photoproduction at different stages of the fireball evolution, location of the critical end point in the nuclear phase diagram through fluctuations, in-medium vorticity and spin transfer through hyperon polarization and non-perturbative efforts such as Lattice QCD to have complementarity in the studies of the phase diagram and the phase transition characterization. Thermal models of heavy-ion collision systems predict a rapid change in the baryon-to-meson ratio as a function of the collision energy. This corresponds to a transition of the hadronic medium from a baryon- to a meson-dominated gas. This is expected to occur at a fireball temperature of about 140 MeV and a baryon chemical potential of around 420 MeV. These physical parameters of the system can be assigned to the corresponding center of mass collision energy of $\sqrt{s_{NN} } =8.2$ GeV, which is well within the expected MPD scan. Recently, the MexNICA group has proposed feasibility studies based on Monte Carlo data~\cite{Ayala:2023lnl} , for measuring the crossing points of the meson and baryon transverse momentum spectra, as a function of centrality and collision energy. 

In the context of relativistic heavy-ion collisions within the NICA energy range, the MexNICA group has studied the evolution with collision energy of the parameters that describe the two-pion correlation function, by performing Monte Carlo simulations, which show that the pion sample produced at the core of the fireball has a large component that comes from the decay of long-lived but slow-moving resonances, as well as a small component of pions coming from primary processes; so the analysis of the relative abundance of pions in the core coming from resonance decays and from primary processes becomes more important as a tool to study signals of criticality within the NICA energy range~\cite{Ayala:2023llp, Ayala:2023sbb, Ayala:2021zst} . Furthermore, in these heavy-ion collisions there are intense magnetic fields produced in the early stages of the collision~\cite{allmag}, which allows the opening of channels for photon production: gluon fusion and gluon splitting. Since at pre-equilibrium, the gluon-dominated occupancy is large, the gluon-driven process for photon production becomes even more relevant. In a series of publications~\cite{Ayala:2024jvc, Ayala:2024ucr, Ayala:2022zhu,Ayala:2019jey}, the MexNICA group collaborating with other Latin American peers, have provided both the theoretical framework and phenomenological impact on relevant yield and flow observables. Using the Linear Sigma Model with quarks, recent efforts of this MexNICA group with Latin American collaborators~\cite{Hernandez:2024nev,Hernandez-Ortiz:2024dci,Ayala:2021tkm} , have focused on the evolution of cumulant ratios describing baryon number fluctuations as a function of the collision energy, on the effects of vorticity on the QCD phase transition, on constraining the critical end point through the speed of sound and on chiral symmetry restoration. This Mexican group has also published a theoretical framework that can provide a way to explain both $\Lambda$ and $\bar\Lambda$ global polarizations in semi-central heavy-ion collisions using the core-corona model where the source of these hyperons is taken as consisting of a high-density core and a less dense corona~\cite{Ayala:2021xrn,Ayala:2020soy} . In connection with these polarization studes, in Refs.~\cite{Ayala:2025kuc,Ayala:2023vgv,Gaspar:2023nqk} , the group reported on results using a field-theoretical calculation to compute the relaxation time that an $s$-quark takes to align its spin with the direction of the angular momentum in a rotating medium at finite temperature and baryon density in order to get a better estimate of the way the vortical motion is transferred to the spin in $\Lambda$s produced in the corona. Finally, the group has made theoretical efforts to use non-perturbative approaches such as lattice QCD simulations, where theoretical models as effective theories for QCD are implemented in Monte Carlo simulations to study alternative versions of the phase diagram~\cite{3dO4,U1cosmo,cool,Fpi}.

Latin American researchers have contributed to the important issue of the anomalous magnetic moment of the muon ($a_\mu$). The anomaly reported in the 2020 White Paper~\cite{Aoyama:2020ynm} disappeared after the 2025 second White Paper~\cite{Aliberti:2025beg}, as suggested by the BMW collaboration result~\cite{Borsanyi:2020mff}, confirmed by other lattice collaborations~\cite{RBC:2018dos,RBC:2024fic,Djukanovic:2024cmq}. Meanwhile, the final FNAL result~\cite{Muong-2:2025xyk} only increased the precision of the experimental number. Three Latin-American scientists were authors of the first White Paper~\cite{Aoyama:2020ynm} (all based in Mexico), which increased to twelve in the second \cite{Aliberti:2025beg} (one Brazilian researcher and the rest from Mexico). Also the number of essential contributions determining the SM prediction to $a_\mu$ with LA autorship grew, from one in \cite{Aoyama:2020ynm} (\cite{Roig:2019reh}) to two in \cite{Aliberti:2025beg} (\cite{Masjuan:2020jsf,Estrada:2024cfy}). We highlight that the last reference was the result of a thesis done in Mexico. Latin-American researchers will continue contributing valuably to this precision program, with the new J-PARC measurement of $a_\mu$~\cite{Saito:2012zz}, with completely different systematics to be considered.


The Latin American Network on Electromagnetic Effects in Strongly Interacting Matter (LANEMqcd) brings together researchers across the region to deepen the understanding of how electromagnetic fields influence QCD matter, particularly the quark–gluon plasma (QGP), in environments ranging from ultra-relativistic heavy-ion collisions to compact astrophysical objects such as neutron stars and possible quark stars. Given that strong magnetic and electric fields naturally emerge in these systems, the study of electromagnetic effects has become essential for interpreting observables related to QGP formation, thermalization, photon production, anisotropies, and transport phenomena, as well as for constraining the dense-matter equation of state relevant to astrophysics.

The LANEMqcd network has 27 members from 20 institutions, belonging to 6 Latin American countries and 4 external members: 

\vspace{0.3cm}

{\bf Brazil} (15 participants)

\begin{itemize}
\item   Universidade Cidade de São Paulo (Unicid)
\item	Instituto de Física Teórica - (IFT-UNESP)
\item	Universidade Estadual de Campinas (UNICAMP)
\item	Universidade Federal do Rio de Janeiro (UFRJ)
\item   Universidade do Estado do Rio de Janeiro (UERJ)
\item	Universidade Federal de Santa Catarina (UFSC)
\item	Universidade Federal de Santa Maria (UFSM)
\item	Universidade Federal do Rio Grande do Sul (UFRGS)
\item   Universidade de São Paulo (USP)
\item   Universidade Federal de Goiás (UFG)
\item   Instituto Tecnológico da Aeronáutica (ITA)
\end{itemize}

\vspace{0.3cm}

{\bf Argentina} (2 participants)
\begin{itemize}
    \item Department of Theoretical Physics, Comisión Nacional de Energía Atómica (CNEA)
\end{itemize}

\vspace{0.3cm}

{\bf Chile} (3 participants)
\begin{itemize}
    \item Pontificia Universidad Católica de Chile (PUC-Chile)
 \item Universidad del Bio-Bio (UBB)
\end{itemize}

\vspace{0.3cm}

{\bf Cuba} (1 participant)
\begin{itemize}
    \item Universidad de La Habana
\end{itemize}

\vspace{0.3cm}

{\bf Colombia} (1 participant)
\begin{itemize}
    \item Universidad del Valle
\end{itemize}

\vspace{0.3cm}

{\bf M\'exico} (5 participants)
\begin{itemize}
    \item Universidad Nacional Autónoma de Mexico (UNAM)
\item	Universidad Autónoma Metropolitana (UAM)
\item	Universidad de Colima (UdeC)
\item	Universidad Michoacana de San Nicolás Hidalgo (UMSNH)
\end{itemize}

{\bf External} (4 participants)
\begin{itemize}
    \item Kent State University
    \item Universitat de Barcelona
    \item University of York
    \item University of Illinois
\end{itemize}

Several members of this network receive individual funding from their respective countries, by agencies CNPq, FAPESP, FAPERJ, Serrapilheira, Conicyt-Chile, Conicet-Argentina and SECIHTI-M\'exico. All the students have fellowships from the same agencies.  The events have been supported by ICTP-SAIFR, CLAF and IANN-QCD. The initiative is part of the Jovem Pesquisador FAPESP grant number 2023/08826-7, received by the coordinator Ana Mizher. 

Recent contributions from Latin American researchers published in 2024 and early 2025 encompass a broad range of theoretical advances. At the level of pre-equilibrium dynamics in heavy-ion collisions, recent work shows that magnetic fields modify gluon pressure anisotropies and accelerate isotropization. Complementary studies demonstrate that strong magnetic fields present during early collision stages can generate additional photon yield and azimuthal anisotropy, motivating the derivation of the two-gluon–one-photon vertex in a magnetized medium~\cite{Ayala:2024jvc,Ayala:2024ucr}. Other contributions address the modification of meson properties under magnetic fields: analyses based on NJL-type models, the linear sigma model, and the Kroll–Lee–Zumino approach show how neutral and charged pion masses, as well as rho-meson screening masses, acquire characteristic magnetic-field dependences, including distinctions between longitudinal and transverse modes once Lorentz symmetry is broken~\cite{Coppola:2023mmq,Ayala:2023llp}. The report also highlights work on fluctuating, or noisy, magnetic backgrounds. In this more realistic scenario, stochastic fields can induce magnetic masses for gauge fields and lead to distinct fermionic spectral widths depending on spin or particle–antiparticle character. The anomalous magnetic moment (AMM) of quarks in extreme fields receives special attention. Recent perturbative calculations reveal asymmetries between spin components of the AMM under strong magnetic backgrounds, while effective-model analyses show that incorporating the AMM can significantly influence the nature of QCD phase transitions~\cite{Hernandez:2025inu,Tavares:2023oln,Tavares:2024myk}. Additional studies investigate how magnetic fields modify strong couplings and form factors. One-loop calculations in the lowest Landau level indicate that QCD couplings grow with field strength, and thermo-magnetic extensions of bag models estimate the pressure of magnetized quark matter~\cite{Fernandez:2024tuk,Valenzuela-Coronado:2024ewi,Braghin:2023ykz}. Corrections to Yukawa potentials and to axial-vector couplings relevant for magnetar physics further illustrate the broad scope of these investigations~\cite{Dominguez:2023bjb,Fraga:2023cef,Fraga:2023lzn}. The report also includes advances in perturbative QCD at extremely high magnetic fields, where two-loop results for the chiral condensate, susceptibility, and pressure offer constraints on quark-magnetar mass–radius relations. The network also reports work on electric-field effects that arise in asymmetric ion collisions, such as Au–Cu. These studies explore pion scattering lengths under electric fields and analyze phase transitions of charged scalar fields at finite electric field and temperature. Finally, the network works on the growing interplay with condensed matter, using Reduced QED (or Pseudo-QED) to study how relativistic-like quasiparticles in materials such as graphene respond to external electromagnetic fields, providing a laboratory analogue to phenomena expected in QCD matter~\cite{Cadiz:2024rzs,Tavares:2024edx,Mizher:2024zag}. 

A central priority of the LANEMqcd network is to translate recent theoretical progress into predictions for experimentally measurable observables in strongly interacting systems under electromagnetic fields. The network continues work in monthly virtual seminars and in-person workshops and to strengthen interactions with experimental groups.

\subsection{HEP and Condensed Matter: Quantum Materials}

\noindent{\bf Introduction, Scientific Goals} 

The main scientific goal is to strengthen the links between condensed matter physics  and high-energy physics, through the study of Dirac materials with a dual focus on fundamental physics and cross-disciplinary feedback with potential applications in the realm of quantum materials. In doing so, a consolidation of the network of researchers is expected in the near future.

Specific scientific goals are:
\begin{enumerate} 
    \item Studying linear and nonlinear transport properties of topological phases using both condensed matter physics tools (Kubo formalism, kinetic theory) and high-energy physics tools (effective actions, vacuum polarization tensors).
    \item     Study the macroscopic consequences of the electromagnetic response of these materials in classical, e.g., waveguides, radiation, etc., and quantum (e.g. interaction with atomic systems, heterostructures, etc.) configurations.
    \item Find analytical solutions for the eigenvalue problem of the effective Hamiltonian that describes strained graphene under general conditions.
    \item Quantum field theoretic treatment of Dirac materials, focusing on anomalies, gauge symmetry, renormalizability, and Green functions.
    \item Studying quantum materials in curved geometries using QFT on curved spacetimes, with the goal of identifying conditions for solid-state analogs of gravitational effects
\end{enumerate} 

\noindent{\bf Technological Impact} 

The contribution of the research carried out by this network is mostly theoretical and with no \textbf{direct}  technological impact. However, a thorough understanding of the new and emerging properties of quantum materials may bridge the gap between fundamental knowledge and application. This theoretical understanding can shed light on:
\begin{enumerate} 
    \item     Next-generation electronic and quantum devices.
    \item Quantum information science, particularly through materials like topological insulators (TIs), Weyl semimetals, and strained graphene.
    \item     Photonic and optoelectronic technologies (e.g., nonreciprocal emitters, harmonic generation).
    \item     Possible solid-state analogs of gravitational phenomena.
\end{enumerate} 
 
The technological implications are promising but speculative. Stronger emphasis on concrete technological pathways or prototypes would enhance the translational relevance of the project.

\noindent{\bf Current Latin American involvement} 

The scientific goals pursued by this network are predicated on
a well-motivated and very active contemporary research frontier, driven by collaborations within this community, with a long and prolific history. Members of this network have begun to study mixed-dimensional field theories (reduced and pseudo-QED) in 1993 \cite{MARINO1993551}  and have continued to collaborate thereafter, e.g., \cite{PhysRevD.102.056020} . In the context of SUSY-QM in Graphene and Dirac Materials, collaborations among this network have been reported since 2007 \cite{7,15,17} , with ongoing efforts to date \cite{20,24,25,26,27} . More recent collaborations have been documented in 2022 \cite{22}  and 2024 \cite{23} . On the topic of Topological Insulators and Axion electrodynamics similar collaborations are reported from 2016 on \cite{PhysRevD.93.045022,PhysRevD.94.085019,PhysRevD.98.056012,PhysRevD.99.116020,MartnRuiz2019,MartnRuiz12019,Silva2020,PhysRevD.107.096024,Franca2022,Medel2023}  and in the related context of anomalies in Lorentz violating QFTs \cite{Arias2007}  and thereafter \cite{Gmez2021, Gmez2022, Gmez2024} . Advances from members of this network on the subject of QFTs in curved spacetimes as an arena to study Dirac materials dates to 2017 \cite{Villarreal,Castro2024} .

\noindent{\bf Funding Profile} 

\begin{itemize} 
  
\item  {\bf Operation Cost:}  
Being a theoretical endeavour, there are no declared operational requirements and costs. However, a posteriori, the network has expressed the need to support a Premium subscription  to Overleaf such that all members of the network can collaborate online on their manuscript preparation
.

\end{itemize} 

\noindent{\bf Areas of expertise} 

\begin{itemize} 
    \item SUSY-QM methods applied to graphene.
    \item Mixed-dimensional gauge theories.
    \item Contributions to electromagnetic modeling of topological insulators and Weyl semimetals.
    \item Analytical techniques in topological matter.
    \item QFT on curved spaces for Dirac materials.
\end{itemize}

\section{Training, outreach, exchange programmes} \label{sec:training} 
There are several activities in HECAP training in the region, described in Chapter 8. Here we focus on the contribution from Central America.

Costa Rica has organized educational events at the undergraduate and high school levels, including masterclasses using \gls{LHCb} open data and a workshop during JOCICI 2023. The country hosted the CERN Latin American School of High-Energy Physics in 2025, which further consolidates regional training efforts. The region is collaborating on the establishment of a joint Central American postgraduate program in HEP. The CAHEP network and regional events, such as CAHEP2020 and the Northern Triangle (El Salvador, Guatemala, Honduras) conference in Copán Ruinas, are important initiatives that foster integration and training.

There is a gap in areas such as GPU programming, ML model deployment on FPGAs, and data-centric algorithm design. While local efforts exist in some institutions in Brazil and Argentina, a continent-wide initiative or regional ML school for HEP would greatly enhance capacity. The authors advocate for integrating such training into the standard HEP education pipeline.

Costa Rica is developing R\&D capabilities in electronics, ML-based uncertainty estimation, and high-performance reconstruction algorithms. Guatemala and Honduras are gaining visibility in theoretical contributions related to BSM physics, scattering amplitudes, and cosmology. These efforts form the basis for leadership roles in both experimental and theoretical domains in the future.

Latin American groups are gaining recognition for their work in ML-driven calorimetry, fast simulation, and trigger development. Examples include \gls{ATLAS} Brazil’s use of deconvolution and AI-based calibration techniques, and Costa Rica’s work on uncertainty quantification in ML models. These areas represent emerging leadership domains for the region in the global HEP community.

\section{Synergies} \label{sec:synergies} 

Central-american countries have formed research networks (e.g., CAHEP) and launched regional initiatives to strengthen capacity. Their shared challenges — limited funding, infrastructure, and human resources — have motivated the development of coordinated strategies in training, research, and international engagement. These synergies are foundational to building a viable HEP research community in the region.

ML research in HEP naturally creates synergies across physics, computer science, and engineering. Collaborative work between institutions on algorithm design, hardware acceleration, and simulation pipelines is already in progress. There is also potential for deeper engagement with the industry, particularly in areas such as edge computing and AI model deployment.

\section{Conclusions} \label{sec:conclusions} 
As can be seen in Table \ref{tab:GreyBook} and in the submitted white papers, there is a large participation of the Latin American community in HEP experiments at CERN. Important developments since the  last PBB include the association of Brazil and Chile to CERN.
In addition, we identified participation in Belle II at SuperKEKB, Jefferson Lab, NICA at JINR and future facilities such as EIC and FCC. A very active theoretical community is also present in the region, with established focused networks such as LANEMqcd. 
The Central American region is becoming an active and coordinated contributor to high-energy physics. With Costa Rica's participation in \gls{LHCb} and the ongoing efforts by Guatemala and Honduras to formalize collaboration, this tri-national effort is positioned to expand Latin America's footprint in global HEP. Continued investment in education, infrastructure, and regional cooperation will be key to sustaining this momentum. Machine learning is transforming the landscape of high-energy physics, from data acquisition to final analysis. Latin American researchers are actively contributing to this shift but face resource and training bottlenecks. Coordinated efforts to address these limitations could enable the region to become a key player in shaping the future of ML-enhanced physics at the \gls{HL-LHC} and next-generation colliders.

Latin America has reached a critical stage in its scientific engagement with global high-energy physics. Across CERN experiments, Belle II, the EIC program, JINR–NICA, and emerging initiatives in detector R{\&}D, the region is no longer a peripheral participant, but a consistent contributor of technical innovation, physics results, and scientists formation and training. The breadth of activities described - from real-time trigger electronics and ultra-fast timing detectors to precision measurements, hadron-structure phenomenology, and machine-learning–driven reconstruction tools - reflects a mature and diversified scientific ecosystem capable of delivering high-impact, mission-critical work. Latin American groups are deeply involved in hardware construction and qualification for next-generation upgrades: LGAD timing systems for ATLAS and CMS, OpenIPMC for ATCA electronics boards for ATLAS, CMS and other experiments, gas mixture R{{\&}D for RPC muon detector systems in CMS, forward calorimetry and muon identification for ALICE3, high-resolution tracking for LHCb Upgrade II, beam-diagnostics instrumentation for Belle II, and a trigger subsystem for the MPD-NICA, to name just a few examples. These contributions are technologically demanding, tied to strict timelines, and essential for the performance of future colliders and experiments. The theoretical and phenomenological programs (spanning QCD under extreme conditions, hadron spectroscopy, electromagnetic effects in strongly interacting matter, and structure functions for the EIC) position Latin America as a relevant source of intellectual leadership in defining the scientific agenda of these facilities.

At the same time, we identify structural challenges that limit the scalability of these achievements. Research teams face irregular funding cycles, insufficient long-term commitments, and gaps in local infrastructure, particularly in high-performance computing and advanced laboratories. These constraints reduce competitiveness and hinder participation in major upgrades, where stable, multi-year investments are essential. Moreover, the lack of sustained postdoctoral programs and mobility funding weakens the training pipeline needed to retain and attract highly skilled personnel. Despite these limitations, the evidence presented underscores a remarkable capacity for impact. With coordinated investment and coherent national policies, Latin America could consolidate itself as an indispensable partner in global high-energy physics over the next two decades, especially as the HL-LHC, ALICE3, Belle II upgrades, the EIC, and future collider initiatives converge. The scientific workforce, its growing instrumentation expertise, and its active engagement in international collaborations make this a strategic moment: targeted support would not only strengthen scientific capabilities but also enhance technological innovation, industrial participation, and advanced training across national systems.

In summary, Latin America’s participation in cutting-edge HEP projects is both scientifically significant and strategically valuable. The region can fully leverage its demonstrated potential and make consistent contributions to the next generation of colliders, with predictable funding, infrastructure development, and financial support for scientists' training and mobility.

\bibliographystyle{unsrt} 
\bibliography{ew-bsm/ew-bsm}

\chapter{Instrumentation and Computing}\label{chapt:instru}
This chapter provides an update on the landscape of instrumentation and computing within the Latin American High Energy Physics, Cosmology, and Astroparticle Physics (HECAP) community, building upon the initial LASF4RI strategic plan from 2020. It draws directly from white papers submitted for the current update process and discussions held at the "III LASF4RI for HECAP Symposium" in August 2024.

It is worth emphasizing that Latin America hosts cutting-edge infrastructures, such as the Vera Rubin Observatory and the Cherenkov Telescope Array Collaboration in Chile, the Pierre Auger Observatory, the QUBIC experiment and the LAMBDA laboratory in Argentina, the CONNIE and ANGRA neutrino experiments, the SIRIUs light source and the future BINGO radiotelescope in Brazil. In addition, the participation of Latin American scientists in world-leading international collaborations in other regions has led to the increased development of instrumentation that has already had important impacts in these collaborations.  

There are also significant effort in training and capacity building in instrumentation and computing that will be addressed in the next chapter.

The goal of this chapter is to point out significant advancements and ongoing efforts across several key areas. Overlap with other chapters of this Briefing Book is unavoidable.

\section{New Developments in Instrumentation Latin American HECAP} \label{sec:instrumentation-computing} 

The region has seen a \textbf{growth and diversification of local instrumentation capabilities}  since the initial 2020 report, with an increasing presence and construction of new local facilities for instrument development, building, testing, and characterization. This showcases a broad diversity of experience and skills developed across different groups and institutions in Latin America.
Below we list (in no particular order) some of these capabilities reported in the submitted White Papers and discussed in the previous chapters.
More details and other contributions can be obtained from the White Papers and references therein.

\begin{itemize}
    \item As mentioned in section 2.3.1, the very wide energy range covered by CTAO-South array necessitates the use of three different
telescope sizes, referred to as Large-, Medium- and Small-Sized Telescopes (LSTs, MSTs and SSTs).
Groups in Brazil are contributing to the development of mechanical components
for the LSTs, to the construction of the support for the camera of the MSTs
and to the construction of an SST prototype called ASTRI, as
discussed in detail in the submitted White Paper.

   \item The prototypes of the horn, transitions, polarizer, magic tees and rectangular-to-coax transitions for the BINGO radiotelescope discussed in section 3.2.4 were designed and fabricated in Brazil and successfully passed the electromagnetic tests.

    \item The {\it Laboratorio Argentino de Mediciones de Bajo umbral de Detección y sus Aplicaciones}  (LAMBDA), detailed in Section 4.2.4, was created in 2022 to investigate and develop the capabilities of Skipper-CCD technology \cite{dm:tiffenberg:2017}  for dark matter searches, neutrino physics experiments, and searches for physics BSM. It is the result of the strong collaboration between the Physics Department of the University of Buenos Aires (UBA) and Fermilab in the USA.

    \item As mentioned in Section 5.4, a Mexican group joined Hyper Kamiokande in 2020 and co-leads the mechanical design, local prototyping, and low-power front-end electronics of multi-PMT optical modules. They will manufacture the first full production batch of them to be installed in 2027.

    \item 
Brazilian groups are leading the construction of the photon detection system, called ARAPUCA, for the DUNE far detector, as mentioned in section 5.8.
Production and assembly will done in Brazil. In addition, Brazilian companies will make a valuable contribution towards the construction of a liquid argon purification system for DUNE.

    \item Chile led the design and manufacturing of 400 high-voltage splitters (HVS) and 200 under water boxes (UWB) for JUNO, successfully completed in late 2022, as described in section 5.9.

    \item An advanced electronic interface system for muon trigger assisted by the hadronic calorimeter
(called TMDB) was designed, produced and commissioned by the ATLAS/Brazil Cluster and has been operating
on ATLAS since Run2 in the context of the Tilemuon project. 
In addition, the group has projected and constructed a dedicated facility 
for the testing Low-Gain Avalanche Detector (LGAD) sensors. 

\item Argentinian institutions are involved in the design, construction and tests of the Global
Trigger component of the first part of the ATLAS trigger, the hardware based Level-0.

    \item The OpenIPMC project, spearheaded by SPRACE in Brazil, has provided open source hardware, firmware, and software for 
more than 1000 electronics boards that were produced and assembled in Brazil and shipped to CERN for the
CMS tracker and high-granularity calorimeter upgrades. Brazilian groups in Rio de Janeiro are developing advanced experimental capabilities for gaseous-detector research, particularly Resistive Plate Chambers (RPCs). Additionally, a group in Rio Grande do Sul is undertaking the study of
artificial diamond as a radiation-hard candidate for fast timing near the beam, emphasizing
their applicability to forward timing in PPS and is developing diamond synthesis, metallization, and sensor
characterization.

    \item Brazilian groups at LHCb work on
detector readout electronics, multi-wire proportional chamber design, solid-state sensors characterization, and
real-time analysis development. They  are involved in the operation of three detector projects of the
current experiment: VELO, SciFi and RTA. 

\item Brazilian groups in ALICE are contributing to its upgrade, including the Time
Projection Chamber (TPC) with GEM detectors and the SAMPA chip, developed in Sao Paulo, which enhances tracking and particle
identification capabilities. 
They are developing the Forward Calorimeter (FoCal) with silicon sensors for
high spatial resolution, and contributing to the readout electronics of the
FoCal system, including the configuration of the HGCROC chip and firmware development for the Common
Readout Unity (CRU). They are also participating in the development of ultra-high timing resolution detectors (around 20 ps) for particle identification, testing advanced technologies like MadPix and LGADs. The Brazilian Center for Research in Energy and Materials (CNPEM) has recently joined the ALICE 3 magnet project with involvement in design and construction.

\item 
The Mexican ICN-UNAM group in ALICE plays a leading role in the conceptual design and
performance studies of a Muon Identification detector  based on plastic scintillator bars with wavelength-shifting (WLS) fibers and SiPM
readout, together with the construction and characterization of small-area prototypes with two orthogonal scintillator layers, 
tested with pion/muon beams at CERN.
    
\end{itemize}

\section{Strategic Advancement in Scientific Computing and Data Infrastructure} \label{subsec:computing-data} 

The increasing data volumes from current and future experiments, particularly those from the High-Luminosity LHC, necessitate continued strategic advances in computing and data infrastructure. 

The Brazilian CMS groups hosts two Tier-2 computing centers with almost 3000 physical cores and a few PB of disk storage space, connected via fast internet connections to the Worldwide LHC Computing Grid.
The SPRACE group is engaged in CMS software and trigger R\&D for the Phase 2 upgrade and led the 
Technical Design Report of the upgraded High-Level Trigger
(HLT). It is part of the Next-Generation
Triggers "Real-time Reconstruction Revolution" (R3) project, which aims to overcome
the current limitations of online reconstruction quality and output bandwidth, leading
the task on optimal calibrations, which seeks to design accelerated calibration
workflows and an improved calibration infrastructure so that HLT can reach offline-
like calibration quality, including the use of online data buffering and predictive AI
techniques. This is being developed in synergy with Run 3 operations, for example
through prototypes of HLT scouting workflows.
SPRACE also participates in CMS planning for computing resources into the
HL-LHC era, including studies of CPU and disk requirements for Runs 3, 4 and 5,
and the impact of R\&D improvements and resource growth scenarios.

The Brazilian ALICE group currently has a cluster with 2408 CPU cores and 1.5 PB of storage. For the next 5 years, it is expected a
flat increase of 15\% annually in the demand for more computer power and storage.

Cosmological projects such as the Legacy Survey of Space and Time (LSST) also involves large amounts of data. LSST will collect 15-20 TB of data per night, totalling more than 5000 PB in its 10 years survey. In Brazil, the Laboratório Interinstitucional de e-Astronomia (LIneA) is building an Independent Data Acess Center (IDAC) for LSST with  1000+ processing cores (~100 Tflops), 5 PB of storage and a 500 TB database with a 40 Gbps connection. Similar efforts are taking place in Mexico, Chile and Argentina.

Key areas of focus include:
\begin{itemize} 
    \item The \textbf{development and implementation of advanced data handling and analysis techniques}.
    \item Progress in \textbf{cloud computing initiatives}, such as the model being developed by the Brazilian INCT-FNAe.
    \item The application of \textbf{Machine Learning (ML) and Artificial Intelligence (AI)}  for data analysis, fast simulations, detector development, and trigger systems. For example, the \gls{ATLAS} collaboration is using GPUs in trigger and offline processing and experimenting with different data flows to enhance physics reach. Argentina is involved in the hardware and firmware implementation of algorithms for the \gls{ATLAS} global trigger system, including electron and photon identification using neural networks, with the goal of providing 23\% of the required modules.
    \item The need for \textbf{robust and reliable data storage and management solutions}.
    \item The development of \textbf{software tools and CI/CD practices}  to support collaborative research.
\end{itemize} 
A dedicated review of scenarios and proposals under the umbrella of a "SC+IT Hub" is underway, showcasing the effective use of hybrid spaces that integrate industrial and academic resources. This includes the introduction of AI models (such as LLMs) into scientific production, and the use of dedicated computing resources, Infrastructure as Code (IaC), and cloud technologies to create labs integrated into classrooms and curricula. These efforts aim to build robust, scalable, and accessible frameworks that support scientific endeavors and capacity-building projects, enhancing the participation of Latin American researchers in the global scientific community.

\section{Fostering Collaboration and Networking} \label{subsec:collaboration-networking} 

The chapter highlights the crucial roles of collaborative efforts, knowledge sharing, and networks and consortia in boosting local efforts. This includes fostering both regional collaborations within Latin America and stronger ties with international partners.

\begin{itemize} 
    \item The \textbf{Latin American Association for High Energy, Cosmology and Astroparticle Physics (LAA-HECAP)} , established in November 2021 and hosted by ICTP-SAIFR, plays a significant role in fostering collaborations within the region. It currently has over 560 members and actively engages with ministries and funding agencies to promote strategic planning.
    \item The LASF4RI process itself has significantly enhanced the \textbf{visibility of the Latin American HECAP community} , as exemplified by a Symmetry Magazine diagram that summarized the Physics Briefing Book's findings.
    \item The \textbf{Colombian Network on High Energy Physics}  emphasizes synergistic efforts in dark matter, cosmology, astroparticle, and neutrino physics, demonstrating the increased interest and capabilities of Colombian researchers.
    \item \textbf{Mexico} is working to consolidate a network of researchers focused on Dirac materials, bridging condensed-matter physics and high-energy physics, and to foster collaborations throughout Latin America.
    \item The Brazilian \textbf{INCT Brasil-CERN} serves as a network of people and institutions with common goals, affiliated with the Ministry of Science and Technology. It aims to promote new transversal collaborations within the community and synergistic interactions with industry.
    \item Latin American groups involved in \textbf{gravitational wave science}  have increased their participation in second-generation detectors and have members in the consortia for third-generation detectors.
\end{itemize} 

\section{Addressing Challenges and Future Directions} \label{subsec:challenges-future} 

The Latin American HECAP community faces several challenges, including securing stable funding, ensuring access to shared infrastructures and tools, and addressing the training, recognition, and retention of human resources.

Future directions should involve a faster development of Artificial Intelligence/Machine learning tools. A dedicated White Paper showed their use in
HEP for simulation, triggering, reconstruction, data quality, data analysis and end-to-end deep learning.

In addition, cloud computing can be explored as a way for sharing resources.

\section{Conclusions} \label{sec:conclusions} 

The Latin American HECAP community has demonstrated \textbf{remarkable growth and diversification in its research efforts}  since the initial LASF4RI strategy, expanding its involvement in numerous prestigious international collaborations and local initiatives. This progress underscores Latin America's growing contributions and engagement in critical areas of scientific inquiry.

Key aspects of this development include:
\begin{itemize} 
    \item \textbf{Broad Participation in International Experiments:}  Latin American researchers are actively involved in major global experiments across various thematic areas, as detailed in the Physics Briefing Book. These include:
    \begin{itemize} 
        \item \textbf{HEP Experiments:}  Continued and expanded participation from several groups in different countries in the Large Hadron Collider (LHC) experiments at CERN, such as ALICE, ATLAS, CMS, and LHCb, with contributions to detector upgrades and data analysis. Mexican participation in the Belle II experiment in SuperKEKB is continuing as well. 
        \item \textbf{Neutrino Physics:}  Strong involvement in neutrino experiments such as the Deep Underground Neutrino Experiment (DUNE), the 
        LA-based the Coherent Neutrino-Nucleus Scattering Experiment (CONNIE), the Jiangmen Underground Neutrino Observatory (JUNO) in China and Hyper-Kamiokande in Japan.
        \item \textbf{Dark Matter and Dark Energy Searches:}  Growing engagement in experiments like the Dark Energy Spectroscopic Instrument (DESI), the Vera Rubin Observatory's Legacy Survey of Space and Time (LSST), and various dark matter dedicated projects such as \gls{SWGO} and efforts related to CTA. The Brazilian dark matter research community has shown remarkable growth since 2011, with over 1000 published papers and theses by mid-2024. Mexico aims to create a control center to process real-time data detected by LIGO, VIRGO, and KAGRA interferometers in the coming years.
        \item \textbf{Astroparticle Physics \& Cosmology:}  Participation in the Cherenkov Telescope Array (CTA), the Latin American Giant Observatory (LAGO), and gravitational wave science, with an expressed interest in developing the South American Gravitational-Wave Observatory (SAGO). Chile will host the \gls{LLAMA} observatory, the first in the Southern Hemisphere.
        \item \textbf{Future Colliders:}  Brazilian groups are interested in joining collaborations for the Electron-Ion Collider (EIC), which is already approved by the DOE. Discussions are also ongoing regarding participation in future circular colliders, such as the FCC.
    \end{itemize} 
    \item \textbf{Instrumentation and Computing Advancements:}  The region has seen a \textbf{growth and diversification of local instrumentation capabilities}, moving towards the construction of new local facilities for instrument development and testing. There's a strong focus on strategic advancements in scientific computing and data infrastructure, including:
    \begin{itemize} 
        \item Development and implementation of \textbf{advanced data handling and analysis techniques}.
        \item Progress in \textbf{cloud computing initiatives}, such as the model by the Brazilian INCT-FNAe.
        \item Application of \textbf{Machine Learning (ML) and Artificial Intelligence (AI)}  for data analysis, fast simulations, detector development, and trigger systems. For example, the \gls{ATLAS} collaboration uses GPUs for trigger and offline processing and is exploring different data flows to enhance physics reach. Argentina is involved in the hardware and firmware implementation of algorithms for the \gls{ATLAS} global trigger system, including electron and photon identification using neural networks, with the aim of providing 23\% of the required modules.
        \item Initiatives to develop \textbf{software tools and CI/CD practices} .
    \end{itemize} 
    \item \textbf{Capacity Building and Training:}  A crucial emphasis is placed on \textbf{developing and retaining highly qualified human resources}  in HECAP. This includes:
    \begin{itemize} 
        \item The \textbf{CERN Schools of Physics}  (Europe, Asia, Latin America), which support Latin American students. The last three schools were held in Mexico, Argentina, and Chile, with the next school planned for Costa Rica in 2025. These are two-week, residential programs for PhD or advanced Master's students, sponsored by CERN and CIEMAT from Spain. CERN also organizes Accelerator and Computing schools, and Brazil is slated to host a Computing School in 2026.
        \item Programs like \textbf{LA-CoNGA Physics}, an open science education collaboration between Latin America and Europe for High Energy Physics, and associated hackathons (e.g., Coafina, which recently completed its third edition).
        \item \textbf{Mobility programs}  for students and professors between Latin American institutions and international labs like CERN and Fermilab. Fermilab's ``Physics Without Frontiers'' program targets undergraduates, and a proposed Mobility Initiative with LAA-HECAP aims to foster collaboration and career paths in neutrino research. Chile offers PhD scholarships open to foreign students, with many recipients from Latin America (128 foreign students selected for the April-March academic year). Ecuador currently lacks universities that grant PhDs in physics, but Master's programs are beginning to emerge as a significant step. Venezuela's efforts focus on access to education and capacity building, utilizing online platforms for events to build community.
    \end{itemize} 
    \item \textbf{Strengthening Collaboration and Funding Advocacy:}  The LASF4RI process itself has significantly fostered regional collaboration and knowledge sharing. There is a recognized need to \textbf{engage proactively with funding agencies and national authorities}  to secure stable and predictable funding for both regional projects and international participation. Challenges include financial constraints, funding instability, and personnel turnover. Brazil, as an associate member of CERN, highlights the need for dedicated funding for researchers' work and mobility, distinct from treaty contributions. Mexico's CONACYT, which is transitioning into the Ministry of Science, Humanities, Technologies, and Innovation, has fundamentally shifted its funding mechanisms to focus on the ``human right of science''. The Brazilian INCT Brasil-CERN actively lobbies funding agencies to secure stable funding for MOUs related to large collaborations. Argentina's participation in the \gls{ATLAS} upgrade relies on continued funding, despite political changes impacting scientific ministries.
\end{itemize} 

The progress made, as highlighted by the submitted white papers and discussions at the LASF4RI symposia, indicates Latin America's growing capacity to play a significant role in the global HECAP landscape, emphasizing both its regional strengths (e.g., geographical advantages for astroparticle observatories) and its increasing impact in international flagship projects.


\chapter{Bridging the HECAP Training Gap}
\label{chapt:training}

\section{Closing the gaps through training} 
Latin America exhibits considerable differences in the HECAP\footnote{High Energy, Cosmology, and Astroparticle Physics} physics research environment. Argentina, Brazil, and México have strong, long-running physics groups. Uruguay and Chile are emerging as new players in the Latin American scientific arena. Colombia, Costa Rica, Ecuador, Paraguay and Perú are still building their science systems. In Central American nations, research centres are rare or non-existent. In most countries of the region, research and development investment is only 0.1\%–0.7\% of GDP, far below the 2–3\% typical in high-income nations\footnote{\url{https://uis.unesco.org/en/news/uis-releases-new-data-sdg-9-5-research-and-development}  and \url{https://www.theglobaleconomy.com/rankings/research_and_development/Latin_America/} } . This funding gap results in a limited number of graduate programs and research opportunities available in the home country.

Closing these gaps through better education and training is vital for the region’s progress. Physics fields, such as high-energy physics, astrophysics, and cosmology, have expanded significantly in recent years, thanks to the support of international partners and internal continental efforts. Still, growth is uneven across countries, and many young scientists do not receive the support they need. Without sufficient, well-trained experts, Latin America relies on external solutions, loses talent to other regions, and has limited influence in shaping global research goals. Training more scientists is not a luxury; it is a necessity, enabling the region to address its problems and fully participate in worldwide science and economic growth.

Financial shortages resulting from limited investments in research and development (R\&D) are the primary obstacle. Because scholarships are few, most graduate students must work full-time and study at night or on weekends. This slows them down and cuts their research time. Only Argentina, Brazil, México, and, recently, Chile offer robust fellowship programs that cover tuition and living expenses, often welcoming students from other Latin American countries. Colombia, Ecuador, and Peru have limited grant opportunities, and high tuition fees usually exclude many talented students. The outlook is worse in Central América, where only Costa Rica has one physics PhD program. Students there must either give up their goals or pay a high price to study abroad. Young scientists often feel they have no choice but to leave the region to continue their careers. This brain drain weakens local programs. Waiting for significant R\&D budget increases is not realistic. The area needs swift, innovative steps that maximise the return on every dollar. Short, intensive schools (lasting 2-3 weeks, with some funding), topic-based workshops (3-5 days), and blended online-and-in-person courses can give high-value training at low cost. By linking and expanding these efforts, Latin America can train and keep its talent at home, even with limited funds. Turning these scattered initiatives into a lasting training pipeline demands coordination and targeted support.

\section{The existing training initiatives in the region} 
\label{traninglandscape} 
The region has various training programs supported by international organisations and local institutes. These initiatives are cost-effective and often subsidised, making them accessible to students from under-resourced institutions. The ICTP South American Institute for Fundamental Research (ICTP-SAIFR) in São Paulo has, since 2012, established itself as the premier hub in Latin America for graduate-level international schools in HECAP with world-leader lecturers. In addition, ICTP-SAIFR has a robust outreach programme that trains high school teachers in both Portuguese and Spanish.

\subsection{Intensive schools} 
Periodically, international physics schools are hosted in Latin America, offering one to two weeks of immersive learning. Crucially, many cover participants' travel and lodging costs, thereby lowering the barriers for students from lower-income backgrounds. These short schools serve as immersive physics bootcamps. Students spend entire days in lectures, hands-on projects, and discussions with renowned scientists, often in remote venues that encourage focused attention.

\subsubsection{CERN Latin-American School of High-Energy Physics}  
The CERN Latin-American School of High-Energy Physics (CLASHEP)\footnote{\url{https://cerncourier.com/a/high-energy-physics-flourishes-in-latin-america/} }  \cite{DuhrMulders}  is a two-week school held every two years in collaboration with CERN. Since 2001, the school has travelled biannually through nine countries, bringing leading physicists to teach Latin American students. CERN waives the registration fee and covers most travel costs for students from the region. Dozens of young researchers join each session to learn, work on projects, and meet experts from around the world. After eleven editions, the school has trained more than 850 early-career physicists; many now hold key roles in Latin American institutions. CLASHEP is therefore a solid foundation for any regional training plan, offering strong courses, hands-on teamwork, and a network that directly links Latin America to CERN’s experiments and technology programmes.

\subsubsection{Joint ICTP-SAIFR/ICTP-Trieste Schools}  
Joint ICTP-SAIFR/ICTP-Trieste Schools pair São Paulo's South American Institute for Fundamental Research with ICTP-Trieste to host alternating two-week summer programs in cosmology and particle physics since 2014\footnote{\url{https://indico.ictp.it/event/10858} } . Each edition combines foundational mini-courses, hands-on tutorials, student talks, and a one-day workshop, waives fees, and offers travel support to broaden participation from Latin America. Ten schools have already trained over a thousand early-career scientists, many of whom are now involved in global sky-survey and collider collaborations. The third particle-physics school took place from June 24 to July 5, 2024, in São Paulo, while the fifth cosmology school happened from July 28 to August 8, 2025. Lecture videos and materials remain freely archived, extending the school's reach and nurturing sustained North-South research collaborations among peers.

\subsubsection{Latin-American School on \gls{CTA} Science} 
ICTP-SAIFR has hosted the First Latin-American School on \gls{CTA} Science back in 2023 \footnote{\url{https://www.ictp-saifr.org/cta2023/} } . The lectures were given by experts on these topics who are familiar with the imaging atmospheric Cherenkov telescopes technique. Dozens of students were familiarized with theoretical aspects of gamma-ray astronomy numerical codes via hands-on. 

\subsubsection{Latin‐American Astroparticle Physics School} 
The CLAF/ICTP-SAIFR Latin American Astroparticle Physics School revives the Latin American School on Cosmic Rays \& Astrophysics\footnote{\url{https://www.ictp-saifr.org/claf-ictp-saifr-laaps/} }, founded in 2004 by Óscar Saavedra, which has had seven very successful series of recurring schools (2004-2017) in various countries of Latin America. The 2025 edition, in São Paulo, delivered intensive training in ground-based gamma-ray and neutrino astronomy, centred on the Cherenkov Telescope Array Observatory and the Southern Wide-field Gamma-ray Observatory (SWGO). Graduate students received a theoretical background in ultra-high-energy cosmic rays, Cherenkov techniques, multi-messenger transients, and active galactic nuclei. Additionally, the participants developed practical analysis skills in \gls{SWGO} simulations and blazar modelling.

\subsubsection{Brazilian School of Cosmology \& Gravitation} 
The Brazilian School of Cosmology \& Gravitation\footnote{\url{https://www.sbfisica.org.br/v1/sbf/xix-brazilian-school-of-cosmology-and-gravitation-bscg/} }, founded at Centro Brasileiro de Pesquisas Físicas in Rio de Janeiro by Mario Novello in 1978, convenes every two to four years for five days of intensive graduate training in relativistic astrophysics, cosmology, and quantum gravity. The 19th edition took place in September 2024. Each session hosts 80–120 Latin American students and offers morning tutorials, afternoon seminars, posters, and roundtables. Most of the refereed proceedings later serve as course texts. Funded by CBPF, CAPES, FAPERJ, and international partners, BSCG has trained over 1,000 physicists, many of whom now lead teams for LSST, J-PAS, and LIGO–Virgo, making it a cornerstone of regional capacity building and advancing regional astrophysics and cosmology today.

\subsubsection{IIP Program on Astroparticle Physics} 
In 2018, the International Institute of Physics created a group in Theoretical Particle and Astroparticle Physics that has been promoting several events devoted to educating both undergraduate and graduate students in the field, namely: 1. Minicourse on Astroparticle Physics (2018), 2. School on Particle and Field (2018), 3. PASCCO-School on Particle, Astroparticle and Cosmology (2020), 4. Minicourse on Astroparticle Physics (2021), 5. Dark Matter School (2021), 6. Particle Physics for High School Students (2022), 7. Minicourse on Dark Matter (2022), 8. PhenoBR (2023), 9. Cosmos in High School (2023), 10. Particle Physics in High School (2024), 11. Mini-course on Dark Matter, Neutrinos and Baryon Asymmetry (2024), 12. School on Primordial Black Holes and Dark Matter (2024), 13. School on Phase Transitions and Gravitational Waves (2024). Thus, more than one event per year. These events have attracted over one thousand students from Latin America.

\subsubsection{IIP Program on Collider Physics} 
IIP has also been promoting more specialized schools in the context of collider physics because UFRN has a signed MoU with the Future Circular Collider at CERN, for instance: School on Particle Colliders (2024), International School on Collider Physics (2024), Minicourse on GEANT and DMG4 (2024)

\subsubsection{IIP Outreach Program in Particle and Astroparticle Physics} 
The IIP has organised several outreach events for high school students on modern physics and basic aspects of particle physics through the Masterclass in Particle Physics and the Research Initiation Program, in which advanced high school students and PhD students supervise undergraduates. Students are offered several lectures in Particle and Astroparticle Physics, learn how to give talks, and produce scientific posters for conferences. The program lasts several months. The final goal of the program is to write an e-book, published with a Digital Object Identifier (DOI), that summarises their acquired knowledge. These e-books are later advertised within their own school. In this way, their role is highlighted both among their peers and within the overall school board.

\subsubsection{Escuela Mexicana de Partículas y Campos} 
Founded in 1984, the biennial Escuela Mexicana de Partículas y Campos (EMPC)\footnote{\url{https://aeifmx.com/xx-escuela-mexicana-de-particulas-y-campos-2023/} } rotates among Mexican institutions to immerse Latin American graduate students and young researchers in frontier particle physics through morning review lectures, afternoon parallel tutorials, and poster forums. This balanced format, which combines theory, experiment, and career panels, encourages active mentorship and collaboration. Over the past four decades, EMPC has developed national and regional expertise that now informs Mexican teams working on major LHC experiments. The 20th edition, hosted by Universidad Autónoma de Yucatán in October 2023, delivered courses on tensor dark matter, machine learning, and quantum computing, complemented by outreach events, and drew more than 120 participants. The next edition is expected in 2026, continuing its role as Latin America's primary gateway to advanced particle physics training and fostering new international scientific collaborations.

\subsubsection{Juan José Giambiagi Winter School} 
The Juan José Giambiagi Winter School of Theoretical Physics\footnote{\url{https://giambiagi2025.df.uba.ar/} } , organised by the Department of Physics at the University of Buenos Aires, is held every July and reached its 26th edition in 2024 and its 27th in 2025. The program features four to six intensive courses delivered by international experts, complemented by poster sessions, tutorials, and open colloquia. Each year, the school awards 20 to 30 travel and accommodation grants for Latin American participants, supported by partnerships with CELFI\footnote{The Centro Latinoamericano de Formación Interdisciplinaria (CELFI) is an Argentine public-sector programme that finances short, high-level training activities across Latin America.  \url{http://www.celfi.gob.ar} }  and CLAF. Topics span precision cosmology, fluid dynamics, artificial intelligence in physics, and quantum-limited mechanical systems. By providing high-level training in regions where postgraduate opportunities are limited, the school helps mitigate brain drain and fosters collaborations that now link institutions from México to Chile.

\subsubsection{International School on High Energy Physics}
International School on High Energy Physics, LISHEP\footnote{\url{https://www.lishep.uerj.br/}}, is an international school created in 1993 to fill a gap in High Energy Physics meetings in Brazil, mainly in Rio de Janeiro, where most Brazilian HEP groups are based. Founded and chaired by Prof. Alberto Franco de Sá Santoro, it initially took its name from the LAFEX Department at CBPF, later renamed CLAFEX. To remain independent of institutional changes, the organisers adopted the generic title “International School on High Energy Physics” while keeping the acronym. LISHEP is promoted by AIAFEX, a non-profit association that connects public and private partners and supports the School’s organisation. Initially structured in three sessions, a fourth was added in 2002; all adopt an experimental and phenomenological focus and have significantly strengthened the HEP community in Brazil and Latin America. The School has received support from Brazilian agencies and international laboratories, and strongly encourages participation from American and European students. The most recent edition took place in 2023, and its activities are widely documented (see  \cite{caruso2000a,alves2000a,caruso2005a,caruso2011b,abreu2018a,abreu2019a} and several entries in INSPIRE \cite{lishepweb}).

\subsubsection{ISYA and ICFA two global intensive schools} 
There are two other scenarios near the HECAP arena; these are global initiatives that impact, with some regularity, the Latin American training landscape

\paragraph{International School for Young Astronomers (ISYA).} 
 Launched in 1967, the International School for Young Astronomers (ISYA) of the International Astronomical Union provides intensive three-week, graduate-level courses in regions with limited access to contemporary astrophysics training.\footnote{\url{https://www.iau.org/OYA/OYA/About-ISYAs.aspx?hkey=6849faaa-17b8-43f3-a201-90b414328eb7} } . Latin American participation began in Argentina (1970) and progressed through editions in Argentina (1974, 2002), Brazil (1977, 1995), Cuba (1989), México (2005, 2023), Trinidad \& Tobago (2009), Honduras (2015), Colombia (2018), and Ecuador (2025).  More than 200 graduates now hold faculty or postdoctoral positions across the continent; follow-up surveys indicate that over 60\% completed their doctorates abroad and returned to establish new research groups.
 Each host institution modernises its instrumentation and updates its curricula in preparation for the school. Through this itinerant model, ISYA injects advanced expertise while fostering lasting institutional growth throughout Latin América.

\paragraph{ICFA School on Instrumentation in Elementary Particle Physics} 
The International Committee for Future Accelerators (ICFA\footnote{\url{https://icfa.hep.net/} } ) created its School on Instrumentation in Elementary Particle Physics to sharpen global expertise in detector technology. In Latin América, the school has catalysed advanced, hands-on training, forged a cross-border community of experimentalists, and boosted the region's contributions to flagship experiments worldwide. The ICFA calendar highlights Latin American editions in Brazil (1990, 2004), México (1997), and Argentina (2010); the documented surge in local detector R\&D confirms the school's lasting, positive impact on the region's high-energy physics ecosystem.

\subsection{Topic-Focused Workshops} 
Short workshops and mini-schools, lasting a few days to a week, focus on specific skills and act as skills accelerators, sparking collaborations on new research problems. They are high-impact and low-cost events, where organisers need only cover a handful of expert speakers and basic local expenses. Grants often cover travel expenses for lecturers and help local students attend. Because the events are brief, participants can join without needing to leave their jobs or coursework for extended periods. The close focus provides trainees with targeted skills, such as data analysis, detector design, and advanced computing, rarely offered at their home institutions.

\subsubsection{Latin American Symposium on High-Energy Physics} 
SILAFAE, the biennial Latin American Symposium on High-Energy Physics, has served as the region's primary meeting point for particle, astroparticle, cosmology, and gravity researchers since 1996\footnote{\url{https://fisindico.uniandes.edu.co/event/18/} } . With 15 editions rotating through 14 host nations, it has expanded from roughly 100 participants to over 180, blending plenary reviews, parallel sessions, AI-driven data workshops, and public outreach events while funding student travel and young scientist tracks. Debates at recent events have spawned two lasting structures: LASF4RI-HECAP, which aligns Latin American bids for extensive research facilities, and LAA-HECAP, the professional body representing the community in global road-mapping exercises. Alums generated links that now embed regional groups in ATLAS, CMS, IceCube, \gls{JUNO} and other experiments, emphasising SILAFAE's role as a collaboration incubator. Following an online pandemic edition, the 2024 meeting in Mexico City reaffirmed its momentum as the strategic Launchpad for Latin American groups to coordinate contributions, train their students, and ensure the region remains an active and creative partner in global HEP. The next edition will be in Colombia in 2026.

\subsubsection{ICTP-SAIFR's Specialised Research Workshops} 
ICTP-SAIFR's specialised workshops\footnote{\url{https://www.ictp-saifr.org/}} are agile, 3–5-day research sprints that gather 20–30 experts and students around one cutting-edge theme—e.g., gravitational-wave data pipelines or observational cosmology. Mornings feature concise plenary talks; afternoons shift to hands-on tutorials, live data challenges, and open discussions, so attendees leave with functioning code or new analytical tools. Participation is free, invited speakers receive local support, and limited travel funds keep the events accessible across Latin America; simultaneous streaming extends reach. Typical outcomes include regional analysis hubs (LIGO/Virgo/KAGRA workshop, 2025) and cross-country collaborations that inform major survey publications. Small cohorts, intense focus, and low barriers accelerate skill transfer, foster networks, and rapidly align Latin American researchers with global physics frontiers.

\subsubsection{CosmoSur/CosmoSul} 
CosmoSur (or CosmoSul)\footnote{\url{https://cosmosul-vii.on.br/} }  is a biennial Southern Cone workshop that, since its 2011 launch in Rio de Janeiro, rotates among Brazilian, Chilean, and Argentine venues to unite cosmology and gravitation groups across the region. The seven editions, the most recent of which took place in Salvador, Brazil, in August 2024, blend plenary reviews, hands-on sessions, and hack days on weak lensing, gravitational-wave cosmology, and \gls{LSST} science, all documented in AIP-published proceedings. Travel bursaries, covering nearly half of student attendees, keep costs low and seed collaborations; ALMA, \gls{LSST} DESC and LIGO-Virgo waveform-modelling projects were planned here, alongside white papers for LASF4RI and ESA Voyage 2050. Its inclusive format, hybrid streaming and emphasis on discussion make CosmoSur the Southern Cone's leading forum for shaping the region's contribution to next-decade, data-rich cosmology.

\subsubsection{Specific Technical Skill Bootcamps} 
Three-to four-day boot camps quickly build key technical skills and connect students with research networks. Sessions focus on detector hardware, scientific computing, and machine learning for physics, often in collaboration with international partners, such as CYTED\footnote{Ciencia y Tecnología para el Desarrollo, \url{https://cyted.org/LAGO-INDICA} }. The Latin American Giant Observatory, for instance, trains participants to build and operate low-cost water Cherenkov detectors.

\subsection{Modular \& Hybrid Graduate Courses} 
Modular, year-long hybrid courses now provide Latin American students with graduate training that minimally disrupts their lives. Offered online or in short on-site blocks, these programs cover everything from detector design to precision cosmology. Their flexible format slashes costs, eliminates the need for relocation or new buildings, and allows students to fit their studies around work. A typical path: attend a summer school for a broad overview, then spend six months in an online module to deepen that knowledge, all while staying at home. Funding more of these hybrids through curriculum grants and better internet access will quickly build a skilled physics workforce without draining local communities.

\subsubsection{Diplomado en Física Teórica UNACH–ICTP} 
The Diplomado en Física Teórica (DFT)\footnote{\url{https://learning.mctp.mx/diplomado-en-fisica-teorica-2024/} }  is the flagship distance-learning diploma of the Mesoamerican Centre for Theoretical Physics, established at Universidad Nacional Autónoma de Chiapas, following the UNACH–ICTP decree. Offered for the first time in October 2021, the programme now runs from September to May and is entirely online, with no charge. A nine-month syllabus delivers two consecutive blocks (Classical, Quantum Mechanics and Electromagnetism, followed by Gravitation, Statistical Physics and Quantum Field Theory) through live lectures, research seminars, and mentored mini-projects, totalling roughly 300 synchronous hours. Registration for the 2025 cohort is open, reaffirming its ``100 \% virtual'' model that allows participants to join from anywhere with internet access. Dozens of advanced undergraduates and recent graduates from at least ten Latin American countries have already completed the course, many of whom are moving on to master's and PhD programmes, making the DFT a low-barrier gateway that is steadily expanding the region's pool of well-prepared theoretical physicists. 

\subsubsection{Physics Without Frontiers \& PhysicsLatam Courses} 
Physics Without Frontiers\footnote{\url{https://www.ictp.it/home/physics-without-frontiers} }  (PWF), launched by ICTP in 2008, dispatches a volunteer network of researchers to deliver short schools, masterclasses and mentoring schemes where graduate infrastructure is scarce. In Latin America, it has conducted high-energy physics masterclasses from Mexico to Argentina, organised the landmark Venezuela–Colombia programme in 2016 at Universidad Industrial de Santander, Bucaramanga, Colombia \cite{Multiverse}, and hosted a School on Machine Learning in Physics at Universidad Técnica Federico Santa María, Chile, in 2025. PhysicsLatam\footnote{\url{https://www.physicslatam.com} } , founded in 2022 by Daniel Galviz with PWF/ICTP backing, complements those on-site efforts through semester-long, free online courses in Spanish and English; its Theoretical Particle Physics course (Sep 2024–Jan 2025) and General Relativity \& Cosmology track (Jun–Aug 2025) stream live lectures, archive recordings and award ICTP-PWF certificates. Together, these initiatives have already trained thousands of Latin American students, funneling many into Trieste Diplomas, European and LA-PhD programs, and local projects. 

\subsubsection{ERASMUS+ Initiatives to train physics in Latin America} 
LA-CoNGA physics (Latin American Collaborative Graduate Network in Physics\footnote{https://laconga.redclara.net/} ) is an ERASMUS- Capacity Building EU-funded partnership of eight universities in Latin America (Venezuela, Colombia, Peru, and Ecuador) and three in Europe (two in France and one in Germany)\cite{PenarodriguezNunez2022} . It revives and augments the CEVALE2VE\cite{CaicedoEtal2017} initiative to remotely train students from Colombia and Venezuela in high-energy physics. The LA-CoNGA physics consortium developed modular master's courses in high-energy physics that cover data science, detector design, and field theory, equipping students with skills valued in both academia and industry. Teaching combines online lectures, hands-on lab sessions with shared kits, and regional hackathons. More than 50 students in the 2021–2023 cohorts finished the full program and earned Bologna-level credits at a low cost.

EL-BONGÓ physics (E-Latin American huB for OpeN Growing cOmmunities in Physics \footnote{\url{https://elbongo.redclara.net/} } ) now scales and extends this model to 11 Central América and Andean countries with four European partner universities. The central idea is to develop four research and learning communities (High-Energy Physics, Multimessenger Astronomy, Seismology \& Earth Hazards, and AI \& Computational Tools). EL-BONGÓ physics adds two new features. First, each course addresses an applied research problem, enabling students to learn by conducting research in the community. Second, a FABLab network trains them in digital fabrication, building strong ``do-it-yourself" skills to develop and maintain low-cost scientific instruments.

Remote labs enable students to operate off-site detectors via low-bandwidth SSH connections. An open e-learning platform stores code, data, and collaboration tools, mirroring the workflow of large observatories and accelerators. This virtual setup provides hands-on, data-driven training, even in areas where local instrumentation facilities are unavailable. It is essential to emphasise that LA-CoNGA physics (and the forthcoming EL-BONGÓ physics program) features a formal ``research apprenticeship" stage, during which students spend three months at a regional lab or on a remote project, applying their skills to a real-world experiment. This cements their training and further integrates the Latin American scientific community through collaboration.

\subsection{Coordinated graduate programs} 
It is worth mentioning the vision and the implementation of the Graduate Program in Astrophysics, Cosmology and Gravitation (Programa de Pós-Graduação em Astrofísica, Cosmologia e Gravitação, PPGCosmo\footnote{\url{https://ppgcosmo.cosmo-ufes.org/} } ). Since its launch in 2016, the Graduate Program in Astrophysics, Cosmology and Gravitation (PPGCosmo) at UFES has become Latin America’s flagship doctoral network in the field. It received an unprecedented initial grade of 5 from Brazil’s evaluation agency, CAPES, and retained that rating in the 2017–2020 cycle. Today, the program unites 16 institutions across nine countries under a dual-supervision scheme that requires every student to spend a whole year with an overseas co-advisor, guaranteeing immersion in diverse research environments.  Its 36-credit curriculum and qualifying exam embed candidates in major collaborations—including LIGO–Virgo, Euclid, LSST, J-PAS, and the Pierre Auger Observatory—providing early access to world-class data sets and leadership roles. As of July 2025, PPGCosmo supports 27 Ph.D. candidates and has graduated 17 researchers now employed across the continent. International calls on platforms such as Hyperspace continually attract applicants worldwide, while the COSMO-UFES group coordinates day-to-day operations and outreach.  By coupling mandatory mobility, joint supervision, and pooled instrumentation, PPGCosmo offers a scalable model that bridges structural training gaps and elevates Latin-American talent to global competitiveness.

\begin{figure} 
    \centering
    \includegraphics[width=.7\linewidth]{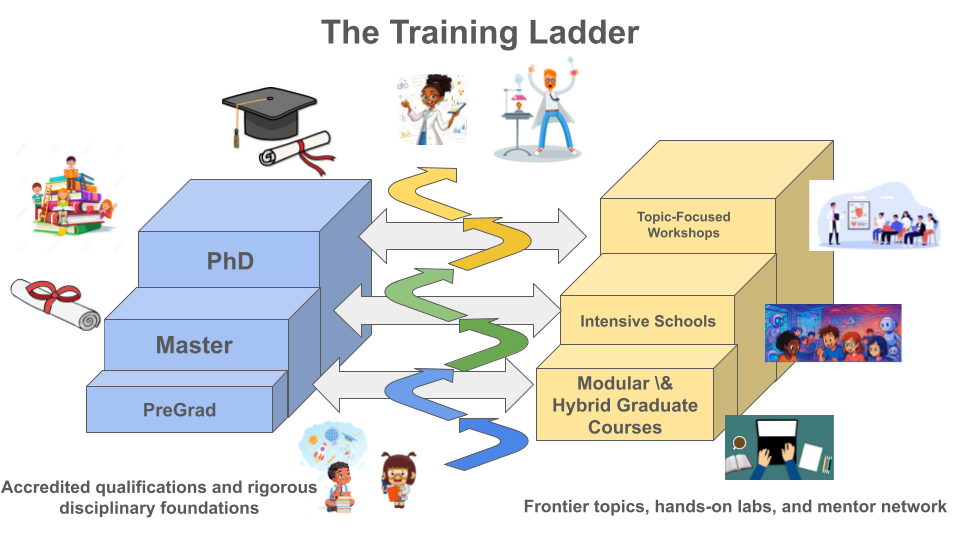} 
\caption{The ``training ladder’’ is a complete path aligned with flexible options (intensive schools, short workshops, and modular courses) to Universities supply accredited degrees and disciplinary depth, while the two-sided ladder complements them by delivering rapid exposure to frontier topics, hands-on detector laboratories, and a continent-wide mentor network. We should coordinate intensive schools, short workshops, and modular courses to align with graduate programs, creating a seamless progression from fundamentals to specialised expertise. In countries with few graduate programs, the ladder bridges existing gaps; where no programs exist, it equips students to undertake degrees elsewhere in the region.} 
    \label{fig:enter-label} 
\end{figure}

\section{Weaving possible training pathways} 
\label{trainingladder} 
There is no one-size-fits-all approach to training new researchers under the limited circumstances in Latin America. Formal graduate programs should mesh with flexible options (intensive schools, short workshops, and modular courses) so they form a seamless training ladder. Latin American universities and research centres must coordinate these offerings to guide students from fundamentals to advanced skills. In countries with few graduate programs, the ladder bridges gaps; where none exist, it equips students to pursue degrees elsewhere in the region.

\begin{enumerate} 
    \item \textbf{Modular advanced coursework} . Students begin with online modules that deliver graduate-level theory and lab skills over one to two years. Programmes such as LA-CoNGA, the upcoming EL-BONGÓ Physics network, and PhysicsLatam let them study detector electronics, data science, or precision cosmology from home on a flexible schedule. Because the courses are hybrid or entirely online, trainees can maintain their day jobs or research assistantships, which are vital for those who cannot pause paid work. A physics teacher, for example, might teach by day and log into an evening quantum-field-theory class. They are better prepared to pursue a formal master's degree, apply for PhD positions, or enter the industry as skilled researchers or engineers.
    \item \textbf{Immersive Schools} . Students—advanced undergraduates or recent graduates—start by attending a fully funded 1–2 week international school. These schools spark interest, offer exposure to cutting-edge topics, and connect participants with leading physicists.  By the end, participants build networks, identify areas of interest, and gain the confidence to pursue physics independently. This first step levels the playing field and motivates students to pursue a career in the field.
    \item \textbf{Skill-Building Workshops.}  A five-day detector-instrumentation bootcamp for experimental particle physics or a data-analysis session for astronomy delivers intensive, practical training and fits easily into a semester break or a few vacation days. One of these events in a year equips students with hands-on skills and plugs them into specialist networks. An alum of a CERN school, for instance, might join a neutrino-physics workshop, learn to analyse telescope data, and meet researchers planning the following experiment. These workshops sharpen technical skills and often open the door to new research collaborations.   
\end{enumerate} 
Synchronising some of the stages creates possible low-cost tracks: students from modest backgrounds move from an introductory school to focused workshops, then to advanced courses and research, all within Latin America's current offerings. The route leverages existing, funded events and online resources, allowing local groups to organise formal training programs. It is merit-driven and inclusive, with open competitions that reward talent and broaden diversity. 

\section{Coordination to stretch scarce funds} 
\label{CoordinationSupport} 
A few strategic spots can turn today's patchwork of schools, workshops, and courses into a single, continent-wide training program. Coordination, shared funding, robust connectivity, and active mentoring are the cornerstones that must be put in place.
\begin{enumerate} 
    \item \textbf{Regional coordination.}  Create a planning body—under CLAF or a new Latin-American Physics Training Consortium—to publish one public calendar, avoid date clashes, and cluster events by location. A typical annual route would be: online pre-course $\rightarrow$ intensive school $\rightarrow$ targeted workshops. With precise sequencing, isolated events merge into a coherent curriculum.
    \item \textbf{Shared scholarships} . Pool government, agency, and donor money into a regional fund. Packages covering one school, one workshop, and an online course for $\approx 50$ top students a year give sustained, flexible support and spread costs among partners. Tracking graduates' output proves the scheme's value. 
    \item \textbf{Robust connectivity} . Collaborate with RedCLARA and national networks to enhance campus bandwidth and establish e-learning hubs, enabling every student to stream lectures and participate in virtual sessions, achieving low costs and high impact.
    \item \textbf{Mentorship network} . Pair each trainee with a senior Latin-American scientist through LASF4RI. Mentors guide event choices, identify curriculum gaps, and keep students engaged, all at minimal cost.

\end{enumerate} 
Together, these measures weave schools, workshops, and courses into an integrated system that reaches more students for every dollar spent.

\section{Concluding remarks} 
Latin América already has a set of ingredients to train physicists, without waiting for a dramatic increase in funding. Intensive schools such as CLASHEP, EMPC, and BSCG offer short three- to five-day research workshops at ICTP-SAIFR and similar hubs, as well as modular hybrid programmes like LA-CoNGA and EL-BONGÓ, collectively forming a cost-efficient ``training ladder". This ladder is designed to complement, not replace, formal university degrees. Universities will continue to provide accredited qualifications and rigorous disciplinary foundations. The ladder fills the experiential gap: offering rapid exposure to frontier topics, hands-on detector labs, and access to a continent-wide mentor network. All these aspects that standard curricula often cannot deliver at scale. 

This discussion should be viewed as a starting point, rather than a fully costed blueprint. Three key design questions require further development:
\begin{itemize} 
    \item A light-touch regional board. Drawn from existing bodies such as LAA-HECAP and CLAF, these structures should oversee a rotating calendar, select host sites, and appoint evaluation committees. Transparent statutes and annual public reports would ensure accountability.
    \item Initial seed funding could come from UNESCO's Participation Programme, OAS academic mobility funds, ERASMUS+ Capacity Building, and private foundations (ICTP, Simons, Sloan). Bundled scholarships covering one school, one workshop, and one online course for approximately 50 students per year would require only about USD 300k annually.
    \item Host universities could award micro-credentials or ECTS-compatible certificates for ladder modules. A shared digital badge system would document competencies and facilitate the recognition of credit across borders.
    \item The success of the PPGCosmo initiative demonstrates how joint supervision, mandatory mobility, and pooled instrumentation can bridge structural training gaps and elevate Latin American talent to a level of global competitiveness.
\end{itemize} 

By coordinating and scaling the initiatives mentioned above, the region could triple its pool of research-ready physicists within a decade. Sequencing these elements, from foundational schools, through skills-focused boot camps, to mentored research apprenticeships, allows students to advance toward the research frontier while staying in the region, helping curb brain drain and maximising national R\&D investments, which still hover at just 0.1–0.7\% of GDP. This pragmatic, complementary scheme could transform limited resources into maximum scientific return, positioning Latin América as an indispensable partner in global high-energy physics, astroparticle studies, and cosmology.

\bibliographystyle{plain} 
\bibliography{Training/Training} 

\chapter{Outlook}
\label{chapt:conclusion}
Over the past decades, Latin America has moved from a largely peripheral presence toward a growing and increasingly structured participation in some of the most ambitious scientific endeavors worldwide. This evolution reflects not only the initial efforts of pioneers, but the sustained efforts of successive generations of scientists, and the gradual maturation of national and regional scientific ecosystems. The next decades will determine whether this progress consolidates into a long-term strategic position.

The global roadmap of fundamental physics is entering a period of unprecedented transformation. New experimental frontiers, ranging from next-generation colliders, precision cosmology, neutrino observatories and gravitational-wave detectors, to advanced gamma-ray and cosmic-ray facilities, and the multi-messenger approach to perform such experiments, will redefine our understanding of the fundamental blocks of nature and the fabric of spacetime. In parallel, scientific practice itself is being reshaped by artificial intelligence, data-intensive discovery, exascale computing, and the convergence of physics, information science and advanced engineering. These developments place extraordinary demands on scientists themselves, but also on infrastructure, governance and long-term financing. For Latin America, they also create an opportunity: to transition from being a powerful participant to becoming a strategic co-designer of the global scientific agenda.

A forward-looking strategy for the region cannot rely solely on continued access to large international collaborations. While participation in major experiments remains indispensable, the next stage of development must be built around a fully integrated scientific ecosystem, connecting experiment, theory, phenomenology, instrumentation, software, data science and high-performance computing. Some of the experiences across the region reported here, demonstrate that lasting scientific impact emerges when these components evolve together in a coordinated manner: a continental-wide coordination framework should enable Latin American countries to define shared priorities, pool strategic investments and present a unified voice in global negotiations for both large and small scale scientific projects. Science at the frontier now operates on timescales of decades and budgets that exceed national capacities for most countries in the region, so only through structured regional cooperation, supported by stable intergovernmental agreements, can Latin America aspire to influence the design, governance and scientific direction of next-generation international facilities.

From the experimental perspective, Latin America must deepen its engagement in the forthcoming generation of flagship projects: future colliders, neutrino facilities, gravitational-wave observatories, large-scale cosmic surveys and multi-messenger astrophysics infrastructures. But equally important is the region’s capacity to host and operate strategic experimental infrastructure on its own soil. Such facilities not only serve the advancement of fundamental knowledge, but also transform local ecosystems through advanced engineering, technology transfer, workforce development and industrial innovation. Hosting frontier infrastructure is no longer a symbolic aspiration; it is a lever for structural development in science-based economies. In theory and phenomenology, Latin America has already achieved international recognition in multiple domains. The coming decades will require a new qualitative leap: sustained programs that are natively integrated with experimental design, detector optimization, simulation and data analysis. Strengthening the integration between theory, computation, machine learning and global data infrastructures, will be essential for the region to retain leadership. Hardware and software development must be explicitly recognized as strategic priorities. The ability to design detectors, front-end electronics, data acquisition systems, firmware, reconstruction algorithms and scientific software platforms constitutes a form of technological sovereignty. High-performance computing, artificial intelligence and large-scale data infrastructures will be among the most decisive bottlenecks of future science. Sustained access to advanced computing resources, coupled to national and regional data strategies, will make Latin America risks an active generator of primary scientific knowledge. Regional high-performance computing centers, shared cloud infrastructures, AI laboratories for scientific discovery and coordinated open-data policies should be understood as critical scientific infrastructure, on par with accelerators and observatories. These capabilities directly impact national innovation systems in sectors such as medical imaging, telecommunications, artificial intelligence, aerospace, energy and security. Investing in scientific infrastructure is therefore not only an investment in knowledge, but in industrial capability, technological diversification and economic resilience.

At the heart of this entire enterprise lies the formation, training, mobility and hiring of scientists. The next generation of Latin American scientists must be trained in environments that are simultaneously global in scope and locally anchored. This requires sustained investment in doctoral and postdoctoral programs, regional schools, long-term mobility schemes, hybrid digital training models, and structured mentorship networks. Diversity, equity and inclusion are not peripheral concerns: they are core conditions for the sustainability and creative vitality of the scientific workforce. A region that retains and attracts global talent is continuously evolving towards scientific leadership.

None of these aspirations can be achieved without long-term political commitment and stable financing mechanisms. Governments must recognize that investments in fundamental physics, astronomy and cosmology are strategic investments in education, innovation, technological autonomy and international cooperation. The returns are not always immediate, but they are cumulative, transformative and intergenerational. Equally important is the construction of a renewed social contract between science and society. Public understanding, transparency, open science practices and meaningful engagement with younger generations are essential to sustain long-term support. Science at the frontier is a collective project that expands knowledge, opportunity and shared futures.

The evidence presented throughout this report (and its continuous updates) demonstrate that the region already possesses the essentials: human talent, emerging infrastructure, international credibility and a growing culture of collaboration. 
It is clear that the many recent developments in the region showcased in this Physics Briefing Book points to a much more active role of the HECAP Latin American in cutting-edge projects around the world.
We look forward to continuing our efforts for Latin America to be a coherent scientific pole with its own long-term strategy, infrastructure and cooperation identity. 

\chapter{Appendix}
\section{List of White Papers}\label{wp}
We list below the White Papers used as a basis for this Physics Briefing Book. They are available \href{https://drive.google.com/drive/folders/18ATto2vVbAKfB3M7hB6ty6wYWQ23f6Xl?usp=sharing}{here}.

\begin{enumerate}
    \item Update on Argentina Experimental HEP Input
    \item Update on the Brazilian Participation in LHCb
    \item GRAND White Paper
    \item ANDES Update
    \item An empirical case and ongoing efforts on Transversal Computer Strategies \& Services for Scientific
and Training efforts for LASF4RI
    \item ASTRI White Paper
    \item Machine Learning in HEP
    \item Brazilian Community Report on Dark Matter
    \item Central American Input for HECAP
    \item CMS-Brazil Input for the PBB
    \item CTA-Brazil White Paper
    \item DESI Extension White Paper
    \item DUNE Colombian Participation Update
    \item Ecuadorian HECAP Groups Update
    \item Latin America and the Electron-Ion Collider
    \item Gravitational Waves in Mexico
    \item The ATLAS/Brazil Cluster Update
    \item Gravitational Waves in Latin America White Paper
    \item LA-CoNGA Update
    \item HIBEAM/NNBAR experiment
    \item LAGO Update
    \item Latin American network on electromagnetic effects in strongly interacting matter
    \item TAMBO 
    \item ALICE-Brazil
    \item BINGO Update
    \item CONNIE
    \item INCT CERN-Brasil
    \item LAMBDA: A World-Class Particle Physics Lab in South America
    \item Southern Wide-field Gamma-ray Observatory (SWGO)
    \item Latin American Contribution to JUNO
    \item Mexican Participation in LSST
    \item The Latin American Participation in LSST
    \item Mexican-led effort to study strongly interacting matter with the MPD of the NICA-JINR complex
    \item Brazilian Effort on Liquid Argon Detectors for Neutrino Physics
    \item Latin American Contributions to the NOvA Experiment
    \item QUBIC Update
    \item A Latin-American network on Astrophysics, Cosmology and Gravitation - PPGCosmo
    \item Future Rare Events Searches with Low-Temperature Detectors and Quantum Sensors in Latin America
    \item The South American Gravitational wave Observatory White Paper
    \item The Group of Theoretical Physics (GI-FT) of the Escuela Politecnica Nacional in Ecuador
    \item Update of the Brazilian Participation in the Next-Generation Collider Experiments
    \item Neutrino Angra Experiment
    \item A Venezuelan input to the LASF4RI
    \item Quantum Materials at the Interface between Condensed Matter and High Energy Physics
    \item Colombian Network on High Energy Physics: Input on Theoretical HEP  
    \item Oscura White Paper
\end{enumerate}

\clearpage
\newpage

\section{Glossary of Experiments}
\printglossaries

\end{document}